\documentclass[
reprint,
superscriptaddress,
frontmatterverbose, 
showpacs,
preprintnumbers,
nofootinbib,
amsmath,
amssymb,
aps,
floatfix,
twocolumn
]{revtex4-2}

\usepackage[inline]{enumitem}
\usepackage[utf8]{inputenc}
\usepackage[normalem]{ulem}
\usepackage{graphicx}
\usepackage{dcolumn}
\usepackage{bm}
\usepackage{color}
\usepackage[dvipsnames]{xcolor}
\usepackage[
    colorlinks = true,
    linkcolor = BlueViolet,
    anchorcolor = purple,
    citecolor = purple,
    filecolor = purple,
    urlcolor = BlueViolet]{hyperref}
    
\usepackage{url}
\usepackage{xspace}
\usepackage{slashed}
\usepackage{multirow,bigstrut}
\usepackage{mathrsfs} 
\usepackage{relsize,amsmath}
\usepackage[geometry]{ifsym}
\usepackage{graphicx}
\usepackage{amssymb}
\usepackage{pifont}
\usepackage{physics}

\usepackage[capitalise]{cleveref}

\usepackage{rotating}
\usepackage{ragged2e}

\usepackage{fontawesome5} 
\definecolor{blue-violet}{rgb}{0.33, 0.17, 0.89}

\def\microboone{MicroBooNE\xspace}

\def\miniboone{MiniBooNE\xspace}

\renewcommand{\phi}{\varphi}

\newcounter{CommentCount}
\definecolor{MH}{rgb}{0.0,0.6,9}
\definecolor{palatinate}{rgb}{0.494, 0.192, 0.482}

\definecolor{teal}{HTML}{008080}

\newcommand{\epluseminus}{$e^+e^-$\xspace}

\newcommand{\darknus}{\texttt{DarkNews}\xspace}
\newcommand{\darknews}{\darknus}

\newcommand{\orcid}[1]{\href{https://orcid.org/#1}{(\textcolor[HTML]{A6CE39}{\faOrcid} #1)}}

\newcommand{\cmark}{\ding{51}}%
\newcommand{\xmark}{\ding{55}}%

\begin{document}

\title{A panorama of new-physics explanations to the MiniBooNE excess}
%

\author{Asli M. Abdullahi}
\email{asli@fnal.gov}
\thanks{\orcid{0000-0002-6122-4986}}
\affiliation{Theoretical Physics Department, Fermi National Accelerator Laboratory, Batavia, IL 60510, USA}

\author{Jaime Hoefken Zink}
\email{jaime.hoefkenzink@ncbj.gov.pl}
\thanks{\orcid{0000-0002-4086-2030}}
\affiliation{National Centre for Nuclear Research, Pasteura 7, Warsaw, PL-02-093, Poland}

\author{Matheus Hostert}
\email{mhostert@g.harvard.edu}
\thanks{\orcid{0000-0002-9584-8877}}
\affiliation{Department of Physics \& Laboratory for Particle Physics and Cosmology, Harvard University, Cambridge, MA 02138, USA}

\author{Daniele Massaro}
\email{daniele.massaro5@unibo.it}
\thanks{\orcid{0000-0002-1013-3953}}
\affiliation{Dipartimento di Fisica e Astronomia, Universit\`a di Bologna, via Irnerio 46, 40126 Bologna, Italy}
\affiliation{INFN, Sezione di Bologna, viale Berti Pichat 6/2, 40127 Bologna, Italy}
\affiliation{Centre for Cosmology, Particle Physics and Phenomenology (CP3), Universit\'e Catholique de Louvain, B-1348 Louvain-la-Neuve, Belgium}

\author{Silvia Pascoli}
\email{silvia.pascoli@unibo.it}
\thanks{\orcid{0000-0002-2958-456X}}
\affiliation{Dipartimento di Fisica e Astronomia, Universit\`a di Bologna, via Irnerio 46, 40126 Bologna, Italy}
\affiliation{INFN, Sezione di Bologna, viale Berti Pichat 6/2, 40127 Bologna, Italy}

\date{\today}

\begin{abstract}
The MiniBooNE low-energy excess stands as an unexplained anomaly in short-baseline neutrino oscillation experiments. 
It has been shown that it can be explained in the context of dark sector models. 
Here, we provide an overview of the possible new-physics solutions based on electron, photon, and dilepton final states. 
We systematically discuss the various production mechanisms for dark particles in neutrino-nucleus scattering.
Our main result is a comprehensive fit to the MiniBooNE energy spectrum in the parameter space of dark neutrino models, where short-lived heavy neutral leptons are produced in neutrino interactions and decay to $e^+e^-$ pairs inside the detector.
For the first time, other experiments will be able to directly confirm or rule out dark neutrino interpretations of the MiniBooNE low-energy excess.
\end{abstract}

\maketitle

\makeatletter
\def\l@subsubsection#1#2{}
\makeatother
\tableofcontents 

\section{Introduction}

Significant anomalies in neutrino experiments have persisted for over two decades, with no conclusive resolution in or outside the Standard Model.
The most statistically significant of these is the apparent $\nu_\mu \to \nu_e$ conversion of neutrinos and antineutrinos at short baselines in the \miniboone experiment~\cite{MiniBooNE:2007uho,MiniBooNE:2008yuf,MiniBooNE:2013uba} and the Liquid Scintillator Neutrino Detector (LSND)~\cite{LSND:2001aii}.
The excess at \miniboone, dubbed the low-energy excess (LEE), is characterized by electron-like events in the energy region between $200$~MeV and $600$~MeV and is coincident in time with the $\langle E_\nu \rangle \sim 800$~MeV neutrino beam.
More detailed background studies and data have become available over the years, increasing the significance of the excess to a total of $4.8\sigma$~\cite{MiniBooNE:2018esg,MiniBooNE:2020pnu}.
The origin of LEE has been debated in the literature, with some authors pointing to combinations of effects that could somewhat reduce its significance~\cite{Brdar:2021ysi,Kelly:2022uaa}.
However, at this time, it is accurate to say that the origin of the LEE remains unknown and that its statistical significance remains strong.

Historically, the most studied beyond-the-Standard-Model explanation to the \miniboone LEE has been the 3+1 oscillation model, where an eV-scale sterile neutrino induces short-distance $\nu_\mu \to \nu_e$ oscillations. 
Other anomalies, including the LSND result and electron-neutrino disappearance gallium experiments~\cite{GALLEX:1994rym,GALLEX:1997lja,Kaether:2010ag,Abdurashitov:1996dp,SAGE:1998fvr,Abdurashitov:2005tb,SAGE:2009eeu}, have also been linked to the LEE under the sterile neutrino hypothesis.
This paradigm, however, is in strong tension with cosmology~\cite{Hamann:2011ge,Archidiacono:2013xxa,Hagstotz:2020ukm} and $\nu_\mu$-disappearance data~\cite{MINOS:2020iqj,IceCube:2020phf,IceCube:2020tka}.
In fact, the internal tension in sterile neutrino global fits is significant and driven not by one but multiple experiments~\cite{Dentler:2018sju,Diaz:2019fwt}.
In addition, the LSND and the gallium experiments rely on different principles of operation than \miniboone, and, therefore, it is not unlikely that the resolution to each of these anomalies is unrelated.
As such, we focus uniquely on the \miniboone LEE and its potential solutions in new physics models.

The \miniboone detector could not distinguish between the Cherenkov emission of electron and photon showers inside the detector.
Therefore, the LEE can be due to multiple kinds of electromagnetic final states, including electrons, photons, and combinations thereof.
One important class of backgrounds at MiniBooNE arises from photon production in neutrino-nucleus neutral-current (NC) interactions through coherent or resonant processes. 
Other proposed solutions rely on non-resonant photons, photon pairs, or electron-positron pairs produced inside the detector in coincidence with the neutrino beam.
When highly collimated or highly energy-asymmetric, the pairs of electromagnetic particles can be misreconstructed as a single electron.
Several beyond-the-SM explanations have exploited this ambiguity in signal reconstruction.
These non-oscillatory explanations are easily embedded in low-scale extensions of the SM, where new dark particles of MeV to GeV masses mediate the production of or decay into electromagnetic activity.
We provide an overview of such new physics interpretations of the \miniboone LEE and discuss their current status.
For a review of broader aspects of short-baseline phenomenology, we point the reader to Ref.~\cite{Acero:2022wqg}.
While our discussion is focused primarily on \miniboone, our list of particle production modes in neutrino scattering can be applied to other neutrino experiments searching for new physics.

To disentangle the possible nature of the LEE, the Short-Baseline Neutrino (SBN) program at Fermilab is underway and has started to probe a few interpretations of the excess~\cite{MicroBooNE:2015bmn,Machado:2019oxb}.
The program consists of three modern detectors in the Booster Neutrino Beam (BNB), where \miniboone was also located.
Thanks to the improved particle-identification (PID) capabilities of the Liquid Argon Time-Projection Chambers (LArTPCs), the SBN program has greater discriminating power between electron and photon-like events.
The first detector to run in the BNB was the \microboone experiment, a $170$~t active-volume detector at $470$~m away from the target.
Two additional detectors are part of SBN: the Icarus detector, currently in operation and with a much larger fiducial mass of $760$~t located $600$~m from the BNB target, and the Short-Baseline Near Detector (SBND), with a $112$-t active volume at a shorter distance, $110$~m away.
Recent data from \microboone shows no excess of $\nu_e$ events~\cite{MicroBooNE:2021rmx,MicroBooNE:2021jwr,MicroBooNE:2021sne,MicroBooNE:2021nxr}, but it could not yet fully exclude the 3+1 oscillation picture \cite{Arguelles:2021meu,MiniBooNE:2022emn,MicroBooNE:2022wdf} (see also \cite{Denton:2021czb}). 
The \microboone experiment has also tested the radiative $\Delta$ decays hypothesis and excluded it at the $94\%$ confidence level (C.L.)~\cite{MicroBooNE:2021zai}. 
The constraint is driven primarily by events with a photon associated with a proton vertex, so non-resonant sources of photons remain largely unconstrained due to the larger backgrounds. 
Furthermore, LEE explanations based on exotic sources of electron-positron pairs~\cite{Bertuzzo:2018itn,Ballett:2018ynz,Ballett:2019pyw,Abdullahi:2020nyr,Abdallah:2020biq,Hammad:2021mpl,Dutta:2020scq,Datta:2020auq,Abdallah:2020vgg,Abdallah:2022grs} or photon pairs~\cite{Datta:2020auq} are still largely untested.

As a characteristic example of a model where $e^+e^-$ final states explain the LEE, we study the so-called dark neutrino explanation~\cite{Bertuzzo:2018itn,Ballett:2018ynz}.
These models are based on low-scale seesaw mechanisms embedded in a dark sector.
The heavy neutral leptons (HNLs) can explain neutrino masses and have new interactions with ordinary matter through light vector or scalar mediators.
The mediators enhance the production of HNLs in neutrino-nucleus upscattering and allow them to quickly decay into visible final states inside neutrino detectors.
We perform a detailed fit to the \miniboone energy distribution for models with a dark photon mediator and one and two HNLs that decay into $e^+e^-$ pairs.
This is the first comprehensive fit to the \miniboone LEE in the context of an $e^+e^-$ interpretation and will allow other neutrino experiments to perform direct tests of this hypothesis.
Our analysis is based on our own simulation of the MiniBooNE experiment in the publicly-available generator \darknews~\cite{Abdullahi:2022cdw}.\footnote{\href{https://github.com/mhostert/DarkNews-generator}{{\large\color{BlueViolet}\faGithub}\,\,{github.com/mhostert/DarkNews-generator}}.}

This article is divided as follows.
In \cref{sec:miniboone}, we review the MiniBooNE LEE and its properties.
In \cref{sec:models}, we discuss the status of different new physics models proposed to explain the excess.
We then turn our focus to dark neutrino models in \cref{sec:fit_darknus}, where we present the results of our fit to the MiniBooNE energy spectrum.
We conclude in \cref{sec:conclusions}.

\section{The \miniboone low-energy excess}
\label{sec:miniboone}

\begin{figure}[t]
    \centering
    \includegraphics[width=0.49\textwidth]{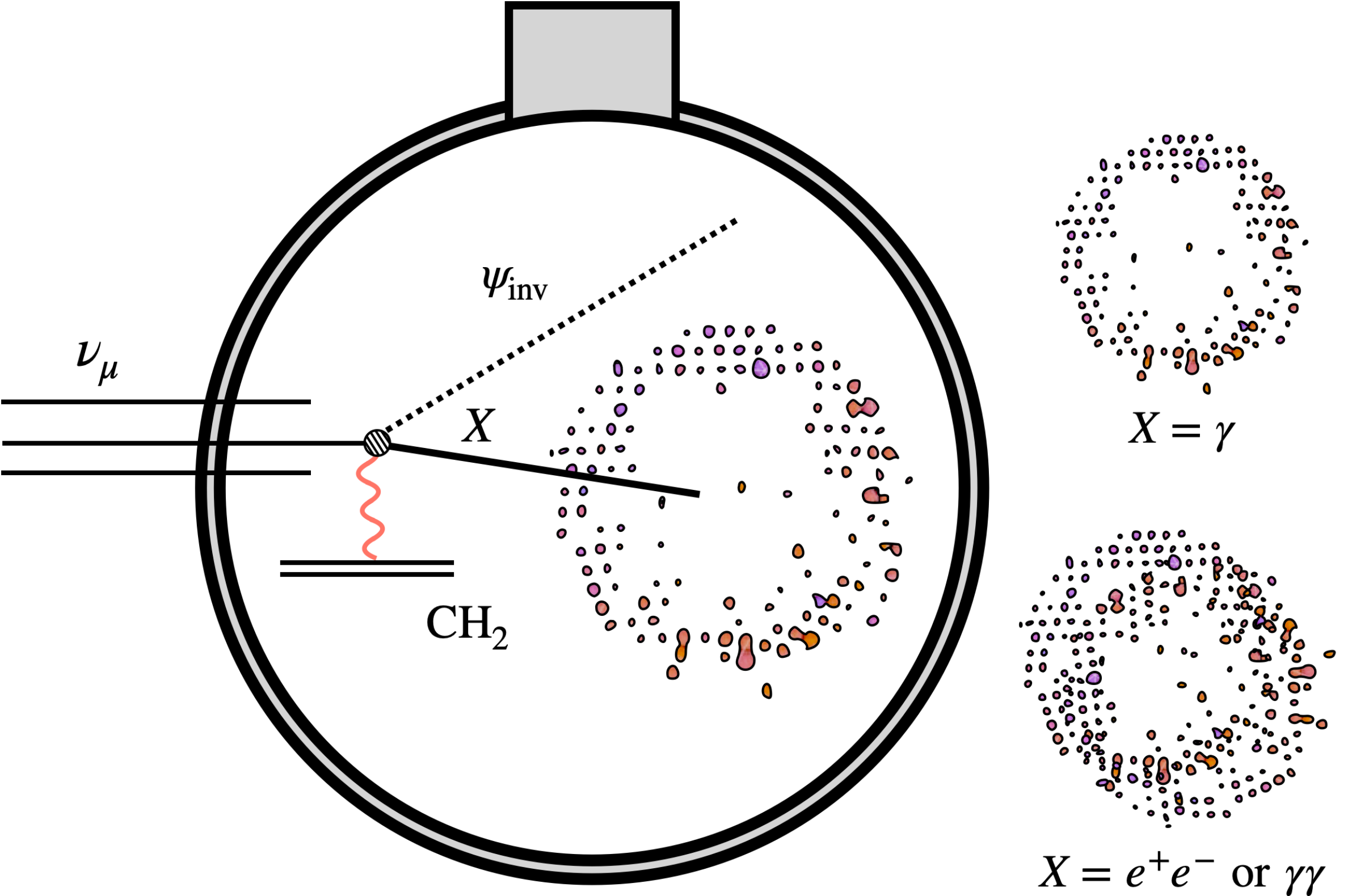}
    \caption{Different scenarios where a new particle $X$ can mimic the $\nu_e$ appearance signal in the \miniboone Cherenkov detector.}
    \label{fig:mb_final_states}
\end{figure}

The \miniboone experiment was located in the Fermilab BNB and employed a $12.2$~m diameter Cherenkov detector filled with 818 tons of pure mineral oil, located $541$~m away from the beryllium target. 
It ran in neutrino mode, with a forward-horn-current (FHC), and anti-neutrino mode, with a reverse-horn-current (RHC). 
Over a span of 17 years, \miniboone observed a large excess of low-energy electron-like events. 
The first excess was reported in neutrino mode between 2007 and 2009. 
For a total of $6.46 \times 10^{20}$ Protons-on-target (POT) in neutrino mode, an excess of $128.8 \pm 43.4$ electron-like events was observed over the background with a significance of $3.0\,\sigma$~\cite{MiniBooNE:2008yuf}. 
The excess was predominantly present in the $200~\mathrm{MeV} < E_\nu^{\mathrm{QE}} < 475$~MeV energy region. 
It subsequently also observed in anti-neutrino mode with a comparable significance of $2.8\sigma$, corresponding to $78.4 \pm 28.5$ excess events observed over the background in the energy range $200~\mathrm{MeV} < E_\nu^{\mathrm{QE}} < 1250$~MeV for a total of $11.27 \times 10^{20}$ POT~\cite{MiniBooNE:2010idf,MiniBooNE:2013uba}. 
The experiment then collected nearly double the amount of POT in neutrino mode, improving the background analysis and reducing systematic uncertainties. 
This led to a substantial increase in the significance of the LEE, with the $\sim 3\,\sigma$ excess rising to $4.7\,\sigma$~\cite{MiniBooNE:2018esg}, and most recently to $4.8\,\sigma$~\cite{MiniBooNE:2020pnu}. 
The combined neutrino and anti-neutrino mode excess currently stands at $638.0 \pm 52.1 (\mbox{stat.}) \pm 122.2$ (syst.) electron-like events.

Charged particles in the \miniboone detector produce directional Cherenkov light and isotropic scintillation light. 
Due to the limited PID capabilities of Cherenkov detectors, the signals of electrons, photons, and collimated $e^+e^-$ and $\gamma\gamma$ pairs are too similar to differentiate. 
Fortunately, the dominant electromagnetic backgrounds at \miniboone are constrained with \emph{in-situ} measurements. 
The beam's intrinsic $\nu_e$ content is constrained by $\nu_\mu$ CC rate measurements.
This is particularly relevant at higher energies, where kaon decays at the target dominate the neutrino flux.
The kaon production rate was directly measured by SciBooNE~\cite{SciBooNE:2011sjq}, which operated at $100$~m from the BNB target for a fraction of the total MiniBooNE lifetime. 
SciBooNE was also used to constrain $\nu_\mu$ disappearance in the BNB beam in combination with MiniBooNE data~\cite{MiniBooNE:2012meu,SciBooNE:2011qyf}.

The $\pi^0$ production rate is inferred by measuring separated $\gamma\gamma$ pairs~\cite{MiniBooNE:2008mmr,MiniBooNE:2009dxl}.
Given the $\pi^0$ kinematics and the Monte-Carlo, this allowed MiniBooNE to directly constrain the number of $\pi^0$s that decayed to collinear or highly energy-asymmetric photons, as well as the number of photon pairs where one of the photons was absorbed or escaped undetected.
In addition, the $\pi^0$ and $\pi^\pm$ rates that are measured by \miniboone constrain the production of the $\Delta(1232)$ resonance, which in turn constrains the rate of single photon events from radiative $\Delta(1232)$ decay.
This constraint, however, is subject to uncertainties on the $\Delta(1232)$ radiative decay branching ratio. 
Nonetheless, an enhancement of this rate to the level necessary to explain the LEE is in conflict with theoretical predictions~\cite{Wang:2014nat} and with recent \microboone data~\cite{MicroBooNE:2021zai}.

In addition to the low-energy nature of the excess, the most prominent features of the LEE are the following:
\begin{enumerate}
    \item \textbf{Angular spectrum}: The LEE shows a mild preference for forward-going events, although it is still broadly distributed in the angular variable $\theta_{\rm beam}$, the angle of the electron-like shower with respect to the neutrino beam.
    Approximately $72\%$ of events have a reconstructed angle satisfying $\cos{\theta_{\rm beam}} \leq 0.9$.
    \item \textbf{Radial profile}: The radial distribution of the excess events is consistent with neutrino interactions inside the tank. The significance of the excess also increases when decreasing the radius of the fiducial volume.
    \item \textbf{Timing}: Most excess events occur within the first $8~\mathrm{ns}$ of bunch timing, coinciding with neutrino scattering events in the detector.
\end{enumerate}
These considerations provide significant constraints on potential LEE mechanisms.
Quantifying the constraining power of the aforementioned distributions is challenging outside the collaboration due to the lack of a correlation matrix, but one can conclude that the radial and timing information indicates a preference for a neutrino-induced excess.
Additional constraints come from the search for sub-GeV dark matter performed by the MiniBooNE-DM collaboration~\cite{MiniBooNEDM:2018cxm}. 
The experiment ran in beam-dump mode, directing protons away from the beryllium target and toward the steel absorber at the end of the decay pipe, $50$~m away. 
Consequently, the beam-dump mode had a smaller and softer neutrino flux, allowing non-neutrino signals to be explored. 
No excess was observed in this search, excluding explanations where new particles are produced in neutral meson decays at the target. 
Due to suppressed off-target production of charged mesons, reciprocal models with charged mesons are still allowed, up to the signal considerations discussed above.
For a detailed discussion of the role of the angular distribution and of the beam-dump mode data on new-physics interpretations of the LEE, see Ref.~\cite{Jordan:2018qiy}.

\renewcommand{\arraystretch}{1.4}
\newcommand\insertdiagram[1]{
    \hspace{-12 ex}
    \begin{minipage}{.1\textwidth}\vspace{1ex}
        \includegraphics[page=#1,height=\textwidth]{table_of_explanations.pdf}
    \end{minipage}
}
\newcommand\biggercell[1]{
    \begin{minipage}{0.2\textwidth}
    \vspace{4ex}
       #1
    \vspace{4ex}
    \end{minipage}
}
\renewcommand{\arraystretch}{1.4}
\newcommand{\nocheck}{\textcolor{red}{\cmark}}
\newcommand{\semicheck}{\textcolor{orange}{\xmark}}
\newcommand{\fullcheck}{\textcolor{NavyBlue}{\cmark}}
\newcommand{\newhline}{\cline{2-8}}
\begin{table*}[t]
    \centering
    \begin{tabular}{
    |c|
    >{\centering\arraybackslash}p{4 cm}|
    >{\centering\arraybackslash}p{1.4 cm}|
    >{\centering\arraybackslash}p{1.3 cm}|
    >{\centering\arraybackslash}p{1.7 cm}|
    >{\centering\arraybackslash}p{1.3 cm}|
    >{\centering\arraybackslash}p{1.7 cm}|
    >{\centering\arraybackslash}p{2.5 cm}|
    }
    \hline\hline
    \multirow{2}{*}{Category} & \multirow{2}{*}{Model}& \multirow{2}{*}{Final state} & \multicolumn{3}{c|}{LEE signal properties} & \multirow{2}{*}{LSND} & \multirow{2}{*}{References} 
    \\ 
    \cline{4-6}
    & &  & energy dist. & angular dist. & timing & &  
    \\
    \hline\hline
\multirow[c]{4}{*}[-3em]{Flavor transitions} 
        & SBL oscillations & $e^-$ & \fullcheck & \fullcheck & \fullcheck & \fullcheck & Reviews and global fits \cite{Dentler:2018sju,Diaz:2019fwt,Boser:2019rta,Dasgupta:2021ies,Acero:2022wqg} 
        \\
        \newhline
        & SBL oscillations with invisible sterile decay & $e^-$ &  \fullcheck &  \fullcheck & \fullcheck & \fullcheck & \cite{Moss:2017pur,Moulai:2019gpi} 
        \\
        \newhline
        & SBL oscillations with anomalous matter effects  & $e^-$ & \fullcheck  & \fullcheck & \fullcheck & \fullcheck & \cite{Akhmedov:2011zza,Bramante:2011uu,Karagiorgi:2012kw,Asaadi:2017bhx,Smirnov:2021zgn,Alves:2022vgn} 
        \\ 
        \newhline
        & neutrino-flavor-changing bremsstrahlung & $e^-$ & \fullcheck & -- & \fullcheck & \fullcheck & \cite{Berryman:2018ogk}  
        \\
        \newhline
\hline\hline
\multirow[c]{2}{*}[-0.5em]{Decays in flight} 
        & SBL oscillations with visible sterile decay & $e^-$ &  \fullcheck&  \fullcheck&  \fullcheck & \fullcheck & \cite{Palomares-Ruiz:2005zbh,Bai:2015ztj,deGouvea:2019qre,Dentler:2019dhz,Hostert:2020oui} 
        \\
        \newhline
        & heavy neutrino decay & $\gamma$, $\gamma \gamma$ \hspace{3ex} $e^+e^-$  & \fullcheck & \semicheck & \semicheck & \semicheck & \cite{Fischer:2019fbw,Chang:2021myh} 
        \\
        \newhline
\hline\hline
\multirow[c]{2}{*}[-1em]{Scattering} 
    & neutrino-induced upscattering & $\gamma$, $\gamma \gamma$ \hspace{3ex} $e^+e^-$ & \fullcheck & \semicheck \, (vector) \fullcheck \, (scalar) \fullcheck \, (TMM) & \fullcheck & \semicheck \, (vector) \fullcheck \, (scalar) \fullcheck \, (TMM) & vector~\cite{Bertuzzo:2018itn,Ballett:2018ynz,Ballett:2019pyw,Abdullahi:2020nyr,Abdallah:2020biq,Hammad:2021mpl}
    scalar~\cite{Datta:2020auq,Dutta:2020scq,Abdallah:2020vgg}
    TMM~\cite{Gninenko:2009ks,Gninenko:2010pr,Gninenko:2012rw,Masip:2012ke,Radionov:2013mca,Magill:2018jla,Schwetz:2020xra,Vergani:2021tgc,Alvarez-Ruso:2021dna,Kamp:2022bpt,Bansal:2022zpi} 
    \\
    \newhline
    & dark particle-induced upscattering  &  $\gamma$, $\gamma \gamma$ \hspace{3ex} $e^+e^-$ & \fullcheck & model dependent & \fullcheck & \semicheck & \cite{Dutta:2021cip} 
    \\
    \hline\hline
    \end{tabular}
    \caption{New physics explanations of the MiniBooNE excess categorized by the origin of the electron-like signature and underlying mechanism. Notation: \fullcheck -- the model can naturally explain the feature,  \semicheck -- the feature is not generally explained by the model. SBL stands for ``short-baseline". 
    We do not make assessments on the viability of models based on other experimental constraints.
    ~\label{tab:big_picture}}
\end{table*}

\section{New physics explanations}
\label{sec:models}

There are four classes of final states that have the potential of explaining the \miniboone low-energy excess: $e^-$, $\gamma$, $e^+ e^-$ and $\gamma \gamma$. These could be produced by different mechanisms within or beyond the SM. 
For each of these final states, we discuss how these signals may appear in new-physics models.
We summarize the new physics models in~\Cref{tab:big_picture}.
It highlights the ability of the model to explain different features of the MiniBooNE LEE, as well as potential compatibility with the LSND excess. 
It does not, however, reflect the current global experimental landscape, as some models are more constrained by other experimental data.
We also summarize explanations based on neutrino scattering processes in \Cref{tab:model_summary}.
It is important to add that previous work~\cite{Abdallah:2020vgg,Abdallah:2022grs} has explored the fact that, in addition to MiniBooNE, one could also explain the LSND excess of inverse beta decay events by using upscattering that knocks out neutrons from the Carbon nucleus in the LSND liquid scintillator. 
Some proposals rely on the DIF (decay in flight) component of the pion source to produce heavier particles.

\subsection{Single electron}

The \miniboone results have been initially interpreted as an excess of $\nu_e$ ($\overline{\nu}_e$) that undergo CC scattering. 
The electrons (positrons) would be detected as single \emph{fuzzy} Cherenkov rings in the \miniboone detector. 
Such an excess of neutrinos in the beam can be due to a mis-modelling of the flux, or from a beyond-the-Standard-Model phenomenon. 
One popular new-physics interpretation of the excess is that of short-baseline neutrino oscillation induced by a new sterile neutrino of eV mass.
This hypothesis gained significant traction due to the natural connection between the LSND anomaly and the \miniboone results, as well as due to other deviations observed in the $\nu_e$ disappearance sector, all of which share similar values of $L/E$. 
The LSND experiment observed a $3.8\sigma$ excess of $\overline{\nu}_e$ in a $\pi$ decay at rest (DAR) neutrino source. 
The antineutrinos were detected via inverse-beta-decay (IBD) in the liquid scintillator of the detector. 
The different operating principles of the two experiments combined with the similar $L/E$ suggested that a common explanation could be found in novel oscillations.
We discuss the recent developments regarding this hypothesis below and briefly review some alternatives to pure oscillatory models. 

The \microboone detector performed the first test of the LEE under the single-electron hypothesis in LAr using the same beam as \miniboone.
The first analysis aimed to constrain the template of the \miniboone LEE with three independent $\nu_e$ event selections~\cite{MicroBooNE:2021rmx,MicroBooNE:2021jwr,MicroBooNE:2021sne,MicroBooNE:2021nxr}.
The template is defined as the difference between the central value of the data minus the central value of backgrounds as a function of the reconstructed neutrino energy.
This, in turn, can be unfolded into an excess flux of $\nu_e$ in the BNB and constrained by \microboone data.
We review the findings of this search and comment on important caveats below.
In the \microboone's LArTPC detector, protons can be observed as tracks, and the signal contributes to either the $1s1t$ or $1s0t$ topologies.\footnote{As both electrons and photons are reconstructed as showers, it is customary to use the notation N$s$M$t$ to specify a topology containing N showers and M tracks.}
Using a deep-learning-based reconstruction approach, one analysis searched for the exclusive $1e1p$ CCQE channel finding a modest deficit with respect to predictions~\cite{MicroBooNE:2021jwr}. 
A second analysis searched for pionless $\nu_e$ scattering using Pandora-based reconstruction in the exclusive channels $1e$N$p0\pi$ (N$>$0) and $1e0p0\pi$~\cite{MicroBooNE:2021sne}. 
While a mild deficit was observed in the $1e$N$p0\pi$ channel, an excess was present in the $1e0p0\pi$ channel. 
The latter, however, has a much lower $\nu_e$ purity due to photon backgrounds, so it is less sensitive to the single electron hypothesis.
The third and highest statistics search was for inclusive $1eX$ $\nu_e$ events using the Wire-Cell reconstruction method~\cite{MicroBooNE:2021nxr}. 
Good agreement between data and prediction is observed, and the fixed LEE template is excluded at $3.75\sigma$.
Combining all aforementioned channels, except the less-sensitive $1e0p0\pi$ one, \microboone rejects the fixed LEE $\nu_e$ template at $>97\%$ C.L. 
This translates into an upper limit on the $\nu_e$ composition of the \miniboone excess, corresponding to $< 51\%$ at $95\%$ C.L. 

Two important caveats to the \microboone model-agnostic template analysis were that i) it used a fixed template for LEE and ii) it did not exclude physical models, opting instead to work with a model-agnostic approach.
The first point, in particular, is critical as the uncertainties on the template's shape can be large due to the significant systematics and correlations in the \miniboone background prediction.
In addition, physical models that attempt to explain the anomaly can also have significantly different shapes from the nominal template and still provide a reasonable fit to the data, thanks to the large systematic uncertainties.
The significance of the shape uncertainties and the exclusion power of \microboone in the parameter space of 3+1 oscillation models were discussed in detail in Ref.~\cite{Arguelles:2021meu}. 
MicroBooNE has subsequently searched for short-baseline oscillations~\cite{MicroBooNE:2022sdp}, excluding a considerable part of the sterile neutrino parameter space that can explain the \miniboone and LSND results, but not fully ruling out this interpretation.
Finally, it has also been pointed out that if the \miniboone excess is predominantly composed of antineutrinos (in neutrino mode), the constraints posed by LAr experiments become significantly weaker due to the smaller antineutrino-argon cross sections~\cite{Kamp:2023mjn}.

We also briefly comment on other measurements of $\nu_e$ events at \microboone. 
In particular, the differential $\nu_e$CC cross section on Argon has been measured using neutrinos from the Booster neutrino beam~\cite{MicroBooNE:2022tdd} as well as from the NuMI beam~\cite{MicroBooNE:2021gfj,MicroBooNE:2021fdt}. 
These measurements still have large statistical and systematic uncertainties, especially in the NuMI neutrino flux prediction, which it can be as large as $20\%$ at the lowest energies ($E_\nu < 300$~MeV). 
The mean energy of NuMI flux in both analyses is similar to that of Booster neutrinos, and the baseline varies between $679$~m, the distance between the NuMI target and \microboone, and $100$~m, the distance between the NuMI absorber and \microboone. 
With increasing statistics, this flux can become an auxiliary tool to explore different $L$ and $E$ configurations, especially relevant for oscillation searches~\cite{MicroBooNE:2022sdp}.

\subsubsection{Sterile-driven short-baseline oscillations}

We start the $\nu_e$ interpretations of the \miniboone excess within the 3+1 model of neutrino oscillations.
A mostly-sterile mass eigenstate with $m_4\gtrsim 1$~eV is introduced, leading to oscillations at $L/E \sim 1\text{ km}/1\text{ GeV}$, where $L$ is the baseline of the experiment and $E$ the neutrino energy.
This interpretation is related to other anomalies in neutrino data, which share similar values of $L/E$.
Of particular relevance is the prediction that the $\nu_\mu \to \nu_e$ appearance signal at \miniboone and LSND should be accompanied by $\nu_e$ and $\nu_\mu$ disappearance in sterile-driven oscillation models.
We briefly review the status of these two channels below.

Some electron-neutrino and antineutrino experiments have observed a deficit of events compatible with sterile-driven neutrino $\nu_e/\overline{\nu}_e$ disappearance. 
An overall deficit of $\overline{\nu}_e$ rates at reactor experiments~\cite{Mention:2011rk}, dubbed the reactor antineutrino anomaly (RAA), was observed when incorporating the Huber-M\"ueller reactor antineutrino flux calculations~\cite{Mueller:2011nm,Huber:2011wv}. 
Since then, newer calculations of the reactor antineutrino flux have diluted the statistical significance of the RAA~\cite{Giunti:2021kab}.
New data on the beta spectrum of $^{235}$U and $^{239}$Pu collected at the Kurchatov Institute~\cite{Kopeikin:2021ugh} showed that data-driven reactor flux models based on data from ILL~\cite{Schreckenbach:1981wlm,VonFeilitzsch:1982jw,Schreckenbach:1985ep,Hahn:1989zr} were overestimating the total number of neutrinos~\cite{Giunti:2021kab,Ricochet:2022pzj}.
This interpretation was recently corroborated by the STEREO reactor experiment~\cite{STEREO:2022nzk} and resolved the RAA.

Reactor experiments that can measure ratios of event rates at different baselines have better control of the flux systematics. 
Dedicated searches for sterile neutrinos were performed by the PROSPECT~\cite{PROSPECT:2018dtt,PROSPECT:2020sxr}, STEREO~\cite{STEREO:2018rfh,STEREO:2019ztb,STEREO:2022nzk}, and DANSS~\cite{DANSS:2018fnn} experiments, as well as by the RENO and NEOS collaborations~\cite{RENO:2020hva}.
All these searches reported null results. 
Neutrino-4~\cite{NEUTRINO-4:2018huq}, on the other hand, claims evidence for sterile neutrinos at a significance of larger than $4\sigma$.
The interpretation of their results has been criticized in the literature~\cite{PROSPECT:2020raz}.
In Ref.~\cite{Giunti:2021iti}, the authors claim that the Neutrino-4 results can only be reproduced when neglecting the detector energy resolution.
Their analysis brings the significance of the excess to as low as $2.2\sigma$, preferring significantly larger and already-excluded mixing angles.

Another anomaly is the deficit of $\nu_e$ observed in gallium experiments when exposing the detectors to radioactive sources~\cite{Acero:2007su,Giunti:2010zu}. 
This anomaly has been observed in a modern setup by the BEST experiment~\cite{Barinov:2021asz}. 
These experiments have low sensitivity to the oscillation frequency ($\Delta m^2_{41}$) due to poor spatial resolution but they observe an overall significant deficit in the expected $\nu_e$ rate.
A sterile-driven $\nu_e$ disappearance explanation, however, requires large mixing angles with electron-neutrinos, around $|U_{e4}|^2 \sim 0.07$, and is in tension with measurements of the total solar neutrino flux~\cite{Goldhagen:2021kxe}.
No consensus has been reached yet on the cause of this anomaly --- for recent reviews on the topic, see Refs.~\cite{Giunti:2022btk,Giunti:2022xat,Brdar:2023cms}.
Global fits to the reactor and gallium experiment data find that the strongest evidence for anomalous $\nu_e$ disappearance comes from BEST and Neutrino-4; no satisfactory agreement among all data in a 3+1 oscillation model was observed~\cite{Berryman:2021yan,Giunti:2022btk}.

In the muon sector, $\nu_\mu$ disappearance is not observed by accelerator experiments~\cite{Dydak:1983zq,SciBooNE:2011qyf,MiniBooNE:2012meu}.
The most significant constraint was placed by the MINOS/MINOS+ collaboration~\cite{MINOS:2020iqj}, surpassing other existing limits by over an order of magnitude.
However, their results, especially their sensitivity to averaged-out oscillations, have been questioned in the literature~\cite{Louis:2018yeg,Diaz:2019fwt}.
Despite this, no independent study has been performed to address the issues raised.
An independent limit on $\nu_\mu$ disappearance comes from atmospheric muon-neutrinos that would oscillate due to a sterile-neutrino-induced parametric resonance inside the Earth. 
This effect is constrained by IceCube~\cite{IceCube:2020phf,IceCube:2020tka} and excludes a significant portion of the sterile neutrino parameter space around $\Delta m^2 \sim 1$~eV$^2$.
In the tau sector, data and constraints on $\nu_\tau$ appearance are scarce and come mostly from atmospheric neutrino experiments.
At accelerators, OPERA set direct limits on $|U_{\tau 4}|^2$ through a search for $\nu_\mu \to \nu_\tau$ appearance~\cite{CHORUS:2007wlo,OPERA:2015zci}.
In the future, IceCube~\cite{Smithers:2021orb} and KM3NET~\cite{KM3NeT:2021uez} could improve on these direct searches.
Indirect limits have been obtained with searches for the disappearance of active neutrinos via neutral-current interactions.
This strategy was pursued at NO$\nu$A~\cite{NOvA:2017geg,NOvA:2021smv} and can also be used in the LAr program at FNAL~\cite{Furmanski:2020smg}.
Other direct probes of sterile neutrinos include searches for anomalous kinks in the beta spectrum of tritium at the KATRIN experiment~\cite{KATRIN:2022ith}, which are sensitive to the $|U_{e4}|^2$ mixing angle.

Light sterile neutrinos are strongly constrained by cosmology as they would have thermalized with the SM bath in the early Universe and count as additional relativistic species at Big Bang Nucleosynthesis~\cite{Planck:2018vyg,Hagstotz:2020ukm,Adams:2020nue}.
Extensions to the minimal 3+1 model have proposed to reconcile the model with cosmology~\cite{Dasgupta:2013zpn,Hannestad:2013ana,Chu:2018gxk,Yaguna:2007wi,Saviano:2013ktj,Giovannini:2002qw,Bezrukov:2017ike,Farzan:2019yvo,Cline:2019seo,Dentler:2019dhz,Archidiacono:2020yey,DiValentino:2021rjj}. 
The general idea consists of suppressing the production of the light sterile neutrinos in the early Universe.
This can be achieved by the so-called secret-interaction mechanism, in which strong self-interactions between sterile neutrinos (or between sterile neutrinos and an ultra-light dark matter background) create a large matter potential for the sterile flavor and suppress their mixing with light neutrinos at early times. 
Other proposals have also been put forward, showing that lowering the reheating temperature to a few MeV can avoid the thermalization of light sterile neutrinos~\cite{Gelmini:2004ah,Yaguna:2007wi}.

It is also possible that short-baseline oscillations are induced by more than just one sterile neutrino.
In addition to the even larger number of relativistic degrees of freedom in the early Universe, this solution is not immune to the tension between appearance and disappearance datasets~\cite{Giunti:2015mwa}.
Global fits still find a significant internal tension in 3+2 model~\cite{Giunti:2011gz,Conrad:2012qt,Kopp:2013vaa,Giunti:2013aea,Diaz:2019fwt}.

In conclusion, sterile neutrino models face significant challenges in explaining global neutrino data, and their existence is also in stark contradiction with standard cosmological models. 
A solution to this conundrum requires additional new physics in the neutrino sector and/or non-standard cosmologies.
More data from the SBN program will be essential to test the sterile neutrino interpretation of \miniboone directly by means of measurements of the $\nu_e$ spectrum at the different baselines of MicroBooNE, SBND, and ICARUS~\cite{MicroBooNE:2015bmn,Cianci:2017okw}.

\subsubsection{Anomalous matter effects}

The non-standard interactions (NSI) framework provides a phenomenological parameterization of deviations from the ordinary matter potential experienced by neutrinos in matter~\cite{Proceedings:2019qno}.
While NSIs alone cannot generate new flavor transitions at short baselines, several works in the literature have studied their impact on sterile neutrino-driven oscillations~\cite{Akhmedov:2010vy}.
NSIs can suppress oscillations at large energies, where $2 E_\nu V > \Delta m^2_{41}$ with $V$ a new matter potential for neutrinos, relaxing constraints from MINOS/MINOS+~\cite{Hu:2020uvx} due to the higher energy of the NuMI beam.
In addition, the energy of the active-sterile transition resonance used to search for sterile neutrinos at IceCube can be lowered.
Even though IceCube is less sensitive to BSM effects at low energies~\cite{Schneider:2021wzs}, the authors of Ref.~\cite{Esmaili:2018qzu} conclude that existing IceCube data already rules this possibility out.
The new matter potential can be associated with new interactions in the $\nu_\mu$ and $\nu_\tau$ sectors~\cite{Liao:2016reh}, with new interactions between sterile neutrinos and ordinary matter~\cite{Liao:2018mbg}, or interactions between sterile neutrinos and a dark matter background~\cite{Denton:2018dqq}.
The typical values of NSIs required for the aforementioned scenarios are large, typically bigger than a few percent of $G_F$. 
They are, therefore, strongly constrained by high-energy neutrino and global data (see~\cite{Proceedings:2019qno}).

Anomalous matter potentials can also induce new resonances at short-baseline experiments, akin to the Mikheyev-Smirnov-Wolfenstein 
effect.
In this case, the appearance signal can be sharply peaked at the resonant energy and would explain the low energy nature of the MiniBooNE excess.
A first proposal was based on an anomalous matter potential sourced by the cosmic neutrino background~\cite{Asaadi:2017bhx}.
However, this idea requires an unrealistic local overdensity of cosmic neutrinos.
Ref.~\cite{Smirnov:2021zgn} also concludes that this scenario is excluded by the different measurements of $\Delta m^2_{31}$ across the different energy scales of T2K, NOVA, MINOS, and atmospheric experiments.
Another possibility is that the matter potential is sourced by ordinary matter and that it exclusively impacts the new sterile neutrino, referred to as a quasi-sterile neutrino in Ref.~\cite{Alves:2022vgn}.
In this case, stronger-than-Weak interactions between quasi-sterile neutrinos and matter particles are required to produce the matter potential.

\subsubsection{Decaying sterile neutrinos}

Heavy, mostly-sterile neutrinos can be produced via mixing in meson decays and subsequently decay in flight into light neutrinos between the source and the detector~\cite{Palomares-Ruiz:2005zbh,Bai:2015ztj,deGouvea:2019qre,Dentler:2019dhz}.
This decay can proceed through a light scalar particle, for instance, $\nu_4 \to \nu_i \phi$, with $i<4$, which can also participate in the secret-interaction mechanism that suppresses $\nu_4$ production in the early Universe.
Typically, $m_\phi + m_{\nu_i} < m_4$ such that $\nu_4$ decays via two-body decays with a lifetime shorter than the short-baseline distances.
In some models, the scalar particle can decay, $\phi \to \nu_i \overline{\nu}_j$, enhancing the final number of neutrinos~\cite{Dentler:2019dhz}.
For Dirac neutrinos, the $\nu_4$ and $\phi$ decays may be into left(right)-handed (anti)neutrinos, in which case it is referred to as visible or into non-interacting right(left)-handed (anti)neutrinos, in which case it is called invisible since the new decays are unobservable. 
The visible and invisible types of decay can coexist, depending on the underlying interactions.
For Majorana neutrinos, the daughter neutrinos and antineutrinos are always interacting.

In the case of invisible decays, an explanation for the short-baseline anomalies still relies on 3+1 oscillations.
However, depending on the experiment energy and baseline, the oscillatory behavior can be damped due to the sterile neutrino decay.
Due to the finite lifetime of the sterile state, limits from searches for a sterile-driven resonance in the earth using $\nu_\mu$ disappearance at IceCube~\cite{Moss:2017pur,Moulai:2019gpi} are relaxed.
In a dedicated analysis, IceCube has searched for the sterile-driven resonance in invisibly-decaying-sterile neutrino models~\cite{Moulai:2021zey,IceCubeCollaboration:2022tso} and found that it actually improves the agreement with data.
The null hypothesis of no 3+1 oscillations is disfavored with a p-value of $2.5\%$, and the hypothesis of 3+1 oscillations without decay is disfavored with a p-value of $\lesssim 1\%$.
The best-fit values prefer larger $\Delta m^2_{41}$ values of $\mathcal{O}(4 - 10)$~eV$^2$, as well as large couplings between the scalar and the sterile neutrino, $\alpha_\phi \simeq 0.6$.
A recent global fit to IceCube and global neutrino data finds that invisible sterile decays alleviate the tension in the 3+1 model~\cite{Hardin:2022muu}.
Theoretically, these solutions can be easily incorporated in models of singlet scalars coupled to the sterile neutrino; however, the coupling constants preferred by data are large, nearing the non-perturbative regime.

For visible decays, the signal of $\nu_\mu \to \nu_e$ appearance can be mimicked by the $\nu_4 \to \nu_F (\phi\to \nu_F \overline{\nu}_F)$ decays, where $\nu_F$ is the low-energy neutrino flavor state, $\ket{\nu_F} = \sum_{i=1}^3 U_{s i}^* \ket{\nu_i}$.
In this case, $\nu_\mu$ and $\nu_e$ disappearance are proportional to $|U_{e4}|^2$ and $|U_{\mu 4}|^2$, respectively, while the effective appearance signal, if all decays are sufficiently short, is proportional to $|U_{\mu 4}|^2$. 
Since both appearance and disappearance probabilities depend on the same power of the mixing elements, the global neutrino data tension is significantly reduced and could disappear altogether.
In the original proposal~\cite{Palomares-Ruiz:2005zbh}, the $\nu_4$ decays do not proceed through mixing, but instead from a dimension-five operator $\frac{g_{e}}{v_h}(\overline{L_e} \tilde{H})\nu_R \phi$.
In that case, appearance signals can still occur without a corresponding $\nu_e$ disappearance channel provided $U_{e4} = 0$.

If $U_{e4} \neq 0$, then $\nu_4$ can be produced in $^8$B decays in the Sun. 
This is strongly constrained when $\nu_4$ decays to active antineutrinos.
Therefore, scenarios where the neutrinos are Majorana or where the scalar is unstable, $\phi \to \nu_F \overline{\nu}_F$, are in tension with solar antineutrino searches~\cite{Hostert:2020oui}.
Dirac neutrinos with a stable $\phi$ particle are not subject to these constraints.

We also comment on an alternative, fine-tuned solution to avoid these limits.
Consider a model with a lepton-number-charged scalar particle $\phi_2$, which carries $L= 2$.
If neutrinos are Dirac and the sterile state is almost degenerate in mass with the scalar, $\epsilon \equiv \frac{m_4 - m_\phi}{m_4} \ll 1$, then the antineutrinos produced in $\nu_4 \to \overline{\nu}_F \phi_2$ decays are not observable due to their small energy, below the inverse-beta-decay threshold.
The subsequent decays of $\phi$ give two interacting visible neutrinos $\phi \to \nu_F \nu_F$, avoiding solar antineutrino limits altogether.
A full analysis is required to determine if the decay couplings can be large enough to overcome the $\epsilon$ suppression to the $\nu_4$ lifetime,
\begin{equation}
\Gamma^{\rm Lab} = \sum_i \frac{m_4^2}{ 4 E_{\nu_4}}|U_{s4} U_{sj}|^2 \alpha_{\phi} \epsilon^2 (2 - \epsilon)^2,
\end{equation}
where $\alpha_\phi = g_\phi^2/4 \pi$.
The model, however, still faces strong limits from cosmology~\cite{Dentler:2019dhz}.

The decaying-sterile neutrino models discussed above have not been directly targeted by \microboone, although the null results of the template analyses and the 3+1 oscillations search can already constrain them.
The main difference with respect to the 3+1 oscillations in the sterile decay signal is the softening of the energy dependence of the oscillation and the suppression of $\nu_\mu$ disappearance.
This can weaken the constraining power of the $\nu_\mu$ sample and wash-out oscillations in the $\nu_e$ appearance signal.
In particular, $\nu_e$ disappearance can be large and suppress the rate of intrinsic $\nu_e$ backgrounds while compensating for the additional $\nu_e$ rate from the effective $\nu_\mu \to \nu_e$ transition.
Given these distinctive features, we encourage a dedicated study of this scenario to assess the sensitivity of \microboone to the decaying-sterile neutrino explanations of the MiniBooNE excess.

\subsubsection{Exotic effects in short-baseline oscillations}

\paragraph{Wavepacket decoherence}
The authors of Ref.~\cite{Arguelles:2022bvt} pointed out that neutrino-wavepacket decoherence could help explain the lack of $\overline{\nu}_e$-disappearance in reactor experiments in 3+1 oscillation models.
The global fit to short-baseline data in Ref.~\cite{Hardin:2022muu} showed that the $4.9\sigma$ internal tension of short-baseline data is reduced to $3.6\sigma$ when including effects due to the finite size of the neutrino wavepacket.
The best-fit result for the size of the wavepacket in that analysis is $\sigma_x \sim  67$~fm, much smaller than the naive expectation of typical inter-atomic distance or the inverse of the detector energy resolution~\cite{Akhmedov:2022bjs}.
For comparison, theoretical estimates of the beta-decay-induced antineutrino wavepacket size~\cite{Jones:2022hme} find $\sigma_x\lesssim 10 - 100$~pm (see also~\cite{Jones:2022cvh,Akhmedov:2022mal}).
For such proposals to be successful, quantum-mechanical decoherence should be more significant than classical averaging from detector resolution.

\paragraph{Energy-dependent mixing angles}
It has been argued that the possibility of having energy-dependent mixing angles and masses can alleviate the tension between appearance and disappearance experiments in 3+1 oscillation models~\cite{Schwetz:2007cd, Babu:2022non}.
As shown in Ref.~\cite{Babu:2022non}, renormalization group evolution can modify disappearance constraints on the square of mixing elements by factors of $\mathcal{O}(2-3)$ because their energy scales are significantly different from those of LSND and MiniBooNE.
The running of the mixing angles also modifies the oscillation at MiniBooNE since the energy scale of detection ($E\sim 3$~GeV) is larger than that of production ($E\sim m_\pi$).
Reactors and solar neutrinos constrain $|U_{e4}|$ at energies of $\mathcal{O}(10)$~MeV or less, leaving room for this parameter to run.
MINOS/MINOS+ and IceCube, however, are sensitive to the same production scale ($E\sim m_\pi$) but constrain $|U_{\mu4}|$ at a larger detection energy of about $3$~GeV and $1$~TeV, respectively.
Oscillation searches at \microboone would not be significantly impacted, although the oscillation maximum at the other SBN detectors could be somewhat modified due to the different baselines.
The benchmark model proposed in Ref.~\cite{Babu:2022non} closely resembles the particle content we discuss in \cref{sec:fit_darknus} but does not require any couplings between the new dark particles and the SM other than to neutrinos themselves.

\paragraph{Space-time modifications}
Altered neutrino dispersion relations have also been discussed to solve the 3+1 oscillation tension.
In large extra dimensions, fluctuations of the brane make the path length of active neutrinos larger than that of sterile ones~\cite{Pas:2005rb,Hollenberg:2009ws,Carena:2017qhd,Doring:2018ncz,Doring:2018cob}.
In some cases, the modified dispersion relations lead to resonances in the neutrino flavor evolution.
Another realization of such scenarios is through Lorentz violation, parameterized in the SM Extension effective theory, which can modify neutrino flavor evolution via Lorentz-violating higher-dimensional operators suppressed by powers of $1/M_{\rm planck}$~\cite{Kostelecky:2004hg,Diaz:2010ft,Diaz:2011ia}.
Although these models solve some of the issues in the 3+1 oscillation paradigm, they still show tension with global neutrino data due to the steep energy dependence~\cite{Barenboim:2019hso}.

\begin{table*}[t]
    \centering
    \begin{tabular}{|c|>{\centering\arraybackslash}p{5 cm}|>{\centering\arraybackslash}p{4 cm}|>{\centering\arraybackslash}p{4 cm}|>{\centering\arraybackslash}p{2 cm}|}
\hline\hline
    \multicolumn{5}{|c|}{New physics in scattering}\\
\hline
    Topology & Model & Diagram & Signal & References\\
\hline\hline
\multirow[t]{2}{*}{single $\gamma$} 
        & neutrino upscattering  & \insertdiagram{2} &  $N\to \nu \gamma$ & \cite{Gninenko:2009ks,Gninenko:2010pr,Gninenko:2012rw,Masip:2012ke,Radionov:2013mca,Magill:2018jla,Schwetz:2020xra,Vergani:2021tgc,Alvarez-Ruso:2021dna,Kamp:2022bpt,Bansal:2022zpi}
        \\
        \cline{2-5}
        & \biggercell{neutrino-induced \\ inverse-Primakoff scattering} & \insertdiagram{3} & $\varphi^* A \to \gamma A$ &  \cite{Bansal:2022zpi}
        \\
        \cline{2-5}
\hline\hline
\multirow[c]{4}{*}{$e^+e^-$} 
        & \biggercell{neutrino upscattering} &
        \multirow{2}{*}{\insertdiagram{1}}  & \biggercell{$N \to \nu e^+e^-$ \\ on-shell $N$} & \cite{Bertuzzo:2018itn,Ballett:2018ynz,Ballett:2019pyw,Datta:2020auq,Dutta:2020scq,Abdullahi:2020nyr,Abdallah:2020biq,Abdallah:2020vgg,Hammad:2021mpl} \quad 
        \Cref{sec:fit_darknus}
        \\
        \cline{2-2}\cline{4-5}
        & \biggercell{neutrino-induced bremsstrahlung} &  &  \biggercell{$Z^\prime \to e^+e^-$ \\ off-shell $N$} & not studied \bigstrut 
        \\
        \cline{2-5}
        &  \biggercell{neutrino-induced \\ Primakoff scattering} &  \insertdiagram{5} & $\varphi \to e^+e^-$ & \cite{Abdallah:2020biq}\bigstrut
        \\
        \cline{2-5}
        &   \biggercell{neutrino-induced \\ inverse-Primakoff scattering} &  \insertdiagram{4} & $Z^\prime \to e^+e^-$ & not studied 
        \\
\hline\hline
\multirow[c]{2}{*}{$\gamma\gamma$} 
        &  \biggercell{neutrino upscattering} &  \insertdiagram{7} & $N \to \nu \gamma \gamma$ & \cite{Datta:2020auq} \bigstrut
        \\
        \cline{2-5}
        &  \biggercell{neutrino-induced \\ Primakoff scattering} &  \insertdiagram{6} & $\varphi \to \gamma \gamma$ & not studied 
        \\
\hline\hline
    \end{tabular}
    \caption{Examples of new particle production in neutrino-nucleus scattering with electromagnetic final states.~\label{tab:model_summary}}
\end{table*}

\subsection{Single photons}

Single photons are an important background in \miniboone at low energies.
The dominant sources of these events come from misidentified $\pi^0$ decays and $\Delta(1232)$ radiative decays~\cite{MiniBooNE:2021bgc}. 
Well-reconstructed two-ring events are measured by \miniboone and directly constrain the $\pi^0$ rate.
While the $\Delta(1232)$ production cross section is also constrained by the $\pi^0$ sample, largely made up of resonant events, the radiative branching ratio $\Delta(1232) \to N \gamma $ ($N = n, p$) is not. 
An enhancement of this branching by $\mathcal{O}(3)$ factors leads to a remarkable agreement with the LEE. 
This possibility is constrained by photoproduction experiments, $\gamma p \to \Delta$, and has recently been directly tested by the \microboone experiment.
Good agreement with \miniboone's estimates was found, excluding this explanation of the LEE at $94\%$ C.L.~\cite{MicroBooNE:2021zai}. 
It is important to note that while this result excludes the SM $\Delta$ hypothesis, it does not necessarily exclude other new physics scenarios that invoke single photons.
In particular, the \microboone analysis was not sufficiently sensitive to single photons not accompanied by a hadronic vertex (e.g., coherent emission) due to the significantly larger backgrounds in the $1\gamma0p$ selection.
Single photons can also contribute to the exclusive $1e0p0\pi$ channel of the \microboone $\nu_e$ search, where, notably, an excess is already observed~\cite{MicroBooNE:2021sne}
In view of that, we proceed to discuss beyond-the-SM sources of single photons.

\subsubsection{Upscattering via transition magnetic moments}
\label{sec:TMM}

Neutrinos interactions with the material inside or outside the \miniboone detector can produce short-lived heavy neutral leptons (HNLs) that subsequently decay visibly inside the fiducial volume.
This scenario is referred to as upscattering.
One model for upscattering where the HNL decays to a single photon is that of a transition magnetic moment~\cite{Gninenko:2009ks,Gninenko:2010pr,Gninenko:2012rw,Magill:2018jla,Vergani:2021tgc,Kamp:2022bpt}.
The low-energy Lagrangian is given by
\begin{equation}\label{eq:TMMlagrangian}
    \mathscr{L} \supset \frac{\mu_{\rm tr}^\alpha}{2} \overline{\nu}_{\alpha} \sigma^{\mu\nu} N_R F_{\mu\nu} + \text{ h.c.},
\end{equation}
where $\alpha$ is the neutrino flavor index.
The interaction above would mediate neutrino-nucleus upscattering that produces the HNL $N$ as well as the subsequent decays of $N \to \nu \gamma$. 
Depending on the details of the model, the massive particle $N$ may be Dirac or Majorana, although the \miniboone angular spectrum prefers the former.
An explanation of the LEE can be achieved with $\mu_{\rm tr}^\mu$ as small as $\mathcal{O}(10^{-9}\mu_B)$, which corresponds to $\mu_{\rm tr}^\mu \sim (1\text{ PeV})^{-1}$.
The muon index indicates that muon neutrinos and antineutrinos, which dominate the flux at MiniBooNE, initiate the upscattering process.
In what follows, we drop the flavor index and assume $\mu_{\rm tr} = \mu_{\rm tr}^\mu$.

In this model, the upscattering cross section and decay rate of the HNL are proportional to $|\mu_{\rm tr}|^2$. 
Once produced, the HNL can decay into either a single photon or, around $0.7\%$ of the time, into a $e^+e^-$ pair. 
The total decay rates for a Dirac HNL are given by~\cite{Arguelles:2021dqn},
\begin{align}
    \Gamma_{N\to \nu \gamma} &= \frac{|\mu_{\rm tr}|^2 m_N^3}{16 \pi},
    \\
    \Gamma_{N\to \nu \ell^+ \ell^-} &\simeq \frac{\alpha |\mu_{\rm tr}|^2}{48\pi^2} m_N^3 \left[ 2\log\left(\frac{m_N}{m_\ell}\right) -3 \right],
\end{align}
where we neglected terms of $\mathcal{O}(m_\ell^2/m_N^2)$ in the dilepton rate. 
The dilepton mode, while subdominant, provides an alternative signature that may be searched for in $K^+,\pi^+ \to \ell^+ N \to \ell^+ \nu e^+e^-$ or at low-density neutrino detectors such as ND280~\cite{Arguelles:2021dqn}.
A detailed fit to the \miniboone excess was performed in \cite{Kamp:2022bpt}, suggesting to explain both the energy and the angular spectrum of the LEE, the HNL should be as massive as $400$~MeV.
Ref.~\cite{Kamp:2022bpt} also derived constraints on the model from the MINERvA neutrino-electron scattering sideband data.
While MINERvA poses strong limits, it is not able to fully exclude the regions of preference, especially at large HNL masses where stringent cuts on the angular spectrum reject most of the signal events.

Several future experiments can probe the TMM scenario, including dark matter direct detection experiments~\cite{Shoemaker:2018vii}, coherent elastic neutrinos-nucleus scattering (CE$\nu$NS) measurements~\cite{Kim:2021lun,Bolton:2021pey}, double-bang searches at IceCube and large volume neutrino detectors~\cite{Coloma:2017ppo,Atkinson:2021rnp,Schwetz:2020xra,Plestid:2020vqf,Huang:2021mki}, and neutrino detectors at the Forward Facilities of the LHC~\cite{Ismail:2021dyp}.

The scale of new physics behind the dimension-5 operator above is model-dependent, but a naive estimate for a TMM induced at the one-loop level can be found using
\begin{equation}
    \mu_{\rm tr} \simeq \frac{y y^\prime e Q_f}{8 \pi^2} \frac{m_f}{M^2_{\rm UV}} \left[ \log{\left(\frac{M_{\rm UV}^2}{m_f^2}\right) - 1} \right],
\end{equation}
where $Q_f$ and $m_f$ stand for the electric charge and the mass of the charged fermion in the loop, respectively, and $y^{(\prime)}$ is the coupling between left-(right-)handed fermions to the new particle of mass $M_{\rm UV}$.
Ref.~\cite{Brdar:2020quo} proposes a leptoquark model as an UV completion to \cref{eq:TMMlagrangian}, where $f = b$-quark.
The leptoquark can be within reach of the LHC with $M_{\rm UV} \gtrsim 1$~TeV for $y=y^\prime=2$ and $\mu_{\rm tr} \sim 1\text{ PeV}^{-1}$.

In general, the UV completion of \cref{eq:TMMlagrangian} would also generate large mass terms, $m_D \overline{\nu}_L N_R$~\cite{Voloshin:1987qy,Barbieri:1988fh,Babu:1989px,Babu:1989wn,Leurer:1989hx} (see also \cite{Babu:2020ivd} for a recent review).
In seesaw models, this is undesirable for two reasons: i) neutrino masses would be too large, 
ii) the mixing between active and heavy neutrinos would be too large. 
While point number i) can be avoided in inverse seesaw models with approximate conservation of lepton number, point ii) poses a much bigger challenge.
The mixing is subject to severe constraints from decay-in-flight and meson peak searches for HNLs.
Schematically, for $f=b$, we get
\begin{equation}
    U_{\alpha N} \sim \frac{\mu_{\rm tr}}{2} \frac{M_{\rm UV}^2}{e Q_f m_N} \sim 3\times 10^{-3} \left(\frac{\mu_{\rm tr}}{1 \text{ PeV}^{-1}}\right) \left(\frac{M_{\rm UV}}{1 \text{ TeV}}\right).
\end{equation}
which is prohibitively large. 
As a result, such models require some fine-tuning to avoid constraints on the mixing angles.

Finally, we note that such transition magnetic moments can also exist between different generations of HNLs~\cite{McKeen:2010rx,Bolton:2021pey}. 
For instance,
\begin{equation}\label{eq:nnprime_tmm}
    \mathscr{L} \supset \frac{\mu_{NN^\prime}}{2} \overline{N_L^\prime} \sigma^{\mu\nu} N_R F_{\mu\nu} + \text{ h.c.},
\end{equation}
where $N^\prime$ may again be a Dirac or Majorana particle. In this case, even relatively light $N$ particles can be produced inside the detector and decay fast enough via $N\to N^\prime \gamma^{(*)}$.    
For instance, for $m_N=10$~MeV and $\mu_{\rm tr}\sim (5\text{ PeV})^{-1}$, the event rate at \miniboone is sufficiently large for $\mu_{NN^\prime}\sim (500\text{ TeV})^{-1}$, provided $N^\prime$ is  light enough. 
Small mass splittings between $N$ and $N^\prime$ can also help ameliorate tensions with the \miniboone angular distribution in this case. 
This is particularly interesting in inverse seesaw models, where lepton number violation is controlled by $M_{N} - M_{N^\prime}$, though more work is needed to find a self-consistent UV completion of \Cref{eq:TMMlagrangian,eq:nnprime_tmm}.

We end this section by noting that HNLs coupled to the SM via mixing and small TMMs can be long-lived and produced in charged meson decays, $\pi, K \to \ell N$.
In this case, the HNL propagates from the target to the detector to decay in flight via $N \to \nu \gamma$~\cite{Fischer:2019fbw}.
This signature is severely constrained by the timing information when the HNL mass is larger than $\sim 50$~MeV.
It is also constrained, although not fully excluded, by the corresponding off-shell photon mode $N \to \nu (\gamma^* \to e^+e^-)$ signature at ND280 and PS-191~\cite{Arguelles:2021dqn}.
The latter, however, does exclude similar models where the dominant branching ratio of the HNL is into $e^+e^-$~\cite{Chang:2021myh}.

\subsubsection{Neutrino-induced inverse-Primakoff scattering}
\label{sec:ips}

We now turn to another possibility for producing single-photon signatures: neutrino-induced inverse Primakoff scattering (IPS). 
In this scenario, the neutrino scatters on a nucleus through a virtual scalar particle that undergoes inverse Primakoff scattering, $\phi^* \, A \to \gamma A$, with $A$ a nuclear target.
The scalar may couple to photons or dark photons ($Z^\prime$), and either mediator may be exchanged with the nucleus.
The MiniBooNE LEE can be explained if a photon is produced in the final state or if a dark photon decays to an $e^+e^-$ pair.
In this section, we focus on the former.
The process $\nu A \to \nu \gamma A$ has been discussed in Ref.~\cite{Bansal:2022zpi} in the context of the dimension-seven Rayleigh operators, $\overline{\nu} \nu F_{\mu\nu}F^{\mu\nu}$ and $\overline{\nu} \nu F_{\mu\nu}\widetilde{F}^{\mu\nu}$.
The scenario we discuss here is one of the completions of the operator above, also discussed by the authors.

We discuss a model of a light scalar particle with loop-induced couplings to SM photons.
For later convenience, we also include couplings to a dark photon, the mediator of a new $U(1)_D$ gauge symmetry. 
The low-energy Lagrangian reads,
\begin{align}\label{eq:phigammagamma}
    \mathscr{L}  &\supset -\frac{1}{4} X_{\mu\nu} X^{\mu\nu} - \frac{\varepsilon}{2 c_W} F_{\mu\nu} X^{\mu\nu} - \frac{1}{2}\partial_\mu \phi\partial^\mu \phi
    \\\nonumber
    & - \frac{\alpha}{8 \pi}  \phi \left( \frac{c_{\gamma\gamma}}{f_{\gamma\gamma}} F_{\mu \nu} F^{\mu \nu} + \frac{c_{\gamma X}}{f_{\gamma X}} F_{\mu \nu} X^{\mu \nu} + \frac{c_{XX}}{f_{XX}} X_{\mu \nu} X^{\mu \nu}\right),
\end{align}
where $A^{\mu\nu} \equiv \partial_\mu A_{\nu} - \partial_\nu A_{\mu}$ for $A = X,\,F$, with $X_\mu$ the $U(1)_D$ mediator before diagonalization of the kinetic and mass terms.
The scales $f_{\gamma \gamma}/c_{\gamma \gamma}$ and $f_{\gamma X}/c_{\gamma X}$ are constrained by direct searches for new charged particles of by the coupling of $\phi$ with SM particles.
If the operators above are generated by charged particles with no $U(1)_D$ charge, we expect $c_{\gamma X} \propto \varepsilon$ and $c_{XX} \propto \varepsilon^2$, upon diagonalization of the gauge kinetic terms. 

The scalar can also couple to the neutrino sector. 
A large direct coupling to SM neutrinos is challenging to achieve within a SU$(2)_L$-invariant model, but a direct coupling to a sterile neutrino $\nu_s$ is less constrained. 
This sterile state $\nu_s$ mixes with SM neutrinos, and in terms of the mass eigenstates, $\ket{\nu_s}=\sum_{i=1}^4 U_{s i}^* \ket{\nu_i}$.
Having this in mind, we can consider
\begin{equation}\label{eq:phinusnus}
    \mathscr{L} \supset c_\nu^{ij} \phi \overline{\nu_i} P_L \nu_j + \text{h.c.},
\end{equation}
where all couplings $c_{\nu}^{ij}$ are proportional to $U_{si}^*U_{s j}$.
For concreteness, we assume neutrinos to be Dirac.

Using the equivalent photon approximation, we estimate the cross-section for the $\nu A \to \nu \gamma A$ process.
In the limit of large $m_{\phi}$, we find
\begin{equation}\label{eq:IPSdxsec}
    \frac{\dd\sigma_{\nu A \to \nu \gamma A}}{\dd Q^2 \dd \hat{s}} \simeq \frac{\alpha^3 Z^2}{64 (4 \pi)^4} \frac{|c_{\gamma\gamma}|^2}{f_{\gamma\gamma}^2} \frac{\hat{s}}{Q^2} \frac{|c_\nu|^2}{m_\phi^4} |F(Q^2)|^2,
\end{equation}
where $Q^2$ is the momentum exchange with the nucleus, $\hat{s} = (k + k_\gamma)^2$ is the center-of-mass energy of the projectile neutrino and the semi-real photon, and $F(Q^2)$ is the nuclear electromagnetic form factor.
The cross section favors large center-of-mass energies thanks to the higher-dimension nature of the interaction and, therefore, produces high-energy photons, in contrast with the low-energy nature of the MiniBooNE excess.
It is also suppressed by large powers of $\alpha$ and phase-space factors.
This is expected since the full process is a one-loop three-body scattering.

It is easy to see that for typical values of allowed couplings and masses the total cross section is far below the SM neutrino cross section.
We have also checked this for small values of $m_\phi$, where the mediator mass no longer suppresses the rate.
This is compatible with the findings of \cite{Bansal:2022zpi}, which quotes limits on the neutrino polarizability operators of order $c_\nu |c_{\gamma\gamma}|/f_{\gamma\gamma} < 3 \times 10^{-3}$~GeV$^{-1}$.
Clearly, $f_{\gamma\gamma}$ is tied to the GeV scale for allowed couplings of $c_\nu$ of $\mathcal{O}(10^{-3})$, and, therefore, requires new large couplings to charged particles below the EW scale, which is highly constrained by direct searches.
The situation is much worse in Majoron models where $c_\nu$ is proportional to neutrino masses.
While a detailed study would be needed to draw definitive conclusions, we deem this model inconsistent with the LEE as i) it produces photons that are too high energy, ii) requires decay constants $f_{\gamma\gamma}$ significantly below the EW scale, signaling a theoretical inconsistency for the couplings required to explain the LEE.

Finally, we mention that IPS can also be initiated by dark particles in the beam.
Ref.~\cite{Dutta:2021cip} considered the production of long-lived scalar particles in charged meson decays, $\pi, K \to \ell \nu \phi$, followed by the IPS of $\phi$ on nuclei inside the detector.
In this case, the scalar particle interacted with matter through a virtual dark photon through the $c_{\gamma X}$ operator to produce a single photon.

\subsection{Electron-positron pairs} 

Another explanation of the excess is an anomalous source of $e^+e^-$ inside the \miniboone detector.
The LEE signal is mimicked if the pairs overlap or are highly asymmetric in energy, so only a single electromagnetic shower is resolved~\cite{Ballett:2018ynz}. 
The $e^+e^-$ explanation of the LEE is unique in that it constitutes both a photon-like (overlapping pairs) and electron-like (energy asymmetric pairs) signal.
Neutrino interactions in MiniBooNE rarely produce $e^+e^-$ pairs directly.
The most common source of these final states is through the production of photons that subsequently convert into overlapping \epluseminus pairs, or the Dalitz decay of $\pi^0 \to e^+e^- \gamma$ with a branching ratio of $\sim 1.2\%$.
We are unaware of a publicly available study of Dalitz decays in MiniBooNE, although the rate is small and most often accompanied by an observable additional photon.\cite{Dutta:2021cip}.

New physics sources of $e^+e^-$ arise naturally in models of dark sectors, i.e., in extensions of the SM below the electroweak scale with small couplings to the SM.
The new light particles can be produced by the interactions of muon-neutrinos and antineutrinos with nuclei in the detector and subsequently decay to \epluseminus.
We will discuss a few examples based on neutrino upscattering to HNLs, neutrino-induced Primakoff scattering, and bremsstrahlung processes.

To date, no dedicated experimental search for an \epluseminus origin of LEE has been carried out, although several constraints have been derived in phenomenological works.
The most constraining data comes from the photon-like sideband of the neutrino-electron scattering measurement of MINERvA~\cite{MINERvA:2015nqi,MINERvA:2019hhc,MINERvA:2022vmb}, which has been used to place limits on dark neutrinos and transition magnetic moments in Refs.~\cite{Arguelles:2018mtc,Kamp:2022bpt}.
It should be noted that MINERvA data does not robustly exclude upscattering explanations of the MiniBooNE excess.
Already in Refs.~\cite{Arguelles:2018mtc,Kamp:2022bpt} it was recognized that the photon and $e^+e^-$ final states appear in a large $dE/dx$ sideband (photon-like) of the MINERvA analysis, which is tuned to data for measurements at low $dE/dx$ (electron-like).
In the absence of theoretical uncertainty bands for the pre-tune photon-like background prediction, the authors of \cite{Arguelles:2018mtc,Kamp:2022bpt} considered an optimistic and a conservative assumption of $30\%$ and $100\%$ overall uncertainty on the background.
In view of the large excess of photon-like events in recent MINERvA analyses~\cite{MINERvA:2023ner}, it is unlikely that the optimistic constraints from \cite{Arguelles:2018mtc,Kamp:2022bpt} MINERvA apply.
While MINERvA could be sensitive to upscattering scenarios with dedicated analysis strategies, it is too early to conclude that the data excludes them.

Another important limit on upscattering scenarios comes from the HNL search in the gaseous Argon TPCs of ND280~\cite{T2K:2019jwa}, the off-axis near detector of T2K.
As part of that analysis, the collaboration searched for \epluseminus final states without any hadronic activity.
With a generic but simplified framework, Ref.~\cite{Brdar:2020tle} repurposed this search to constrain MiniBooNE explanations, concluding that upscattering models where the parent HNL has a lifetime greater than $c\tau^0 > 10$~cm are excluded by T2K data.
In Ref.~\cite{Arguelles:2022lzs}, the authors performed a more detailed simulation of the signature inside the multi-component detector, further strengthening the limits for displaced decays.
While a comparison to the MiniBooNE region of preference was possible in only a few cases, it is fair to say that it would be difficult to reconcile MiniBooNE explanations based on long-lived particles that decay primarily to $e^+e^-$ with T2K data.
This is primarily due to the large amounts of lead and iron in front of the low-density TPCs of ND280 detector, which enhances the rate of coherent upscattering events, and the low background nature of the analysis.
Scenarios with photon and two photon final states are less constrained as the photons would not convert inside the gaseous TPCs.

\subsubsection{Neutrino-induced boson fusion}
\label{sec:models:VBF}

As another possibility for producing new particles by neutrino-nucleus scattering inside MiniBooNE, we discuss the neutrino-induced vector boson fusion (VBF) reactions.
Like the IPS process discussed in \Cref{sec:ips}, this is a $2 \to 3$ scattering mediated by light particles.
The dark photon interactions in \Cref{eq:phigammagamma} and the neutrino-scalar interaction of \Cref{eq:phinusnus} could lead to the on-shell production of the scalar particle $\phi$ or of the dark photon $Z^\prime$, which subsequently decay to $e^+e^-$.
The topology of the three-body process need not arise from the higher-dimensional operators in IPS.
When considering a dark photon, the IPS diagram can proceed through the direct interaction between $\phi$ and $Z^\prime$, as considered in Ref.~\cite{Abdallah:2020biq}.
We leave a detailed study of this cross section to future literature.
In what follows, we comment on the total cross section and argue that it can be sufficiently large for the allowed parameter space. 

For dark photon masses below the $\mathcal{O}(100)$~MeV scale, the equivalent photon approximation can give a crude estimate of the cross-section for the $\nu A \to \nu Z^\prime A$ process.
In the limit of small mediator masses, we find
\begin{equation}\label{eq:VBFdxsec}
    \frac{\dd\sigma_{\nu A \to \nu Z^\prime A}}{\dd Q^2 \dd \hat{s}} \simeq |c_\nu|^2\alpha_D  \frac{ \alpha\varepsilon^2 Z^2}{4 \pi} \frac{m_{Z^\prime}^2}{\hat{s}^3 Q^2} |F(Q^2)|^2,
\end{equation}
where $Q^2$ is the momentum exchange with the nucleus, $\hat{s} = 2 k_1 \cdot k_\gamma$ is the center-of-mass energy of the projectile neutrino and the semi-real photon, and $F(Q^2)$ is the nuclear electromagnetic form factor.
Contrary to \cref{eq:IPSdxsec}, dark photon production prefers low energy exchange and, therefore, can, in principle, lead to low-energy events.
We also note that the rate is proportional to the dark photon mass, which is a reflection of the fact that the scalar coupling to the dark photon is proportional to $g_D^2 v_\varphi$, where $v_\varphi$ is the dark scalar vacuum expectation value.

It is easy to see that for allowed parameters the cross section can be as large as a few percent of the total neutrino cross section in the SM, indicating that as far as the total rate is concerned, it can successfully reproduce the number of events at MiniBooNE.
A detailed study of the energy and angular spectra is required to draw any additional conclusions.

\subsubsection{Upscattering to heavy neutrinos}
\label{sec:models:upscattering}

In \cref{sec:TMM}, we discussed the production of HNLs through a transition magnetic moment. 
Now, we will focus on models where HNLs are generated by new light mediators and decay into dilepton pairs instead of a single photon.
These scenarios are particularly simple from a model-building point of view and can be linked to consistent low-energy extensions of the SM.
The first proposals in Refs.~\cite{Bertuzzo:2018itn,Ballett:2018ynz} were based on dark photon models with a single HNL.
The signature is given by
\begin{equation}\label{eq:upscattering_process}
    \nu_\mu A \to (N \to \nu Z^\prime{}^{(*)} \to \nu e^+e^-) A,
\end{equation}
where $A$ is some nuclear target, $Z^\prime$ the dark photon, and $N$ an HNL.
As indicated, the dark photon may be produced on or off its mass shell.
If on-shell, the decay chain is typically prompt, and the final states are more forward-going.
If the dark photon is off-shell, the HNL will decay via a three-body process and be longer-lived. 
In that case, production from the dirt upstream of the detector can be important.

Later studies considered vector mediators of a $B-L$ gauge symmetry~\cite{Hammad:2021mpl} and extended scalar sectors~\cite{Datta:2020auq,Dutta:2020scq}, where new light scalars play the role of the dark photon in \cref{eq:upscattering_process}.
Another logical possibility is that a scalar particle mediates the scattering process, but the decay proceeds through a light dark photon.
An advantage of scalar mediators is that the upscattering is predominantly helicity-flipping, leading to less forward production of the HNL.
This can lead to less forward angular distributions at MiniBooNE when compared to vector mediator of the upscattering process.

The HNL models that can give the upscattering signature in \cref{eq:upscattering_process} typically require multiple states for anomaly cancellation and neutrino mass generation.
This was explored in Refs.~\cite{Ballett:2019pyw,Abdullahi:2020nyr} that generalized the dark neutrino signature to a cascade of decays in the dark sector,
\begin{equation}\label{eq:upscattering_process_2}
    \nu_\mu A \to (N_i \to N_j Z^\prime{}^{(*)} \to N_j e^+e^-) A,
\end{equation}
where $N_j$ is the lighter of the two HNLs and can be much longer-lived than $N_i$.
Phenomenologically, these models are called $3+n$ models, where $n$ stands for the number of HNLs considered. 
In \cref{sec:fit_darknus}, we study 3+1 and 3+2 models in the context of the \miniboone excess.
Due to the larger dark photon coupling to heavy neutrinos, the parent HNL decay in 3+2 models can be much faster than in 3+1 models as it does not involve the coupling of the dark photon to light neutrinos.

\paragraph{Dark photon models} The UV completions of dark neutrino sectors with dark Abelian gauge symmetry $U(1)_{D}$ 
can be separated according to the origin of the coupling between the dark leptons, $\nu_D$, and active neutrinos, $\nu_\alpha$ --- we refer to this coupling as the dark neutrino portal.
In Ref.~\cite{Bertuzzo:2018ftf}, a new scalar SU$(2)_L$ doublet charged under the $U(1)_{D}$, $H_D$, realizes the inverse seesaw mechanism in the neutral lepton sector with a vector-like dark neutrino, $\nu_D$.
The dark neutrino portal is then given by the Yukawa coupling between active and dark neutrinos, $\overline{L} H_D \nu_D$.
Another possibility studied by Refs.~\cite{Ballett:2019pyw,Abdullahi:2020nyr} was to consider an SU$(2)_L$-singlet complex scalar $\Phi$ as well as a singlet sterile neutrino $\nu_N$.
In this case, the dark neutrino portal is generated in a two-step process: the complex scalar couples the dark neutrinos to the sterile states, $\overline{\nu_N} \nu_D \Phi$, while the latter couples to active neutrinos via the usual neutrino portal, $\overline{L} H \nu_N$.
Alternatively, when the sterile neutrino is heavy, one may integrate it out to find the higher-dimensional portal coupling $M_N^{-1}(\overline{L} H) (\nu_D \Phi)$, where $M_N$ is the mass of the sterile neutrino.

The new gauge boson $X_\mu$ interacts with the SM particles via kinetic mixing with the SM hypercharge boson $B_\mu$. 
The dark photon mass may be generated by the Stueckelberg or Higgs mechanisms. 
The latter implies the existence of new scalar particles, such as a single dark Higgs field, $\Phi$, charged under $U(1)_{D}$.
Heavy neutrino fields charged under the new gauge symmetry, referred to as dark neutrinos, can then simultaneously explain the origin of neutrino masses via the seesaw mechanism and the \miniboone signature.
The latter is possible thanks to their mixing with SM light neutrinos and the interaction with the dark photon.

\begin{figure}[t]
    \centering
    \includegraphics[width=0.49\textwidth]{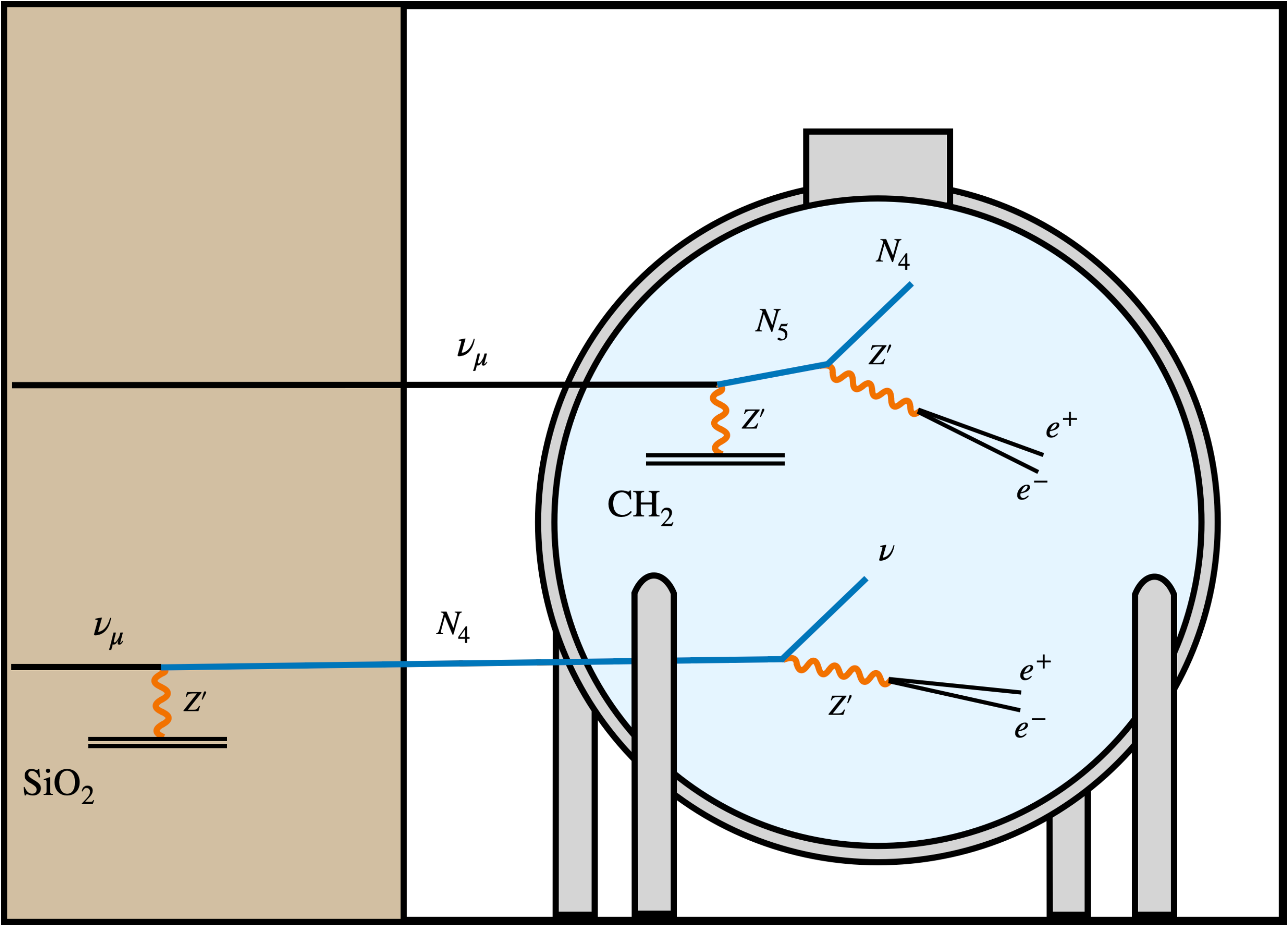}
    \caption{The dark neutrino upscattering signature at MiniBooNE for 3+1 models ($N_4 \to \nu e^+e^-$) and 3+2 models ($N_5 \to N_4 e^+e^-$).
    As an example, we show a scattering inside the detector for 3+2 and in the dirt upstream of the detector for 3+1.
    }
    \label{fig:miniboone_diagram_2}
\end{figure}

For concreteness, we present the full model of Ref.~\cite{Abdullahi:2020nyr}, which can accommodate the simpler phenomenological model used in \cref{sec:fit_darknus}.
The Lagrangian is given by,
\begin{equation}
\label{eq:upscattering_model_minimal}
\begin{split}
    \mathscr{L} \supset &-\frac{1}{4} X^{\mu\nu} X_{\mu\nu} - \frac{\varepsilon}{2 c_W} X_{\mu\nu} B^{\mu\nu} + (\mathcal{D}_\mu \Phi)^\dagger (\mathcal{D}^{\mu} \Phi) \\
    & - V(\Phi, H) + \overline{\nu_N} i \slashed{\partial} \nu_N + \overline{\nu_D} i \slashed{\mathcal{D}} \nu_D \\
    & - \biggl[ (\overline{L} \tilde{H}) Y \nu_N^c + \frac{1}{2} \nu_N M_N \nu_N^c + \overline{\nu}_D M_X \nu_D \\
    &  \qquad + \overline{\nu_N} ( Y_L \nu_{D_L}^c \Phi + Y_R \nu_{D_R} \Phi^\ast) + \mathrm{h.c.} \biggr]
\end{split}
\end{equation}
where: $\widetilde{H} \equiv i \sigma_2 H^*$, $\mathcal{D}_\mu \Phi \equiv \partial_\mu - i g_D X_\mu$, and $g_D$ is the coupling constant of the new force. 
The neutral lepton sector contains a number $d$ of vector-like dark neutrinos $\nu_D = \nu_{D_L} + \nu_{D_R}$ and $n$ sterile states $\nu_N$.
Upon diagonalization of the kinetic and mass terms, the gauge boson sector is comprised of the photon $A$, the SM $Z$ boson, and the dark photon $Z^\prime$. 



One advantage of dark neutrino models is their rich connection to other low-energy phenomenology.
For instance, the model of \cref{eq:upscattering_model_minimal} offers a viable solution to the $(g-2)_\mu$ anomaly~\cite{Muong-2:2021ojo}.
Kinetic mixing can explain the deviation from the Standard Model~\cite{Pospelov:2008zw}, but this solution is already excluded if the dark photon decays predominantly to $e^+e^-$ or fully invisibly. 
In the invisible case, BaBar~\cite{BaBar:2017tiz} and NA64~\cite{Banerjee:2019pds} exclude the region of preference.
However, if the dark photon can decay to several short-lived HNLs, as in dark neutrino models, it becomes a semi-visible particle.
That is, it decays to both visible and invisible particles, such as in the reaction $Z^\prime \to (N_i \to N_j e^+e^-)( N_k \to N_l e^+e^-)$, relaxing the constraints above~\cite{Mohlabeng:2019vrz,Abdullahi:2023tyk}.
A simultaneous explanation of the MiniBooNE and $(g-2)_\mu$ anomalies is possible in 3+n models with $n>1$ and requires $\sim 1$~GeV, dark photons.
Another important connection is to kaon decays, where HNLs can produce multi-lepton signatures, such as $K^+ \to \ell^+ (N_i \to N_j e^+e^-)$~\cite{Ballett:2019pyw}.
For on-shell dark photons, this decay cascade represents a double bump hunt, where $(p_K - p_\ell)^2 = M_{N_i}^2$ and $(p_{e^+}+p_{e^-})^2 = m_{Z^\prime}$.
Finally, similarly to the transition magnetic moment case, large-volume experiments can target the double-bang feature of the upscattering signal~\cite{Coloma:2017ppo,Atkinson:2021rnp,Schwetz:2020xra,Plestid:2020vqf,Huang:2021mki}.

\paragraph{Scalar models}  
The upscattering can also be mediated by a scalar particle. 
This can be achieved, for example, in Higgs portal models, where a dark scalar mixes with the Higgs, or in two Higgs doublet models, where one of the scalars in the extended sector has a large Yukawa coupling to electrons. 
The latter is preferable since the decay rate of the HNL into electron-positron pairs ($N \to \nu (\phi^{(*)} \to e^+e^-)$) is not necessarily suppressed by $(m_e/v_{\rm EW})^2$.
In addition, Higgs portal models are subject to strong constraints from Higgs decays at CMS~\cite{CMS:2022qva} and ATLAS~\cite{ATLAS:2019cid}. 

\begin{figure*}
    \centering
    \includegraphics[width=0.49\textwidth]{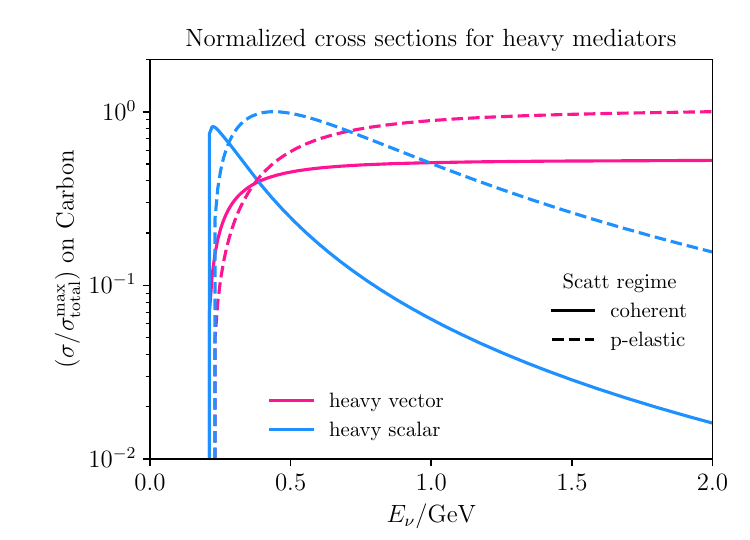}
    \includegraphics[width=0.49\textwidth]{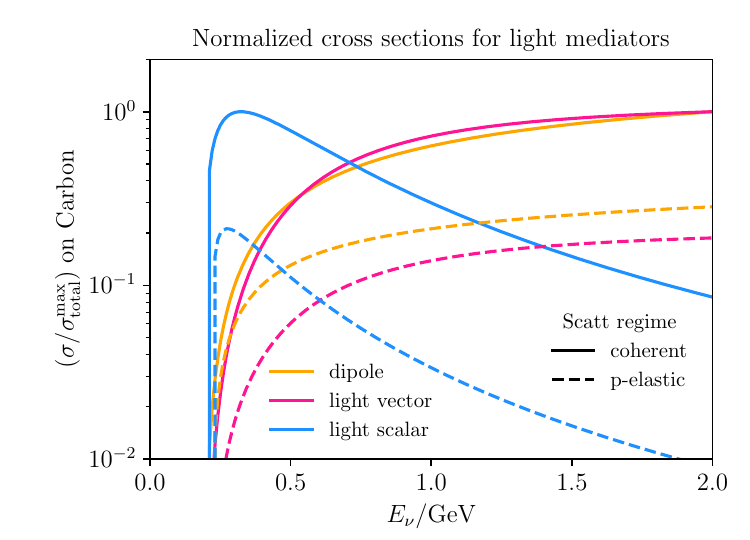}
    \caption{Neutrino upscattering cross sections for the vector mediator, scalar mediator, and magnetic moment upscattering.
    On the left, we show the cases for $m_{Z^\prime} = m_{h^\prime} = 1.25$~GeV and on the right $m_{Z^\prime} = m_{h^\prime} = 30$~MeV.
    In all cases, we fix $m_4 = 200$~MeV.
    }
    \label{fig:scalar_vs_vector}
\end{figure*}

One of the main advantages to considering a scalar mediator scenario is that the the cross section for upscattering is suppressed at high energies, as can be seen in \cref{fig:scalar_vs_vector} and as discussed in Refs.~\cite{Abdallah:2020biq,Abdallah:2022grs}. 
In the coherent and nucleon-elastic scattering regimes, the vector and transition magnetic moments cross sections grow slightly.
On the other hand, the scalar ones decrease for $E_\nu$ much larger than the HNL mass.
This maintains the upscattering events to low energies, in agreement with the MiniBooNE data.
The helicity-flipping dominance of the process also helps widen the angular distribution of the events, especially for heavy scalar mediators where nucleon elastic contributions become dominant. 
Finally, these models are much less constrained by high-energy data from, e.g., CHARM II and MINERvA~\cite{Arguelles:2018mtc}.
Despite the attractive features of these scenarios, we leave a detailed fit to the excess to future literature.

\begin{figure*}
    \centering
    \includegraphics[width=0.49\textwidth]{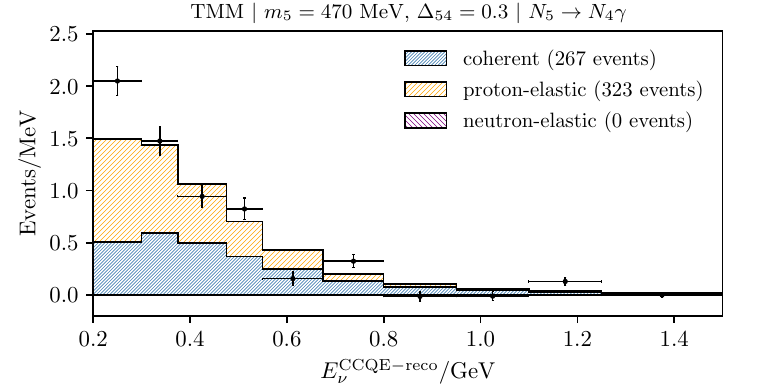}
    \includegraphics[width=0.49\textwidth]{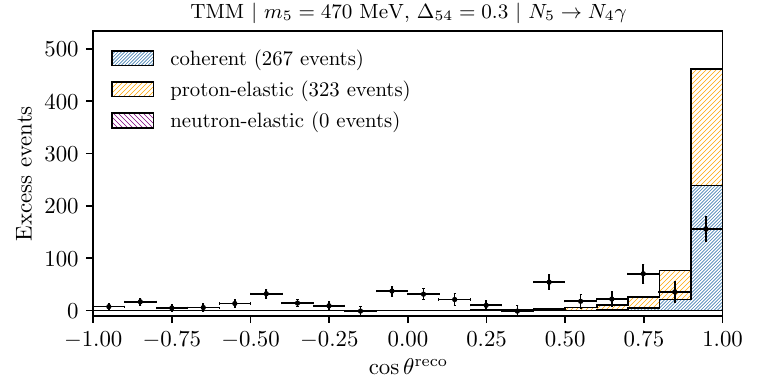}
    \includegraphics[width=0.49\textwidth]{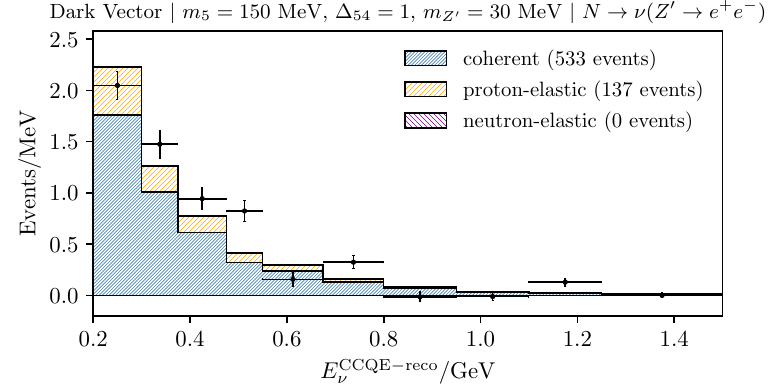}
    \includegraphics[width=0.49\textwidth]{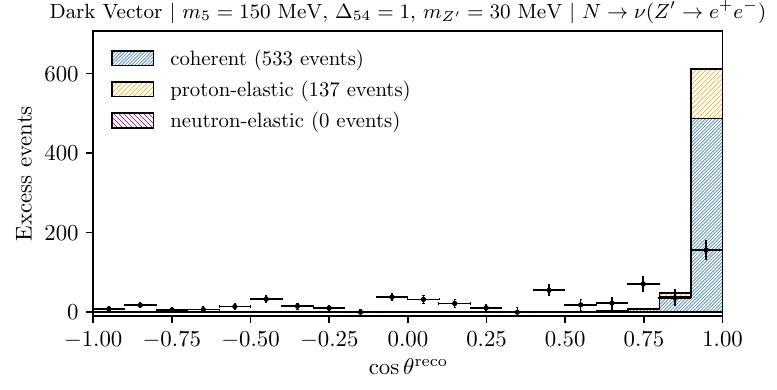}
    \includegraphics[width=0.49\textwidth]{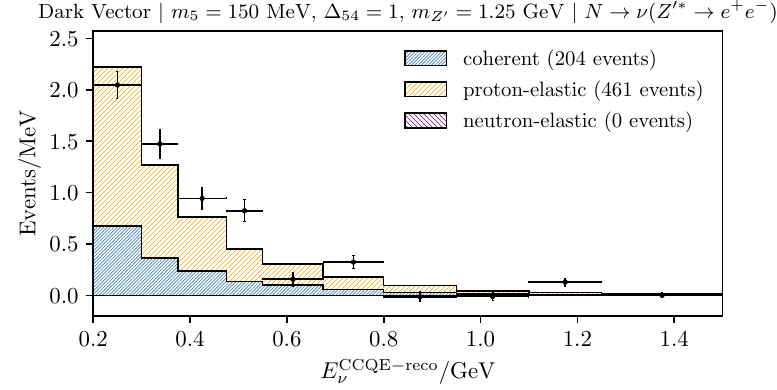}
    \includegraphics[width=0.49\textwidth]{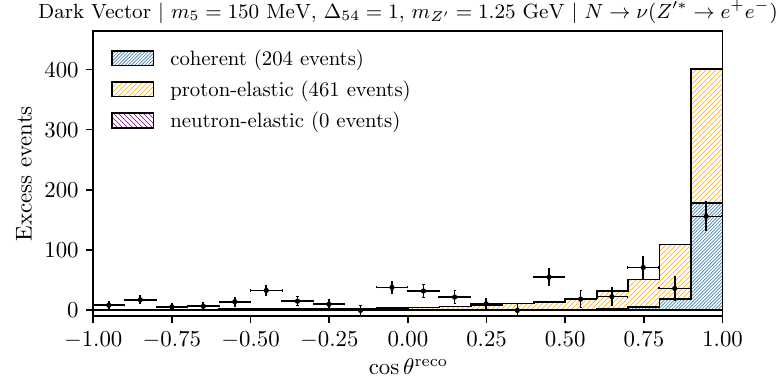}
    \includegraphics[width=0.49\textwidth]{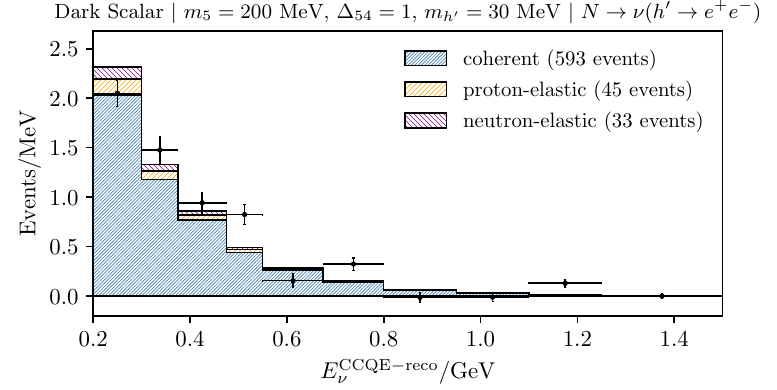}
    \includegraphics[width=0.49\textwidth]{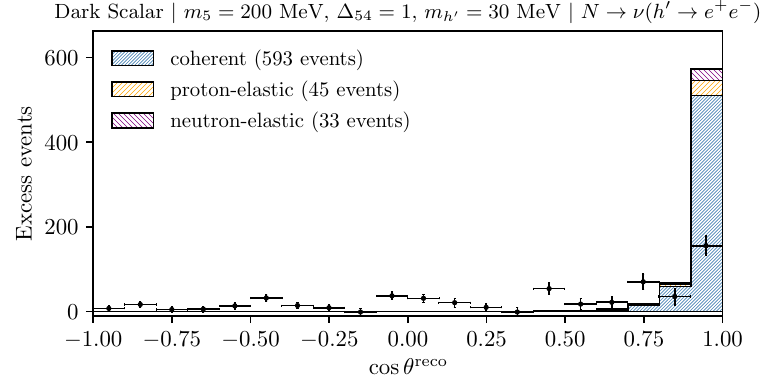}
    \includegraphics[width=0.49\textwidth]{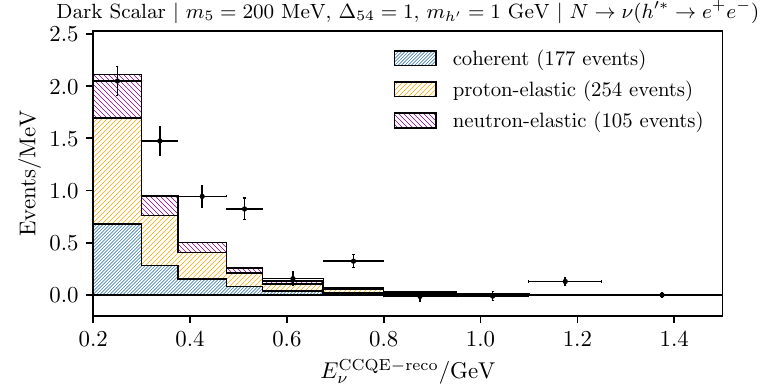}
    \includegraphics[width=0.49\textwidth]{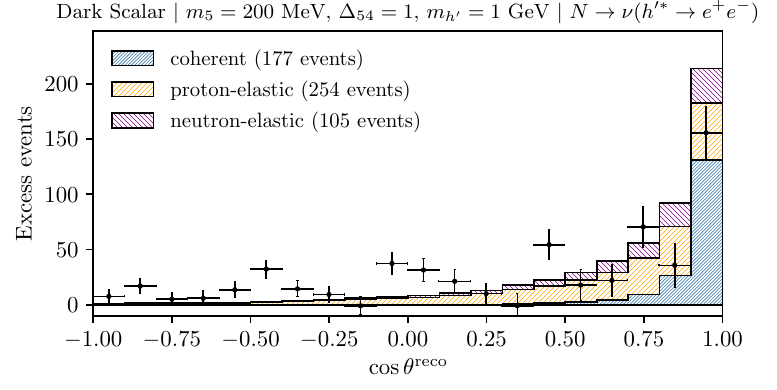}
    \caption{
    The MiniBooNE excess and the new physics prediction in reconstructed neutrino energy (left) and the cosine of the angle with the beam (right) in different models of upscattering.
    Each row corresponds to a different assumption of the mediating particle.
    Here, $\Delta_{54} = (m_5 - m_4)/m_4$. 
    }
    \label{fig:spectra_benchmarks}
\end{figure*}

We end this section by comparing different upscattering models against MiniBooNE data for a few representative points in \cref{fig:spectra_benchmarks}.
To study the shape differences between different upscattering scenarios, we pick different values of HNL and mediator masses in $3+2$ models and minimize the MiniBooNE $\chi^2$ (for more details, see \cref{sec:fit_darknus}) to find the best-fit for the normalization.
We show both the reconstructed neutrino energy and angular distributions, separating the coherent and nucleon-elastic scattering regimes.
While in TMM and dark photon models the scattering takes place on electric charge, in the scalar model, the upscattering can also be significant on neutrons (for more details on the treatment of the scalar cross sections, see \cite{Abdullahi:2022cdw}).
Scenarios with a heavy mediator are better able to explain the angular distribution thanks to the larger momentum transfer to the hadronic target.

\subsubsection{Dark bremsstrahlung}

Neutrinos can also bremsstrahlung dark photons or scalars upon scattering with nuclei.
These light bosons can then subsequently decay to $e^+e^-$. 
The same boson can also mediate the interaction with nuclei, and if it is light, the cross section can be significantly enhanced, similar to the upscattering case. 
Light bosons will produce collimated \epluseminus pairs to be misreconstructed as single electron events. 
Ref.~\cite{deGouvea:2018cfv} considered a similar process initiated by dark matter particles.
Here, we are interested in the possibility of bremsstrahlung initiated by neutrinos.
The kinetic mixing parameter between the SM and the dark photon is much more constrained.

For concreteness, let us consider a dark photon, $Z^\prime$.
The coupling to matter will be proportional to charge, while the coupling to neutrinos will be proportional to a dark coupling $g_D U_{Di}^* U_{D j}$.
We can obtain a naive estimate of the dark bremsstrahlung cross section relative to the upscattering one.
For dark photons and HNLs much lighter than the energy transfer in the scattering process, we naively expect,
\begin{equation}
    \frac{\sigma_{\rm brem}}{\sigma_{\rm ups}} \simeq \frac{\alpha_D}{4 \pi} \log\left( \frac{s}{m_{Z^\prime}^2} \right),
\end{equation}
which is typically smaller than the upscattering cross sections used in \cref{sec:models:upscattering}.
Nevertheless, when $\alpha_D$ is large and $m_{Z^\prime}$ is lighter than 100~MeV, the emission rate could still be large enough to match the number of events of the LEE.
The emission cross section is peaked in the forward direction, indicating that it may be challenging to reconcile this model with the angular distribution of the MiniBooNE LEE.

\subsection{Photon pairs} 

A less explored option is to have dark particles produced in neutrino scattering decay to photon pairs. 
If the photons are highly collimated, or one of the photons exits the detector before pair converting, the signal appears as an electron-like event at \miniboone. 
This can happen if a scalar boson is produced in the dirt or detector and undergoes $\phi \to \gamma \gamma $ inside the fiducial volume.
The production of the boson can take place either via upscattering, replacing the dark photon of \cref{sec:models:upscattering} for a scalar, or via IPS and VBF processes with the production of an on-shell scalar particle.
In all cases, the scalar decay proceeds via the non-renormalizable interaction between $\phi$ and the two photons in \cref{eq:phigammagamma}, giving
\begin{align}\label{eq:scalar_decay}
    \Gamma_{\phi \to \gamma \gamma} &=  \frac{\alpha^2}{256 \pi^3} \frac{|c_{\gamma \gamma}|^2 m_\phi^3}{f_{\gamma \gamma}^2} 
    \\ \nonumber
    &= \frac{1}{5\text{ cm}} \left(\frac{m_\phi}{200 \text{ MeV}} \right)^3\left(\frac{100\text{ GeV}}{f_{\gamma \gamma}/|c_{\gamma \gamma}|} \right)^2.
\end{align}
For the relatively large scalar mass of 200~MeV, the decay constant cannot be significantly larger than $100$~GeV, as otherwise direct constraints from beam dumps become prohibitively strong~\cite{Dobrich:2015jyk}.
This, in turn, implies the need for a relatively low-scale UV completion of the two-photon operator, adding further strain on this interpretation.

In the case of upscattering, the same scalar boson may mediate the upscattering process.
The HNL produced would decay via $N_4 \to \nu_\ell (\phi \to \gamma\gamma)$.
The scalar should be produced on-shell; otherwise, the HNL decay rate will be too small.
In that case, the HNL decay can be regarded as effectively prompt.
Like the dark photon decay to $e^+e^-$, the scalar decay is isotropic, and so, as far as the decay kinematics is concerned, the two scenarios are similar.
One difference, however, is the small fraction of events where one of the photons may escape the fiducial volume before converting to a visible \epluseminus pair, leading to a genuine single photon signal.
This upscattering scenario was explored in Ref.~\cite{Datta:2020auq} and was linked to the $(g-2)_\mu$ anomaly as well; however, in this case, the contribution to the anomalous magnetic moment took place through the Barr-Zee diagram, connecting the decay rate of the $\phi \to \gamma \gamma$ to the $(g-2)_\mu$ anomaly.

Another possibility to produce $\gamma \gamma $ pairs in MiniBooNE is through the decay of dark particles into neutral pions.
One minimal example would a HNL produced via upscattering through a dark boson with axial-vector couplings to quarks.
Such scenarios can appear in $Z^\prime$ models where the dark boson mixes with the SM $Z$ via mass-mixing.
Contrary to the vectorially coupled dark photon, the axial-vector can mediate $N \to \nu \pi^0$ decays.
This scenario requires the $\pi^0$ to have significantly different kinematics from the SM $\pi^0$ production in order to reproduce the excess, as otherwise it is normalized away by the in-situ $\pi^0$ constraint.
We leave a detailed study of these two-photon scenarios to future literature.

\section{Dark neutrino MiniBooNE fit}
\label{sec:fit_darknus}

\renewcommand{\arraystretch}{1.2}
\begin{table*}[t]
    \centering
    \begin{tabular}{|c|c|c|c|c|c|c|c|c|c|}
    \hline
        Number of HNLs & Parent HNL $m_{3+n}$ & Mass splitting $\Delta$ & Mediator $m_{Z^\prime}$ & $|V_{\mu h}|^2$ & $\varepsilon$ & $p_{\rm val}$  & $p_{\rm val}$ & $p_{\rm val}$  \\
        -- & fit & fixed & fixed & fit & fixed & $\nu$ mode & $\overline{\nu}$ mode & combined \\
        \hline\hline
        $n=1$ & $444$~MeV & -- & $30$~MeV & $3.8\times 10^{-8}$ & $8\times 10^{-4}$ & $46\%$ & $7.6\%$ & $6.7\%$\\
        $n=1$ & $326$~MeV & -- & $100$~MeV & $7.5\times 10^{-8}$ & $8\times 10^{-4}$ & $27\%$ & $5.7\%$ & $1.1\%$\\
        \hline \hline
        $n=2$ & $723$~MeV & $0.3$ & $100$~MeV & $3.5\times 10^{-6}$ & $8\times 10^{-4}$ & $46\%$ & $14\%$ & $13\%$
        \\
        $n=2$ & $723$~MeV & $0.5$ & $500$~MeV & $1.3\times 10^{-4}$ & $8\times 10^{-4}$ & $37\%$ & $9.6\%$ & $11\%$
        \\
        $n=2$ & $587$~MeV & $1.0$ & $1.25$~GeV & $1.2\times 10^{-3}$ & $8\times 10^{-4}$ & $32\%$ & $5.0\%$ & $5.5\%$
        \\
        $n=2$ & $322$~MeV & $3.0$ & $1.25$~GeV & $2.3\times 10^{-4}$ & $8\times 10^{-4}$ & $30\%$ & $3.5\%$ & $2.3\%$
        \\

    \hline
    \end{tabular}
    \caption{Best-fit values to the $E_\nu^{\rm CCQE-reco}$ spectrum at \miniboone for the combined fit to $\nu$ and $\overline{\nu}$ modes in    \cref{fig:3p1_all_mzs,fig:3p2_all_mzs}.
    For 3+1 models, $h=4$ and for 3+2 models, $h=5$.
    The number of degrees of freedom for the $\chi^2$ probability $p_{\rm val}$ is $6.8$, $6.9$, and $15.6$ $\nu$ mode, $\overline{\nu}$ mode, and the combination, respectively~\cite{MiniBooNE:2012maf}.
    \label{tab:BPs}}
\end{table*}

Having surveyed the MiniBooNE explanations, we now turn to the specific case of dark neutrinos introduced in \cref{sec:models}. 
This class of models is largely unexplored experimentally, can be embedded in self-consistent low-energy extensions of the SM, and does not contradict cosmological constraints.
While they do not explain other anomalies in short-baseline experiments, like the gallium and LSND anomalies, they provide an excellent fit to the \miniboone energy spectrum.
The signal consists of the decays of short-lived HNLs produced by neutrinos in the dirt or inside the detector.
A schematic of the signature at \miniboone is shown in \cref{fig:miniboone_diagram_2}.
The production requires a new mediator particle to enhance production and shorten the lifetime of the HNLs.
In this work, we focus on a dark photon due to the simplicity of the underlying model.
Our study covers a similar physics to the one proposed in scalar-mediator and similar models~\cite{Dutta:2020scq,Datta:2020auq,Abdallah:2020biq,Abdallah:2020vgg,Hammad:2021mpl,Abdallah:2022grs}.

\begin{figure*}[t]
    \centering
    \includegraphics[width=0.49\textwidth]{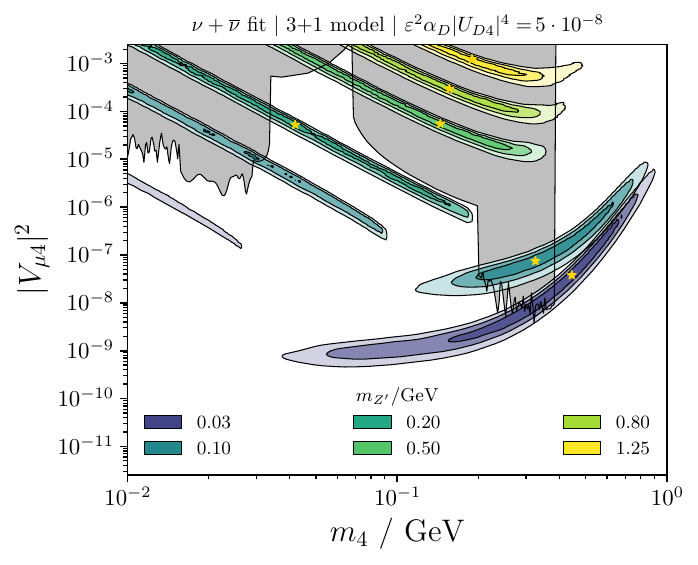}
    \includegraphics[width=0.49\textwidth]{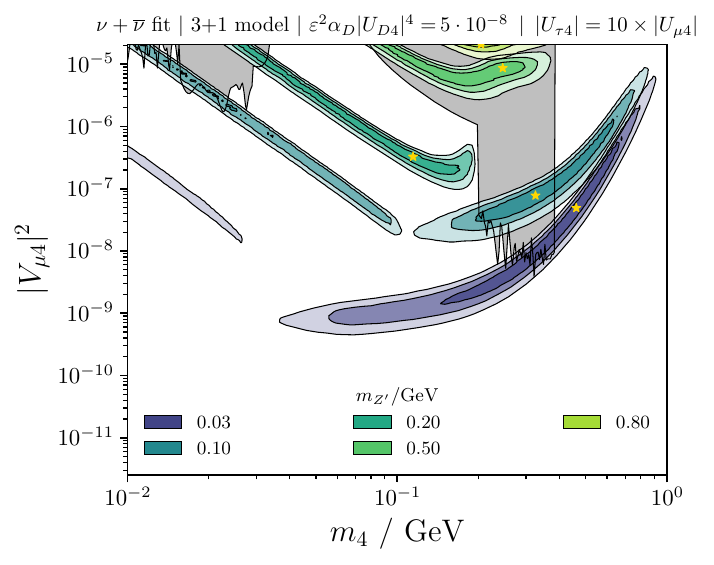}
   \caption{
   The MiniBooNE $E_\nu^{\rm CCQE-reco}$ best-fit regions in the $|V_{\mu4}|^2$ and $m_4$ plane for the 3+1 model.
   On the left, we set $|U_{\tau 4}| = 0$, and on the right $|U_{\tau 4}| = 10 \times |U_{\mu 4}|$.
   Each color represents a given fixed value of $m_{Z^\prime}$, and the different shading corresponds to the $1\sigma$, $2\sigma$, and $3\sigma$ CL regions (2 d.o.f.).
   Model-independent limits on heavy neutrinos exclude the shadowed region~\cite{Fernandez-Martinez:2023phj}.
   Other constraints from neutrino scattering and meson decays are not shown.
   \label{fig:3p1_all_mzs}}
\end{figure*}

\begin{figure*}[t]
    \centering
    \includegraphics[width=0.49\textwidth]{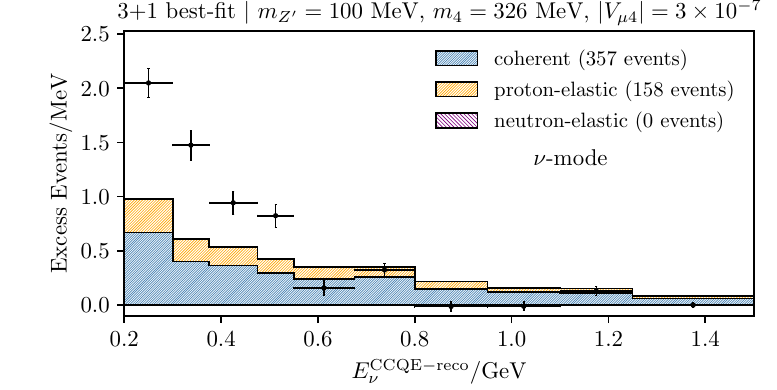}
    \includegraphics[width=0.49\textwidth]{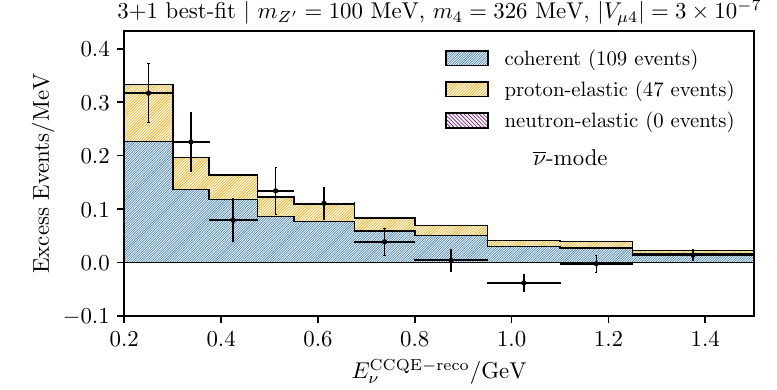} 
    \caption{
    The excess of events as a function of the reconstructed neutrino energy $E_\nu^{\rm CCQE-reco}$ at MiniBooNE for the FHC ($\nu$) mode (left panel) and the RHC ($\overline{\nu}$) mode (right panel). 
    The prediction of the best-fit point in the 3+1-model ($|V_{\mu 4}|^2 = 3 \times 10^{-7}$, $|V_{\tau 4}| = 0$, $m_4 = 326$~MeV, $m_{Z^\prime} = 100$~MeV) is shown as different shades in the histogram separating events from coherent neutrino-nucleus scattering, proton-elastic, and neutron-elastic neutrino scattering.
    \label{fig:3plus1Dist}}
\end{figure*}

\begin{figure}[t]
    \centering
    \includegraphics[width=0.49\textwidth]{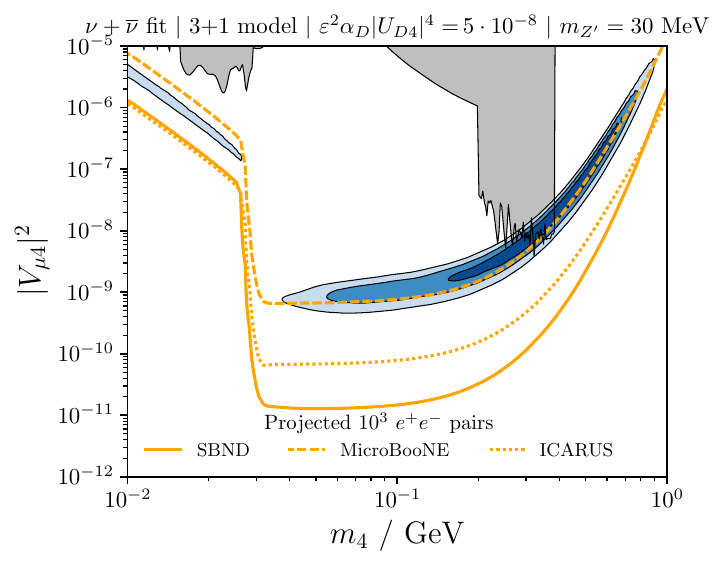}
    \caption{
    Same as \cref{fig:3p1_all_mzs} showing only the case where $m_{Z^\prime} = 30$~MeV and $|U_{\tau 4}| = 0$.
    The orange lines indicate where the \microboone (dashed), Icarus (dotted), and SBND (solid) detectors can expect $10^{3}$ $e^+e^-$ pairs from $N_5$ decays to be produced inside their respective fiducial volumes (before any reconstruction efficiencies).
    \label{fig:3p1_sbn}
    }
\end{figure}

\subsection{Phenomenological model}
\label{sec:pheno_model}

We introduce a minimal phenomenological model to describe HNL production and decay via dark photon interactions.
The interactions and particle content we introduce are based on a seesaw extension of the SM with a new $U(1)_D$ dark gauge symmetry.
Examples of UV completions can be found in Refs.~\cite{Bertuzzo:2018ftf,Ballett:2019pyw,Abdullahi:2020nyr}.

The minimal particle content consists of a massive dark photon that kinetically mixes with the SM hypercharge and the $n$ heavy neutral leptons that interact through the dark force and mix with light neutrinos.
Upon diagonalization of the kinetic and mass terms, the dark photon interactions can be written as
\begin{align}\label{eq:couplings_to_currents}
    \mathscr{L}_{\rm int.} &\supset Z^\prime_{\mu} \left( g_D  \mathcal{J}^{\mu}_D - e \varepsilon  \mathcal{J}^{\mu}_{\rm EM} - \varepsilon  t_{\rm W} \frac{m_{Z^\prime}^2}{m_Z^2} \frac{g}{2 c_{\rm W}}\mathcal{J}^{\mu}_{\rm NC}\right),
\end{align}
where it is assumed that $m_{Z^\prime}^2 \ll m_Z^2$. 
The dark current $\mathcal{J}_{\rm D}^\mu$ can be expressed in general terms based on the neutral lepton mass states,
\begin{equation}\label{eq:generic_dark_current}
\mathcal{J}_{D}^\mu \equiv \sum_{i,j = 1}^{3+n} V_{ij} \overline{\nu_i} \gamma^\mu \nu_j,
\end{equation}
where $n$ is the total number of heavy neutrino mass states, $V_{ij} \equiv \sum_k Q^D_k U^*_{ki} U_{kj}$ is the interaction vertex and $k$ runs over the number of generations of dark states in the theory. 
$Q^D_k$ is the dark charge of those, which we will assume to be equal to $+1$ ($-1$) for neutrinos (antineutrinos), and $U$ is the neutrino mixing matrix.
We adopt a similar notation to Ref.~\cite{Arguelles:2022lzs}, extending it to models with two heavy neutrinos.
For convenience, we define the dark photon couplings between the low-energy flavor neutrino state, $\ket{\hat{\nu}_\alpha} = \sum_{i=1}^3 \ket{\nu_i}$, and the upscattered HNLs $N_{h}$ with $h = 4,5$.
In 3+2 models, assuming unitarity,
\begin{align}\label{eq:defVmuN}
    V_{\alpha h} &\equiv  U_{D h} \frac{U_{\alpha 4}  U_{D4}^* + U_{\alpha 5} U_{D5}^* }{\sqrt{1 - \sum_{\beta} (|U_{\beta 4}|^2 + |U_{\beta 5}|^2)}},
\end{align}
where $\alpha, \beta \in \{e, \mu, \tau\}$.
The equivalent expression for 3+1 models can be recovered by setting all mixing elements $U_{\alpha 5}$ to zero.
For simplicity and having the LEE in mind, unless otherwise specified, we will assume that the HNLs mix with the muon neutrino flavor only.
In addition, we always assume that $|U_{\mu 4}| = |U_{\mu 5}|$ and $|U_{D 4}| = |U_{D 5}|$.
In our simulations we set $|U_{D4}|$ and $|U_{D5}|$ are taken as $1/\sqrt{2}$, neglecting the small correction from dark-light mixing angles, $|U_{D4}|^2 + |U_{D5}|^2 = 1 - \sum_{i = 1}^{3}|U_{Di}|^2$.
In this way, $|V_{\mu 4(5)}|$ can be used as a proxy for the parameter $|U_{\mu 4(5)}|$, which is directly constrained by several model-independent HNL limits.
Finally, we also define the coupling of HNLs to all light neutrinos, collectively denoted by $\nu_\ell$,
\begin{align}\label{eq:defVellN}
    |V_{\ell h}|^2 &\equiv \sum_{i = 1}^3 |V_{h i}|^2 \simeq |U_{Dh}|^2 \sum_{\alpha = e}^\tau |U_{\alpha 4}|^2.
\end{align}
In 3+2 models, there must also be two dark flavors, in which case we assume they mix equally with light neutrinos.

We will vary the kinetic mixing parameter, $\varepsilon$, as well as the masses $m_4$, $m_5$, and $m_{Z^\prime}$.
In the 3+2 model, it will be convenient to define the mass splitting parameter,
\begin{equation}
    \Delta \equiv \frac{m_5 - m_4}{m_4}.
\end{equation}
This parameter controls the mass gap between the two HNL states, and therefore, also the energy release in the $N_5 \to N_4 e^+e^-$ decays.
In what follows, we will assume the physical mass eigenstates to be of pseudo-Dirac nature. 
This leads to less energetic and less-forward $e^+e^-$ pairs than in the Majorana case.

For large mass splitting in the 3+2 model, the parent HNL can decay invisibly via $N_5 \to N_4 N_4 N_4 (\nu_i)$ decays.
While this channel is forbidden for $\Delta < 2 (1)$, it can also be suppressed by the dark photon coupling matrix, $V_{ij}$.
To thoroughly explore the parameter space of the 3+2 model, we make the assumption that $|V_{4i}| \ll |V_{44}|, |V_{55}| \ll e \varepsilon$, for $i < 4$, allowing us to consider arbitrarily large $\Delta$ values without worrying about the invisible decays of $N_5$. 
This assumption may seem strong at first, but it can be thought of as the result of a conserved $C$-symmetry in the dark sector, as discussed in detail in Ref.~\cite{Abdullahi:2023tyk}.
Light neutrinos ($C= -1$) and $N_4$ ($C= -1$) are both odd under $C$ and cannot interact with each other via the dark photon ($C= -1$).
On the other hand, $N_5$ ($C= + 1$) is even and can interact with the lighter odd neutral lepton states via the dark photon.
In summary, the $C$ symmetry allows the dark photon to mediate transitions between $N_5$ and the lighter neutral leptons, but not between $\nu_i$ and $N_4$.
We further assume that the scalar interactions are sufficiently weak so as not to spoil this structure.
Therefore, in what follows, it is understood that $N_5$ always decays via $N_5 \to N_4 \ell^+ \ell^-$ and $N_5 \to N_4 \pi^+ \pi^-$, even for values of $\Delta > 2$, where $m_5 > 3 m_4$.

\begin{figure*}[t]
    \centering
    \includegraphics[width=0.49\textwidth]{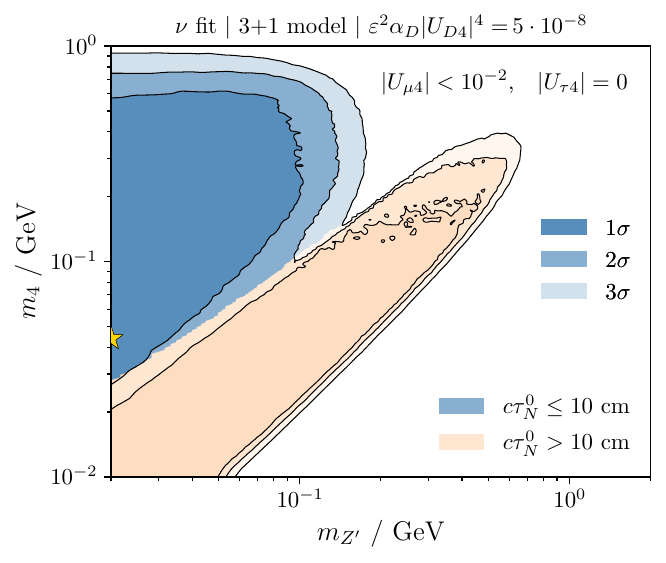}
    \includegraphics[width=0.49\textwidth]{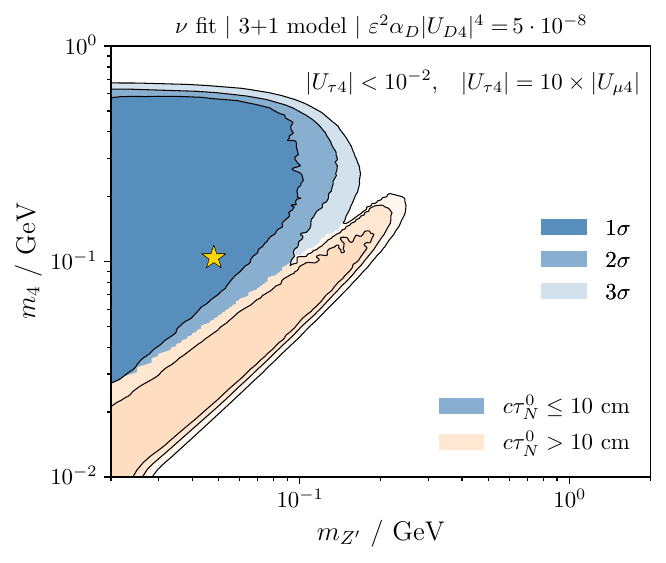}
    \caption{
    The MiniBooNE $E_\nu^{\rm CCQE-reco}$ best-fit regions in the $m_4$ and $m_{Z^\prime}$ plane for the 3+1 model without (left panel) and with (right panel) mixing between $N_4$ and tau neutrinos.
    The different regions show the $1\sigma$, $2\sigma$, and $3\sigma$ best-fit regions.
    The coupling $|U_{\mu 4}|^2$ is profiled over with an upper bound of $|U_{\mu 4}|^2 < 10^{-4}$ (left) and $|U_{\mu 4}|^2 < 10^{-4}/10^2$ (right).
    The dark coupling and kinetic mixing are fixed, as indicated at the top of the panels.
    We divide the regions into a short-lived and a long-lived regime, where the HNLs decay rapidly, $c\tau^0_N \leq 10$~cm, and a region where dirt events and displaced vertices are expected, $c\tau^0_N > 10$~cm, where $c\tau^0_N$ is the HNL proper lifetime.
    \label{fig:3p1_general_fit}
    }
\end{figure*}

The relevant decay widths of HNLs into the respective final states are given by,
\begin{align}
    \Gamma_{N_4 \rightarrow \nu Z'} = \frac{\alpha_D |V_{\ell 4}|^2}{4} \frac{m_4^3}{m^2_{Z'}} \Bigg(1 - \frac{m^2_{Z'}}{m^2_{4}} \Bigg)^2 \Bigg(\frac{1}{2} + \frac{m^2_{Z'}}{m^2_{4}} \Bigg),
    \\ \nonumber
    \Gamma_{Z^\prime \to \ell^+\ell^-} = \frac{\alpha \varepsilon^2}{3} m_{Z^\prime} \sqrt{1- \frac{4 m_\ell^2}{m_{Z^\prime}^2}}\left(1 + \frac{2 m_\ell^2}{m_{Z^\prime}^2}\right)         
\end{align}
in the on-shell dark photon case. 
In the off-shell dark photon case, the HNLs tend to travel longer distances than in the on-shell case due to the three-body nature of the decay.
For off-shell cases, we find,
\begin{align}
    \Gamma_{N_4 \rightarrow \nu e^+ e^-} = \frac{\alpha \alpha_D \epsilon^2 |V_{\ell 4}|^2}{48 \pi} \frac{m_4^5}{m^4_{Z'}} L\Big(m^2_4 / m^2_{Z'} \Big)
    \\ \nonumber
    \Gamma_{N_5 \rightarrow N_4 e^+ e^-} = \frac{\alpha \alpha_D \epsilon^2 |V_{54}|^2}{48 \pi} \frac{m_5^5}{m^4_{Z'}} F\Big(m^2_4 / m^2_5 \Big)
\end{align}
where $L(x) = \frac{12}{x^4}\Big(x - \frac{x^2}{2} - \frac{x^3}{6} - (1-x)\log \frac{1}{1-x}\Big)$ and $F(x) = 1 + 2x  - 8x^2 + 18x^3- 18 x^5 + 8 x^6 - 2 x^7 - x^8 + 24 x^3 (1 - x + x^2) \log(x)$.
In both cases, we neglect the charged lepton masses.
In the second rate, we neglected higher order terms in $m_4/m_{Z^\prime}$.
Note that for small $\Delta$ values (large $m_4/m_5$), neglecting the lepton masses is a bad approximation, and the full width should be calculated.
In our analysis, we use the full decay rate provided in \darknews, including $Z^\prime$ and SM $Z/W$ bosons contributions.

\subsection{DarkNews simulation}

We use \darknews~\cite{Abdullahi:2022cdw} to simulate the production of heavy neutrinos by neutrino upscattering in the dirt and inside the \miniboone and SBN detectors.
The \miniboone detector is modeled as a sphere of $6.1$~m radius, filled with CH$_2$.
The dirt is modeled as a truncated cone of uniform density filled with SiO$_2$.
The cone z-axis (height) is aligned with the direction of the neutrino beam and passes through the center of the \miniboone detector. The cone minor and major radii equal $1.047$ and $10.28$~m, respectively.
The total length of the dirt cone is $474$~m, extending from the wall of the \miniboone vault to the beam absorber.
The distance between the MiniBooNE vault wall and the outer shell of the \miniboone detector is $6.35$~m. 
The beam flux is assumed not to change with respect to the azimuthal angle.
We use the FHC and RHC mode neutrino fluxes from \cite{MiniBooNE:2008hfu}, and simulate events using the $\nu_\mu$ and $\overline{\nu}_\mu$ components of the beam for both modes.
The FHC (RHC) data corresponds to $18.75\, (11.27) \times 10^{20}$ POT.

We calculate the probability of decay inside the \miniboone fiducial volume for every heavy neutrino produced to reweight the events accordingly.
If their travel direction misses the detector altogether, this probability vanishes.
The fiducial volume is defined as a sphere of $5$~m radius at the center of the detector.
To model reconstruction effects in the detector, we smear the electron-positron pairs using a Gaussian with the true energy of the electrons or positrons as the mean and a $\sigma$ equals to $0.12 \times \sqrt{E_e^{\mathrm{true}}} + 0.01$ GeV~\cite{Patterson:2009ki}.

\begin{figure*}[t]
    \centering
    \includegraphics[width=0.49\textwidth]{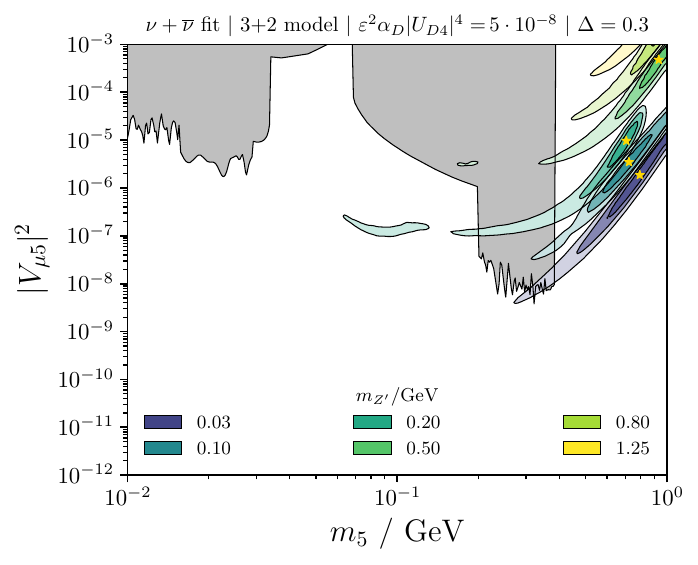}
    \includegraphics[width=0.49\textwidth]{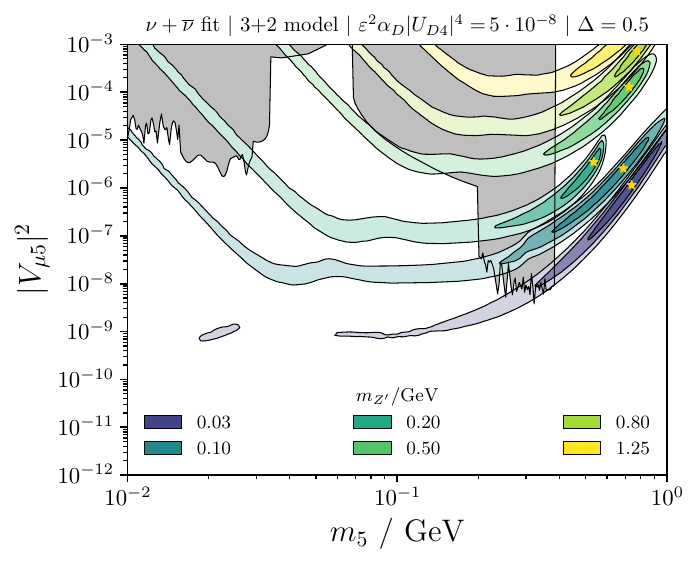}
    \\
    \includegraphics[width=0.49\textwidth]{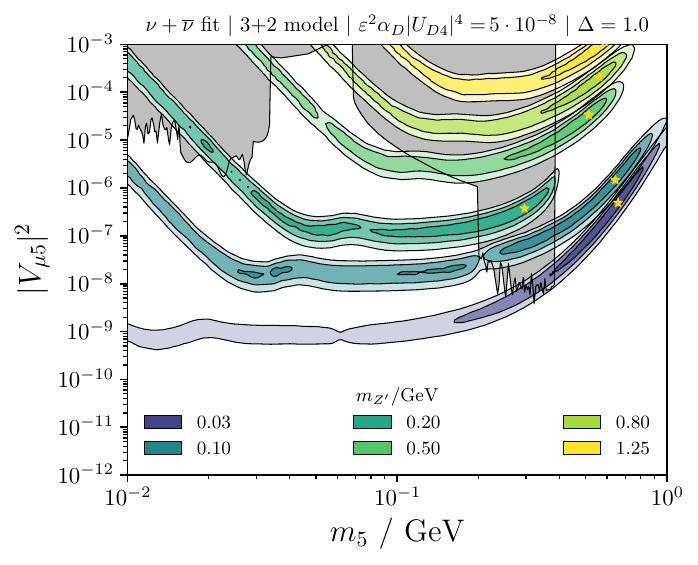}
    \includegraphics[width=0.49\textwidth]{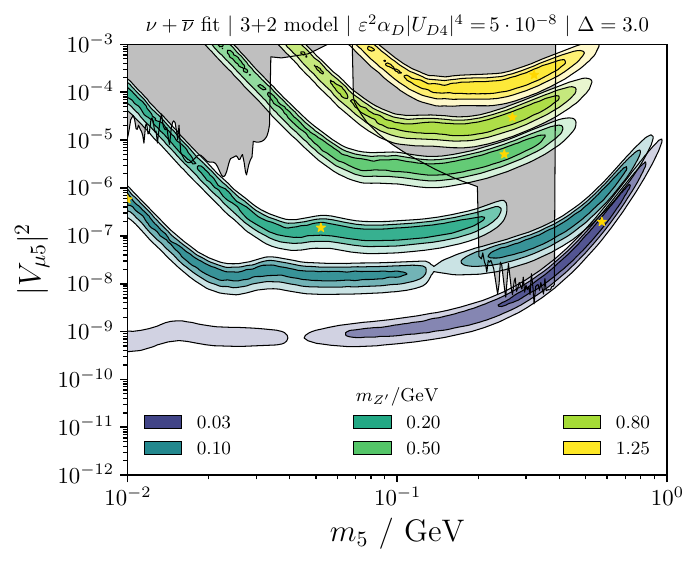}
   \caption{ 
   The MiniBooNE $E_\nu^{\rm CCQE-reco}$ best-fit regions in the $|V_{\mu5}|^2$ and $m_5$ plane for the 3+2 model.
   Each panel corresponds to a different value of $\Delta$, as indicated at the top of the panel.
   Each color represents a given fixed value of $m_{Z^\prime}$, and the different shading corresponds to the $1\sigma$, $2\sigma$, and $3\sigma$ CL regions (2 d.o.f.).
   The $1\sigma$ regions in the low-mass and large-coupling regime appear disconnected due to an interpolation artifact.
   This is due to the strong dependence of the event rate on the HNL lifetime when the signal is dominated by dirt upscattering.
   Model-independent limits on heavy neutrinos exclude the shadowed region~\cite{Fernandez-Martinez:2023phj}.
   Other constraints from neutrino scattering and meson decays are not shown.
    \label{fig:3p2_all_mzs}
    }
\end{figure*}

\begin{figure*}[t]
    \centering
    \includegraphics[width=0.49\textwidth]{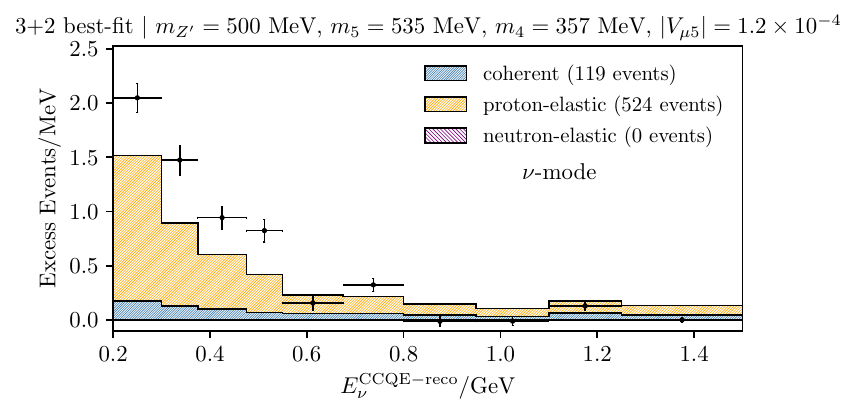}
    \includegraphics[width=0.49\textwidth]{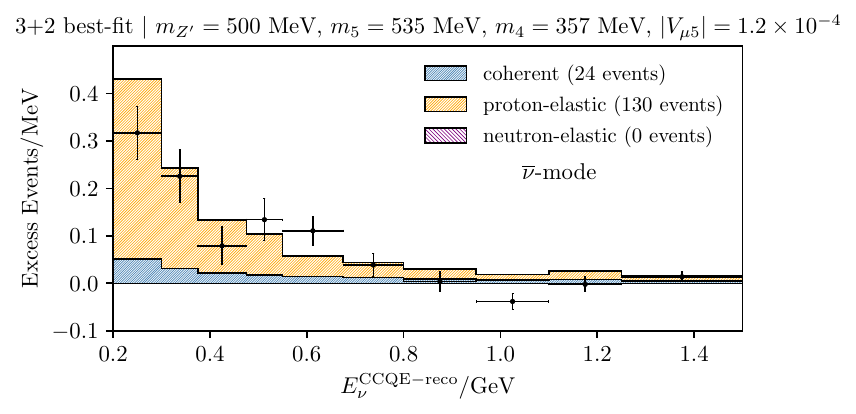}
    \caption{
    The excess of events as a function of the reconstructed neutrino energy $E_\nu^{\rm CCQE-reco}$ at MiniBooNE for the FHC ($\nu$) mode (left panel) and the RHC ($\overline{\nu}$) mode (right panel). 
    The prediction of the best-fit point in the 3+2-model ($m_{Z^\prime} = 100$~MeV, $m_5 = 615$~MeV, $\Delta = 0.3$, $|V_{\mu 4}|^2 = |V_{\mu 5}|^2 = 1.1\times 10^{-6}$, $\varepsilon = 8 \times 10^{-4}$) is shown as different shades in the histogram separating events from coherent neutrino-nucleus scattering, proton-elastic, and neutron-elastic neutrino scattering.
    \label{fig:3plus2Dist}
    }
\end{figure*}

\begin{figure*}[t]
    \centering
    \includegraphics[width=0.49\textwidth]{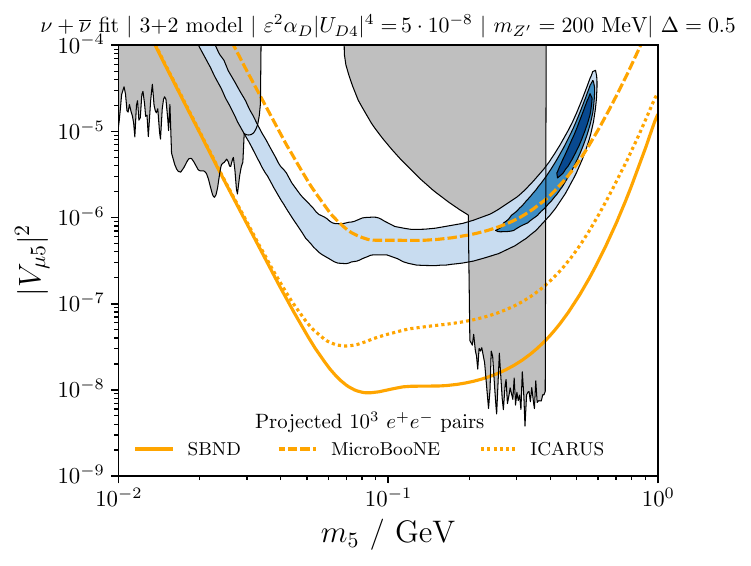}
    \includegraphics[width=0.49\textwidth]{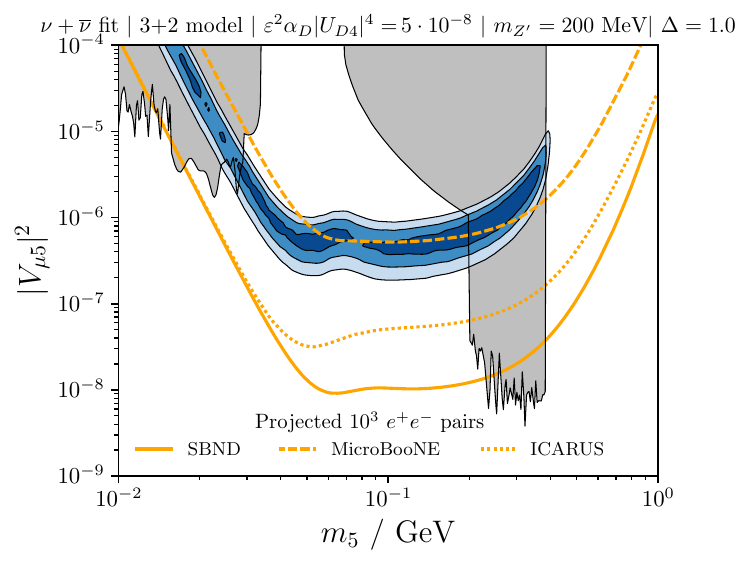}
    \caption{
    Same as \cref{fig:3p2_all_mzs} but for $\Delta = 0.5$ (left panel) and $\Delta = 1$ (right panel).
    The mediator mass has been fixed to $m_{Z^\prime} = 200$~MeV.
    The orange lines indicate where the \microboone (dashed), Icarus (dotted), and SBND (solid) detectors can expect $10^{3}$ $e^+e^-$ pairs from $N_5$ decays to be produced inside their respective fiducial volumes (before any reconstruction efficiencies).
    \label{fig:3p2_sbn}
    }
\end{figure*}

Our event selection is detailed in \cref{app:selection}.
To mimic the LEE signature, the \epluseminus must be reconstructed as a single shower.
Overlapping and asymmetric events are interpreted as single showers with a ``visible" four-momentum $p_{e^+} + p_{e^-}$.
We adopt the same criterion for overlapping and energy-asymmetry as the $\pi^0$ study of Ref.~\cite{Kelly:2022uaa}.
With the visible energy $E_{\rm vis}$ of the misreconstructed single shower and its angle $\theta_{\rm beam}$ with respect to the neutrino beam, we can calculate $E_\nu^{\rm CCQE-reco}$, defined as the reconstructed neutrino energy under the hypothesis of CCQE scattering,
\begin{equation}
    E_\nu^{\rm CCQE-reco} = \frac{1}{2} \frac{2m_n E_{\rm vis} - (m_n^2 + m_e^2 - m_p^2)}{m_n - E_{\rm vis} + p_{\rm vis}\cos \theta_{\rm beam}}
\end{equation}
where $m_p$, $m_n$, $m_e$ stand for the proton, neutron, and electron mass, and $p_{\rm vis} = \sqrt{E_{\rm vis}^2 - m_e^2}$.
As a last step, we multiply our final efficiencies by the official single-photon reconstruction efficiency taken from~\cite{MiniBooNE:2012maf}, shown in \cref{fig:miniboone_effs}.
This includes the fiducialization of events, requiring $R<500$~cm, with a $55\%$ efficiency.
Since we have already performed the fiducialization in our own simulation, we divide the official efficiencies by this number.

A few examples of our reconstruction efficiencies are shown in \cref{fig:miniboone_effs_3p1,fig:miniboone_effs_3p2} for different values of the parent HNL and dark photon masses in the 3+1 and 3+2 models.
These include only reconstruction effects but no geometrical acceptance.
The transition between on and off-shell decays is visible in the plots as a sharp drop in efficiency. 
For on-shell decays, the efficiencies tend to be smaller at large dark photon masses due to the larger separation angles between the \epluseminus pairs for less energetic dark photons.

We also show the total number of $e^+e^-$ pairs expected at the SBN detectors. 
For that, we use the \darknews implementation of the MicroBooNE, SBND, and ICARUS detectors and the corresponding dirt volumes upstream.
Our projections for $6.8\times 10^{20}$ POT at MicroBooNE and $15.6\times 10^{20}$ POT at SBND and ICARUS~\cite{SBNpots}.

\subsection{Statistical procedure}

We calculate the \miniboone $\chi^2$ based on the distribution of $E_\nu^{\rm CCQE-reco}$ using the covariance matrices provided by the MiniBooNE collaboration in~\cite{MiniBooNE:2021bgc}. 
The $\chi^2$ surfaces are obtained by repeating this calculation across two-dimensional grids of model parameters.
We do this in two ways:
i) for plots of coupling versus mass, we fix all parameters that do not appear in the $x$ and $y$ axes and compute the $\chi^2$ in the $\nu$ and $\overline{\nu}$ mode, while ii) for plots of mass versus mass, such as $m_4$ versus $m_{Z^\prime}$, we fix all parameters except for one mixing angle which we profile over: $U_{\mu 4}$ in the 3+1 model and $U_{\mu 4} = U_{\mu 5}$ in the 3+2 model.
This allows us to show the mass parameters that best fit the shape of the LEE, independently of the coupling needed. 
In this case, we compute the $\chi^2$ using $\nu$ mode data only. 
As in Ref.~\cite{MiniBooNE:2012maf}, when quoting the $\chi^2$ probability $p_{\rm val}$ of our best-fit points, we use $6.8$, $6.9$, and $15.6$ degrees of freedom for the neutrino mode, antineutrino mode, and their combination, respectively. This assumes two independent fit parameters, typically $|V_{\mu 4/5}|^2$ and $m_{4/5}$.

In more detail, we vary $m_4$ and $m_{Z\prime}$ for the 3+1 model, while in the 3+2 one we vary $m_5$ and $\Delta = (m_5 - m_4)/m_4$ while fixing $m_{Z\prime}$ or vary $m_5$ and $m_{Z\prime}$ while fixing $m_{4}$.
In both cases, we profiled over a mixing parameter, setting a maximum value of $10^{-2}$ for $|U_{\mu 4}|$ (3+1) or for $|U_{\mu 4}|$ and $|U_{\mu 5}|$ (3+2). 
For the remaining parameters, we use the following default values: $g_D = 2$, $U_{D5} = U_{D4} = 1/\sqrt{2}$, and $\varepsilon = 8 \times 10^{-4}$ ($\varepsilon = 10^{-2}$) for 3+1 (3+2). 
With our assumptions, the couplings of the dark photon to the HNLs always appear in the combination $\alpha_D^2 |U_{D4}|^4 \varepsilon^2$.
Therefore, for different values of each one of these parameters, as long as $|U_{D4}| = |U_{D5}|$, the best-fit regions can be trivially rescaled.

\subsection{Results}

\begin{figure*}[t]
    \centering
    \includegraphics[width=0.49\textwidth]{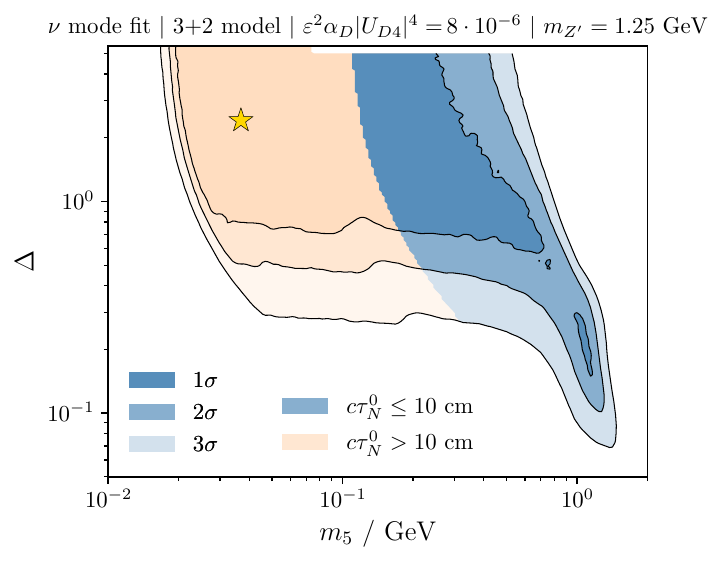}
    \includegraphics[width=0.49\textwidth]{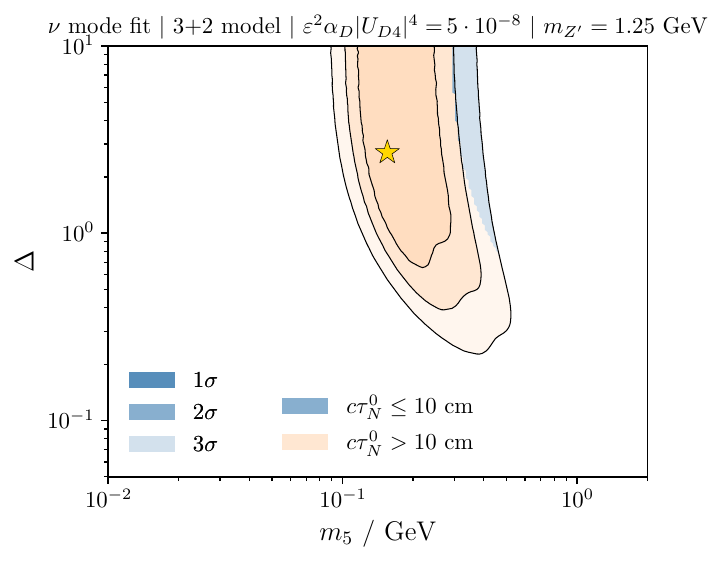}
    \includegraphics[width=0.49\textwidth]{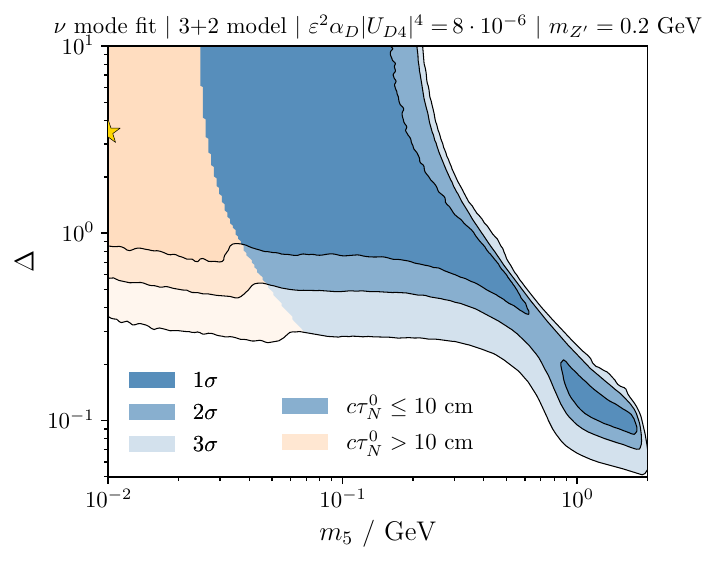}
    \includegraphics[width=0.49\textwidth]{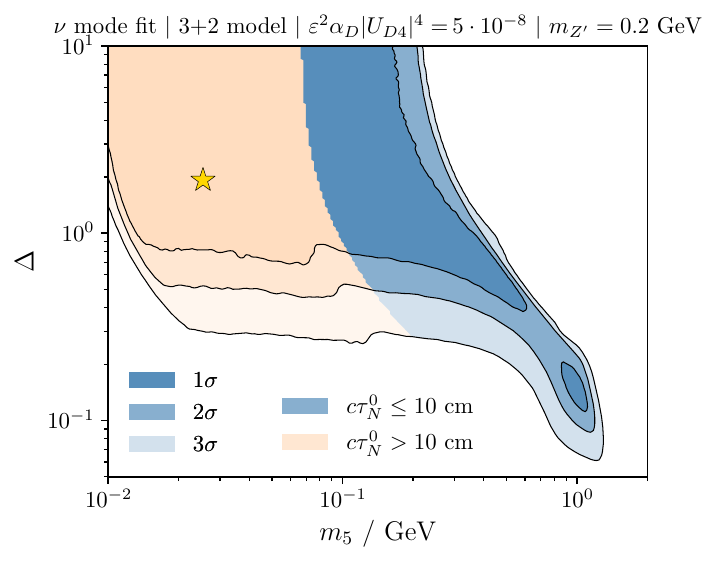}
    \caption{
    The MiniBooNE $E_\nu^{\rm CCQE-reco}$ best-fit regions in the $m_5$ and $\Delta$ plane for the 3+2 model for two choices of couplings.
    The different regions show the $1\sigma$, $2\sigma$, and $3\sigma$ best-fit regions.
    The couplings $|U_{\mu 4}|^2=|U_{\mu 5}|^2$ are profiled over with an upper bound of $|U_{\mu 4}|^2 < 10^{-2}$, which can be tightened by choosing larger dark couplings and kinetic mixing.
    All the plots consider $m_{Z\prime} = 1.25 \text{ GeV}$ and couplings as indicated at the top of each panel.   
    We divide the regions into a short-lived and a long-lived regime, where the HNLs decay rapidly, $c\tau^0_N \leq 10$~cm, and a region where dirt events and displaced vertices are expected, $c\tau^0_N > 10$~cm, where $c\tau^0_N$ is the HNL proper lifetime.
    \label{fig:3p2_general_fit_delta}
    }
\end{figure*}

\begin{figure*}[t]
    \includegraphics[width=0.95\textwidth]{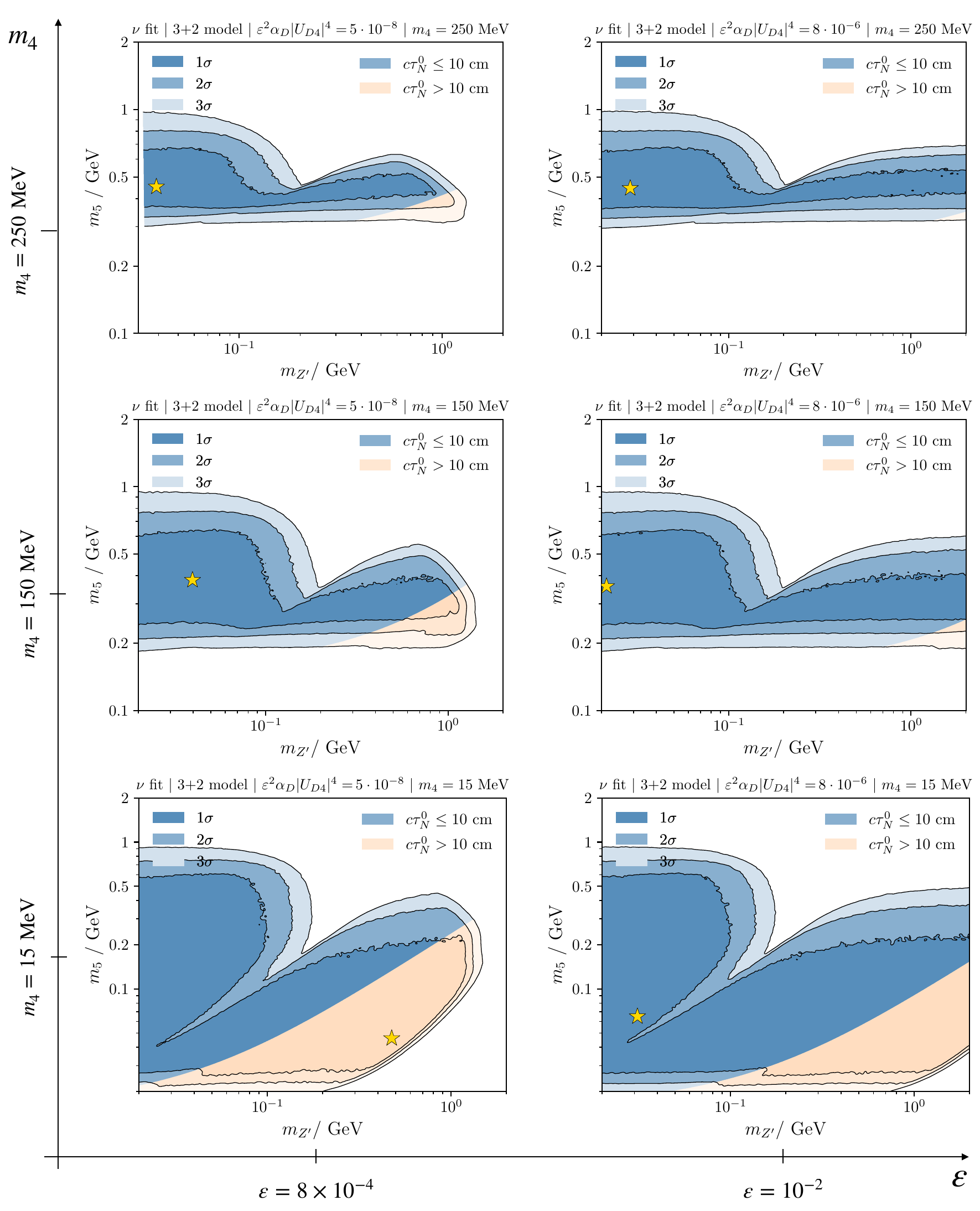}
    \caption{
    The MiniBooNE $E_\nu^{\rm CCQE-reco}$ best-fit regions in the $\Delta$ versus $m_5$ plane for the 3+2 model with $m_{Z\prime} = 1.25$~GeV.
    The different regions show the $1\sigma$, $2\sigma$, and $3\sigma$ CL regions (2 d.o.f.).
    From the bottom to the upper rows, we set $m_4$ equal to $15$, $150$, and $250$ MeV, respectively. 
    In the left and right columns we set $\epsilon$ equal to $8\times 10^{-4}$ and $10^{-2}$, respectively.
    The couplings $|U_{\mu 4}|^2 = |U_{\mu 5}|^2$ are profiled over with an upper bound of $|U_{\mu 4}|^2 < 10^{-4}$.
    We divide the regions into a short-lived and a long-lived regime, where the HNLs decay rapidly, $c\tau^0_N \leq 10$~cm, and a region where dirt events and displaced vertices are expected, $c\tau^0_N > 10$~cm, where $c\tau^0_N$ is the HNL proper lifetime.
    \label{fig:3p2_general_fit_m4}}
\end{figure*}

\subsubsection{3+1 scenario}

\cref{fig:3p1_all_mzs} shows the regions of preference in the combined $\nu$ and $\overline{\nu}$ mode \miniboone fit for different values of $m_4$ and $|V_{\mu 4}|^2$.
Both plots set $\varepsilon^2 \alpha_D |U_{D4}|^4 = 5 \times 10^{-8}$ in our benchmark parameters. 
On the left panel, $|U_{\tau 4}| = 0 $, while in the right panel, $|U_{\tau 4}| = 10 |U_{\mu 4}|$.
The latter helps decrease the HNL lifetime as the coupling $|V_{\ell 4}|^2$ that controls the decay rate is now larger than $|V_{\mu 4}|^2$, which controls the upscattering cross section.
Where a discontinuity is present, it can be attributed to the change of on-shell to off-shell dark photon regimes.

High values of $m_4$ need higher couplings to produce enough events due to the high threshold for HNL production. 
On the other hand, when $m_4 < m_{Z^\prime}$ and the dark photon is off-shell in the HNL decay, large couplings are required due to the long decay length of the HNLs. 
In the off-shell regime, most of the signal stems from HNLs produced in the dirt upstream of the detector.
This regime typically produces less energetic \epluseminus pairs than the prompt-decay regime, as the probability of decay inside the \miniboone tank is larger for slower HNLs.
Furthermore, heavy dark photons tend to produce less forward \epluseminus pairs, further reducing their energy. 

\Cref{fig:3p1_sbn} shows the same regions of preference of the left panel of \cref{fig:3p1_all_mzs}, but exclusively for $m_{Z^\prime} = 30$ MeV.
It is then compared with the predictions for the event rates at the three SBN detectors: \microboone, Icarus, and SBND.
The orange curves in different dash styles correspond to the parameters where $10^{3}$ \epluseminus pairs would be produced inside the fiducial volume of each detector, considering upscattering in the upstream dirt and inside the detector.
This number is presented without any reconstruction or selection efficiencies.
The real sensitivity of each experiment will depend on the backgrounds and selection strategy adopted.
We leave a detailed study to future work and experimental collaborations.

The best fit of the 3+1 model for $|U_{\tau 4}| = 0$ is given by $|V_{\mu 4}|^2 = 4.8 \times 10^{-7}$, $m_4 = 20$~MeV, and $m_{Z^\prime} = 30$~MeV.
We show the prediction for the reconstructed neutrino energy spectrum for this best fit in \cref{fig:3plus1Dist} for neutrino and antineutrino modes.
Most events come from coherent scattering on nuclei, although proton-elastic interactions are responsible for about $10\%$ of the total signal.
It can also be seen that the rate in neutrino mode is lower than the data, while in antineutrino mode, it is larger than the data.
This is due to the vector nature of the dark photon couplings to quarks.
It predicts that the neutrino and antineutrino cross sections on nuclei are the same.
This is in mild tension with the MiniBooNE observation that the antineutrino excess is comparatively smaller than the neutrino mode one.
The tension is also visible in \cref{tab:BPs}.
The goodness-of-fit is significantly better for neutrino than antineutrino data, so combining the two provides an overall smaller $\chi^2$ probability.
Since the excess in neutrino mode is more significant, this mode drives the fit and leads to an overprediction for the number of events in antineutrino mode.
In the 3+1 model, the agreement with the data is typically better for long-lived HNL regimes, but it is still at a $\gtrsim 2\sigma$ tension with the combination of the neutrino and antineutrino excess.

Finally, we emphasize that the reconstructed neutrino energy fit has a subleading dependence on the angular distribution and cannot adequately quantify the agreement with the \miniboone angular spectrum.
A quantitative estimate of this tension, taking into account the correlations between angular bins, is not possible outside the collaboration.
Nevertheless, in 3+1 models, we find that virtually all points predict no events in the $\cos\theta \lesssim 0.9$ region.
This is due to two main effects:
i) for off-shell dark photons, the signal is dominated by dirt production of HNL, so the geometrical acceptance of the detector biases the angular spectrum to be more forward, and
i) in the on-shell dark photon regime, the signal is dominated by HNLs produced inside the detector, but the upscattering cross section is predominantly coherent, leading to forward HNL production due to the low momentum exchange with the nucleus.
In the next section, we explore $3+2$ scenarios, where HNLs can be short-lived even for off-shell mediators and, therefore, lead to slightly less forward signatures.

Finally, we also explore the effect of simultaneously varying the HNL and mediator masses in \cref{fig:3p1_general_fit}.
The darkest to lightest filled regions show the one, two, and three $\sigma$ regions of preference, respectively, for the 3+1 model in the plane of $m_4$ versus $m_{Z^\prime}$.
The parameter $|U_{\mu 4}|$ is profiled over with a hard upper bound of $|U_{\mu 4}|^2 < 10^{-4}$.
The plot on the right includes one extra mixing, $|U_{\tau 4}|$, which is set to $10 \times |U_{\mu 4}|$. 
In this case, the upper bound is set on the tau mixing parameter, $|U_{\tau 4}|^2 < 10^{-4}$, explaining why the long-lived region is significantly smaller than in the previous case.
The plots are divided in two: a short-lived HNL region ($c\tau^0_N \leq 10$~cm) shown in shades of blue and a long-lived HNL region ($c\tau^0_N > 10$~cm) shown in shades of beige, where $c\tau^0_N$ as the decay length of the parent HNL in its rest frame. 
The two regions correspond to the on-shell and off-shell dark photon cases, respectively.
Although we color the regions differently, the fit is done over the entire plane, showing that both on and off-shell regimes can lead to an acceptable fit to the \miniboone neutrino energy spectrum.

\subsubsection{3+2 scenario}

Similarly to the 3+1 scenario, we show the resulting regions of preference in the coupling $|V_{\mu 5}|^2$ versus mass $m_5$ plane in \cref{fig:3p2_all_mzs}.
Each panel corresponds to a fixed value of $\Delta \equiv m_5/m_4 - 1 = 0.3$, $0.5$, $1$ and $3$, going from mildly degenerate to mildly hierarchical masses.
In turn, for each panel, we show the resulting preference regions for different fixed values of $m_{Z\prime}$, corresponding to $0.03$, $0.06$, $0.1$, $0.2$, $0.5$, $0.8$ and $1.25$ GeV.
In these plots, higher values of $\Delta$ allow $N_5$ to be shorter-lived and for it to release more energy into the \epluseminus system.
This improves the fit to the excess, up until $\Delta$ becomes too large and the decays resemble the 3+1 scenario. 
The shape of the regions of preference can be understood by noticing that the downward trend at low $m_5$ values is caused by long lifetime of $N_5$, which is typically produced in the dirt and decays more often inside MiniBooNE for larger $m_5$ due to the steep dependence of the decay rate, $\Gamma \propto m_5^5$. 
Once $N_5$ becomes too short-lived to propagate the distance from the dirt to the detector, the regions of preference turn over to then stay flat up to the point where the $N_5$ production energy threshold is too large.
The transition from off-shell to on-shell regions is less noticeable in these plots since HNLs can be short-lived even for off-shell mediators.
Nevertheless, it can still be observed as a gap between the two $1\sigma$ and $2\sigma$ closed regions.

In \cref{fig:3plus2Dist}, we can see the distribution of the reconstructed neutrino energy of the best-fit point in the 3+2 model, namely $m_{Z^\prime} = 100$~MeV, $m_5 = 615$~MeV, $\Delta = 0.3$, $|V_{\mu 4}|^2 = |V_{\mu 5}|^2 = 1.1\times 10^{-6}$, and $\varepsilon = 8 \times 10^{-4}$.
Once more, the signal prediction for the neutrino mode undershoots the excess, while the antineutrino signal overshoots it.
Overall, \cref{tab:BPs} shows that 3+2 models better fit the MiniBooNE energy spectrum, especially for smaller $\Delta$ values where the HNL decays produce lower-energy \epluseminus pairs.

Our projections for event rates at the SBN program are shown in \cref{fig:3p2_sbn} for two cases: $m_{Z^\prime} = 30$ and $200$ MeV, both for $\Delta = 1$.
As before, the contours show the parameter space where we project that $10^{3}$ pairs of $e^+ e^-$ will be produced inside the SBND, \microboone, and ICARUS fiducial volumes. 
As in the 3+1 model, when the signal is dominated by HNL production in the dirt, ICARUS is expected to see as many events as SBND.
The smaller neutrino flux at the detector's location is compensated by the greater extent of dirt upstream.
It is easy to see that the SBND detector could observe as many as $10^{5}$ \epluseminus pairs with its full exposure.

Turning to mass versus mass plots, \cref{fig:3p2_general_fit_delta} shows a fit to neutrino mode data in the plane of $\Delta$ versus $m_5$.
We fix $\varepsilon^2 \alpha_D |U_{D5}|^4 = 8 \times 10^{-6}$ and $5 \times 10^{-8}$, and impose an upper bound on $|U_{\mu 4}|^2=|U_{\mu 5}|^2 < 10^{-4}$. 
The plot exhibits two different shaded colors. 
The beige color corresponds to $N_5$ proper decay lengths greater than $10$ cm, while the blue one is smaller than $10$ cm. 
The fit is mostly insensitive to the values of $\Delta$ once larger than $\Delta \gtrsim 1$.
In that case, we recover a scenario similar to the 3+1 model, albeit with a larger coupling on the HNL decay process.
Small values of $\Delta$ are disfavored due to the longer lifetime and the significant suppression of the energy released in $N_5 \to N_4 e^+e^-$ decays.
Nevertheless, a feature can still be observed at large $N_5$ masses, where the HNL production occurs by the highest energy neutrinos in the beam at the cost of requiring larger $|V_{\mu 5}|^2$ couplings.
We recall that when $1 < \Delta < 2$, the decay process $N_5 \rightarrow \nu N_4 N_4$ could, in principle, take place.
Similarly for $N_5 \rightarrow N_4 N_4 N_4$ when $\Delta > 2$. 
As discussed in \cref{sec:pheno_model}, we assume both of these channels to be subdominant due to the structure of the dark photon couplings, which can be easily achieved with symmetry arguments.


The second mass fit is shown in \cref{fig:3p2_general_fit_m4}.
For each panel in the vertical direction, we vary the mass of $N_4$ to $15$, $150$, and $250$ MeV, while in the horizontal direction, we consider values of $\varepsilon$ of $8 \times 10^{-4}$ and $10^{-2}$. 
We also float $|U_{\mu4 }|^2 = |U_{\mu 5}|^2$, with a hard upper bound of $10^{-4}$.
As the kinetic mixing increases, more parameter space opens up at large $m_{Z^\prime}$, where the lifetime of the HNL can still be sufficiently short to induce a signal in \miniboone. 
A similar effect happens for lighter $N_4$, as the mass splitting $\Delta$ increases and the $N_5$ decays faster. 
The change in regimes from on-shell to off-shell dark photons can be seen in all panels as a feature at intermediate $m_{Z^\prime}$ values.

In summary, the 3+2 model can provide a better fit to the LEE than the 3+1 model when $\Delta \lesssim 1$ due to the lower energy emitted in HNL decays.
It still, however, faces two main challenges: the LEE angular distribution and the tension between neutrino and antineutrino data.
While we do show the angular distributions here, we find that 3+2 models tend to predict more events outside the $\cos{\theta_{\rm beam}} > 0.9$ region, although the prediction still significantly overshoots the data on the most forward bin.
The values of kinetic mixing, $|V_{\mu 4}|$, and $|V_{\mu 5}|$ can be significantly smaller in this model, even for heavy dark photons, indicating that this scenario is less stringently constrained by other indirect searches for dark photons and HNLs.

\section{Conclusions}
\label{sec:conclusions}

We have reviewed proposed solutions to the low-energy excess of electron-like events at \miniboone, specifically in the context of new-physics models. 
Most explanations exploit the limitations in the particle-identification capabilities of the \miniboone detector as it cannot distinguish between single $e^\pm$, single $\gamma$, and collimated or energy-asymmetric \epluseminus or $\gamma\gamma$ pairs.
New-physics models that lead to an excess of $\nu_e$ and $\overline{\nu}_e$ in the beam, like the popular eV sterile neutrino oscillations, are associated with the single $e^-$ and single $e^+$ hypothesis and have already been constrained by several experiments, including \microboone, which operated in the same beam as \miniboone.
Other explanations based on single $\gamma$ final states, like upscattering to heavy neutrinos with transition magnetic moments or inverse-Primakoff scattering, are far less constrained.
The previous single photon search at \microboone is less sensitive to these explanations as it exclusively targets the radiative decays of the $\Delta(1232)$ resonance.
Unlike the resonant channel, the single-$\gamma$ dark sector models predict a dominant coherent-scattering component, which is subject to a larger background due to the absence of a hadronic vertex.
We have also discussed proposals based on dark particle decays into $e^+e^-$ and $\gamma \gamma$ pairs inside the detector. 
These include dark neutrino models, in which neutrinos interact with nuclei in the detector to produce short-lived heavy neutral leptons.
We also comment on a few less explored alternatives to upscattering, including inverse-Primakoff scattering, neutrino-induced vector boson fusion, and dark bremsstrahlung, all of which warrant further study.

After this detailed overview, we have focused on a representative model that advocates HNLs upscattering and subsequent fast decays and have performed a comprehensive fit to the LEE neutrino energy spectrum. 
We consider both cases of one HNL (3+1) and two HNLs (3+2) in the spectrum.
The 3+2 model is less constrained and can accommodate solutions to the LEE at large mediator masses, where the HNL decays $N_5 \to N_4 e^+e^-$ proceed via off-shell mediators.
In addition, the mass splitting between the parent and daughter HNLs can lead to even lower-energy events, providing a better fit to the LEE.
In 3+2 models, we find better fits to the LEE than the best fit in standard sterile-neutrino-driven oscillation hypotheses, with an overall $\chi^2$ probability of $14\%$.
The goodness-of-fit is primarily limited by the tension between the relative number of excess events in neutrino and antineutrino modes.
\miniboone is compatible with the SM expectation that neutrino cross sections are larger than antineutrino ones; in contrast, the vectorial nature of dark photon interactions predicts they are the same.
Another challenge for dark photon models is reproducing the LEE angular spectrum.
For dark neutrinos to produce events that are not fully concentrated in the region $\cos{\theta_{\rm beam}} > 0.9$, the model requires off-shell mediators and large HNL masses, $m_N \gtrsim 400$~MeV.
In addition to the uncertainties on the \epluseminus reconstruction, quantifying the agreement with the angular spectrum and its bin-to-bin correlations is not possible with publicly available information, although the prediction significantly overshoots the excess in the most-forward bin, where only $\sim 32\%$ of the total excess is concentrated.
Improvements to the angular spectrum and the neutrino-to-antineutrino ratio can be obtained in models where the upscattering is mediated by a scalar or an axial-vector coupling to quarks.
Both possibilities, however, are more severely constrained by other direct searches.

For the first time, our results allow a thorough comparison of the dark neutrino interpretation of \miniboone with other experiments' data.  
Extending existing constraints on dark neutrino models to the slices of parameter space we show here will be crucial to understanding this LEE interpretation status. 
Looking forward, the SBN program at Fermilab will be able to test the dark neutrino LEE hypothesis using the same beamline in which \miniboone operated.
As a crude estimate of the reach of each detector, we show the signal event rate expected before reconstruction efficiencies.
In the \miniboone regions of preference, the number of \epluseminus pairs produced inside the fiducial volume of \microboone can be of $\mathcal{O}(10^{3})$, while for SBND and ICARUS, this number can be well above $\mathcal{O}(10^{4})$.
Despite being located further away from the target, ICARUS can observe as many events as SBND thanks to the larger amount of dirt upstream of the detector, offering the ideal conditions to constrain the parameter space where HNLs are long-lived.
A full sensitivity study that contextualizes these event rates on top of SM backgrounds and reconstruction efficiencies in each detector is in order.

We note that our dark neutrino fit is subject to uncertainties in the treatment of the upscattering cross section.
These uncertainties are small for light mediators since the signal is dominated by the cleaner coherent neutrino-nucleus scattering channel.
However, our \darknews simulation shows that the signal can contain a significant portion of proton-elastic events for heavier mediators, indicating that larger energy transfer regimes can dominate the upscattering.
The tools available to simulate such processes lack a detailed treatment of the nuclear response in these regimes.
Future efforts with automated tools like \textsc{Achilles}~\cite{Isaacson:2022cwh} and more comprehensive implementations of dark neutrinos in tools like GENIE~\cite{Andreopoulos:2009rq} will be crucial to support searches for dark particles at next-generation neutrino experiments.

\acknowledgments
We acknowledge discussions with Mark Ross-Lonergan and Georgia Karagiorgi on the capabilities of the MicroBooNE detector and thank Kevin Kelly for providing the $\pi^0$ misidentification probability in \cref{fig:efficiency_reco_pi0}.
We also thank Pedro Machado and William Louis for their comments on the draft.
The research of M.H. is supported by the Perimeter Institute for Theoretical Physics. 
Research at Perimeter Institute is supported in part by the Government of Canada through the Department of Innovation, Science and Economic Development and by the Province of Ontario through the Ministry of Colleges and Universities.
The research of S.P. and J.H.Z. has received funding / support from the European Union’s Horizon 2020 research and innovation programme under the Marie Skłodowska-Curie grant agreement No 860881-HIDDeN. J.H.Z is supported by the National Science Centre, Poland (research grant No. 2021/42/E/ST2/00031).

\appendix

\section{Signal selection}
\label{app:selection}

The MiniBooNE signal selection focused on a $\nu_\mu \to \nu_e$ signal from oscillations.
The data released by the collaboration~\cite{MiniBooNE:2021bgc} includes data and background distributions in reconstructed quantities like visible shower energy, shower angles, and reconstructed neutrino energy under the assumption of CCQE scattering.
The collaboration also released Monte-Carlo events with true and reconstructed variables and selection efficiencies for electron and photon final states.
These data releases are best suited for studies of single-electron, positron, or photon final states.
Other models based on misidentifying multiple electromagnetic showers into one require greater care since they depend on the detector response and signal reconstruction specifics.
The two main challenges are i) obtaining realistic distributions of reconstructed quantities without a detailed detector simulation and ii) evaluating the signal efficiency without access to high-level reconstructed variables.
For instance, part of the signal selection in MiniBooNE is performed using particle identification likelihoods, which quantify the compatibility of an observed event with the hypothesis that it consists of an electron, muon, or pion final state.
Without a detector simulation, obtaining the distributions of these abstract quantities and the selection efficiency of the respective cuts in these variables is not realistic. 
As a result, phenomenological studies of models that rely on the misidentification of final states are subject to more uncertainties than those based on oscillation signatures, for example.

In the absence of dedicated studies of exotic $e^+e^-$ and $\gamma\gamma$ final states by the collaboration, we turn to simplified signal selection criteria.
As the main benchmark of this article, we focus on the dark neutrino model of\cref{sec:fit_darknus}, where the main source of uncertainty in the analysis is the probability that an $e^+e^-$ final state is selected as an electron-like event in the LEE region.
We also turn to the study in Ref.~\cite{Kelly:2022uaa}, where the authors attempted to reproduce the MiniBooNE $\pi^0$ signal selection without a complete detector simulation.
In our own fits, we implement a procedure analogous to theirs and compare the final signal efficiency obtained under different selection and reconstruction assumptions.

\begin{figure}[t]
    \centering
    \includegraphics[width=0.49\textwidth]{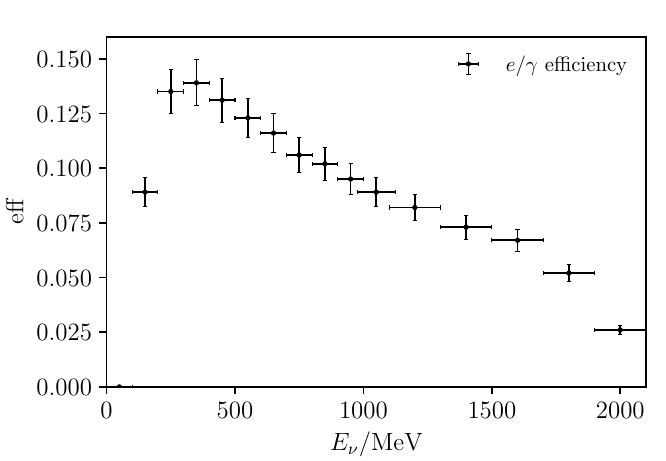}
    \caption{MiniBooNE single electron/photon efficiencies as a function of energy~\cite{MiniBooNE:2012maf} used in this work.
    \label{fig:miniboone_effs}}
\end{figure}

\begin{figure*}[t]
    \centering
    \includegraphics[width=0.49\textwidth]{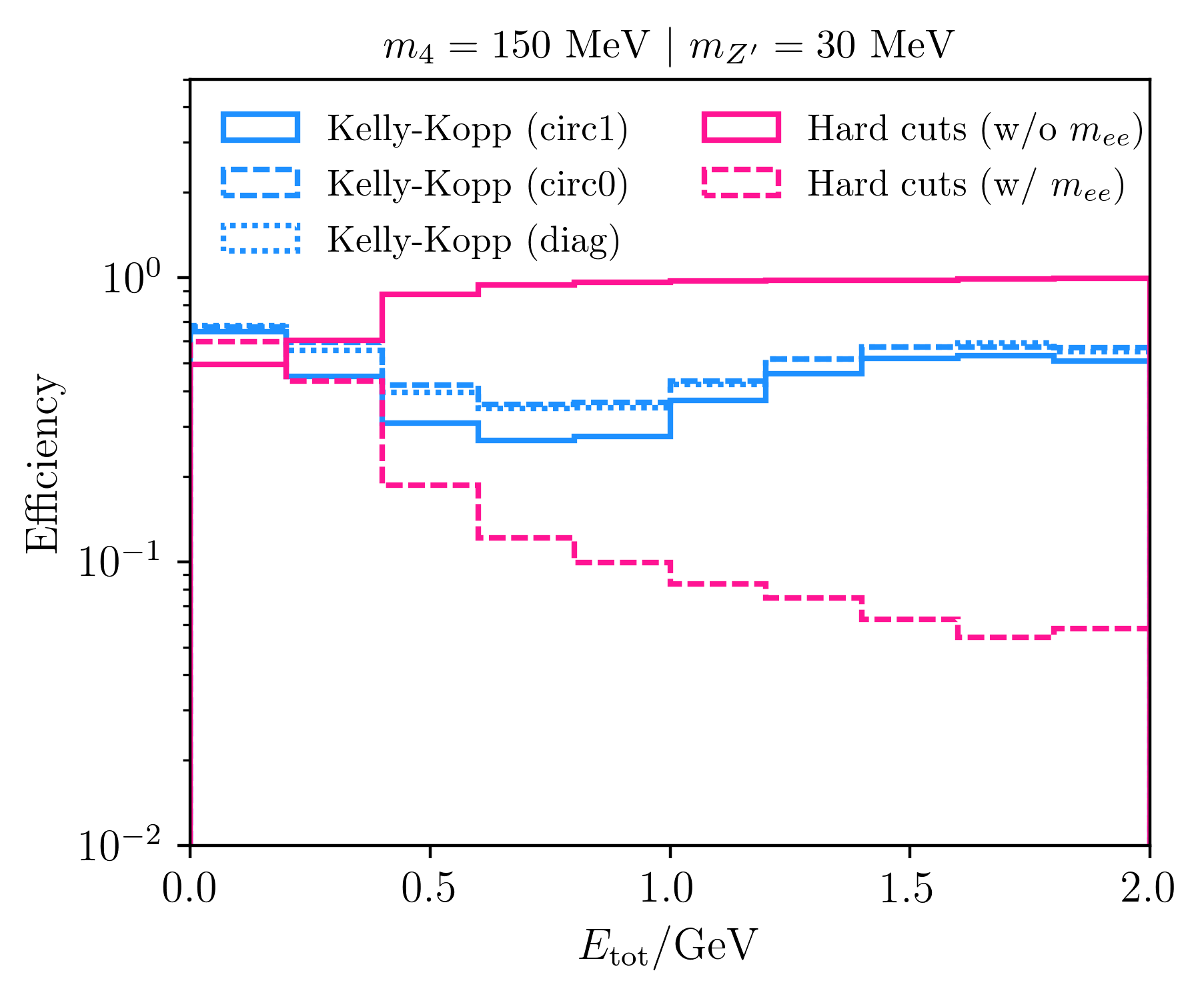}
    \includegraphics[width=0.49\textwidth]{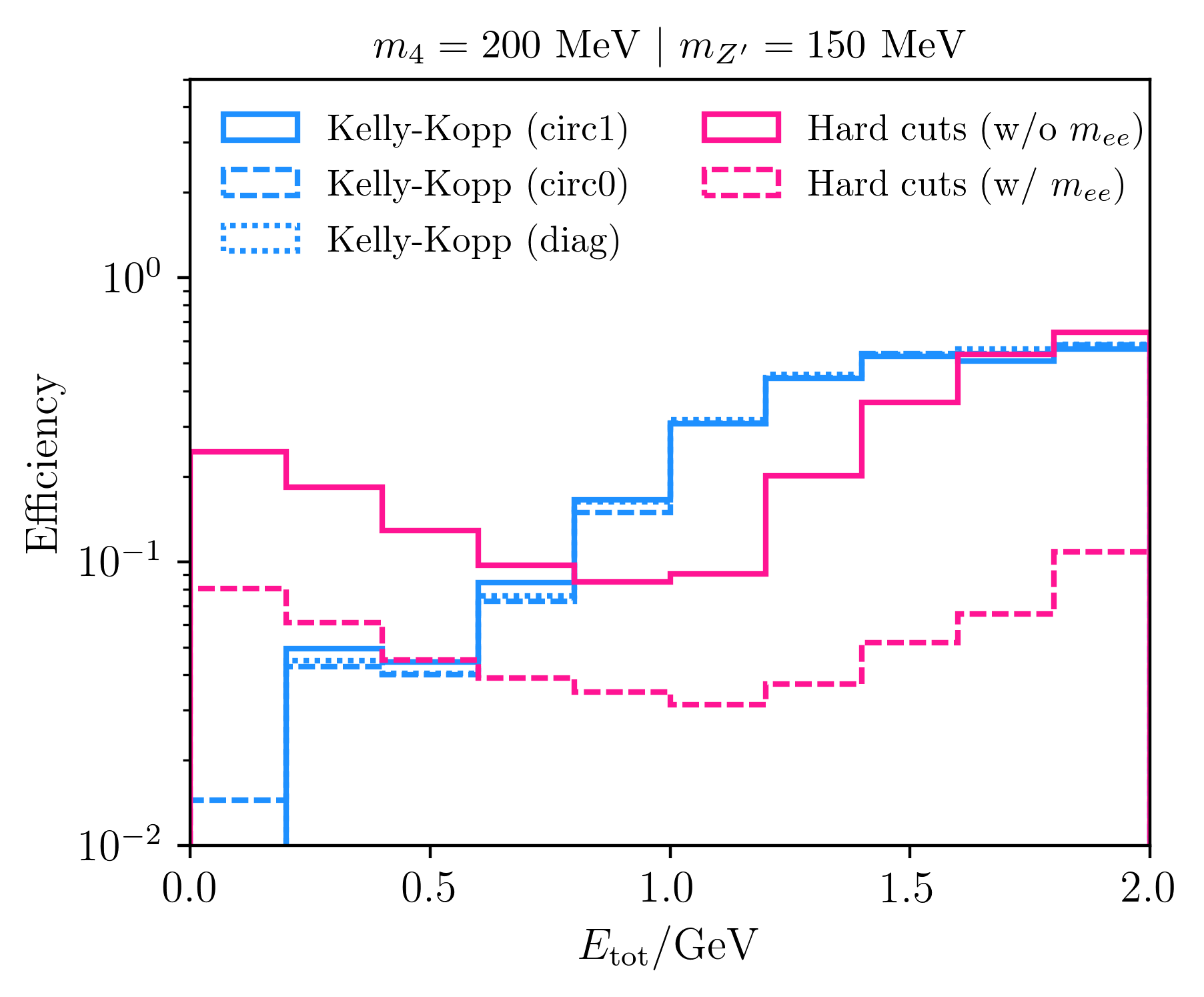}
    \includegraphics[width=0.49\textwidth]{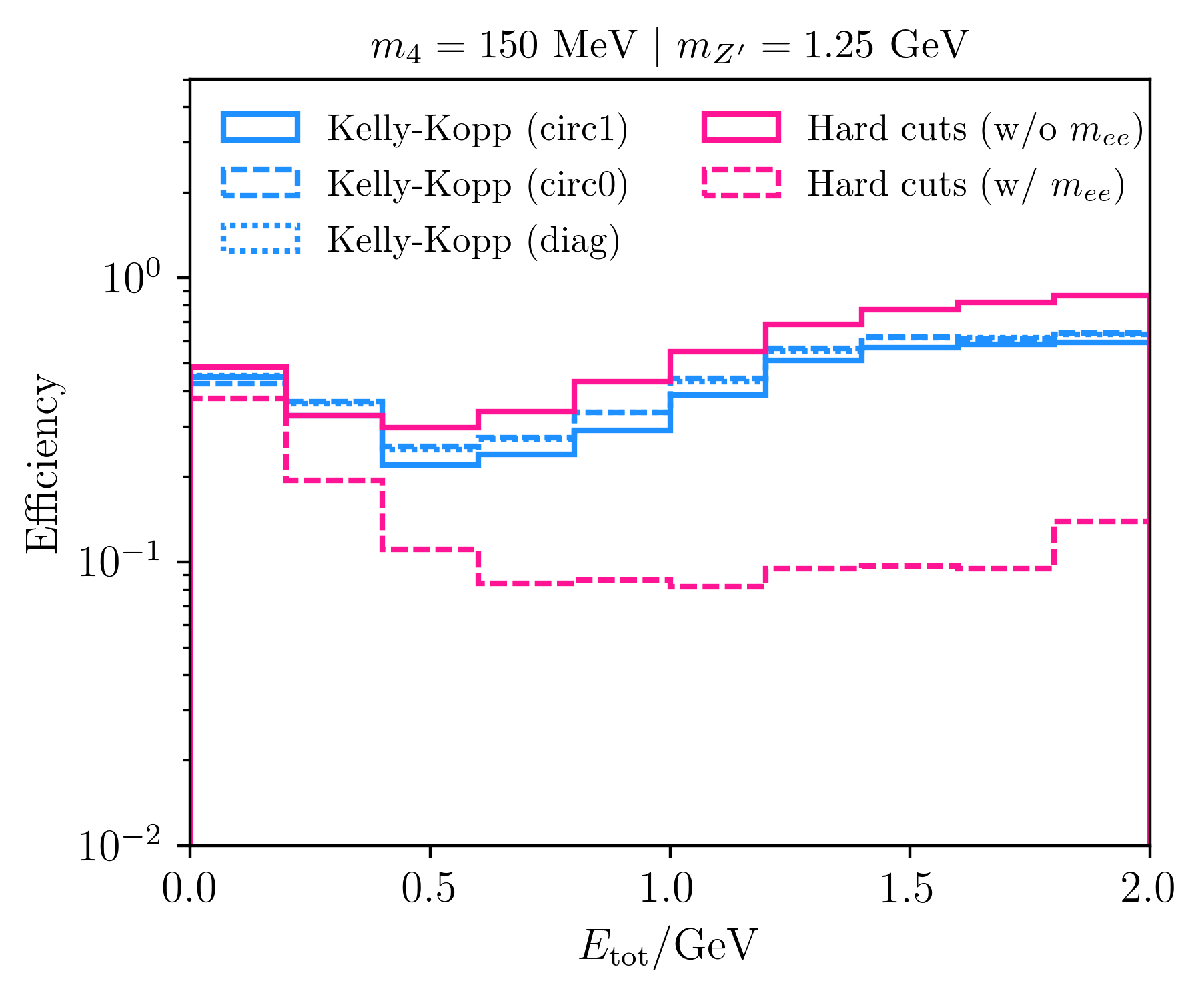}
    \includegraphics[width=0.49\textwidth]{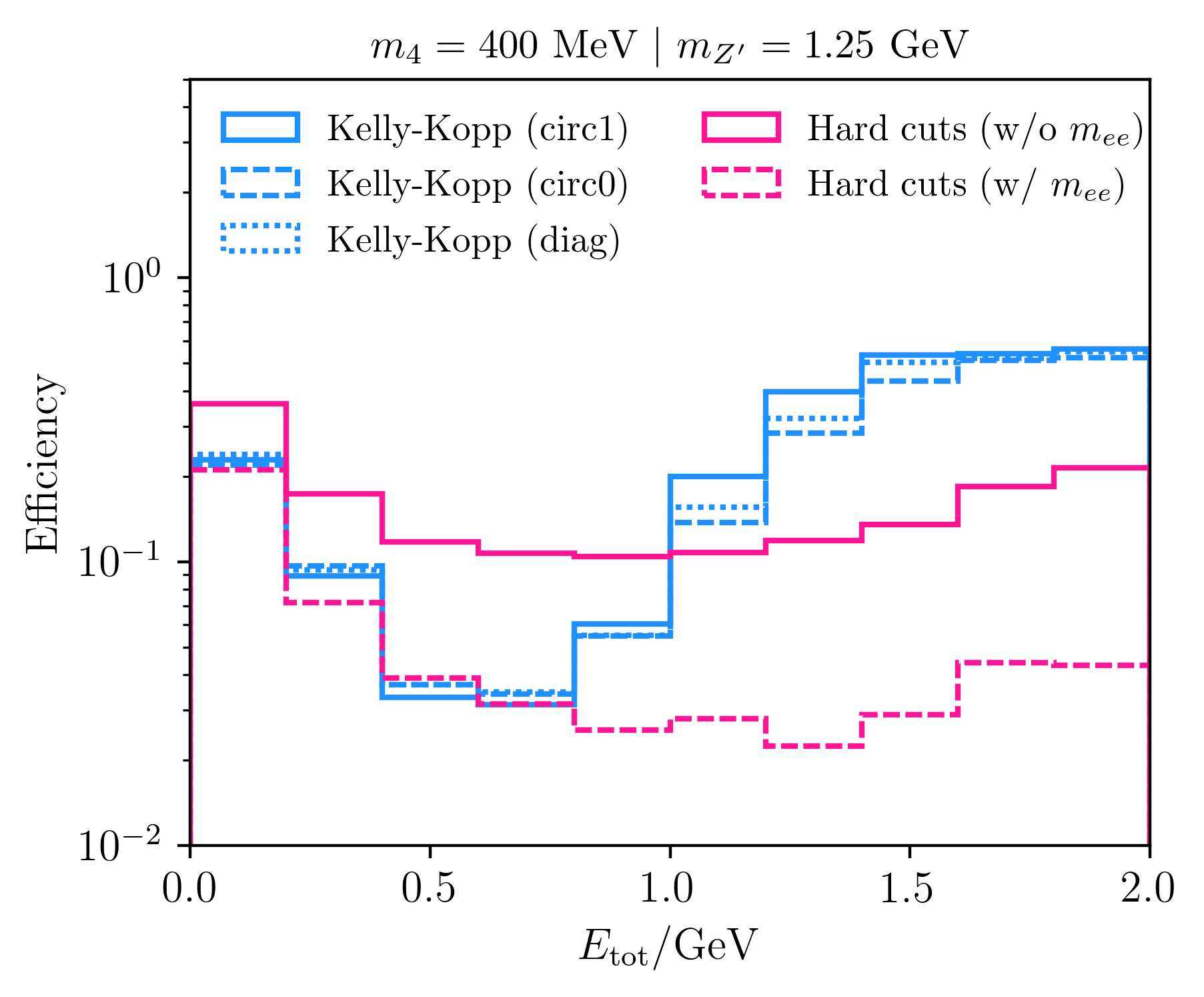}
    \caption{The final signal efficiency using different pre-selection methods. 
    The true-variable method, discussed in \cref{sec:true-var}, is shown with and without applying the cut on the \epluseminus invariant mass $m_{ee}$.
    The simplified $\pi^0$ method, discussed in \cref{sec:simple-pi0}, is referred to as Kelly-Kopp.
    We show the different methods based on the choice of the $r$ variable, as defined in \cite{Kelly:2022uaa}.
    In all cases, we generate upscattering inside the detector and enforce $N_4$ to decay promptly.
    The fits in the main text are based on the Kelly-Kopp (circ1) method.
    \label{fig:comparison_preselection}}
\end{figure*}

\begin{figure*}[ht]
    \centering
    \includegraphics[width=0.49\textwidth]{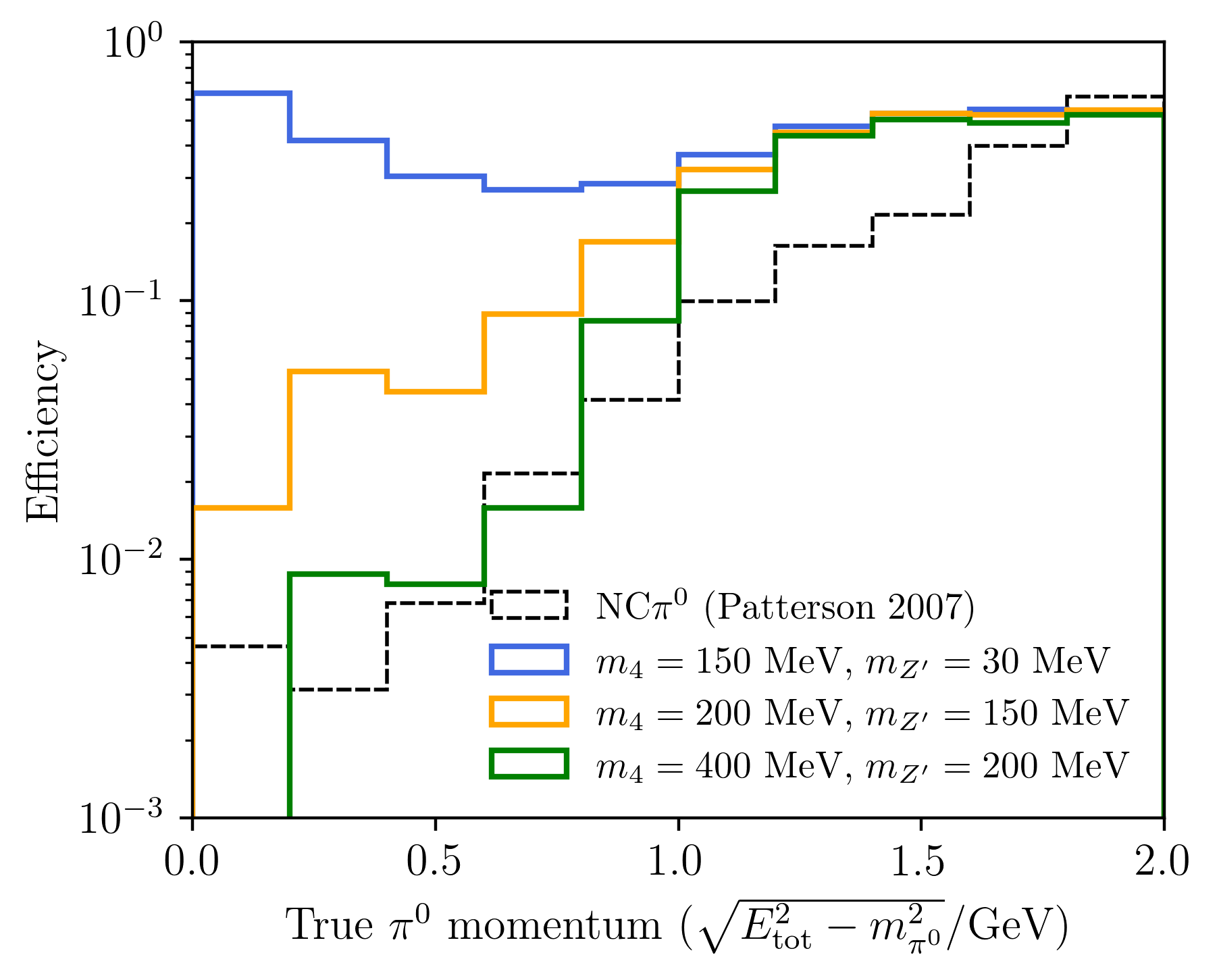}
    \includegraphics[width=0.49\textwidth]{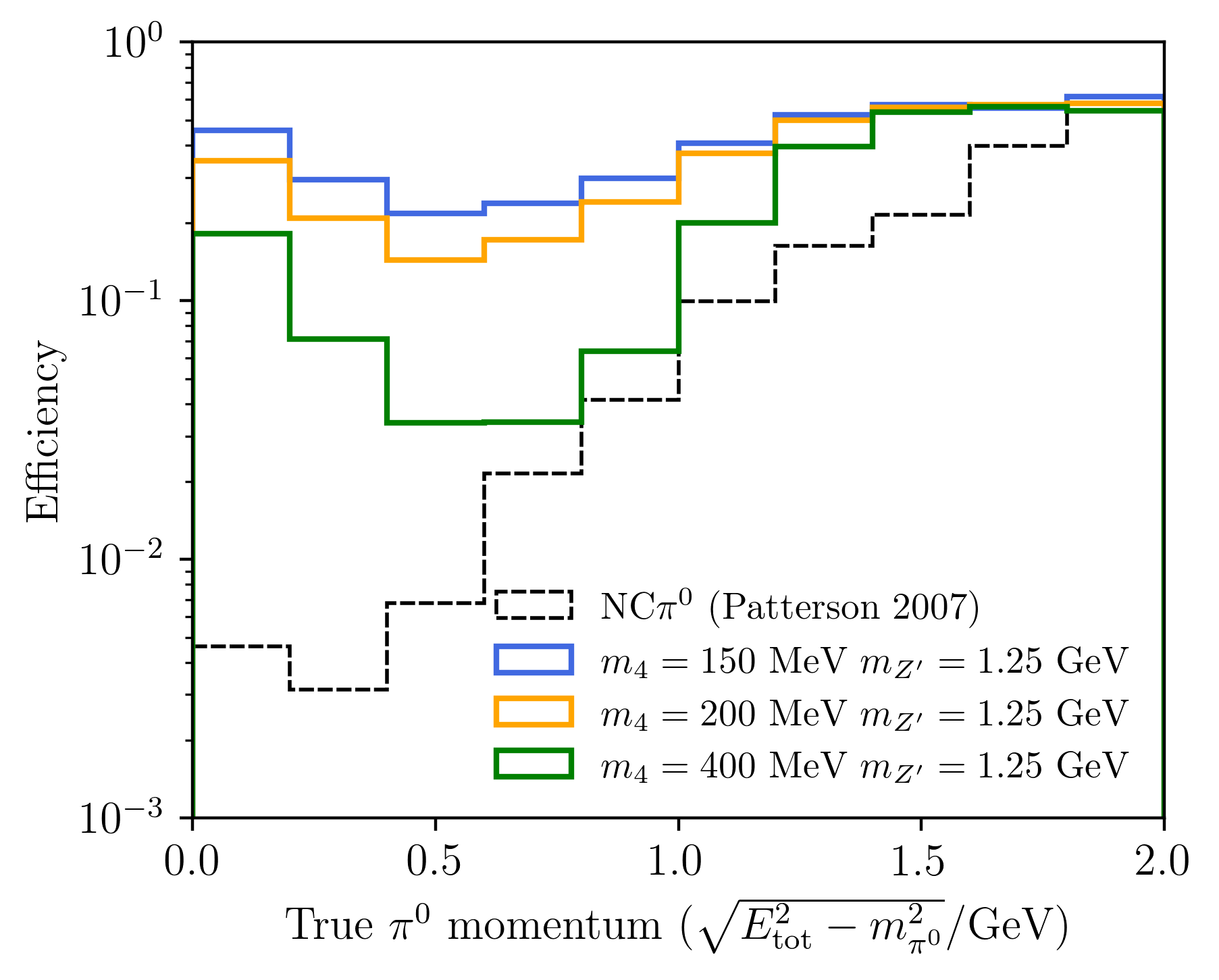}
    \caption{A comparison between the MiniBooNE signal selection efficiency for dark neutrino events and the misidentification probability for $\pi^0$ background events (from \cite{Patterson:2007zz}) in the true $\pi^0$ momentum variable.
    The latter is defined as $\sqrt{E_{\rm tot}^2 - m_{\pi^0}^2}$ for \epluseminus events.
    \label{fig:efficiency_true_pi0}}
\end{figure*}

\begin{figure*}[ht]
    \centering
    \includegraphics[width=0.49\textwidth]{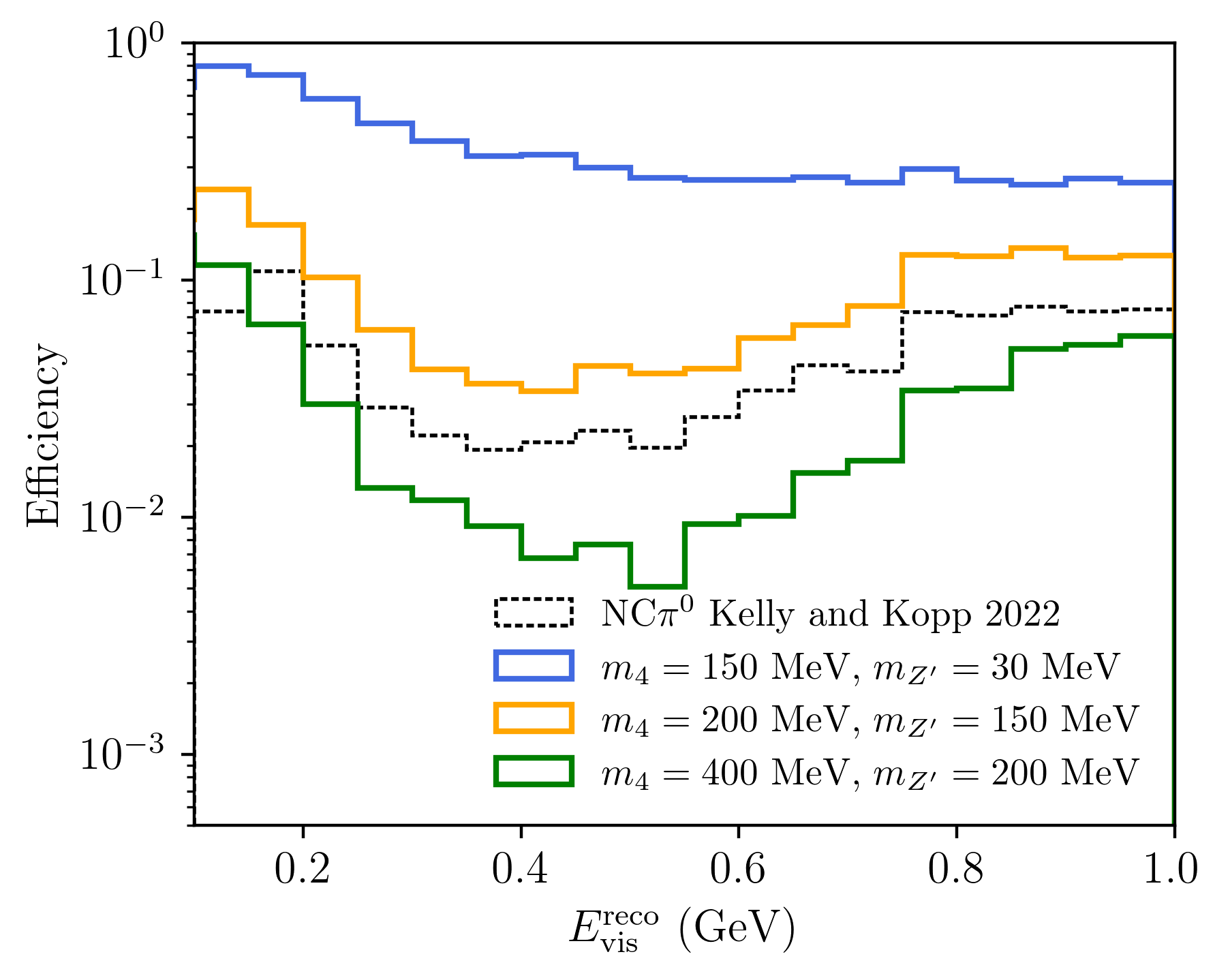}
    \includegraphics[width=0.49\textwidth]{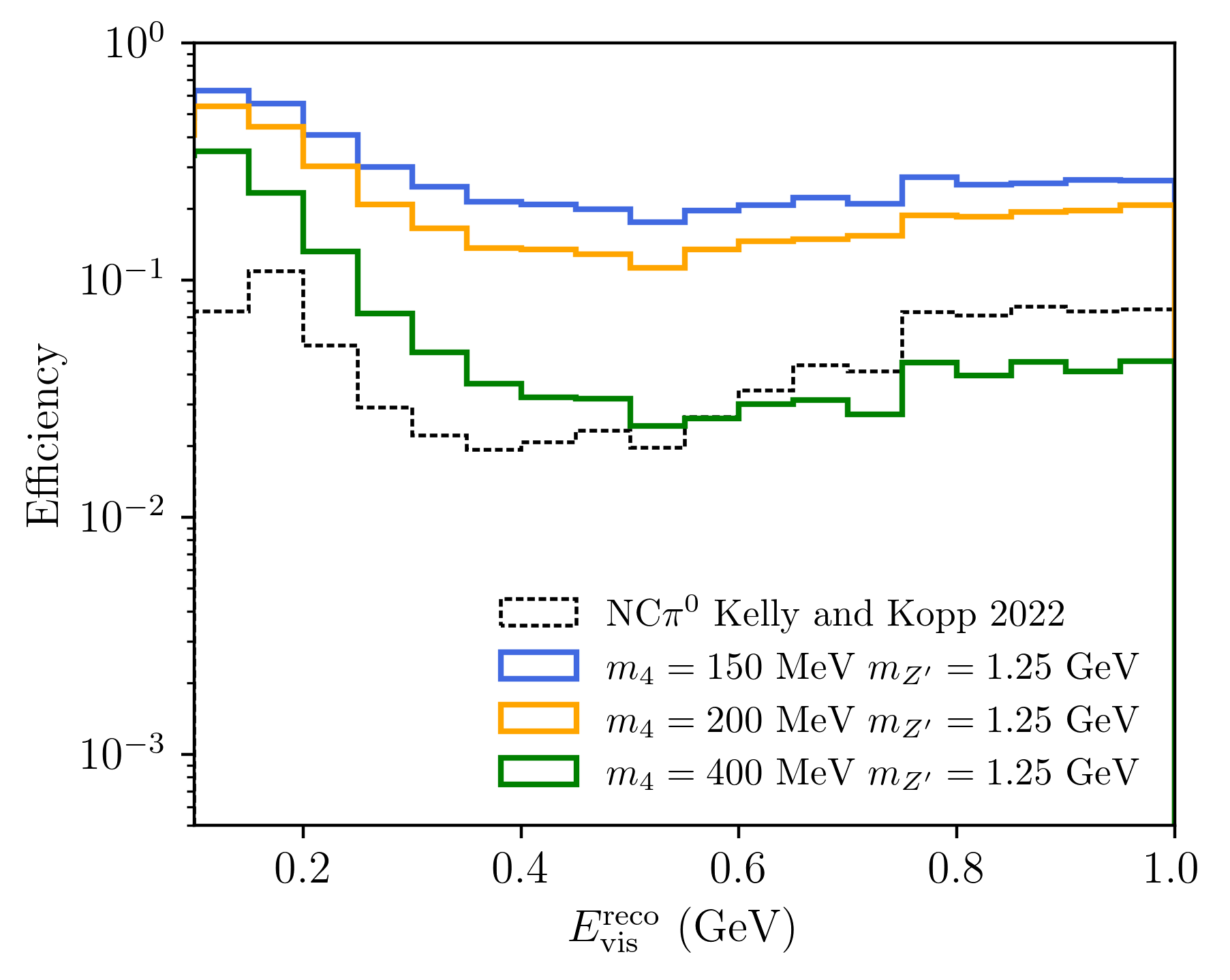}
    \caption{A comparison between the pre-selection (circ1) efficiency for dark neutrino events and the misidentification probability for $\pi^0$ background events (from \cite{Kelly:2022uaa}) in the reconstructed visible energy variable. \label{fig:efficiency_reco_pi0}}
\end{figure*}

\begin{figure*}[t]
    \centering
    \includegraphics[width=0.48\textwidth]{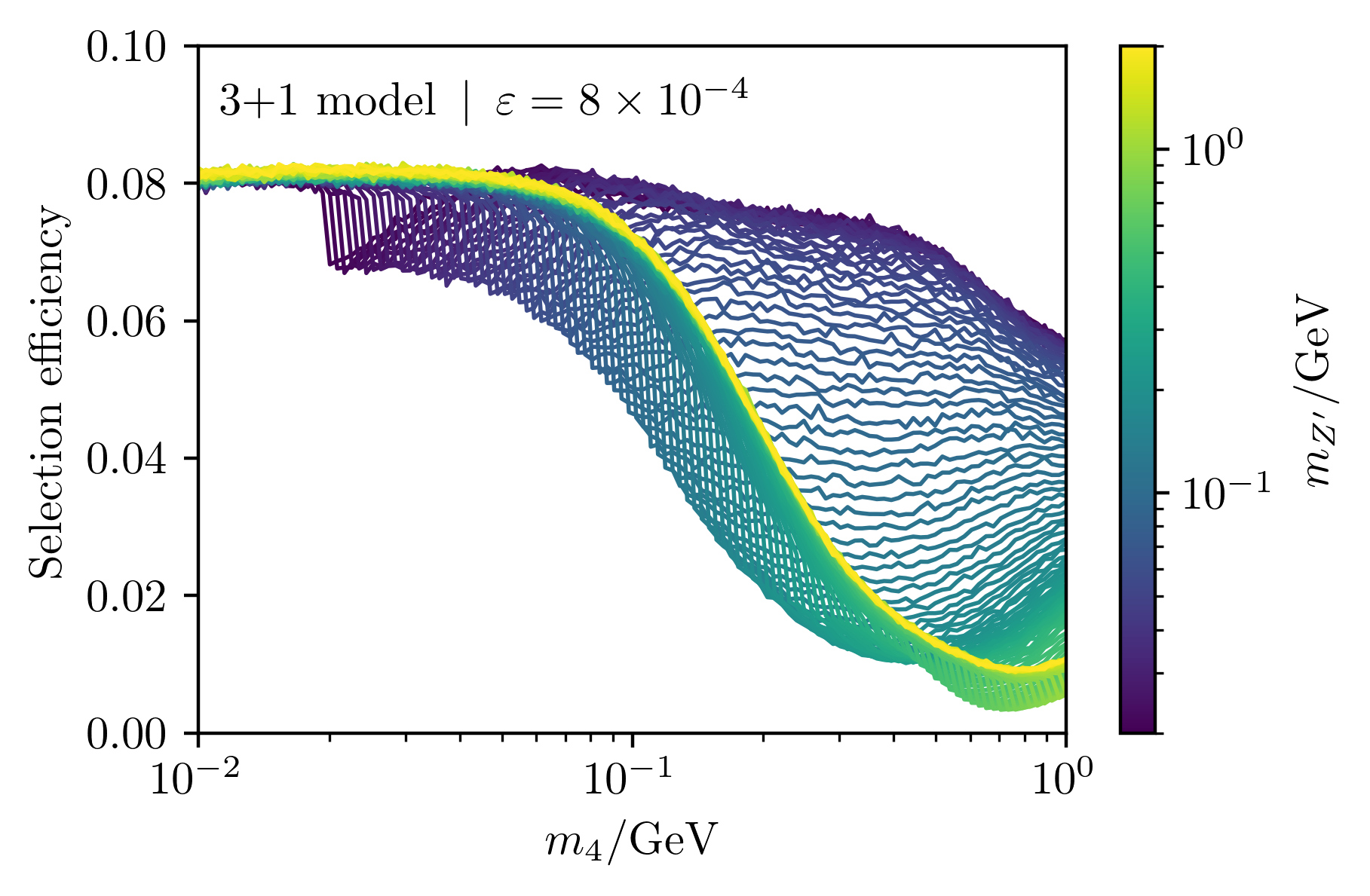}
    \includegraphics[width=0.48\textwidth]{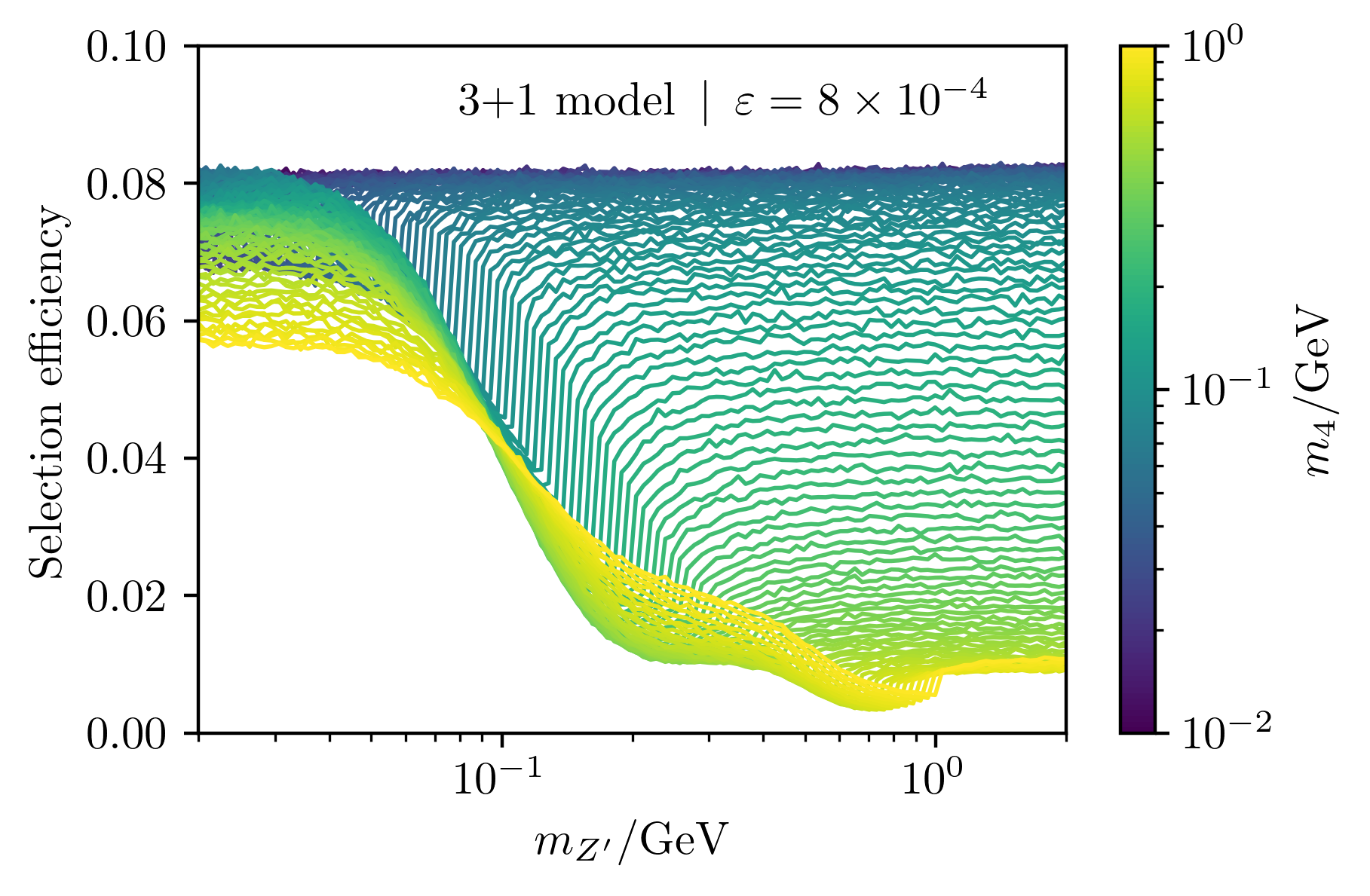}
    \caption{
    The signal selection efficiency of our simulation for dark neutrino events in MiniBooNE for the 3+1 model.
    On the left, we show the efficiency as a function of the heavy neutrino mass $m_4$ for fixed values of the mediator mass, $m_{Z^\prime}$.
    On the right, we show the same efficiency, now as a function of $m_{Z^\prime}$ for fixed values of $m_4$.
    The geometric acceptance is not included.
    \label{fig:miniboone_effs_3p1}}
\end{figure*}

\begin{figure*}[t]
    \centering
    \includegraphics[width=0.48\textwidth]{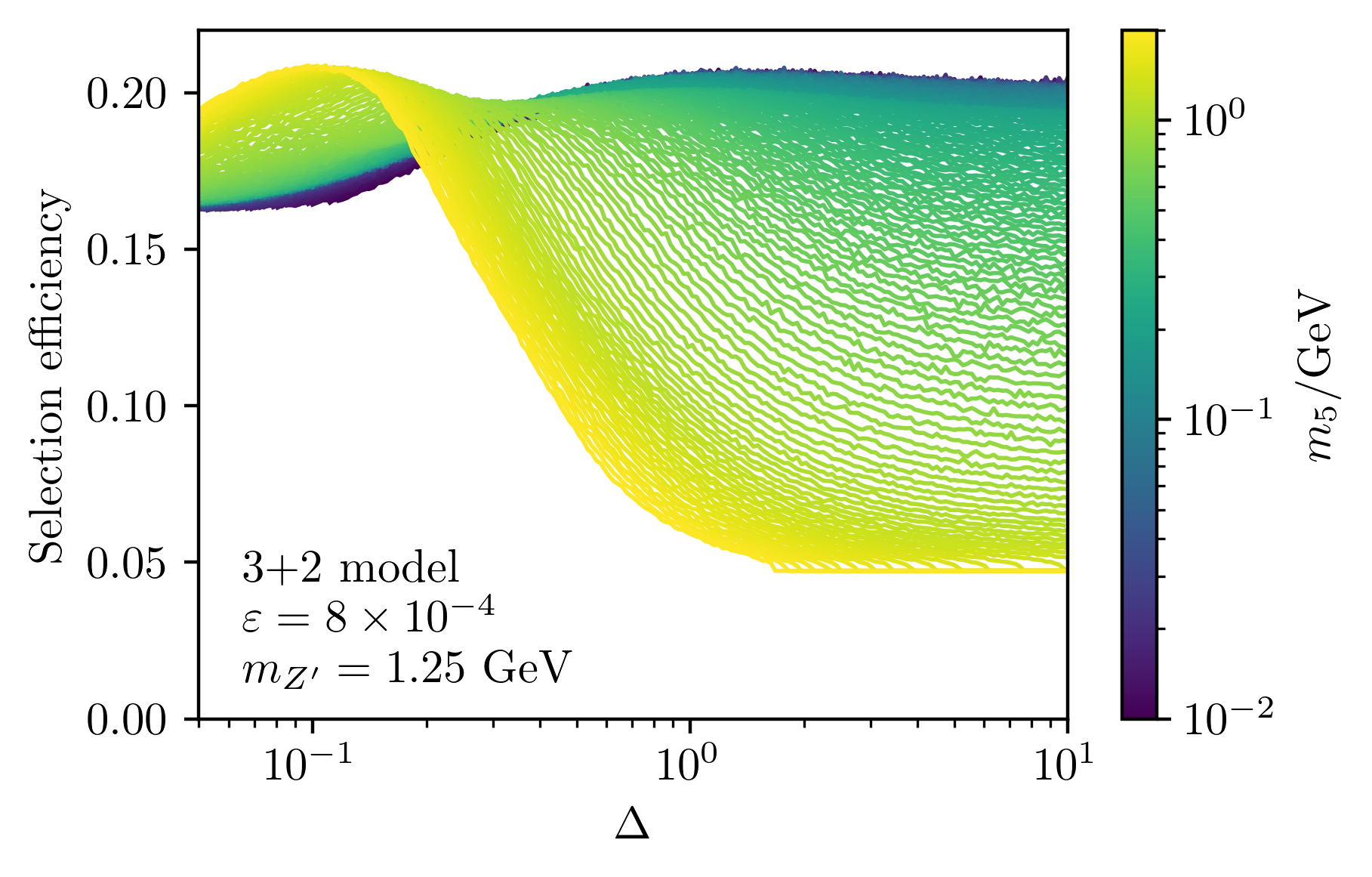}
    \includegraphics[width=0.48\textwidth]{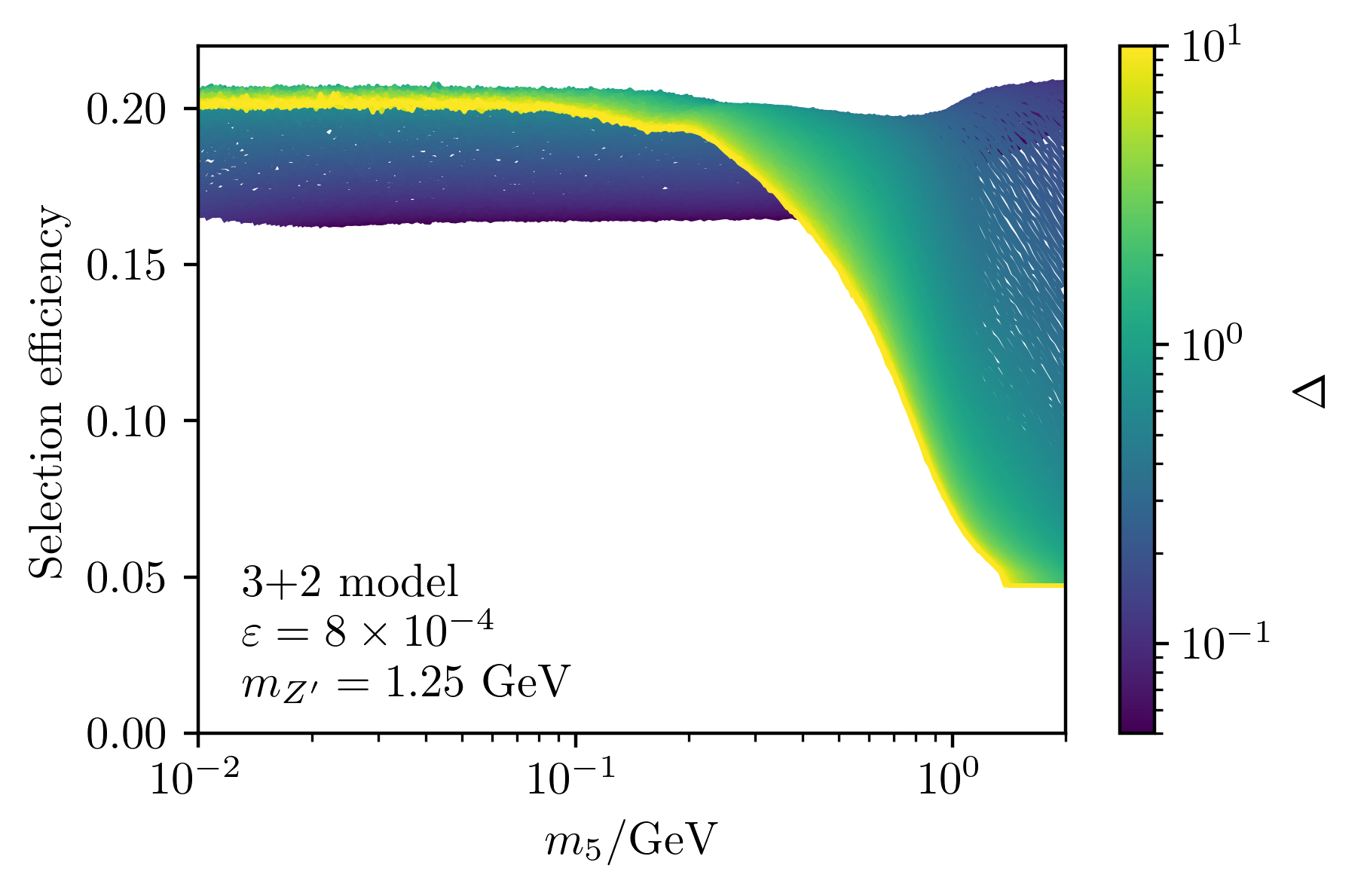}
    \\
    \includegraphics[width=0.48\textwidth]{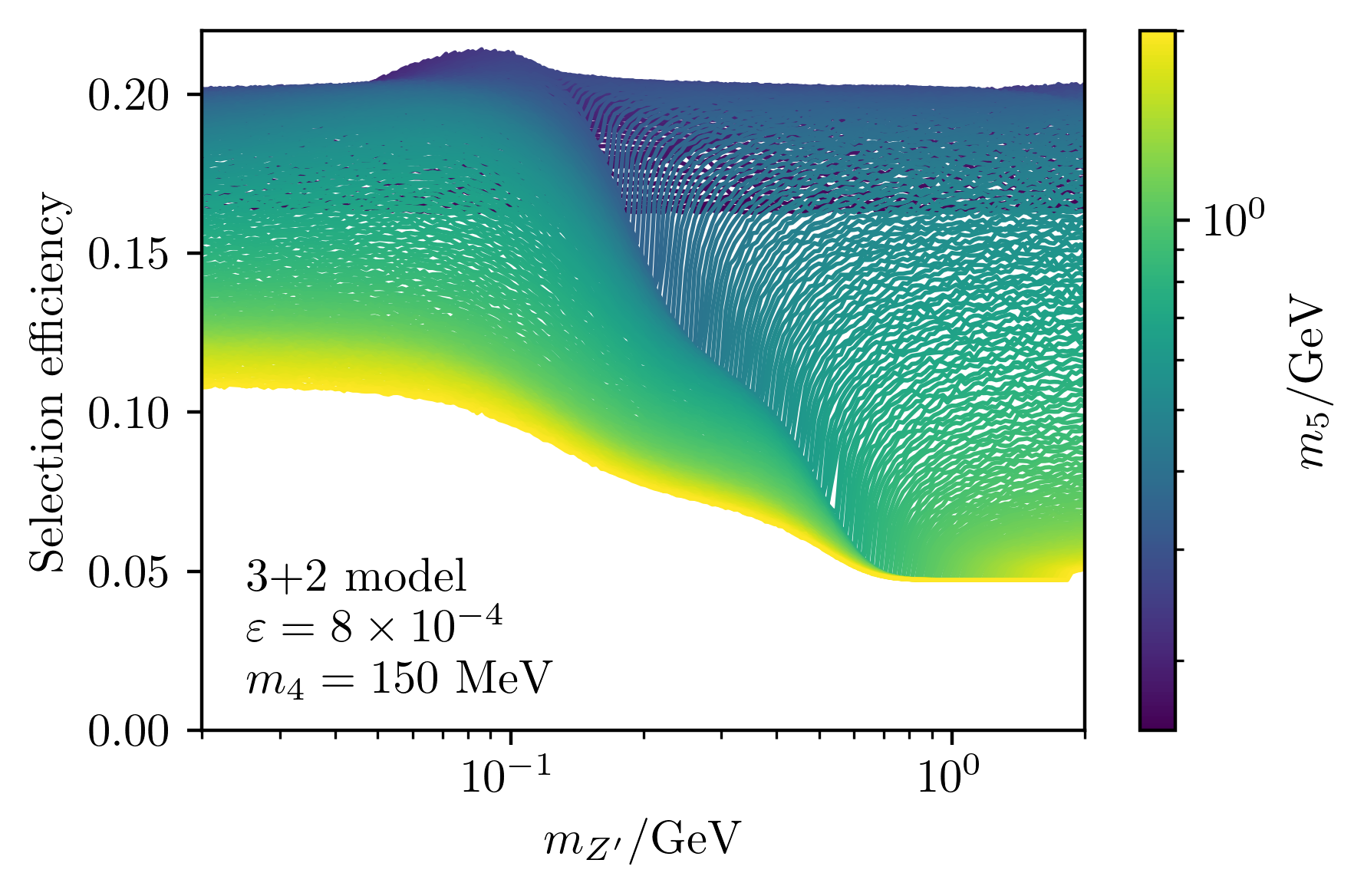}
    \includegraphics[width=0.48\textwidth]{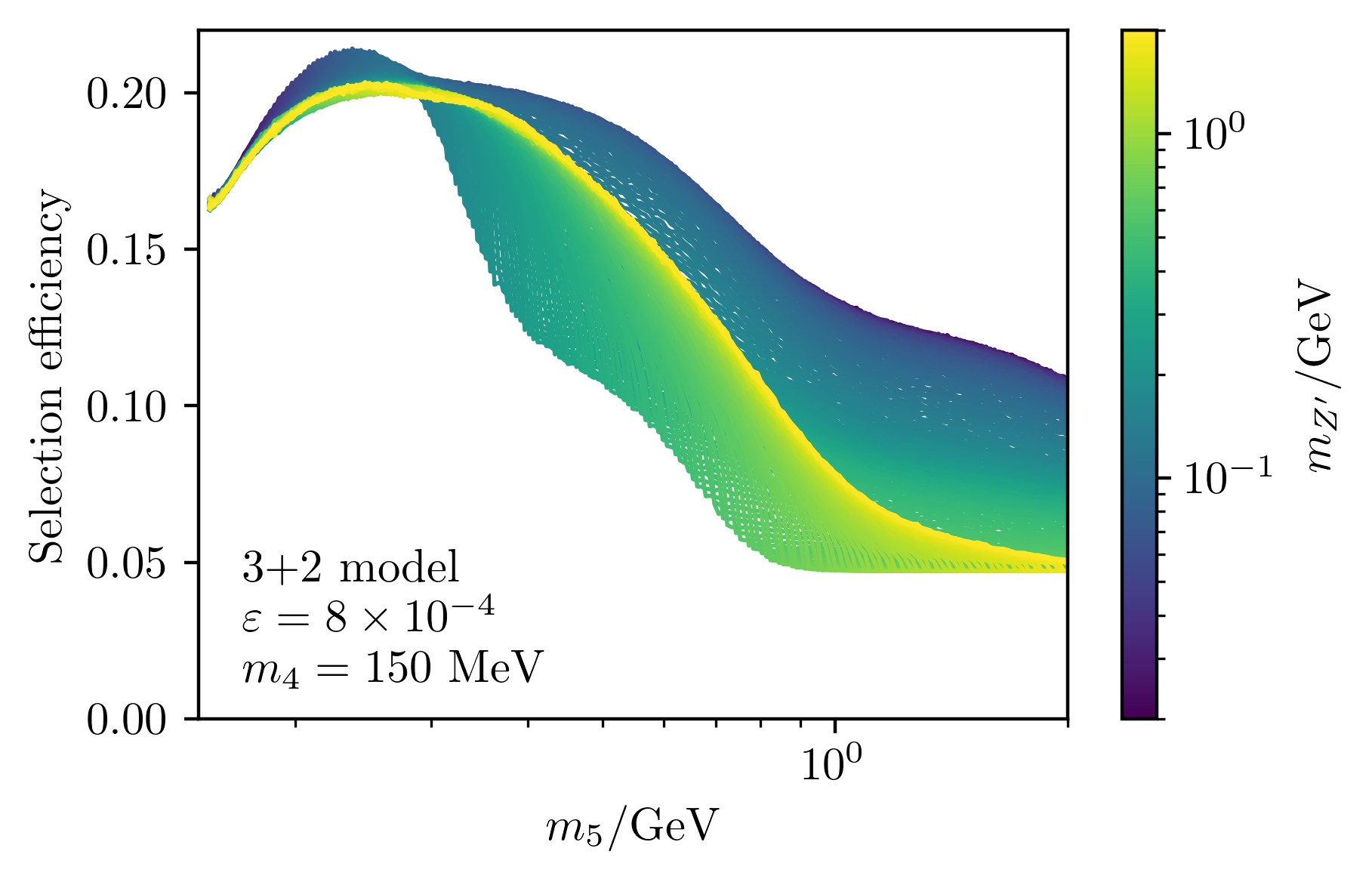}
    \caption{
    The signal selection efficiency of our simulation for dark neutrino events in MiniBooNE for the 3+2 model.
    The upper panels show the efficiency as a function of the HNL mass splitting $\Delta$ (left) and the parent HNL mass $m_5$ (right) for a single fixed value of $m_{Z^\prime}$.
    The lower panels show the efficiency as a function of the mediator mass $m_{Z^\prime}$ (left) and the parent HNL mass $m_5$ (right) for a single fixed value of $m_4$.
    The geometric acceptance is not included.
    \label{fig:miniboone_effs_3p2}}
\end{figure*}
\begin{figure}[t]
    \centering
    \includegraphics[width=0.49\textwidth]{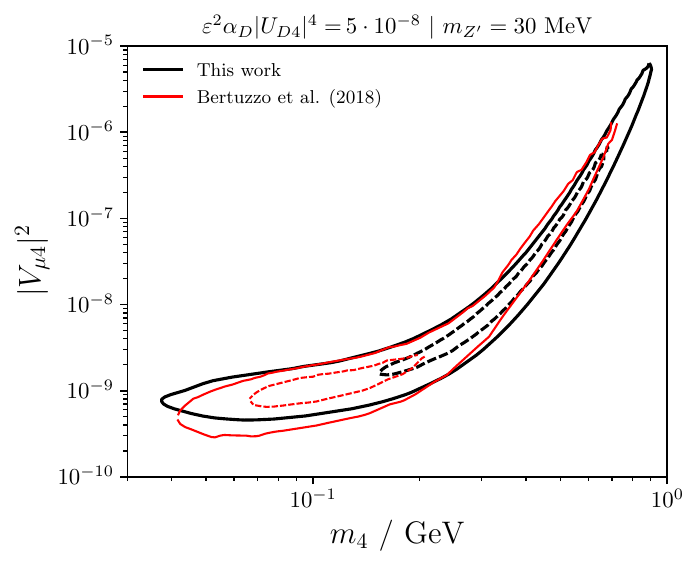}
    \caption{A comparison of the LEE neutrino energy spectrum fit in this work and that of Ref.~\cite{Bertuzzo:2018itn} for the choice of $m_{Z^\prime} = 30$~MeV.
    The $1\sigma$ and $2\sigma$ regions are shown are shown as smaller and larger closed regions, respectively.
    We assume that the results in Ref.~\cite{Bertuzzo:2018itn} are quoted for $|U_{D4}|^2 = 1$.
    \label{fig:}}
\end{figure}

We divide the selection criteria into two steps.
\begin{itemize}
    \item Pre-selection: requirements that the \epluseminus pair is reconstructed as a single electromagnetic shower.
    \item Selection: fiducialization, PID, and energy requirements. 
\end{itemize}
To evaluate the efficiency of the second step, we apply the official MiniBooNE reconstruction efficiencies shown in \cref{fig:miniboone_effs} for each event given its $E_{\rm vis}$.
Since we apply fiducial volume cuts ourselves, we divide the official MiniBooNE efficiencies by $55\%$~\cite{MiniBooNE:2007uho}.
The efficiencies also include the cut on $E_{\rm vis} > 140$~MeV, so we extend our prediction down to $E_{\rm vis} < 100$~MeV.
This is also an approximation since the shape of CCQE events and dark-neutrino events within this low-energy bin can differ.
Also implicit is the assumption that the PID cuts have the same efficiency for CCQE events and \epluseminus ones.
These cuts account for approximately a factor of two reduction in the overall efficiency~\cite{MiniBooNE:2007uho}. 
This number may vary for \epluseminus events due to the different topology, but it cannot be estimated here without access to the reconstructed electron, muon, and pion likelihoods for each event.
Finally, other cuts aimed at reducing external (dirt and wall) events and comics are also included in \cref{fig:miniboone_effs} and are assumed to have the same effect on \epluseminus events.

We now describe two simplified pre-selection procedures and compare their efficiencies in \cref{fig:efficiency_true_pi0,fig:efficiency_reco_pi0}.

\subsection{Simplified hard cuts pre-selection}
\label{sec:true-var}

A dark neutrino event can be reconstructed as a single shower for sufficiently energy-asymmetric or collimated \epluseminus pairs.
A simplified approach adopted in Refs.~\cite{Ballett:2018ynz,Abdullahi:2020nyr} was to define an energy-asymmetric \epluseminus pair as one where the energy of one of the particles was below $30$~MeV (after smearing).
Events were classified as overlapping when the opening angle between the leptons was at most $13^\circ$ (after smearing).
In addition, a cut on the \emph{true} invariant mass of all events was performed, simulating the selection criterion in Table 12.1 of Ref.~\cite{Patterson:2007zz},
\begin{align}\label{eq:invmass_cut}
    m_{e^+e^-}^{\rm true} &< \left[32.03
     + 7.417\left(\frac{E_{e^+} + E_{e^-}}{\text{ GeV}}\right)\right.
    \\\nonumber
    &\qquad \qquad \left. + 27.38\left(\frac{E_{e^+} + E_{e^-}}{\text{ GeV}}\right)^2\right] \text{ MeV}.
\end{align}
In the \miniboone analysis, this cut is applied to the invariant mass obtained under the two-photon reconstruction hypothesis.
Without such a reconstructed variable, we apply the cut to the true invariant mass of the \epluseminus pair.
This implicitly assumes that the MiniBooNE two-photon reconstruction correctly reconstructs the $e^+e^-$ invariant mass, which is less likely to be the case for events that appear more single-shower-like.
While a crude approximation, this cut ensures that heavy parent particles do not contribute significantly to the LEE, since the daughter \epluseminus pair will typically be well separated.
As we will see, this cut tends to produce much smaller efficiencies than the method implemented in the analysis of the main text.

\subsection{Simplified $\pi^0$-based pre-selection (Kelly-Kopp)}
\label{sec:simple-pi0}

This method is the one we adopt for our dark neutrino fits and is adapted from Ref.~\cite{Kelly:2022uaa}.
The authors implement an energy-dependent cut on the plane of the $\pi^0$ kinematics: $\cos{\theta_{\gamma \gamma}}$ vs $E_{\rm max}/E_{\rm vis}$.
This is achieved with a cut on $r$, the distance between points in the $\pi^0$ kinematic plane and the most-overlapping and most-energy-asymmetric region. 
The selection criterion is designed to reproduce the correct $\pi^0$-background distribution, and it was shown to be mostly independent of the exact choice of the abstract variable $r$.
Due to the similarity between the topology of dark neutrino and $\pi^0$ events, we readily adapt the selection for \epluseminus events.
For concreteness, we choose to work with the variable $r_{\rm circ1}$, defined here as
\begin{equation}
    r_{\rm circ1}^2 = \left(\frac{1 - \cos{\theta_{ee}}}{2} \right)^2 + \left( 1 - \frac{E_{\rm max}}{E_{\rm vis}}\right)^2.
\end{equation}
We then apply an energy-dependent cut on $r_{\rm circ1}$ following the data release of Ref.~\cite{Kelly:2022uaa}.
The three different $r$ variables defined in Ref.~\cite{Kelly:2022uaa}, namely, $r_{\rm circ1}$, $r_{\rm circ0}$, and $r_{\rm diag}$, lead to very similar results.

A comparison of the efficiency for the two pre-selection methods discussed above is shown in \cref{fig:comparison_preselection}.
We generate upscattering events exclusively inside the detector and enforce the $N_4$ decays to be prompt. 
In this way, the shape of the efficiencies as a function of the reconstructed visible energy $E_{\rm vis}^{\rm reco}$ depends uniquely on the kinematics and not on the HNL lifetime.
The fiducial volume cut is included in our pre-selection efficiency but does not depend on the kinematics.
The comparison shows that the cut in \cref{eq:invmass_cut} significantly impacts the true-variable selection method. 
In addition, the methods of \cref{sec:true-var} and \cref{sec:simple-pi0} can differ significantly in shape and overall normalization. 
The discrepancy is also significant for $m_{Z^\prime} \simeq m_\pi$, indicating that the true-variable method would most likely not reproduce the correct $\pi^0$ spectrum at \miniboone if applied to single $\pi^0$ background events.
This further motivates us to work with the simplified $\pi^0$ method of Ref.~\cite{Kelly:2022uaa}.

We also compare our pre-selection method (circ1) with the misidentification of $\pi^0$ backgrounds at \miniboone.
In \cref{fig:efficiency_true_pi0}, we show the pre-selection efficiencies for a few dark neutrino models and compare them with the misidentification probabilities for genuine $\pi^0$ events quoted in Fig.~11.8 of Ref.\cite{Patterson:2007zz}.
Although we refer to the latter as an efficiency, it should not be confused with the efficiency of reconstructing well-separated $2\gamma$ event from $\pi^0$s.
We convert the true \epluseminus total energy $E_{\rm tot}$ into a mock $\pi^0$ momentum variable according to $p_\pi = \sqrt{E_{\rm tot}^2 - m_{\pi^0}^2}$.
The differences in shapes are due to the different boost factors of the parent particle.
Light mediators produced on-shell are more likely to be misidentified as a single shower at much lower momenta than $\pi^0$ simply due to their larger boosts.
In the case of off-shell mediators, the HNLs are misidentified as a single shower more often across the entire energy region.
Finally, we also compare our preselection with the efficiency in \cite{Kelly:2022uaa}, as a function of the reconstructed visible energy $E_{\rm vis}^{\rm reco}$.

A comparison of the final signal selection efficiencies for various points in the parameter space of dark neutrino events is shown in \cref{fig:miniboone_effs_3p1,fig:miniboone_effs_3p2}.
We do not include the geometrical acceptance in order to isolate the effects of the kinematics on the efficiency.
However, we do include the official MiniBooNE $e/\gamma$ efficiencies.

For ease of comparison with previous literature, we compare our own dark neutrino fit for 3+1 model with $m_{Z^\prime} = 30$~MeV with the fit in Ref.~\cite{Bertuzzo:2018itn}.
While the set of phenomenological parameters used in their work is not the same, we assume that $|U_{D4}| = 1$, instead of our own value of $|U_{D4}| = 1/\sqrt{2}$.
Good agreement is observed between the two cases.

\bibliographystyle{apsrev4-1}
\bibliography{lib}{}

\begin{thebibliography}{220}%
\makeatletter
\providecommand \@ifxundefined [1]{%
 \@ifx{#1\undefined}
}%
\providecommand \@ifnum [1]{%
 \ifnum #1\expandafter \@firstoftwo
 \else \expandafter \@secondoftwo
 \fi
}%
\providecommand \@ifx [1]{%
 \ifx #1\expandafter \@firstoftwo
 \else \expandafter \@secondoftwo
 \fi
}%
\providecommand \natexlab [1]{#1}%
\providecommand \enquote  [1]{``#1''}%
\providecommand \bibnamefont  [1]{#1}%
\providecommand \bibfnamefont [1]{#1}%
\providecommand \citenamefont [1]{#1}%
\providecommand \href@noop [0]{\@secondoftwo}%
\providecommand \href [0]{\begingroup \@sanitize@url \@href}%
\providecommand \@href[1]{\@@startlink{#1}\@@href}%
\providecommand \@@href[1]{\endgroup#1\@@endlink}%
\providecommand \@sanitize@url [0]{\catcode `\\12\catcode `\$12\catcode
  `\&12\catcode `\#12\catcode `\^12\catcode `\_12\catcode `\%12\relax}%
\providecommand \@@startlink[1]{}%
\providecommand \@@endlink[0]{}%
\providecommand \url  [0]{\begingroup\@sanitize@url \@url }%
\providecommand \@url [1]{\endgroup\@href {#1}{\urlprefix }}%
\providecommand \urlprefix  [0]{URL }%
\providecommand \Eprint [0]{\href }%
\providecommand \doibase [0]{http://dx.doi.org/}%
\providecommand \selectlanguage [0]{\@gobble}%
\providecommand \bibinfo  [0]{\@secondoftwo}%
\providecommand \bibfield  [0]{\@secondoftwo}%
\providecommand \translation [1]{[#1]}%
\providecommand \BibitemOpen [0]{}%
\providecommand \bibitemStop [0]{}%
\providecommand \bibitemNoStop [0]{.\EOS\space}%
\providecommand \EOS [0]{\spacefactor3000\relax}%
\providecommand \BibitemShut  [1]{\csname bibitem#1\endcsname}%
\let\auto@bib@innerbib\@empty
\bibitem [{\citenamefont {Aguilar-Arevalo}\ \emph {et~al.}(2007)\citenamefont
  {Aguilar-Arevalo} \emph {et~al.}}]{MiniBooNE:2007uho}%
  \BibitemOpen
  \bibfield  {author} {\bibinfo {author} {\bibfnamefont {A.~A.}\ \bibnamefont
  {Aguilar-Arevalo}} \emph {et~al.} (\bibinfo {collaboration} {MiniBooNE}),\
  }\href {\doibase 10.1103/PhysRevLett.98.231801} {\bibfield  {journal}
  {\bibinfo  {journal} {Phys. Rev. Lett.}\ }\textbf {\bibinfo {volume} {98}},\
  \bibinfo {pages} {231801} (\bibinfo {year} {2007})},\ \Eprint
  {http://arxiv.org/abs/0704.1500} {arXiv:0704.1500 [hep-ex]} \BibitemShut
  {NoStop}%
\bibitem [{\citenamefont {Aguilar-Arevalo}\ \emph
  {et~al.}(2009{\natexlab{a}})\citenamefont {Aguilar-Arevalo} \emph
  {et~al.}}]{MiniBooNE:2008yuf}%
  \BibitemOpen
  \bibfield  {author} {\bibinfo {author} {\bibfnamefont {A.~A.}\ \bibnamefont
  {Aguilar-Arevalo}} \emph {et~al.} (\bibinfo {collaboration} {MiniBooNE}),\
  }\href {\doibase 10.1103/PhysRevLett.102.101802} {\bibfield  {journal}
  {\bibinfo  {journal} {Phys. Rev. Lett.}\ }\textbf {\bibinfo {volume} {102}},\
  \bibinfo {pages} {101802} (\bibinfo {year} {2009}{\natexlab{a}})},\ \Eprint
  {http://arxiv.org/abs/0812.2243} {arXiv:0812.2243 [hep-ex]} \BibitemShut
  {NoStop}%
\bibitem [{\citenamefont {Aguilar-Arevalo}\ \emph {et~al.}(2013)\citenamefont
  {Aguilar-Arevalo} \emph {et~al.}}]{MiniBooNE:2013uba}%
  \BibitemOpen
  \bibfield  {author} {\bibinfo {author} {\bibfnamefont {A.~A.}\ \bibnamefont
  {Aguilar-Arevalo}} \emph {et~al.} (\bibinfo {collaboration} {MiniBooNE}),\
  }\href {\doibase 10.1103/PhysRevLett.110.161801} {\bibfield  {journal}
  {\bibinfo  {journal} {Phys. Rev. Lett.}\ }\textbf {\bibinfo {volume} {110}},\
  \bibinfo {pages} {161801} (\bibinfo {year} {2013})},\ \Eprint
  {http://arxiv.org/abs/1303.2588} {arXiv:1303.2588 [hep-ex]} \BibitemShut
  {NoStop}%
\bibitem [{\citenamefont {Aguilar}\ \emph {et~al.}(2001)\citenamefont {Aguilar}
  \emph {et~al.}}]{LSND:2001aii}%
  \BibitemOpen
  \bibfield  {author} {\bibinfo {author} {\bibfnamefont {A.}~\bibnamefont
  {Aguilar}} \emph {et~al.} (\bibinfo {collaboration} {LSND}),\ }\href
  {\doibase 10.1103/PhysRevD.64.112007} {\bibfield  {journal} {\bibinfo
  {journal} {Phys. Rev. D}\ }\textbf {\bibinfo {volume} {64}},\ \bibinfo
  {pages} {112007} (\bibinfo {year} {2001})},\ \Eprint
  {http://arxiv.org/abs/hep-ex/0104049} {arXiv:hep-ex/0104049} \BibitemShut
  {NoStop}%
\bibitem [{\citenamefont {Aguilar-Arevalo}\ \emph
  {et~al.}(2018{\natexlab{a}})\citenamefont {Aguilar-Arevalo} \emph
  {et~al.}}]{MiniBooNE:2018esg}%
  \BibitemOpen
  \bibfield  {author} {\bibinfo {author} {\bibfnamefont {A.~A.}\ \bibnamefont
  {Aguilar-Arevalo}} \emph {et~al.} (\bibinfo {collaboration} {MiniBooNE}),\
  }\href {\doibase 10.1103/PhysRevLett.121.221801} {\bibfield  {journal}
  {\bibinfo  {journal} {Phys. Rev. Lett.}\ }\textbf {\bibinfo {volume} {121}},\
  \bibinfo {pages} {221801} (\bibinfo {year} {2018}{\natexlab{a}})},\ \Eprint
  {http://arxiv.org/abs/1805.12028} {arXiv:1805.12028 [hep-ex]} \BibitemShut
  {NoStop}%
\bibitem [{\citenamefont {Aguilar-Arevalo}\ \emph
  {et~al.}(2021{\natexlab{a}})\citenamefont {Aguilar-Arevalo} \emph
  {et~al.}}]{MiniBooNE:2020pnu}%
  \BibitemOpen
  \bibfield  {author} {\bibinfo {author} {\bibfnamefont {A.~A.}\ \bibnamefont
  {Aguilar-Arevalo}} \emph {et~al.} (\bibinfo {collaboration} {MiniBooNE}),\
  }\href {\doibase 10.1103/PhysRevD.103.052002} {\bibfield  {journal} {\bibinfo
   {journal} {Phys. Rev. D}\ }\textbf {\bibinfo {volume} {103}},\ \bibinfo
  {pages} {052002} (\bibinfo {year} {2021}{\natexlab{a}})},\ \Eprint
  {http://arxiv.org/abs/2006.16883} {arXiv:2006.16883 [hep-ex]} \BibitemShut
  {NoStop}%
\bibitem [{\citenamefont {Brdar}\ and\ \citenamefont
  {Kopp}(2022)}]{Brdar:2021ysi}%
  \BibitemOpen
  \bibfield  {author} {\bibinfo {author} {\bibfnamefont {V.}~\bibnamefont
  {Brdar}}\ and\ \bibinfo {author} {\bibfnamefont {J.}~\bibnamefont {Kopp}},\
  }\href {\doibase 10.1103/PhysRevD.105.115024} {\bibfield  {journal} {\bibinfo
   {journal} {Phys. Rev. D}\ }\textbf {\bibinfo {volume} {105}},\ \bibinfo
  {pages} {115024} (\bibinfo {year} {2022})},\ \Eprint
  {http://arxiv.org/abs/2109.08157} {arXiv:2109.08157 [hep-ph]} \BibitemShut
  {NoStop}%
\bibitem [{\citenamefont {Kelly}\ and\ \citenamefont
  {Kopp}(2023)}]{Kelly:2022uaa}%
  \BibitemOpen
  \bibfield  {author} {\bibinfo {author} {\bibfnamefont {K.~J.}\ \bibnamefont
  {Kelly}}\ and\ \bibinfo {author} {\bibfnamefont {J.}~\bibnamefont {Kopp}},\
  }\href {\doibase 10.1007/JHEP05(2023)113} {\bibfield  {journal} {\bibinfo
  {journal} {JHEP}\ }\textbf {\bibinfo {volume} {05}},\ \bibinfo {pages} {113}
  (\bibinfo {year} {2023})},\ \Eprint {http://arxiv.org/abs/2210.08021}
  {arXiv:2210.08021 [hep-ph]} \BibitemShut {NoStop}%
\bibitem [{\citenamefont {Anselmann}\ \emph {et~al.}(1995)\citenamefont
  {Anselmann} \emph {et~al.}}]{GALLEX:1994rym}%
  \BibitemOpen
  \bibfield  {author} {\bibinfo {author} {\bibfnamefont {P.}~\bibnamefont
  {Anselmann}} \emph {et~al.} (\bibinfo {collaboration} {GALLEX}),\ }\href
  {\doibase 10.1016/0370-2693(94)01586-2} {\bibfield  {journal} {\bibinfo
  {journal} {Phys. Lett. B}\ }\textbf {\bibinfo {volume} {342}},\ \bibinfo
  {pages} {440} (\bibinfo {year} {1995})}\BibitemShut {NoStop}%
\bibitem [{\citenamefont {Hampel}\ \emph {et~al.}(1998)\citenamefont {Hampel}
  \emph {et~al.}}]{GALLEX:1997lja}%
  \BibitemOpen
  \bibfield  {author} {\bibinfo {author} {\bibfnamefont {W.}~\bibnamefont
  {Hampel}} \emph {et~al.} (\bibinfo {collaboration} {GALLEX}),\ }\href
  {\doibase 10.1016/S0370-2693(97)01562-1} {\bibfield  {journal} {\bibinfo
  {journal} {Phys. Lett. B}\ }\textbf {\bibinfo {volume} {420}},\ \bibinfo
  {pages} {114} (\bibinfo {year} {1998})}\BibitemShut {NoStop}%
\bibitem [{\citenamefont {Kaether}\ \emph {et~al.}(2010)\citenamefont
  {Kaether}, \citenamefont {Hampel}, \citenamefont {Heusser}, \citenamefont
  {Kiko},\ and\ \citenamefont {Kirsten}}]{Kaether:2010ag}%
  \BibitemOpen
  \bibfield  {author} {\bibinfo {author} {\bibfnamefont {F.}~\bibnamefont
  {Kaether}}, \bibinfo {author} {\bibfnamefont {W.}~\bibnamefont {Hampel}},
  \bibinfo {author} {\bibfnamefont {G.}~\bibnamefont {Heusser}}, \bibinfo
  {author} {\bibfnamefont {J.}~\bibnamefont {Kiko}}, \ and\ \bibinfo {author}
  {\bibfnamefont {T.}~\bibnamefont {Kirsten}},\ }\href {\doibase
  10.1016/j.physletb.2010.01.030} {\bibfield  {journal} {\bibinfo  {journal}
  {Phys. Lett. B}\ }\textbf {\bibinfo {volume} {685}},\ \bibinfo {pages} {47}
  (\bibinfo {year} {2010})},\ \Eprint {http://arxiv.org/abs/1001.2731}
  {arXiv:1001.2731 [hep-ex]} \BibitemShut {NoStop}%
\bibitem [{\citenamefont {Abdurashitov}\ \emph {et~al.}(1996)\citenamefont
  {Abdurashitov} \emph {et~al.}}]{Abdurashitov:1996dp}%
  \BibitemOpen
  \bibfield  {author} {\bibinfo {author} {\bibfnamefont {D.~N.}\ \bibnamefont
  {Abdurashitov}} \emph {et~al.},\ }\href {\doibase
  10.1103/PhysRevLett.77.4708} {\bibfield  {journal} {\bibinfo  {journal}
  {Phys. Rev. Lett.}\ }\textbf {\bibinfo {volume} {77}},\ \bibinfo {pages}
  {4708} (\bibinfo {year} {1996})}\BibitemShut {NoStop}%
\bibitem [{\citenamefont {Abdurashitov}\ \emph {et~al.}(1999)\citenamefont
  {Abdurashitov} \emph {et~al.}}]{SAGE:1998fvr}%
  \BibitemOpen
  \bibfield  {author} {\bibinfo {author} {\bibfnamefont {J.~N.}\ \bibnamefont
  {Abdurashitov}} \emph {et~al.} (\bibinfo {collaboration} {SAGE}),\ }\href
  {\doibase 10.1103/PhysRevC.59.2246} {\bibfield  {journal} {\bibinfo
  {journal} {Phys. Rev. C}\ }\textbf {\bibinfo {volume} {59}},\ \bibinfo
  {pages} {2246} (\bibinfo {year} {1999})},\ \Eprint
  {http://arxiv.org/abs/hep-ph/9803418} {arXiv:hep-ph/9803418} \BibitemShut
  {NoStop}%
\bibitem [{\citenamefont {Abdurashitov}\ \emph {et~al.}(2006)\citenamefont
  {Abdurashitov} \emph {et~al.}}]{Abdurashitov:2005tb}%
  \BibitemOpen
  \bibfield  {author} {\bibinfo {author} {\bibfnamefont {J.~N.}\ \bibnamefont
  {Abdurashitov}} \emph {et~al.},\ }\href {\doibase 10.1103/PhysRevC.73.045805}
  {\bibfield  {journal} {\bibinfo  {journal} {Phys. Rev. C}\ }\textbf {\bibinfo
  {volume} {73}},\ \bibinfo {pages} {045805} (\bibinfo {year} {2006})},\
  \Eprint {http://arxiv.org/abs/nucl-ex/0512041} {arXiv:nucl-ex/0512041}
  \BibitemShut {NoStop}%
\bibitem [{\citenamefont {Abdurashitov}\ \emph {et~al.}(2009)\citenamefont
  {Abdurashitov} \emph {et~al.}}]{SAGE:2009eeu}%
  \BibitemOpen
  \bibfield  {author} {\bibinfo {author} {\bibfnamefont {J.~N.}\ \bibnamefont
  {Abdurashitov}} \emph {et~al.} (\bibinfo {collaboration} {SAGE}),\ }\href
  {\doibase 10.1103/PhysRevC.80.015807} {\bibfield  {journal} {\bibinfo
  {journal} {Phys. Rev. C}\ }\textbf {\bibinfo {volume} {80}},\ \bibinfo
  {pages} {015807} (\bibinfo {year} {2009})},\ \Eprint
  {http://arxiv.org/abs/0901.2200} {arXiv:0901.2200 [nucl-ex]} \BibitemShut
  {NoStop}%
\bibitem [{\citenamefont {Hamann}\ \emph {et~al.}(2011)\citenamefont {Hamann},
  \citenamefont {Hannestad}, \citenamefont {Raffelt},\ and\ \citenamefont
  {Wong}}]{Hamann:2011ge}%
  \BibitemOpen
  \bibfield  {author} {\bibinfo {author} {\bibfnamefont {J.}~\bibnamefont
  {Hamann}}, \bibinfo {author} {\bibfnamefont {S.}~\bibnamefont {Hannestad}},
  \bibinfo {author} {\bibfnamefont {G.~G.}\ \bibnamefont {Raffelt}}, \ and\
  \bibinfo {author} {\bibfnamefont {Y.~Y.~Y.}\ \bibnamefont {Wong}},\ }\href
  {\doibase 10.1088/1475-7516/2011/09/034} {\bibfield  {journal} {\bibinfo
  {journal} {JCAP}\ }\textbf {\bibinfo {volume} {09}},\ \bibinfo {pages} {034}
  (\bibinfo {year} {2011})},\ \Eprint {http://arxiv.org/abs/1108.4136}
  {arXiv:1108.4136 [astro-ph.CO]} \BibitemShut {NoStop}%
\bibitem [{\citenamefont {Archidiacono}\ \emph {et~al.}(2013)\citenamefont
  {Archidiacono}, \citenamefont {Fornengo}, \citenamefont {Giunti},
  \citenamefont {Hannestad},\ and\ \citenamefont
  {Melchiorri}}]{Archidiacono:2013xxa}%
  \BibitemOpen
  \bibfield  {author} {\bibinfo {author} {\bibfnamefont {M.}~\bibnamefont
  {Archidiacono}}, \bibinfo {author} {\bibfnamefont {N.}~\bibnamefont
  {Fornengo}}, \bibinfo {author} {\bibfnamefont {C.}~\bibnamefont {Giunti}},
  \bibinfo {author} {\bibfnamefont {S.}~\bibnamefont {Hannestad}}, \ and\
  \bibinfo {author} {\bibfnamefont {A.}~\bibnamefont {Melchiorri}},\ }\href
  {\doibase 10.1103/PhysRevD.87.125034} {\bibfield  {journal} {\bibinfo
  {journal} {Phys. Rev. D}\ }\textbf {\bibinfo {volume} {87}},\ \bibinfo
  {pages} {125034} (\bibinfo {year} {2013})},\ \Eprint
  {http://arxiv.org/abs/1302.6720} {arXiv:1302.6720 [astro-ph.CO]} \BibitemShut
  {NoStop}%
\bibitem [{\citenamefont {Hagstotz}\ \emph {et~al.}(2021)\citenamefont
  {Hagstotz}, \citenamefont {de~Salas}, \citenamefont {Gariazzo}, \citenamefont
  {Gerbino}, \citenamefont {Lattanzi}, \citenamefont {Vagnozzi}, \citenamefont
  {Freese},\ and\ \citenamefont {Pastor}}]{Hagstotz:2020ukm}%
  \BibitemOpen
  \bibfield  {author} {\bibinfo {author} {\bibfnamefont {S.}~\bibnamefont
  {Hagstotz}}, \bibinfo {author} {\bibfnamefont {P.~F.}\ \bibnamefont
  {de~Salas}}, \bibinfo {author} {\bibfnamefont {S.}~\bibnamefont {Gariazzo}},
  \bibinfo {author} {\bibfnamefont {M.}~\bibnamefont {Gerbino}}, \bibinfo
  {author} {\bibfnamefont {M.}~\bibnamefont {Lattanzi}}, \bibinfo {author}
  {\bibfnamefont {S.}~\bibnamefont {Vagnozzi}}, \bibinfo {author}
  {\bibfnamefont {K.}~\bibnamefont {Freese}}, \ and\ \bibinfo {author}
  {\bibfnamefont {S.}~\bibnamefont {Pastor}},\ }\href {\doibase
  10.1103/PhysRevD.104.123524} {\bibfield  {journal} {\bibinfo  {journal}
  {Phys. Rev. D}\ }\textbf {\bibinfo {volume} {104}},\ \bibinfo {pages}
  {123524} (\bibinfo {year} {2021})},\ \Eprint
  {http://arxiv.org/abs/2003.02289} {arXiv:2003.02289 [astro-ph.CO]}
  \BibitemShut {NoStop}%
\bibitem [{\citenamefont {Adamson}\ \emph {et~al.}(2020)\citenamefont {Adamson}
  \emph {et~al.}}]{MINOS:2020iqj}%
  \BibitemOpen
  \bibfield  {author} {\bibinfo {author} {\bibfnamefont {P.}~\bibnamefont
  {Adamson}} \emph {et~al.} (\bibinfo {collaboration} {MINOS+, Daya Bay}),\
  }\href {\doibase 10.1103/PhysRevLett.125.071801} {\bibfield  {journal}
  {\bibinfo  {journal} {Phys. Rev. Lett.}\ }\textbf {\bibinfo {volume} {125}},\
  \bibinfo {pages} {071801} (\bibinfo {year} {2020})},\ \Eprint
  {http://arxiv.org/abs/2002.00301} {arXiv:2002.00301 [hep-ex]} \BibitemShut
  {NoStop}%
\bibitem [{\citenamefont {Aartsen}\ \emph
  {et~al.}(2020{\natexlab{a}})\citenamefont {Aartsen} \emph
  {et~al.}}]{IceCube:2020phf}%
  \BibitemOpen
  \bibfield  {author} {\bibinfo {author} {\bibfnamefont {M.~G.}\ \bibnamefont
  {Aartsen}} \emph {et~al.} (\bibinfo {collaboration} {IceCube}),\ }\href
  {\doibase 10.1103/PhysRevLett.125.141801} {\bibfield  {journal} {\bibinfo
  {journal} {Phys. Rev. Lett.}\ }\textbf {\bibinfo {volume} {125}},\ \bibinfo
  {pages} {141801} (\bibinfo {year} {2020}{\natexlab{a}})},\ \Eprint
  {http://arxiv.org/abs/2005.12942} {arXiv:2005.12942 [hep-ex]} \BibitemShut
  {NoStop}%
\bibitem [{\citenamefont {Aartsen}\ \emph
  {et~al.}(2020{\natexlab{b}})\citenamefont {Aartsen} \emph
  {et~al.}}]{IceCube:2020tka}%
  \BibitemOpen
  \bibfield  {author} {\bibinfo {author} {\bibfnamefont {M.~G.}\ \bibnamefont
  {Aartsen}} \emph {et~al.} (\bibinfo {collaboration} {IceCube}),\ }\href
  {\doibase 10.1103/PhysRevD.102.052009} {\bibfield  {journal} {\bibinfo
  {journal} {Phys. Rev. D}\ }\textbf {\bibinfo {volume} {102}},\ \bibinfo
  {pages} {052009} (\bibinfo {year} {2020}{\natexlab{b}})},\ \Eprint
  {http://arxiv.org/abs/2005.12943} {arXiv:2005.12943 [hep-ex]} \BibitemShut
  {NoStop}%
\bibitem [{\citenamefont {Dentler}\ \emph {et~al.}(2018)\citenamefont
  {Dentler}, \citenamefont {Hern\'andez-Cabezudo}, \citenamefont {Kopp},
  \citenamefont {Machado}, \citenamefont {Maltoni}, \citenamefont
  {Martinez-Soler},\ and\ \citenamefont {Schwetz}}]{Dentler:2018sju}%
  \BibitemOpen
  \bibfield  {author} {\bibinfo {author} {\bibfnamefont {M.}~\bibnamefont
  {Dentler}}, \bibinfo {author} {\bibfnamefont {A.}~\bibnamefont
  {Hern\'andez-Cabezudo}}, \bibinfo {author} {\bibfnamefont {J.}~\bibnamefont
  {Kopp}}, \bibinfo {author} {\bibfnamefont {P.~A.~N.}\ \bibnamefont
  {Machado}}, \bibinfo {author} {\bibfnamefont {M.}~\bibnamefont {Maltoni}},
  \bibinfo {author} {\bibfnamefont {I.}~\bibnamefont {Martinez-Soler}}, \ and\
  \bibinfo {author} {\bibfnamefont {T.}~\bibnamefont {Schwetz}},\ }\href
  {\doibase 10.1007/JHEP08(2018)010} {\bibfield  {journal} {\bibinfo  {journal}
  {JHEP}\ }\textbf {\bibinfo {volume} {08}},\ \bibinfo {pages} {010} (\bibinfo
  {year} {2018})},\ \Eprint {http://arxiv.org/abs/1803.10661} {arXiv:1803.10661
  [hep-ph]} \BibitemShut {NoStop}%
\bibitem [{\citenamefont {Diaz}\ \emph {et~al.}(2020)\citenamefont {Diaz},
  \citenamefont {Arg\"uelles}, \citenamefont {Collin}, \citenamefont {Conrad},\
  and\ \citenamefont {Shaevitz}}]{Diaz:2019fwt}%
  \BibitemOpen
  \bibfield  {author} {\bibinfo {author} {\bibfnamefont {A.}~\bibnamefont
  {Diaz}}, \bibinfo {author} {\bibfnamefont {C.~A.}\ \bibnamefont
  {Arg\"uelles}}, \bibinfo {author} {\bibfnamefont {G.~H.}\ \bibnamefont
  {Collin}}, \bibinfo {author} {\bibfnamefont {J.~M.}\ \bibnamefont {Conrad}},
  \ and\ \bibinfo {author} {\bibfnamefont {M.~H.}\ \bibnamefont {Shaevitz}},\
  }\href {\doibase 10.1016/j.physrep.2020.08.005} {\bibfield  {journal}
  {\bibinfo  {journal} {Phys. Rept.}\ }\textbf {\bibinfo {volume} {884}},\
  \bibinfo {pages} {1} (\bibinfo {year} {2020})},\ \Eprint
  {http://arxiv.org/abs/1906.00045} {arXiv:1906.00045 [hep-ex]} \BibitemShut
  {NoStop}%
\bibitem [{\citenamefont {Acero}\ \emph {et~al.}(2022)\citenamefont {Acero}
  \emph {et~al.}}]{Acero:2022wqg}%
  \BibitemOpen
  \bibfield  {author} {\bibinfo {author} {\bibfnamefont {M.~A.}\ \bibnamefont
  {Acero}} \emph {et~al.},\ }\href@noop {} {\  (\bibinfo {year} {2022})},\
  \Eprint {http://arxiv.org/abs/2203.07323} {arXiv:2203.07323 [hep-ex]}
  \BibitemShut {NoStop}%
\bibitem [{\citenamefont {Acciarri}\ \emph {et~al.}(2015)\citenamefont
  {Acciarri} \emph {et~al.}}]{MicroBooNE:2015bmn}%
  \BibitemOpen
  \bibfield  {author} {\bibinfo {author} {\bibfnamefont {R.}~\bibnamefont
  {Acciarri}} \emph {et~al.} (\bibinfo {collaboration} {MicroBooNE, LAr1-ND,
  ICARUS-WA104}),\ }\href@noop {} {\  (\bibinfo {year} {2015})},\ \Eprint
  {http://arxiv.org/abs/1503.01520} {arXiv:1503.01520 [physics.ins-det]}
  \BibitemShut {NoStop}%
\bibitem [{\citenamefont {Machado}\ \emph {et~al.}(2019)\citenamefont
  {Machado}, \citenamefont {Palamara},\ and\ \citenamefont
  {Schmitz}}]{Machado:2019oxb}%
  \BibitemOpen
  \bibfield  {author} {\bibinfo {author} {\bibfnamefont {P.~A.}\ \bibnamefont
  {Machado}}, \bibinfo {author} {\bibfnamefont {O.}~\bibnamefont {Palamara}}, \
  and\ \bibinfo {author} {\bibfnamefont {D.~W.}\ \bibnamefont {Schmitz}},\
  }\href {\doibase 10.1146/annurev-nucl-101917-020949} {\bibfield  {journal}
  {\bibinfo  {journal} {Ann. Rev. Nucl. Part. Sci.}\ }\textbf {\bibinfo
  {volume} {69}},\ \bibinfo {pages} {363} (\bibinfo {year} {2019})},\ \Eprint
  {http://arxiv.org/abs/1903.04608} {arXiv:1903.04608 [hep-ex]} \BibitemShut
  {NoStop}%
\bibitem [{\citenamefont {Abratenko}\ \emph
  {et~al.}(2022{\natexlab{a}})\citenamefont {Abratenko} \emph
  {et~al.}}]{MicroBooNE:2021rmx}%
  \BibitemOpen
  \bibfield  {author} {\bibinfo {author} {\bibfnamefont {P.}~\bibnamefont
  {Abratenko}} \emph {et~al.} (\bibinfo {collaboration} {MicroBooNE}),\ }\href
  {\doibase 10.1103/PhysRevLett.128.241801} {\bibfield  {journal} {\bibinfo
  {journal} {Phys. Rev. Lett.}\ }\textbf {\bibinfo {volume} {128}},\ \bibinfo
  {pages} {241801} (\bibinfo {year} {2022}{\natexlab{a}})},\ \Eprint
  {http://arxiv.org/abs/2110.14054} {arXiv:2110.14054 [hep-ex]} \BibitemShut
  {NoStop}%
\bibitem [{\citenamefont {Abratenko}\ \emph
  {et~al.}(2022{\natexlab{b}})\citenamefont {Abratenko} \emph
  {et~al.}}]{MicroBooNE:2021jwr}%
  \BibitemOpen
  \bibfield  {author} {\bibinfo {author} {\bibfnamefont {P.}~\bibnamefont
  {Abratenko}} \emph {et~al.} (\bibinfo {collaboration} {MicroBooNE}),\ }\href
  {\doibase 10.1103/PhysRevD.105.112003} {\bibfield  {journal} {\bibinfo
  {journal} {Phys. Rev. D}\ }\textbf {\bibinfo {volume} {105}},\ \bibinfo
  {pages} {112003} (\bibinfo {year} {2022}{\natexlab{b}})},\ \Eprint
  {http://arxiv.org/abs/2110.14080} {arXiv:2110.14080 [hep-ex]} \BibitemShut
  {NoStop}%
\bibitem [{\citenamefont {Abratenko}\ \emph
  {et~al.}(2022{\natexlab{c}})\citenamefont {Abratenko} \emph
  {et~al.}}]{MicroBooNE:2021sne}%
  \BibitemOpen
  \bibfield  {author} {\bibinfo {author} {\bibfnamefont {P.}~\bibnamefont
  {Abratenko}} \emph {et~al.} (\bibinfo {collaboration} {(The MicroBooNE
  Collaboration)*, MicroBooNE}),\ }\href {\doibase 10.1103/PhysRevD.105.112004}
  {\bibfield  {journal} {\bibinfo  {journal} {Phys. Rev. D}\ }\textbf {\bibinfo
  {volume} {105}},\ \bibinfo {pages} {112004} (\bibinfo {year}
  {2022}{\natexlab{c}})},\ \Eprint {http://arxiv.org/abs/2110.14065}
  {arXiv:2110.14065 [hep-ex]} \BibitemShut {NoStop}%
\bibitem [{\citenamefont {Abratenko}\ \emph
  {et~al.}(2022{\natexlab{d}})\citenamefont {Abratenko} \emph
  {et~al.}}]{MicroBooNE:2021nxr}%
  \BibitemOpen
  \bibfield  {author} {\bibinfo {author} {\bibfnamefont {P.}~\bibnamefont
  {Abratenko}} \emph {et~al.} (\bibinfo {collaboration} {MicroBooNE}),\ }\href
  {\doibase 10.1103/PhysRevD.105.112005} {\bibfield  {journal} {\bibinfo
  {journal} {Phys. Rev. D}\ }\textbf {\bibinfo {volume} {105}},\ \bibinfo
  {pages} {112005} (\bibinfo {year} {2022}{\natexlab{d}})},\ \Eprint
  {http://arxiv.org/abs/2110.13978} {arXiv:2110.13978 [hep-ex]} \BibitemShut
  {NoStop}%
\bibitem [{\citenamefont {Arg\"uelles}\ \emph
  {et~al.}(2022{\natexlab{a}})\citenamefont {Arg\"uelles}, \citenamefont
  {Esteban}, \citenamefont {Hostert}, \citenamefont {Kelly}, \citenamefont
  {Kopp}, \citenamefont {Machado}, \citenamefont {Martinez-Soler},\ and\
  \citenamefont {Perez-Gonzalez}}]{Arguelles:2021meu}%
  \BibitemOpen
  \bibfield  {author} {\bibinfo {author} {\bibfnamefont {C.~A.}\ \bibnamefont
  {Arg\"uelles}}, \bibinfo {author} {\bibfnamefont {I.}~\bibnamefont
  {Esteban}}, \bibinfo {author} {\bibfnamefont {M.}~\bibnamefont {Hostert}},
  \bibinfo {author} {\bibfnamefont {K.~J.}\ \bibnamefont {Kelly}}, \bibinfo
  {author} {\bibfnamefont {J.}~\bibnamefont {Kopp}}, \bibinfo {author}
  {\bibfnamefont {P.~A.~N.}\ \bibnamefont {Machado}}, \bibinfo {author}
  {\bibfnamefont {I.}~\bibnamefont {Martinez-Soler}}, \ and\ \bibinfo {author}
  {\bibfnamefont {Y.~F.}\ \bibnamefont {Perez-Gonzalez}},\ }\href {\doibase
  10.1103/PhysRevLett.128.241802} {\bibfield  {journal} {\bibinfo  {journal}
  {Phys. Rev. Lett.}\ }\textbf {\bibinfo {volume} {128}},\ \bibinfo {pages}
  {241802} (\bibinfo {year} {2022}{\natexlab{a}})},\ \Eprint
  {http://arxiv.org/abs/2111.10359} {arXiv:2111.10359 [hep-ph]} \BibitemShut
  {NoStop}%
\bibitem [{\citenamefont {Aguilar-Arevalo}\ \emph {et~al.}(2022)\citenamefont
  {Aguilar-Arevalo} \emph {et~al.}}]{MiniBooNE:2022emn}%
  \BibitemOpen
  \bibfield  {author} {\bibinfo {author} {\bibfnamefont {A.~A.}\ \bibnamefont
  {Aguilar-Arevalo}} \emph {et~al.} (\bibinfo {collaboration} {MiniBooNE}),\
  }\href {\doibase 10.1103/PhysRevLett.129.201801} {\bibfield  {journal}
  {\bibinfo  {journal} {Phys. Rev. Lett.}\ }\textbf {\bibinfo {volume} {129}},\
  \bibinfo {pages} {201801} (\bibinfo {year} {2022})},\ \Eprint
  {http://arxiv.org/abs/2201.01724} {arXiv:2201.01724 [hep-ex]} \BibitemShut
  {NoStop}%
\bibitem [{\citenamefont {Abratenko}\ \emph
  {et~al.}(2023{\natexlab{a}})\citenamefont {Abratenko} \emph
  {et~al.}}]{MicroBooNE:2022wdf}%
  \BibitemOpen
  \bibfield  {author} {\bibinfo {author} {\bibfnamefont {P.}~\bibnamefont
  {Abratenko}} \emph {et~al.} (\bibinfo {collaboration} {MicroBooNE}),\ }\href
  {\doibase 10.1103/PhysRevLett.130.011801} {\bibfield  {journal} {\bibinfo
  {journal} {Phys. Rev. Lett.}\ }\textbf {\bibinfo {volume} {130}},\ \bibinfo
  {pages} {011801} (\bibinfo {year} {2023}{\natexlab{a}})},\ \Eprint
  {http://arxiv.org/abs/2210.10216} {arXiv:2210.10216 [hep-ex]} \BibitemShut
  {NoStop}%
\bibitem [{\citenamefont {Denton}(2022)}]{Denton:2021czb}%
  \BibitemOpen
  \bibfield  {author} {\bibinfo {author} {\bibfnamefont {P.~B.}\ \bibnamefont
  {Denton}},\ }\href {\doibase 10.1103/PhysRevLett.129.061801} {\bibfield
  {journal} {\bibinfo  {journal} {Phys. Rev. Lett.}\ }\textbf {\bibinfo
  {volume} {129}},\ \bibinfo {pages} {061801} (\bibinfo {year} {2022})},\
  \Eprint {http://arxiv.org/abs/2111.05793} {arXiv:2111.05793 [hep-ph]}
  \BibitemShut {NoStop}%
\bibitem [{\citenamefont {Abratenko}\ \emph
  {et~al.}(2022{\natexlab{e}})\citenamefont {Abratenko} \emph
  {et~al.}}]{MicroBooNE:2021zai}%
  \BibitemOpen
  \bibfield  {author} {\bibinfo {author} {\bibfnamefont {P.}~\bibnamefont
  {Abratenko}} \emph {et~al.} (\bibinfo {collaboration} {MicroBooNE}),\ }\href
  {\doibase 10.1103/PhysRevLett.128.111801} {\bibfield  {journal} {\bibinfo
  {journal} {Phys. Rev. Lett.}\ }\textbf {\bibinfo {volume} {128}},\ \bibinfo
  {pages} {111801} (\bibinfo {year} {2022}{\natexlab{e}})},\ \Eprint
  {http://arxiv.org/abs/2110.00409} {arXiv:2110.00409 [hep-ex]} \BibitemShut
  {NoStop}%
\bibitem [{\citenamefont {Bertuzzo}\ \emph {et~al.}(2018)\citenamefont
  {Bertuzzo}, \citenamefont {Jana}, \citenamefont {Machado},\ and\
  \citenamefont {Zukanovich~Funchal}}]{Bertuzzo:2018itn}%
  \BibitemOpen
  \bibfield  {author} {\bibinfo {author} {\bibfnamefont {E.}~\bibnamefont
  {Bertuzzo}}, \bibinfo {author} {\bibfnamefont {S.}~\bibnamefont {Jana}},
  \bibinfo {author} {\bibfnamefont {P.~A.~N.}\ \bibnamefont {Machado}}, \ and\
  \bibinfo {author} {\bibfnamefont {R.}~\bibnamefont {Zukanovich~Funchal}},\
  }\href {\doibase 10.1103/PhysRevLett.121.241801} {\bibfield  {journal}
  {\bibinfo  {journal} {Phys. Rev. Lett.}\ }\textbf {\bibinfo {volume} {121}},\
  \bibinfo {pages} {241801} (\bibinfo {year} {2018})},\ \Eprint
  {http://arxiv.org/abs/1807.09877} {arXiv:1807.09877 [hep-ph]} \BibitemShut
  {NoStop}%
\bibitem [{\citenamefont {Ballett}\ \emph {et~al.}(2019)\citenamefont
  {Ballett}, \citenamefont {Pascoli},\ and\ \citenamefont
  {Ross-Lonergan}}]{Ballett:2018ynz}%
  \BibitemOpen
  \bibfield  {author} {\bibinfo {author} {\bibfnamefont {P.}~\bibnamefont
  {Ballett}}, \bibinfo {author} {\bibfnamefont {S.}~\bibnamefont {Pascoli}}, \
  and\ \bibinfo {author} {\bibfnamefont {M.}~\bibnamefont {Ross-Lonergan}},\
  }\href {\doibase 10.1103/PhysRevD.99.071701} {\bibfield  {journal} {\bibinfo
  {journal} {Phys. Rev. D}\ }\textbf {\bibinfo {volume} {99}},\ \bibinfo
  {pages} {071701} (\bibinfo {year} {2019})},\ \Eprint
  {http://arxiv.org/abs/1808.02915} {arXiv:1808.02915 [hep-ph]} \BibitemShut
  {NoStop}%
\bibitem [{\citenamefont {Ballett}\ \emph {et~al.}(2020)\citenamefont
  {Ballett}, \citenamefont {Hostert},\ and\ \citenamefont
  {Pascoli}}]{Ballett:2019pyw}%
  \BibitemOpen
  \bibfield  {author} {\bibinfo {author} {\bibfnamefont {P.}~\bibnamefont
  {Ballett}}, \bibinfo {author} {\bibfnamefont {M.}~\bibnamefont {Hostert}}, \
  and\ \bibinfo {author} {\bibfnamefont {S.}~\bibnamefont {Pascoli}},\ }\href
  {\doibase 10.1103/PhysRevD.101.115025} {\bibfield  {journal} {\bibinfo
  {journal} {Phys. Rev. D}\ }\textbf {\bibinfo {volume} {101}},\ \bibinfo
  {pages} {115025} (\bibinfo {year} {2020})},\ \Eprint
  {http://arxiv.org/abs/1903.07589} {arXiv:1903.07589 [hep-ph]} \BibitemShut
  {NoStop}%
\bibitem [{\citenamefont {Abdullahi}\ \emph {et~al.}(2021)\citenamefont
  {Abdullahi}, \citenamefont {Hostert},\ and\ \citenamefont
  {Pascoli}}]{Abdullahi:2020nyr}%
  \BibitemOpen
  \bibfield  {author} {\bibinfo {author} {\bibfnamefont {A.}~\bibnamefont
  {Abdullahi}}, \bibinfo {author} {\bibfnamefont {M.}~\bibnamefont {Hostert}},
  \ and\ \bibinfo {author} {\bibfnamefont {S.}~\bibnamefont {Pascoli}},\ }\href
  {\doibase 10.1016/j.physletb.2021.136531} {\bibfield  {journal} {\bibinfo
  {journal} {Phys. Lett. B}\ }\textbf {\bibinfo {volume} {820}},\ \bibinfo
  {pages} {136531} (\bibinfo {year} {2021})},\ \Eprint
  {http://arxiv.org/abs/2007.11813} {arXiv:2007.11813 [hep-ph]} \BibitemShut
  {NoStop}%
\bibitem [{\citenamefont {Abdallah}\ \emph {et~al.}(2020)\citenamefont
  {Abdallah}, \citenamefont {Gandhi},\ and\ \citenamefont
  {Roy}}]{Abdallah:2020biq}%
  \BibitemOpen
  \bibfield  {author} {\bibinfo {author} {\bibfnamefont {W.}~\bibnamefont
  {Abdallah}}, \bibinfo {author} {\bibfnamefont {R.}~\bibnamefont {Gandhi}}, \
  and\ \bibinfo {author} {\bibfnamefont {S.}~\bibnamefont {Roy}},\ }\href
  {\doibase 10.1007/JHEP12(2020)188} {\bibfield  {journal} {\bibinfo  {journal}
  {JHEP}\ }\textbf {\bibinfo {volume} {12}},\ \bibinfo {pages} {188} (\bibinfo
  {year} {2020})},\ \Eprint {http://arxiv.org/abs/2006.01948} {arXiv:2006.01948
  [hep-ph]} \BibitemShut {NoStop}%
\bibitem [{\citenamefont {Hammad}\ \emph {et~al.}(2022)\citenamefont {Hammad},
  \citenamefont {Rashed},\ and\ \citenamefont {Moretti}}]{Hammad:2021mpl}%
  \BibitemOpen
  \bibfield  {author} {\bibinfo {author} {\bibfnamefont {A.}~\bibnamefont
  {Hammad}}, \bibinfo {author} {\bibfnamefont {A.}~\bibnamefont {Rashed}}, \
  and\ \bibinfo {author} {\bibfnamefont {S.}~\bibnamefont {Moretti}},\ }\href
  {\doibase 10.1016/j.physletb.2022.136945} {\bibfield  {journal} {\bibinfo
  {journal} {Phys. Lett. B}\ }\textbf {\bibinfo {volume} {827}},\ \bibinfo
  {pages} {136945} (\bibinfo {year} {2022})},\ \Eprint
  {http://arxiv.org/abs/2110.08651} {arXiv:2110.08651 [hep-ph]} \BibitemShut
  {NoStop}%
\bibitem [{\citenamefont {Dutta}\ \emph {et~al.}(2020)\citenamefont {Dutta},
  \citenamefont {Ghosh},\ and\ \citenamefont {Li}}]{Dutta:2020scq}%
  \BibitemOpen
  \bibfield  {author} {\bibinfo {author} {\bibfnamefont {B.}~\bibnamefont
  {Dutta}}, \bibinfo {author} {\bibfnamefont {S.}~\bibnamefont {Ghosh}}, \ and\
  \bibinfo {author} {\bibfnamefont {T.}~\bibnamefont {Li}},\ }\href {\doibase
  10.1103/PhysRevD.102.055017} {\bibfield  {journal} {\bibinfo  {journal}
  {Phys. Rev. D}\ }\textbf {\bibinfo {volume} {102}},\ \bibinfo {pages}
  {055017} (\bibinfo {year} {2020})},\ \Eprint
  {http://arxiv.org/abs/2006.01319} {arXiv:2006.01319 [hep-ph]} \BibitemShut
  {NoStop}%
\bibitem [{\citenamefont {Datta}\ \emph {et~al.}(2020)\citenamefont {Datta},
  \citenamefont {Kamali},\ and\ \citenamefont {Marfatia}}]{Datta:2020auq}%
  \BibitemOpen
  \bibfield  {author} {\bibinfo {author} {\bibfnamefont {A.}~\bibnamefont
  {Datta}}, \bibinfo {author} {\bibfnamefont {S.}~\bibnamefont {Kamali}}, \
  and\ \bibinfo {author} {\bibfnamefont {D.}~\bibnamefont {Marfatia}},\ }\href
  {\doibase 10.1016/j.physletb.2020.135579} {\bibfield  {journal} {\bibinfo
  {journal} {Phys. Lett. B}\ }\textbf {\bibinfo {volume} {807}},\ \bibinfo
  {pages} {135579} (\bibinfo {year} {2020})},\ \Eprint
  {http://arxiv.org/abs/2005.08920} {arXiv:2005.08920 [hep-ph]} \BibitemShut
  {NoStop}%
\bibitem [{\citenamefont {Abdallah}\ \emph {et~al.}(2021)\citenamefont
  {Abdallah}, \citenamefont {Gandhi},\ and\ \citenamefont
  {Roy}}]{Abdallah:2020vgg}%
  \BibitemOpen
  \bibfield  {author} {\bibinfo {author} {\bibfnamefont {W.}~\bibnamefont
  {Abdallah}}, \bibinfo {author} {\bibfnamefont {R.}~\bibnamefont {Gandhi}}, \
  and\ \bibinfo {author} {\bibfnamefont {S.}~\bibnamefont {Roy}},\ }\href
  {\doibase 10.1103/PhysRevD.104.055028} {\bibfield  {journal} {\bibinfo
  {journal} {Phys. Rev. D}\ }\textbf {\bibinfo {volume} {104}},\ \bibinfo
  {pages} {055028} (\bibinfo {year} {2021})},\ \Eprint
  {http://arxiv.org/abs/2010.06159} {arXiv:2010.06159 [hep-ph]} \BibitemShut
  {NoStop}%
\bibitem [{\citenamefont {Abdallah}\ \emph {et~al.}(2022)\citenamefont
  {Abdallah}, \citenamefont {Gandhi},\ and\ \citenamefont
  {Roy}}]{Abdallah:2022grs}%
  \BibitemOpen
  \bibfield  {author} {\bibinfo {author} {\bibfnamefont {W.}~\bibnamefont
  {Abdallah}}, \bibinfo {author} {\bibfnamefont {R.}~\bibnamefont {Gandhi}}, \
  and\ \bibinfo {author} {\bibfnamefont {S.}~\bibnamefont {Roy}},\ }\href
  {\doibase 10.1007/JHEP06(2022)160} {\bibfield  {journal} {\bibinfo  {journal}
  {JHEP}\ }\textbf {\bibinfo {volume} {06}},\ \bibinfo {pages} {160} (\bibinfo
  {year} {2022})},\ \Eprint {http://arxiv.org/abs/2202.09373} {arXiv:2202.09373
  [hep-ph]} \BibitemShut {NoStop}%
\bibitem [{\citenamefont {Abdullahi}\ \emph {et~al.}(2022)\citenamefont
  {Abdullahi}, \citenamefont {Hoefken~Zink}, \citenamefont {Hostert},
  \citenamefont {Massaro},\ and\ \citenamefont {Pascoli}}]{Abdullahi:2022cdw}%
  \BibitemOpen
  \bibfield  {author} {\bibinfo {author} {\bibfnamefont {A.~M.}\ \bibnamefont
  {Abdullahi}}, \bibinfo {author} {\bibfnamefont {J.}~\bibnamefont
  {Hoefken~Zink}}, \bibinfo {author} {\bibfnamefont {M.}~\bibnamefont
  {Hostert}}, \bibinfo {author} {\bibfnamefont {D.}~\bibnamefont {Massaro}}, \
  and\ \bibinfo {author} {\bibfnamefont {S.}~\bibnamefont {Pascoli}},\
  }\href@noop {} {\  (\bibinfo {year} {2022})},\ \Eprint
  {http://arxiv.org/abs/2207.04137} {arXiv:2207.04137 [hep-ph]} \BibitemShut
  {NoStop}%
\bibitem [{\citenamefont {Aguilar-Arevalo}\ \emph
  {et~al.}(2010{\natexlab{a}})\citenamefont {Aguilar-Arevalo} \emph
  {et~al.}}]{MiniBooNE:2010idf}%
  \BibitemOpen
  \bibfield  {author} {\bibinfo {author} {\bibfnamefont {A.~A.}\ \bibnamefont
  {Aguilar-Arevalo}} \emph {et~al.} (\bibinfo {collaboration} {MiniBooNE}),\
  }\href {\doibase 10.1103/PhysRevLett.105.181801} {\bibfield  {journal}
  {\bibinfo  {journal} {Phys. Rev. Lett.}\ }\textbf {\bibinfo {volume} {105}},\
  \bibinfo {pages} {181801} (\bibinfo {year} {2010}{\natexlab{a}})},\ \Eprint
  {http://arxiv.org/abs/1007.1150} {arXiv:1007.1150 [hep-ex]} \BibitemShut
  {NoStop}%
\bibitem [{\citenamefont {Cheng}\ \emph {et~al.}(2011)\citenamefont {Cheng}
  \emph {et~al.}}]{SciBooNE:2011sjq}%
  \BibitemOpen
  \bibfield  {author} {\bibinfo {author} {\bibfnamefont {G.}~\bibnamefont
  {Cheng}} \emph {et~al.} (\bibinfo {collaboration} {SciBooNE}),\ }\href
  {\doibase 10.1103/PhysRevD.84.012009} {\bibfield  {journal} {\bibinfo
  {journal} {Phys. Rev. D}\ }\textbf {\bibinfo {volume} {84}},\ \bibinfo
  {pages} {012009} (\bibinfo {year} {2011})},\ \Eprint
  {http://arxiv.org/abs/1105.2871} {arXiv:1105.2871 [hep-ex]} \BibitemShut
  {NoStop}%
\bibitem [{\citenamefont {Cheng}\ \emph {et~al.}(2012)\citenamefont {Cheng}
  \emph {et~al.}}]{MiniBooNE:2012meu}%
  \BibitemOpen
  \bibfield  {author} {\bibinfo {author} {\bibfnamefont {G.}~\bibnamefont
  {Cheng}} \emph {et~al.} (\bibinfo {collaboration} {MiniBooNE, SciBooNE}),\
  }\href {\doibase 10.1103/PhysRevD.86.052009} {\bibfield  {journal} {\bibinfo
  {journal} {Phys. Rev. D}\ }\textbf {\bibinfo {volume} {86}},\ \bibinfo
  {pages} {052009} (\bibinfo {year} {2012})},\ \Eprint
  {http://arxiv.org/abs/1208.0322} {arXiv:1208.0322 [hep-ex]} \BibitemShut
  {NoStop}%
\bibitem [{\citenamefont {Mahn}\ \emph {et~al.}(2012)\citenamefont {Mahn} \emph
  {et~al.}}]{SciBooNE:2011qyf}%
  \BibitemOpen
  \bibfield  {author} {\bibinfo {author} {\bibfnamefont {K.~B.~M.}\
  \bibnamefont {Mahn}} \emph {et~al.} (\bibinfo {collaboration} {SciBooNE,
  MiniBooNE}),\ }\href {\doibase 10.1103/PhysRevD.85.032007} {\bibfield
  {journal} {\bibinfo  {journal} {Phys. Rev. D}\ }\textbf {\bibinfo {volume}
  {85}},\ \bibinfo {pages} {032007} (\bibinfo {year} {2012})},\ \Eprint
  {http://arxiv.org/abs/1106.5685} {arXiv:1106.5685 [hep-ex]} \BibitemShut
  {NoStop}%
\bibitem [{\citenamefont {Aguilar-Arevalo}\ \emph {et~al.}(2008)\citenamefont
  {Aguilar-Arevalo} \emph {et~al.}}]{MiniBooNE:2008mmr}%
  \BibitemOpen
  \bibfield  {author} {\bibinfo {author} {\bibfnamefont {A.~A.}\ \bibnamefont
  {Aguilar-Arevalo}} \emph {et~al.} (\bibinfo {collaboration} {MiniBooNE}),\
  }\href {\doibase 10.1016/j.physletb.2008.05.006} {\bibfield  {journal}
  {\bibinfo  {journal} {Phys. Lett. B}\ }\textbf {\bibinfo {volume} {664}},\
  \bibinfo {pages} {41} (\bibinfo {year} {2008})},\ \Eprint
  {http://arxiv.org/abs/0803.3423} {arXiv:0803.3423 [hep-ex]} \BibitemShut
  {NoStop}%
\bibitem [{\citenamefont {Aguilar-Arevalo}\ \emph
  {et~al.}(2010{\natexlab{b}})\citenamefont {Aguilar-Arevalo} \emph
  {et~al.}}]{MiniBooNE:2009dxl}%
  \BibitemOpen
  \bibfield  {author} {\bibinfo {author} {\bibfnamefont {A.~A.}\ \bibnamefont
  {Aguilar-Arevalo}} \emph {et~al.} (\bibinfo {collaboration} {MiniBooNE}),\
  }\href {\doibase 10.1103/PhysRevD.81.013005} {\bibfield  {journal} {\bibinfo
  {journal} {Phys. Rev. D}\ }\textbf {\bibinfo {volume} {81}},\ \bibinfo
  {pages} {013005} (\bibinfo {year} {2010}{\natexlab{b}})},\ \Eprint
  {http://arxiv.org/abs/0911.2063} {arXiv:0911.2063 [hep-ex]} \BibitemShut
  {NoStop}%
\bibitem [{\citenamefont {Wang}\ \emph {et~al.}(2015)\citenamefont {Wang},
  \citenamefont {Alvarez-Ruso},\ and\ \citenamefont {Nieves}}]{Wang:2014nat}%
  \BibitemOpen
  \bibfield  {author} {\bibinfo {author} {\bibfnamefont {E.}~\bibnamefont
  {Wang}}, \bibinfo {author} {\bibfnamefont {L.}~\bibnamefont {Alvarez-Ruso}},
  \ and\ \bibinfo {author} {\bibfnamefont {J.}~\bibnamefont {Nieves}},\ }\href
  {\doibase 10.1016/j.physletb.2014.11.025} {\bibfield  {journal} {\bibinfo
  {journal} {Phys. Lett. B}\ }\textbf {\bibinfo {volume} {740}},\ \bibinfo
  {pages} {16} (\bibinfo {year} {2015})},\ \Eprint
  {http://arxiv.org/abs/1407.6060} {arXiv:1407.6060 [hep-ph]} \BibitemShut
  {NoStop}%
\bibitem [{\citenamefont {Aguilar-Arevalo}\ \emph
  {et~al.}(2018{\natexlab{b}})\citenamefont {Aguilar-Arevalo} \emph
  {et~al.}}]{MiniBooNEDM:2018cxm}%
  \BibitemOpen
  \bibfield  {author} {\bibinfo {author} {\bibfnamefont {A.~A.}\ \bibnamefont
  {Aguilar-Arevalo}} \emph {et~al.} (\bibinfo {collaboration} {MiniBooNE DM}),\
  }\href {\doibase 10.1103/PhysRevD.98.112004} {\bibfield  {journal} {\bibinfo
  {journal} {Phys. Rev. D}\ }\textbf {\bibinfo {volume} {98}},\ \bibinfo
  {pages} {112004} (\bibinfo {year} {2018}{\natexlab{b}})},\ \Eprint
  {http://arxiv.org/abs/1807.06137} {arXiv:1807.06137 [hep-ex]} \BibitemShut
  {NoStop}%
\bibitem [{\citenamefont {Jordan}\ \emph {et~al.}(2019)\citenamefont {Jordan},
  \citenamefont {Kahn}, \citenamefont {Krnjaic}, \citenamefont {Moschella},\
  and\ \citenamefont {Spitz}}]{Jordan:2018qiy}%
  \BibitemOpen
  \bibfield  {author} {\bibinfo {author} {\bibfnamefont {J.~R.}\ \bibnamefont
  {Jordan}}, \bibinfo {author} {\bibfnamefont {Y.}~\bibnamefont {Kahn}},
  \bibinfo {author} {\bibfnamefont {G.}~\bibnamefont {Krnjaic}}, \bibinfo
  {author} {\bibfnamefont {M.}~\bibnamefont {Moschella}}, \ and\ \bibinfo
  {author} {\bibfnamefont {J.}~\bibnamefont {Spitz}},\ }\href {\doibase
  10.1103/PhysRevLett.122.081801} {\bibfield  {journal} {\bibinfo  {journal}
  {Phys. Rev. Lett.}\ }\textbf {\bibinfo {volume} {122}},\ \bibinfo {pages}
  {081801} (\bibinfo {year} {2019})},\ \Eprint
  {http://arxiv.org/abs/1810.07185} {arXiv:1810.07185 [hep-ph]} \BibitemShut
  {NoStop}%
\bibitem [{\citenamefont {B\"oser}\ \emph {et~al.}(2020)\citenamefont
  {B\"oser}, \citenamefont {Buck}, \citenamefont {Giunti}, \citenamefont
  {Lesgourgues}, \citenamefont {Ludhova}, \citenamefont {Mertens},
  \citenamefont {Schukraft},\ and\ \citenamefont {Wurm}}]{Boser:2019rta}%
  \BibitemOpen
  \bibfield  {author} {\bibinfo {author} {\bibfnamefont {S.}~\bibnamefont
  {B\"oser}}, \bibinfo {author} {\bibfnamefont {C.}~\bibnamefont {Buck}},
  \bibinfo {author} {\bibfnamefont {C.}~\bibnamefont {Giunti}}, \bibinfo
  {author} {\bibfnamefont {J.}~\bibnamefont {Lesgourgues}}, \bibinfo {author}
  {\bibfnamefont {L.}~\bibnamefont {Ludhova}}, \bibinfo {author} {\bibfnamefont
  {S.}~\bibnamefont {Mertens}}, \bibinfo {author} {\bibfnamefont
  {A.}~\bibnamefont {Schukraft}}, \ and\ \bibinfo {author} {\bibfnamefont
  {M.}~\bibnamefont {Wurm}},\ }\href {\doibase 10.1016/j.ppnp.2019.103736}
  {\bibfield  {journal} {\bibinfo  {journal} {Prog. Part. Nucl. Phys.}\
  }\textbf {\bibinfo {volume} {111}},\ \bibinfo {pages} {103736} (\bibinfo
  {year} {2020})},\ \Eprint {http://arxiv.org/abs/1906.01739} {arXiv:1906.01739
  [hep-ex]} \BibitemShut {NoStop}%
\bibitem [{\citenamefont {Dasgupta}\ and\ \citenamefont
  {Kopp}(2021)}]{Dasgupta:2021ies}%
  \BibitemOpen
  \bibfield  {author} {\bibinfo {author} {\bibfnamefont {B.}~\bibnamefont
  {Dasgupta}}\ and\ \bibinfo {author} {\bibfnamefont {J.}~\bibnamefont
  {Kopp}},\ }\href {\doibase 10.1016/j.physrep.2021.06.002} {\bibfield
  {journal} {\bibinfo  {journal} {Phys. Rept.}\ }\textbf {\bibinfo {volume}
  {928}},\ \bibinfo {pages} {1} (\bibinfo {year} {2021})},\ \Eprint
  {http://arxiv.org/abs/2106.05913} {arXiv:2106.05913 [hep-ph]} \BibitemShut
  {NoStop}%
\bibitem [{\citenamefont {Moss}\ \emph {et~al.}(2018)\citenamefont {Moss},
  \citenamefont {Moulai}, \citenamefont {Arg\"uelles},\ and\ \citenamefont
  {Conrad}}]{Moss:2017pur}%
  \BibitemOpen
  \bibfield  {author} {\bibinfo {author} {\bibfnamefont {Z.}~\bibnamefont
  {Moss}}, \bibinfo {author} {\bibfnamefont {M.~H.}\ \bibnamefont {Moulai}},
  \bibinfo {author} {\bibfnamefont {C.~A.}\ \bibnamefont {Arg\"uelles}}, \ and\
  \bibinfo {author} {\bibfnamefont {J.~M.}\ \bibnamefont {Conrad}},\ }\href
  {\doibase 10.1103/PhysRevD.97.055017} {\bibfield  {journal} {\bibinfo
  {journal} {Phys. Rev. D}\ }\textbf {\bibinfo {volume} {97}},\ \bibinfo
  {pages} {055017} (\bibinfo {year} {2018})},\ \Eprint
  {http://arxiv.org/abs/1711.05921} {arXiv:1711.05921 [hep-ph]} \BibitemShut
  {NoStop}%
\bibitem [{\citenamefont {Moulai}\ \emph {et~al.}(2020)\citenamefont {Moulai},
  \citenamefont {Arg\"uelles}, \citenamefont {Collin}, \citenamefont {Conrad},
  \citenamefont {Diaz},\ and\ \citenamefont {Shaevitz}}]{Moulai:2019gpi}%
  \BibitemOpen
  \bibfield  {author} {\bibinfo {author} {\bibfnamefont {M.~H.}\ \bibnamefont
  {Moulai}}, \bibinfo {author} {\bibfnamefont {C.~A.}\ \bibnamefont
  {Arg\"uelles}}, \bibinfo {author} {\bibfnamefont {G.~H.}\ \bibnamefont
  {Collin}}, \bibinfo {author} {\bibfnamefont {J.~M.}\ \bibnamefont {Conrad}},
  \bibinfo {author} {\bibfnamefont {A.}~\bibnamefont {Diaz}}, \ and\ \bibinfo
  {author} {\bibfnamefont {M.~H.}\ \bibnamefont {Shaevitz}},\ }\href {\doibase
  10.1103/PhysRevD.101.055020} {\bibfield  {journal} {\bibinfo  {journal}
  {Phys. Rev. D}\ }\textbf {\bibinfo {volume} {101}},\ \bibinfo {pages}
  {055020} (\bibinfo {year} {2020})},\ \Eprint
  {http://arxiv.org/abs/1910.13456} {arXiv:1910.13456 [hep-ph]} \BibitemShut
  {NoStop}%
\bibitem [{\citenamefont {Akhmedov}\ and\ \citenamefont
  {Schwetz}(2011)}]{Akhmedov:2011zza}%
  \BibitemOpen
  \bibfield  {author} {\bibinfo {author} {\bibfnamefont {E.~K.}\ \bibnamefont
  {Akhmedov}}\ and\ \bibinfo {author} {\bibfnamefont {T.}~\bibnamefont
  {Schwetz}},\ }\href {\doibase 10.1016/j.nuclphysbps.2011.04.106} {\bibfield
  {journal} {\bibinfo  {journal} {Nucl. Phys. B Proc. Suppl.}\ }\textbf
  {\bibinfo {volume} {217}},\ \bibinfo {pages} {217} (\bibinfo {year}
  {2011})}\BibitemShut {NoStop}%
\bibitem [{\citenamefont {Bramante}(2013)}]{Bramante:2011uu}%
  \BibitemOpen
  \bibfield  {author} {\bibinfo {author} {\bibfnamefont {J.}~\bibnamefont
  {Bramante}},\ }\href {\doibase 10.1142/S0217751X1350067X} {\bibfield
  {journal} {\bibinfo  {journal} {Int. J. Mod. Phys. A}\ }\textbf {\bibinfo
  {volume} {28}},\ \bibinfo {pages} {1350067} (\bibinfo {year} {2013})},\
  \Eprint {http://arxiv.org/abs/1110.4871} {arXiv:1110.4871 [hep-ph]}
  \BibitemShut {NoStop}%
\bibitem [{\citenamefont {Karagiorgi}\ \emph {et~al.}(2012)\citenamefont
  {Karagiorgi}, \citenamefont {Shaevitz},\ and\ \citenamefont
  {Conrad}}]{Karagiorgi:2012kw}%
  \BibitemOpen
  \bibfield  {author} {\bibinfo {author} {\bibfnamefont {G.}~\bibnamefont
  {Karagiorgi}}, \bibinfo {author} {\bibfnamefont {M.~H.}\ \bibnamefont
  {Shaevitz}}, \ and\ \bibinfo {author} {\bibfnamefont {J.~M.}\ \bibnamefont
  {Conrad}},\ }\href@noop {} {\  (\bibinfo {year} {2012})},\ \Eprint
  {http://arxiv.org/abs/1202.1024} {arXiv:1202.1024 [hep-ph]} \BibitemShut
  {NoStop}%
\bibitem [{\citenamefont {Asaadi}\ \emph {et~al.}(2018)\citenamefont {Asaadi},
  \citenamefont {Church}, \citenamefont {Guenette}, \citenamefont {Jones},\
  and\ \citenamefont {Szelc}}]{Asaadi:2017bhx}%
  \BibitemOpen
  \bibfield  {author} {\bibinfo {author} {\bibfnamefont {J.}~\bibnamefont
  {Asaadi}}, \bibinfo {author} {\bibfnamefont {E.}~\bibnamefont {Church}},
  \bibinfo {author} {\bibfnamefont {R.}~\bibnamefont {Guenette}}, \bibinfo
  {author} {\bibfnamefont {B.~J.~P.}\ \bibnamefont {Jones}}, \ and\ \bibinfo
  {author} {\bibfnamefont {A.~M.}\ \bibnamefont {Szelc}},\ }\href {\doibase
  10.1103/PhysRevD.97.075021} {\bibfield  {journal} {\bibinfo  {journal} {Phys.
  Rev. D}\ }\textbf {\bibinfo {volume} {97}},\ \bibinfo {pages} {075021}
  (\bibinfo {year} {2018})},\ \Eprint {http://arxiv.org/abs/1712.08019}
  {arXiv:1712.08019 [hep-ph]} \BibitemShut {NoStop}%
\bibitem [{\citenamefont {Smirnov}\ and\ \citenamefont
  {Valera}(2021)}]{Smirnov:2021zgn}%
  \BibitemOpen
  \bibfield  {author} {\bibinfo {author} {\bibfnamefont {A.~Y.}\ \bibnamefont
  {Smirnov}}\ and\ \bibinfo {author} {\bibfnamefont {V.~B.}\ \bibnamefont
  {Valera}},\ }\href {\doibase 10.1007/JHEP09(2021)177} {\bibfield  {journal}
  {\bibinfo  {journal} {JHEP}\ }\textbf {\bibinfo {volume} {09}},\ \bibinfo
  {pages} {177} (\bibinfo {year} {2021})},\ \Eprint
  {http://arxiv.org/abs/2106.13829} {arXiv:2106.13829 [hep-ph]} \BibitemShut
  {NoStop}%
\bibitem [{\citenamefont {Alves}\ \emph {et~al.}(2022)\citenamefont {Alves},
  \citenamefont {Louis},\ and\ \citenamefont {deNiverville}}]{Alves:2022vgn}%
  \BibitemOpen
  \bibfield  {author} {\bibinfo {author} {\bibfnamefont {D.~S.~M.}\
  \bibnamefont {Alves}}, \bibinfo {author} {\bibfnamefont {W.~C.}\ \bibnamefont
  {Louis}}, \ and\ \bibinfo {author} {\bibfnamefont {P.~G.}\ \bibnamefont
  {deNiverville}},\ }\href {\doibase 10.1007/JHEP08(2022)034} {\bibfield
  {journal} {\bibinfo  {journal} {JHEP}\ }\textbf {\bibinfo {volume} {08}},\
  \bibinfo {pages} {034} (\bibinfo {year} {2022})},\ \Eprint
  {http://arxiv.org/abs/2201.00876} {arXiv:2201.00876 [hep-ph]} \BibitemShut
  {NoStop}%
\bibitem [{\citenamefont {Berryman}\ \emph {et~al.}(2018)\citenamefont
  {Berryman}, \citenamefont {De~Gouv\^ea}, \citenamefont {Kelly},\ and\
  \citenamefont {Zhang}}]{Berryman:2018ogk}%
  \BibitemOpen
  \bibfield  {author} {\bibinfo {author} {\bibfnamefont {J.~M.}\ \bibnamefont
  {Berryman}}, \bibinfo {author} {\bibfnamefont {A.}~\bibnamefont
  {De~Gouv\^ea}}, \bibinfo {author} {\bibfnamefont {K.~J.}\ \bibnamefont
  {Kelly}}, \ and\ \bibinfo {author} {\bibfnamefont {Y.}~\bibnamefont
  {Zhang}},\ }\href {\doibase 10.1103/PhysRevD.97.075030} {\bibfield  {journal}
  {\bibinfo  {journal} {Phys. Rev. D}\ }\textbf {\bibinfo {volume} {97}},\
  \bibinfo {pages} {075030} (\bibinfo {year} {2018})},\ \Eprint
  {http://arxiv.org/abs/1802.00009} {arXiv:1802.00009 [hep-ph]} \BibitemShut
  {NoStop}%
\bibitem [{\citenamefont {Palomares-Ruiz}\ \emph {et~al.}(2005)\citenamefont
  {Palomares-Ruiz}, \citenamefont {Pascoli},\ and\ \citenamefont
  {Schwetz}}]{Palomares-Ruiz:2005zbh}%
  \BibitemOpen
  \bibfield  {author} {\bibinfo {author} {\bibfnamefont {S.}~\bibnamefont
  {Palomares-Ruiz}}, \bibinfo {author} {\bibfnamefont {S.}~\bibnamefont
  {Pascoli}}, \ and\ \bibinfo {author} {\bibfnamefont {T.}~\bibnamefont
  {Schwetz}},\ }\href {\doibase 10.1088/1126-6708/2005/09/048} {\bibfield
  {journal} {\bibinfo  {journal} {JHEP}\ }\textbf {\bibinfo {volume} {09}},\
  \bibinfo {pages} {048} (\bibinfo {year} {2005})},\ \Eprint
  {http://arxiv.org/abs/hep-ph/0505216} {arXiv:hep-ph/0505216} \BibitemShut
  {NoStop}%
\bibitem [{\citenamefont {Bai}\ \emph {et~al.}(2016)\citenamefont {Bai},
  \citenamefont {Lu}, \citenamefont {Lu}, \citenamefont {Salvado},\ and\
  \citenamefont {Stefanek}}]{Bai:2015ztj}%
  \BibitemOpen
  \bibfield  {author} {\bibinfo {author} {\bibfnamefont {Y.}~\bibnamefont
  {Bai}}, \bibinfo {author} {\bibfnamefont {R.}~\bibnamefont {Lu}}, \bibinfo
  {author} {\bibfnamefont {S.}~\bibnamefont {Lu}}, \bibinfo {author}
  {\bibfnamefont {J.}~\bibnamefont {Salvado}}, \ and\ \bibinfo {author}
  {\bibfnamefont {B.~A.}\ \bibnamefont {Stefanek}},\ }\href {\doibase
  10.1103/PhysRevD.93.073004} {\bibfield  {journal} {\bibinfo  {journal} {Phys.
  Rev. D}\ }\textbf {\bibinfo {volume} {93}},\ \bibinfo {pages} {073004}
  (\bibinfo {year} {2016})},\ \Eprint {http://arxiv.org/abs/1512.05357}
  {arXiv:1512.05357 [hep-ph]} \BibitemShut {NoStop}%
\bibitem [{\citenamefont {de~Gouv\^ea}\ \emph {et~al.}(2020)\citenamefont
  {de~Gouv\^ea}, \citenamefont {Peres}, \citenamefont {Prakash},\ and\
  \citenamefont {Stenico}}]{deGouvea:2019qre}%
  \BibitemOpen
  \bibfield  {author} {\bibinfo {author} {\bibfnamefont {A.}~\bibnamefont
  {de~Gouv\^ea}}, \bibinfo {author} {\bibfnamefont {O.~L.~G.}\ \bibnamefont
  {Peres}}, \bibinfo {author} {\bibfnamefont {S.}~\bibnamefont {Prakash}}, \
  and\ \bibinfo {author} {\bibfnamefont {G.~V.}\ \bibnamefont {Stenico}},\
  }\href {\doibase 10.1007/JHEP07(2020)141} {\bibfield  {journal} {\bibinfo
  {journal} {JHEP}\ }\textbf {\bibinfo {volume} {07}},\ \bibinfo {pages} {141}
  (\bibinfo {year} {2020})},\ \Eprint {http://arxiv.org/abs/1911.01447}
  {arXiv:1911.01447 [hep-ph]} \BibitemShut {NoStop}%
\bibitem [{\citenamefont {Dentler}\ \emph {et~al.}(2020)\citenamefont
  {Dentler}, \citenamefont {Esteban}, \citenamefont {Kopp},\ and\ \citenamefont
  {Machado}}]{Dentler:2019dhz}%
  \BibitemOpen
  \bibfield  {author} {\bibinfo {author} {\bibfnamefont {M.}~\bibnamefont
  {Dentler}}, \bibinfo {author} {\bibfnamefont {I.}~\bibnamefont {Esteban}},
  \bibinfo {author} {\bibfnamefont {J.}~\bibnamefont {Kopp}}, \ and\ \bibinfo
  {author} {\bibfnamefont {P.}~\bibnamefont {Machado}},\ }\href {\doibase
  10.1103/PhysRevD.101.115013} {\bibfield  {journal} {\bibinfo  {journal}
  {Phys. Rev. D}\ }\textbf {\bibinfo {volume} {101}},\ \bibinfo {pages}
  {115013} (\bibinfo {year} {2020})},\ \Eprint
  {http://arxiv.org/abs/1911.01427} {arXiv:1911.01427 [hep-ph]} \BibitemShut
  {NoStop}%
\bibitem [{\citenamefont {Hostert}\ and\ \citenamefont
  {Pospelov}(2021)}]{Hostert:2020oui}%
  \BibitemOpen
  \bibfield  {author} {\bibinfo {author} {\bibfnamefont {M.}~\bibnamefont
  {Hostert}}\ and\ \bibinfo {author} {\bibfnamefont {M.}~\bibnamefont
  {Pospelov}},\ }\href {\doibase 10.1103/PhysRevD.104.055031} {\bibfield
  {journal} {\bibinfo  {journal} {Phys. Rev. D}\ }\textbf {\bibinfo {volume}
  {104}},\ \bibinfo {pages} {055031} (\bibinfo {year} {2021})},\ \Eprint
  {http://arxiv.org/abs/2008.11851} {arXiv:2008.11851 [hep-ph]} \BibitemShut
  {NoStop}%
\bibitem [{\citenamefont {Fischer}\ \emph {et~al.}(2020)\citenamefont
  {Fischer}, \citenamefont {Hern\'andez-Cabezudo},\ and\ \citenamefont
  {Schwetz}}]{Fischer:2019fbw}%
  \BibitemOpen
  \bibfield  {author} {\bibinfo {author} {\bibfnamefont {O.}~\bibnamefont
  {Fischer}}, \bibinfo {author} {\bibfnamefont {A.}~\bibnamefont
  {Hern\'andez-Cabezudo}}, \ and\ \bibinfo {author} {\bibfnamefont
  {T.}~\bibnamefont {Schwetz}},\ }\href {\doibase 10.1103/PhysRevD.101.075045}
  {\bibfield  {journal} {\bibinfo  {journal} {Phys. Rev. D}\ }\textbf {\bibinfo
  {volume} {101}},\ \bibinfo {pages} {075045} (\bibinfo {year} {2020})},\
  \Eprint {http://arxiv.org/abs/1909.09561} {arXiv:1909.09561 [hep-ph]}
  \BibitemShut {NoStop}%
\bibitem [{\citenamefont {Chang}\ \emph {et~al.}(2021)\citenamefont {Chang},
  \citenamefont {Chen}, \citenamefont {Ho},\ and\ \citenamefont
  {Tseng}}]{Chang:2021myh}%
  \BibitemOpen
  \bibfield  {author} {\bibinfo {author} {\bibfnamefont {C.-H.~V.}\
  \bibnamefont {Chang}}, \bibinfo {author} {\bibfnamefont {C.-R.}\ \bibnamefont
  {Chen}}, \bibinfo {author} {\bibfnamefont {S.-Y.}\ \bibnamefont {Ho}}, \ and\
  \bibinfo {author} {\bibfnamefont {S.-Y.}\ \bibnamefont {Tseng}},\ }\href
  {\doibase 10.1103/PhysRevD.104.015030} {\bibfield  {journal} {\bibinfo
  {journal} {Phys. Rev. D}\ }\textbf {\bibinfo {volume} {104}},\ \bibinfo
  {pages} {015030} (\bibinfo {year} {2021})},\ \Eprint
  {http://arxiv.org/abs/2102.05012} {arXiv:2102.05012 [hep-ph]} \BibitemShut
  {NoStop}%
\bibitem [{\citenamefont {Gninenko}(2009)}]{Gninenko:2009ks}%
  \BibitemOpen
  \bibfield  {author} {\bibinfo {author} {\bibfnamefont {S.~N.}\ \bibnamefont
  {Gninenko}},\ }\href {\doibase 10.1103/PhysRevLett.103.241802} {\bibfield
  {journal} {\bibinfo  {journal} {Phys. Rev. Lett.}\ }\textbf {\bibinfo
  {volume} {103}},\ \bibinfo {pages} {241802} (\bibinfo {year} {2009})},\
  \Eprint {http://arxiv.org/abs/0902.3802} {arXiv:0902.3802 [hep-ph]}
  \BibitemShut {NoStop}%
\bibitem [{\citenamefont {Gninenko}(2011)}]{Gninenko:2010pr}%
  \BibitemOpen
  \bibfield  {author} {\bibinfo {author} {\bibfnamefont {S.~N.}\ \bibnamefont
  {Gninenko}},\ }\href {\doibase 10.1103/PhysRevD.83.015015} {\bibfield
  {journal} {\bibinfo  {journal} {Phys. Rev. D}\ }\textbf {\bibinfo {volume}
  {83}},\ \bibinfo {pages} {015015} (\bibinfo {year} {2011})},\ \Eprint
  {http://arxiv.org/abs/1009.5536} {arXiv:1009.5536 [hep-ph]} \BibitemShut
  {NoStop}%
\bibitem [{\citenamefont {Gninenko}(2012)}]{Gninenko:2012rw}%
  \BibitemOpen
  \bibfield  {author} {\bibinfo {author} {\bibfnamefont {S.~N.}\ \bibnamefont
  {Gninenko}},\ }\href {\doibase 10.1016/j.physletb.2012.02.071} {\bibfield
  {journal} {\bibinfo  {journal} {Phys. Lett. B}\ }\textbf {\bibinfo {volume}
  {710}},\ \bibinfo {pages} {86} (\bibinfo {year} {2012})},\ \Eprint
  {http://arxiv.org/abs/1201.5194} {arXiv:1201.5194 [hep-ph]} \BibitemShut
  {NoStop}%
\bibitem [{\citenamefont {Masip}\ \emph {et~al.}(2013)\citenamefont {Masip},
  \citenamefont {Masjuan},\ and\ \citenamefont {Meloni}}]{Masip:2012ke}%
  \BibitemOpen
  \bibfield  {author} {\bibinfo {author} {\bibfnamefont {M.}~\bibnamefont
  {Masip}}, \bibinfo {author} {\bibfnamefont {P.}~\bibnamefont {Masjuan}}, \
  and\ \bibinfo {author} {\bibfnamefont {D.}~\bibnamefont {Meloni}},\ }\href
  {\doibase 10.1007/JHEP01(2013)106} {\bibfield  {journal} {\bibinfo  {journal}
  {JHEP}\ }\textbf {\bibinfo {volume} {01}},\ \bibinfo {pages} {106} (\bibinfo
  {year} {2013})},\ \Eprint {http://arxiv.org/abs/1210.1519} {arXiv:1210.1519
  [hep-ph]} \BibitemShut {NoStop}%
\bibitem [{\citenamefont {Radionov}(2013)}]{Radionov:2013mca}%
  \BibitemOpen
  \bibfield  {author} {\bibinfo {author} {\bibfnamefont {A.}~\bibnamefont
  {Radionov}},\ }\href {\doibase 10.1103/PhysRevD.88.015016} {\bibfield
  {journal} {\bibinfo  {journal} {Phys. Rev. D}\ }\textbf {\bibinfo {volume}
  {88}},\ \bibinfo {pages} {015016} (\bibinfo {year} {2013})},\ \Eprint
  {http://arxiv.org/abs/1303.4587} {arXiv:1303.4587 [hep-ph]} \BibitemShut
  {NoStop}%
\bibitem [{\citenamefont {Magill}\ \emph {et~al.}(2018)\citenamefont {Magill},
  \citenamefont {Plestid}, \citenamefont {Pospelov},\ and\ \citenamefont
  {Tsai}}]{Magill:2018jla}%
  \BibitemOpen
  \bibfield  {author} {\bibinfo {author} {\bibfnamefont {G.}~\bibnamefont
  {Magill}}, \bibinfo {author} {\bibfnamefont {R.}~\bibnamefont {Plestid}},
  \bibinfo {author} {\bibfnamefont {M.}~\bibnamefont {Pospelov}}, \ and\
  \bibinfo {author} {\bibfnamefont {Y.-D.}\ \bibnamefont {Tsai}},\ }\href
  {\doibase 10.1103/PhysRevD.98.115015} {\bibfield  {journal} {\bibinfo
  {journal} {Phys. Rev. D}\ }\textbf {\bibinfo {volume} {98}},\ \bibinfo
  {pages} {115015} (\bibinfo {year} {2018})},\ \Eprint
  {http://arxiv.org/abs/1803.03262} {arXiv:1803.03262 [hep-ph]} \BibitemShut
  {NoStop}%
\bibitem [{\citenamefont {Schwetz}\ \emph {et~al.}(2020)\citenamefont
  {Schwetz}, \citenamefont {Zhou},\ and\ \citenamefont
  {Zhu}}]{Schwetz:2020xra}%
  \BibitemOpen
  \bibfield  {author} {\bibinfo {author} {\bibfnamefont {T.}~\bibnamefont
  {Schwetz}}, \bibinfo {author} {\bibfnamefont {A.}~\bibnamefont {Zhou}}, \
  and\ \bibinfo {author} {\bibfnamefont {J.-Y.}\ \bibnamefont {Zhu}},\ }\href
  {\doibase 10.1007/JHEP07(2021)200} {\bibfield  {journal} {\bibinfo  {journal}
  {JHEP}\ }\textbf {\bibinfo {volume} {21}},\ \bibinfo {pages} {200} (\bibinfo
  {year} {2020})},\ \Eprint {http://arxiv.org/abs/2105.09699} {arXiv:2105.09699
  [hep-ph]} \BibitemShut {NoStop}%
\bibitem [{\citenamefont {Vergani}\ \emph {et~al.}(2021)\citenamefont
  {Vergani}, \citenamefont {Kamp}, \citenamefont {Diaz}, \citenamefont
  {Arg\"uelles}, \citenamefont {Conrad}, \citenamefont {Shaevitz},\ and\
  \citenamefont {Uchida}}]{Vergani:2021tgc}%
  \BibitemOpen
  \bibfield  {author} {\bibinfo {author} {\bibfnamefont {S.}~\bibnamefont
  {Vergani}}, \bibinfo {author} {\bibfnamefont {N.~W.}\ \bibnamefont {Kamp}},
  \bibinfo {author} {\bibfnamefont {A.}~\bibnamefont {Diaz}}, \bibinfo {author}
  {\bibfnamefont {C.~A.}\ \bibnamefont {Arg\"uelles}}, \bibinfo {author}
  {\bibfnamefont {J.~M.}\ \bibnamefont {Conrad}}, \bibinfo {author}
  {\bibfnamefont {M.~H.}\ \bibnamefont {Shaevitz}}, \ and\ \bibinfo {author}
  {\bibfnamefont {M.~A.}\ \bibnamefont {Uchida}},\ }\href {\doibase
  10.1103/PhysRevD.104.095005} {\bibfield  {journal} {\bibinfo  {journal}
  {Phys. Rev. D}\ }\textbf {\bibinfo {volume} {104}},\ \bibinfo {pages}
  {095005} (\bibinfo {year} {2021})},\ \Eprint
  {http://arxiv.org/abs/2105.06470} {arXiv:2105.06470 [hep-ph]} \BibitemShut
  {NoStop}%
\bibitem [{\citenamefont {Alvarez-Ruso}\ and\ \citenamefont
  {Saul-Sala}(2021)}]{Alvarez-Ruso:2021dna}%
  \BibitemOpen
  \bibfield  {author} {\bibinfo {author} {\bibfnamefont {L.}~\bibnamefont
  {Alvarez-Ruso}}\ and\ \bibinfo {author} {\bibfnamefont {E.}~\bibnamefont
  {Saul-Sala}},\ }\href {\doibase 10.1140/epjs/s11734-021-00293-9} {\bibfield
  {journal} {\bibinfo  {journal} {Eur. Phys. J. ST}\ }\textbf {\bibinfo
  {volume} {230}},\ \bibinfo {pages} {4373} (\bibinfo {year} {2021})},\ \Eprint
  {http://arxiv.org/abs/2111.02504} {arXiv:2111.02504 [hep-ph]} \BibitemShut
  {NoStop}%
\bibitem [{\citenamefont {Kamp}\ \emph
  {et~al.}(2023{\natexlab{a}})\citenamefont {Kamp}, \citenamefont {Hostert},
  \citenamefont {Schneider}, \citenamefont {Vergani}, \citenamefont
  {Arg\"uelles}, \citenamefont {Conrad}, \citenamefont {Shaevitz},\ and\
  \citenamefont {Uchida}}]{Kamp:2022bpt}%
  \BibitemOpen
  \bibfield  {author} {\bibinfo {author} {\bibfnamefont {N.~W.}\ \bibnamefont
  {Kamp}}, \bibinfo {author} {\bibfnamefont {M.}~\bibnamefont {Hostert}},
  \bibinfo {author} {\bibfnamefont {A.}~\bibnamefont {Schneider}}, \bibinfo
  {author} {\bibfnamefont {S.}~\bibnamefont {Vergani}}, \bibinfo {author}
  {\bibfnamefont {C.~A.}\ \bibnamefont {Arg\"uelles}}, \bibinfo {author}
  {\bibfnamefont {J.~M.}\ \bibnamefont {Conrad}}, \bibinfo {author}
  {\bibfnamefont {M.~H.}\ \bibnamefont {Shaevitz}}, \ and\ \bibinfo {author}
  {\bibfnamefont {M.~A.}\ \bibnamefont {Uchida}},\ }\href {\doibase
  10.1103/PhysRevD.107.055009} {\bibfield  {journal} {\bibinfo  {journal}
  {Phys. Rev. D}\ }\textbf {\bibinfo {volume} {107}},\ \bibinfo {pages}
  {055009} (\bibinfo {year} {2023}{\natexlab{a}})},\ \Eprint
  {http://arxiv.org/abs/2206.07100} {arXiv:2206.07100 [hep-ph]} \BibitemShut
  {NoStop}%
\bibitem [{\citenamefont {Bansal}\ \emph {et~al.}(2023)\citenamefont {Bansal},
  \citenamefont {Paz}, \citenamefont {Petrov}, \citenamefont {Tammaro},\ and\
  \citenamefont {Zupan}}]{Bansal:2022zpi}%
  \BibitemOpen
  \bibfield  {author} {\bibinfo {author} {\bibfnamefont {S.}~\bibnamefont
  {Bansal}}, \bibinfo {author} {\bibfnamefont {G.}~\bibnamefont {Paz}},
  \bibinfo {author} {\bibfnamefont {A.}~\bibnamefont {Petrov}}, \bibinfo
  {author} {\bibfnamefont {M.}~\bibnamefont {Tammaro}}, \ and\ \bibinfo
  {author} {\bibfnamefont {J.}~\bibnamefont {Zupan}},\ }\href {\doibase
  10.1007/JHEP05(2023)142} {\bibfield  {journal} {\bibinfo  {journal} {JHEP}\
  }\textbf {\bibinfo {volume} {05}},\ \bibinfo {pages} {142} (\bibinfo {year}
  {2023})},\ \Eprint {http://arxiv.org/abs/2210.05706} {arXiv:2210.05706
  [hep-ph]} \BibitemShut {NoStop}%
\bibitem [{\citenamefont {Dutta}\ \emph {et~al.}(2022)\citenamefont {Dutta},
  \citenamefont {Kim}, \citenamefont {Thompson}, \citenamefont {Thornton},\
  and\ \citenamefont {Van~de Water}}]{Dutta:2021cip}%
  \BibitemOpen
  \bibfield  {author} {\bibinfo {author} {\bibfnamefont {B.}~\bibnamefont
  {Dutta}}, \bibinfo {author} {\bibfnamefont {D.}~\bibnamefont {Kim}}, \bibinfo
  {author} {\bibfnamefont {A.}~\bibnamefont {Thompson}}, \bibinfo {author}
  {\bibfnamefont {R.~T.}\ \bibnamefont {Thornton}}, \ and\ \bibinfo {author}
  {\bibfnamefont {R.~G.}\ \bibnamefont {Van~de Water}},\ }\href {\doibase
  10.1103/PhysRevLett.129.111803} {\bibfield  {journal} {\bibinfo  {journal}
  {Phys. Rev. Lett.}\ }\textbf {\bibinfo {volume} {129}},\ \bibinfo {pages}
  {111803} (\bibinfo {year} {2022})},\ \Eprint
  {http://arxiv.org/abs/2110.11944} {arXiv:2110.11944 [hep-ph]} \BibitemShut
  {NoStop}%
\bibitem [{\citenamefont {Abratenko}\ \emph
  {et~al.}(2023{\natexlab{b}})\citenamefont {Abratenko} \emph
  {et~al.}}]{MicroBooNE:2022sdp}%
  \BibitemOpen
  \bibfield  {author} {\bibinfo {author} {\bibfnamefont {P.}~\bibnamefont
  {Abratenko}} \emph {et~al.} (\bibinfo {collaboration} {MicroBooNE}),\ }\href
  {\doibase 10.1103/PhysRevLett.130.011801} {\bibfield  {journal} {\bibinfo
  {journal} {Phys. Rev. Lett.}\ }\textbf {\bibinfo {volume} {130}},\ \bibinfo
  {pages} {011801} (\bibinfo {year} {2023}{\natexlab{b}})},\ \Eprint
  {http://arxiv.org/abs/2210.10216} {arXiv:2210.10216 [hep-ex]} \BibitemShut
  {NoStop}%
\bibitem [{\citenamefont {Kamp}\ \emph
  {et~al.}(2023{\natexlab{b}})\citenamefont {Kamp}, \citenamefont {Hostert},
  \citenamefont {Arg\"uelles}, \citenamefont {Conrad},\ and\ \citenamefont
  {Shaevitz}}]{Kamp:2023mjn}%
  \BibitemOpen
  \bibfield  {author} {\bibinfo {author} {\bibfnamefont {N.~W.}\ \bibnamefont
  {Kamp}}, \bibinfo {author} {\bibfnamefont {M.}~\bibnamefont {Hostert}},
  \bibinfo {author} {\bibfnamefont {C.~A.}\ \bibnamefont {Arg\"uelles}},
  \bibinfo {author} {\bibfnamefont {J.~M.}\ \bibnamefont {Conrad}}, \ and\
  \bibinfo {author} {\bibfnamefont {M.~H.}\ \bibnamefont {Shaevitz}},\ }\href
  {\doibase 10.1103/PhysRevD.107.092002} {\bibfield  {journal} {\bibinfo
  {journal} {Phys. Rev. D}\ }\textbf {\bibinfo {volume} {107}},\ \bibinfo
  {pages} {092002} (\bibinfo {year} {2023}{\natexlab{b}})},\ \Eprint
  {http://arxiv.org/abs/2301.12573} {arXiv:2301.12573 [hep-ph]} \BibitemShut
  {NoStop}%
\bibitem [{\citenamefont {Abratenko}\ \emph
  {et~al.}(2022{\natexlab{f}})\citenamefont {Abratenko} \emph
  {et~al.}}]{MicroBooNE:2022tdd}%
  \BibitemOpen
  \bibfield  {author} {\bibinfo {author} {\bibfnamefont {P.}~\bibnamefont
  {Abratenko}} \emph {et~al.} (\bibinfo {collaboration} {MicroBooNE}),\ }\href
  {\doibase 10.1103/PhysRevD.106.L051102} {\bibfield  {journal} {\bibinfo
  {journal} {Phys. Rev. D}\ }\textbf {\bibinfo {volume} {106}},\ \bibinfo
  {pages} {L051102} (\bibinfo {year} {2022}{\natexlab{f}})},\ \Eprint
  {http://arxiv.org/abs/2208.02348} {arXiv:2208.02348 [hep-ex]} \BibitemShut
  {NoStop}%
\bibitem [{\citenamefont {Abratenko}\ \emph {et~al.}(2021)\citenamefont
  {Abratenko} \emph {et~al.}}]{MicroBooNE:2021gfj}%
  \BibitemOpen
  \bibfield  {author} {\bibinfo {author} {\bibfnamefont {P.}~\bibnamefont
  {Abratenko}} \emph {et~al.} (\bibinfo {collaboration} {MicroBooNE}),\ }\href
  {\doibase 10.1103/PhysRevD.104.052002} {\bibfield  {journal} {\bibinfo
  {journal} {Phys. Rev. D}\ }\textbf {\bibinfo {volume} {104}},\ \bibinfo
  {pages} {052002} (\bibinfo {year} {2021})},\ \Eprint
  {http://arxiv.org/abs/2101.04228} {arXiv:2101.04228 [hep-ex]} \BibitemShut
  {NoStop}%
\bibitem [{\citenamefont {Abratenko}\ \emph
  {et~al.}(2022{\natexlab{g}})\citenamefont {Abratenko} \emph
  {et~al.}}]{MicroBooNE:2021fdt}%
  \BibitemOpen
  \bibfield  {author} {\bibinfo {author} {\bibfnamefont {P.}~\bibnamefont
  {Abratenko}} \emph {et~al.} (\bibinfo {collaboration} {MicroBooNE}),\ }\href
  {\doibase 10.1103/PhysRevD.105.L051102} {\bibfield  {journal} {\bibinfo
  {journal} {Phys. Rev. D}\ }\textbf {\bibinfo {volume} {105}},\ \bibinfo
  {pages} {L051102} (\bibinfo {year} {2022}{\natexlab{g}})},\ \Eprint
  {http://arxiv.org/abs/2109.06832} {arXiv:2109.06832 [hep-ex]} \BibitemShut
  {NoStop}%
\bibitem [{\citenamefont {Mention}\ \emph {et~al.}(2011)\citenamefont
  {Mention}, \citenamefont {Fechner}, \citenamefont {Lasserre}, \citenamefont
  {Mueller}, \citenamefont {Lhuillier}, \citenamefont {Cribier},\ and\
  \citenamefont {Letourneau}}]{Mention:2011rk}%
  \BibitemOpen
  \bibfield  {author} {\bibinfo {author} {\bibfnamefont {G.}~\bibnamefont
  {Mention}}, \bibinfo {author} {\bibfnamefont {M.}~\bibnamefont {Fechner}},
  \bibinfo {author} {\bibfnamefont {T.}~\bibnamefont {Lasserre}}, \bibinfo
  {author} {\bibfnamefont {T.~A.}\ \bibnamefont {Mueller}}, \bibinfo {author}
  {\bibfnamefont {D.}~\bibnamefont {Lhuillier}}, \bibinfo {author}
  {\bibfnamefont {M.}~\bibnamefont {Cribier}}, \ and\ \bibinfo {author}
  {\bibfnamefont {A.}~\bibnamefont {Letourneau}},\ }\href {\doibase
  10.1103/PhysRevD.83.073006} {\bibfield  {journal} {\bibinfo  {journal} {Phys.
  Rev. D}\ }\textbf {\bibinfo {volume} {83}},\ \bibinfo {pages} {073006}
  (\bibinfo {year} {2011})},\ \Eprint {http://arxiv.org/abs/1101.2755}
  {arXiv:1101.2755 [hep-ex]} \BibitemShut {NoStop}%
\bibitem [{\citenamefont {Mueller}\ \emph {et~al.}(2011)\citenamefont {Mueller}
  \emph {et~al.}}]{Mueller:2011nm}%
  \BibitemOpen
  \bibfield  {author} {\bibinfo {author} {\bibfnamefont {T.~A.}\ \bibnamefont
  {Mueller}} \emph {et~al.},\ }\href {\doibase 10.1103/PhysRevC.83.054615}
  {\bibfield  {journal} {\bibinfo  {journal} {Phys. Rev. C}\ }\textbf {\bibinfo
  {volume} {83}},\ \bibinfo {pages} {054615} (\bibinfo {year} {2011})},\
  \Eprint {http://arxiv.org/abs/1101.2663} {arXiv:1101.2663 [hep-ex]}
  \BibitemShut {NoStop}%
\bibitem [{\citenamefont {Huber}(2011)}]{Huber:2011wv}%
  \BibitemOpen
  \bibfield  {author} {\bibinfo {author} {\bibfnamefont {P.}~\bibnamefont
  {Huber}},\ }\href {\doibase 10.1103/PhysRevC.85.029901} {\bibfield  {journal}
  {\bibinfo  {journal} {Phys. Rev. C}\ }\textbf {\bibinfo {volume} {84}},\
  \bibinfo {pages} {024617} (\bibinfo {year} {2011})},\ \bibinfo {note}
  {[Erratum: Phys.Rev.C 85, 029901 (2012)]},\ \Eprint
  {http://arxiv.org/abs/1106.0687} {arXiv:1106.0687 [hep-ph]} \BibitemShut
  {NoStop}%
\bibitem [{\citenamefont {Giunti}\ \emph
  {et~al.}(2022{\natexlab{a}})\citenamefont {Giunti}, \citenamefont {Li},
  \citenamefont {Ternes},\ and\ \citenamefont {Xin}}]{Giunti:2021kab}%
  \BibitemOpen
  \bibfield  {author} {\bibinfo {author} {\bibfnamefont {C.}~\bibnamefont
  {Giunti}}, \bibinfo {author} {\bibfnamefont {Y.~F.}\ \bibnamefont {Li}},
  \bibinfo {author} {\bibfnamefont {C.~A.}\ \bibnamefont {Ternes}}, \ and\
  \bibinfo {author} {\bibfnamefont {Z.}~\bibnamefont {Xin}},\ }\href {\doibase
  10.1016/j.physletb.2022.137054} {\bibfield  {journal} {\bibinfo  {journal}
  {Phys. Lett. B}\ }\textbf {\bibinfo {volume} {829}},\ \bibinfo {pages}
  {137054} (\bibinfo {year} {2022}{\natexlab{a}})},\ \Eprint
  {http://arxiv.org/abs/2110.06820} {arXiv:2110.06820 [hep-ph]} \BibitemShut
  {NoStop}%
\bibitem [{\citenamefont {Kopeikin}\ \emph {et~al.}(2021)\citenamefont
  {Kopeikin}, \citenamefont {Skorokhvatov},\ and\ \citenamefont
  {Titov}}]{Kopeikin:2021ugh}%
  \BibitemOpen
  \bibfield  {author} {\bibinfo {author} {\bibfnamefont {V.}~\bibnamefont
  {Kopeikin}}, \bibinfo {author} {\bibfnamefont {M.}~\bibnamefont
  {Skorokhvatov}}, \ and\ \bibinfo {author} {\bibfnamefont {O.}~\bibnamefont
  {Titov}},\ }\href {\doibase 10.1103/PhysRevD.104.L071301} {\bibfield
  {journal} {\bibinfo  {journal} {Phys. Rev. D}\ }\textbf {\bibinfo {volume}
  {104}},\ \bibinfo {pages} {L071301} (\bibinfo {year} {2021})},\ \Eprint
  {http://arxiv.org/abs/2103.01684} {arXiv:2103.01684 [nucl-ex]} \BibitemShut
  {NoStop}%
\bibitem [{\citenamefont {Schreckenbach}\ \emph {et~al.}(1981)\citenamefont
  {Schreckenbach}, \citenamefont {Faust}, \citenamefont {von Feilitzsch},
  \citenamefont {Hahn}, \citenamefont {Hawerkamp},\ and\ \citenamefont
  {Vuilleumier}}]{Schreckenbach:1981wlm}%
  \BibitemOpen
  \bibfield  {author} {\bibinfo {author} {\bibfnamefont {K.}~\bibnamefont
  {Schreckenbach}}, \bibinfo {author} {\bibfnamefont {H.~R.}\ \bibnamefont
  {Faust}}, \bibinfo {author} {\bibfnamefont {F.}~\bibnamefont {von
  Feilitzsch}}, \bibinfo {author} {\bibfnamefont {A.~A.}\ \bibnamefont {Hahn}},
  \bibinfo {author} {\bibfnamefont {K.}~\bibnamefont {Hawerkamp}}, \ and\
  \bibinfo {author} {\bibfnamefont {J.~L.}\ \bibnamefont {Vuilleumier}},\
  }\href {\doibase 10.1016/0370-2693(81)91120-5} {\bibfield  {journal}
  {\bibinfo  {journal} {Phys. Lett. B}\ }\textbf {\bibinfo {volume} {99}},\
  \bibinfo {pages} {251} (\bibinfo {year} {1981})}\BibitemShut {NoStop}%
\bibitem [{\citenamefont {Von~Feilitzsch}\ \emph {et~al.}(1982)\citenamefont
  {Von~Feilitzsch}, \citenamefont {Hahn},\ and\ \citenamefont
  {Schreckenbach}}]{VonFeilitzsch:1982jw}%
  \BibitemOpen
  \bibfield  {author} {\bibinfo {author} {\bibfnamefont {F.}~\bibnamefont
  {Von~Feilitzsch}}, \bibinfo {author} {\bibfnamefont {A.~A.}\ \bibnamefont
  {Hahn}}, \ and\ \bibinfo {author} {\bibfnamefont {K.}~\bibnamefont
  {Schreckenbach}},\ }\href {\doibase 10.1016/0370-2693(82)90622-0} {\bibfield
  {journal} {\bibinfo  {journal} {Phys. Lett. B}\ }\textbf {\bibinfo {volume}
  {118}},\ \bibinfo {pages} {162} (\bibinfo {year} {1982})}\BibitemShut
  {NoStop}%
\bibitem [{\citenamefont {Schreckenbach}\ \emph {et~al.}(1985)\citenamefont
  {Schreckenbach}, \citenamefont {Colvin}, \citenamefont {Gelletly},\ and\
  \citenamefont {Von~Feilitzsch}}]{Schreckenbach:1985ep}%
  \BibitemOpen
  \bibfield  {author} {\bibinfo {author} {\bibfnamefont {K.}~\bibnamefont
  {Schreckenbach}}, \bibinfo {author} {\bibfnamefont {G.}~\bibnamefont
  {Colvin}}, \bibinfo {author} {\bibfnamefont {W.}~\bibnamefont {Gelletly}}, \
  and\ \bibinfo {author} {\bibfnamefont {F.}~\bibnamefont {Von~Feilitzsch}},\
  }\href {\doibase 10.1016/0370-2693(85)91337-1} {\bibfield  {journal}
  {\bibinfo  {journal} {Phys. Lett. B}\ }\textbf {\bibinfo {volume} {160}},\
  \bibinfo {pages} {325} (\bibinfo {year} {1985})}\BibitemShut {NoStop}%
\bibitem [{\citenamefont {Hahn}\ \emph {et~al.}(1989)\citenamefont {Hahn},
  \citenamefont {Schreckenbach}, \citenamefont {Colvin}, \citenamefont
  {Krusche}, \citenamefont {Gelletly},\ and\ \citenamefont
  {Von~Feilitzsch}}]{Hahn:1989zr}%
  \BibitemOpen
  \bibfield  {author} {\bibinfo {author} {\bibfnamefont {A.~A.}\ \bibnamefont
  {Hahn}}, \bibinfo {author} {\bibfnamefont {K.}~\bibnamefont {Schreckenbach}},
  \bibinfo {author} {\bibfnamefont {G.}~\bibnamefont {Colvin}}, \bibinfo
  {author} {\bibfnamefont {B.}~\bibnamefont {Krusche}}, \bibinfo {author}
  {\bibfnamefont {W.}~\bibnamefont {Gelletly}}, \ and\ \bibinfo {author}
  {\bibfnamefont {F.}~\bibnamefont {Von~Feilitzsch}},\ }\href {\doibase
  10.1016/0370-2693(89)91598-0} {\bibfield  {journal} {\bibinfo  {journal}
  {Phys. Lett. B}\ }\textbf {\bibinfo {volume} {218}},\ \bibinfo {pages} {365}
  (\bibinfo {year} {1989})}\BibitemShut {NoStop}%
\bibitem [{\citenamefont {Augier}\ \emph {et~al.}(2023)\citenamefont {Augier}
  \emph {et~al.}}]{Ricochet:2022pzj}%
  \BibitemOpen
  \bibfield  {author} {\bibinfo {author} {\bibfnamefont {C.}~\bibnamefont
  {Augier}} \emph {et~al.} (\bibinfo {collaboration} {Ricochet}),\ }\href
  {\doibase 10.1140/epjc/s10052-022-11150-x} {\bibfield  {journal} {\bibinfo
  {journal} {Eur. Phys. J. C}\ }\textbf {\bibinfo {volume} {83}},\ \bibinfo
  {pages} {20} (\bibinfo {year} {2023})},\ \Eprint
  {http://arxiv.org/abs/2208.01760} {arXiv:2208.01760 [astro-ph.IM]}
  \BibitemShut {NoStop}%
\bibitem [{\citenamefont {Almaz\'an}\ \emph {et~al.}(2023)\citenamefont
  {Almaz\'an} \emph {et~al.}}]{STEREO:2022nzk}%
  \BibitemOpen
  \bibfield  {author} {\bibinfo {author} {\bibfnamefont {H.}~\bibnamefont
  {Almaz\'an}} \emph {et~al.} (\bibinfo {collaboration} {STEREO}),\ }\href
  {\doibase 10.1038/s41586-022-05568-2} {\bibfield  {journal} {\bibinfo
  {journal} {Nature}\ }\textbf {\bibinfo {volume} {613}},\ \bibinfo {pages}
  {257} (\bibinfo {year} {2023})},\ \Eprint {http://arxiv.org/abs/2210.07664}
  {arXiv:2210.07664 [hep-ex]} \BibitemShut {NoStop}%
\bibitem [{\citenamefont {Ashenfelter}\ \emph {et~al.}(2018)\citenamefont
  {Ashenfelter} \emph {et~al.}}]{PROSPECT:2018dtt}%
  \BibitemOpen
  \bibfield  {author} {\bibinfo {author} {\bibfnamefont {J.}~\bibnamefont
  {Ashenfelter}} \emph {et~al.} (\bibinfo {collaboration} {PROSPECT}),\ }\href
  {\doibase 10.1103/PhysRevLett.121.251802} {\bibfield  {journal} {\bibinfo
  {journal} {Phys. Rev. Lett.}\ }\textbf {\bibinfo {volume} {121}},\ \bibinfo
  {pages} {251802} (\bibinfo {year} {2018})},\ \Eprint
  {http://arxiv.org/abs/1806.02784} {arXiv:1806.02784 [hep-ex]} \BibitemShut
  {NoStop}%
\bibitem [{\citenamefont {Andriamirado}\ \emph {et~al.}(2021)\citenamefont
  {Andriamirado} \emph {et~al.}}]{PROSPECT:2020sxr}%
  \BibitemOpen
  \bibfield  {author} {\bibinfo {author} {\bibfnamefont {M.}~\bibnamefont
  {Andriamirado}} \emph {et~al.} (\bibinfo {collaboration} {PROSPECT}),\ }\href
  {\doibase 10.1103/PhysRevD.103.032001} {\bibfield  {journal} {\bibinfo
  {journal} {Phys. Rev. D}\ }\textbf {\bibinfo {volume} {103}},\ \bibinfo
  {pages} {032001} (\bibinfo {year} {2021})},\ \Eprint
  {http://arxiv.org/abs/2006.11210} {arXiv:2006.11210 [hep-ex]} \BibitemShut
  {NoStop}%
\bibitem [{\citenamefont {Almaz\'an}\ \emph {et~al.}(2018)\citenamefont
  {Almaz\'an} \emph {et~al.}}]{STEREO:2018rfh}%
  \BibitemOpen
  \bibfield  {author} {\bibinfo {author} {\bibfnamefont {H.}~\bibnamefont
  {Almaz\'an}} \emph {et~al.} (\bibinfo {collaboration} {STEREO}),\ }\href
  {\doibase 10.1103/PhysRevLett.121.161801} {\bibfield  {journal} {\bibinfo
  {journal} {Phys. Rev. Lett.}\ }\textbf {\bibinfo {volume} {121}},\ \bibinfo
  {pages} {161801} (\bibinfo {year} {2018})},\ \Eprint
  {http://arxiv.org/abs/1806.02096} {arXiv:1806.02096 [hep-ex]} \BibitemShut
  {NoStop}%
\bibitem [{\citenamefont {Almaz\'an}\ \emph {et~al.}(2020)\citenamefont
  {Almaz\'an} \emph {et~al.}}]{STEREO:2019ztb}%
  \BibitemOpen
  \bibfield  {author} {\bibinfo {author} {\bibfnamefont {H.}~\bibnamefont
  {Almaz\'an}} \emph {et~al.} (\bibinfo {collaboration} {STEREO}),\ }\href
  {\doibase 10.1103/PhysRevD.102.052002} {\bibfield  {journal} {\bibinfo
  {journal} {Phys. Rev. D}\ }\textbf {\bibinfo {volume} {102}},\ \bibinfo
  {pages} {052002} (\bibinfo {year} {2020})},\ \Eprint
  {http://arxiv.org/abs/1912.06582} {arXiv:1912.06582 [hep-ex]} \BibitemShut
  {NoStop}%
\bibitem [{\citenamefont {Alekseev}\ \emph {et~al.}(2018)\citenamefont
  {Alekseev} \emph {et~al.}}]{DANSS:2018fnn}%
  \BibitemOpen
  \bibfield  {author} {\bibinfo {author} {\bibfnamefont {I.}~\bibnamefont
  {Alekseev}} \emph {et~al.} (\bibinfo {collaboration} {DANSS}),\ }\href
  {\doibase 10.1016/j.physletb.2018.10.038} {\bibfield  {journal} {\bibinfo
  {journal} {Phys. Lett. B}\ }\textbf {\bibinfo {volume} {787}},\ \bibinfo
  {pages} {56} (\bibinfo {year} {2018})},\ \Eprint
  {http://arxiv.org/abs/1804.04046} {arXiv:1804.04046 [hep-ex]} \BibitemShut
  {NoStop}%
\bibitem [{\citenamefont {Atif}\ \emph {et~al.}(2022)\citenamefont {Atif} \emph
  {et~al.}}]{RENO:2020hva}%
  \BibitemOpen
  \bibfield  {author} {\bibinfo {author} {\bibfnamefont {Z.}~\bibnamefont
  {Atif}} \emph {et~al.} (\bibinfo {collaboration} {RENO, NEOS}),\ }\href
  {\doibase 10.1103/PhysRevD.105.L111101} {\bibfield  {journal} {\bibinfo
  {journal} {Phys. Rev. D}\ }\textbf {\bibinfo {volume} {105}},\ \bibinfo
  {pages} {L111101} (\bibinfo {year} {2022})},\ \Eprint
  {http://arxiv.org/abs/2011.00896} {arXiv:2011.00896 [hep-ex]} \BibitemShut
  {NoStop}%
\bibitem [{\citenamefont {Serebrov}\ \emph {et~al.}(2019)\citenamefont
  {Serebrov} \emph {et~al.}}]{NEUTRINO-4:2018huq}%
  \BibitemOpen
  \bibfield  {author} {\bibinfo {author} {\bibfnamefont {A.~P.}\ \bibnamefont
  {Serebrov}} \emph {et~al.} (\bibinfo {collaboration} {NEUTRINO-4}),\ }\href
  {\doibase 10.1134/S0021364019040040} {\bibfield  {journal} {\bibinfo
  {journal} {Pisma Zh. Eksp. Teor. Fiz.}\ }\textbf {\bibinfo {volume} {109}},\
  \bibinfo {pages} {209} (\bibinfo {year} {2019})},\ \Eprint
  {http://arxiv.org/abs/1809.10561} {arXiv:1809.10561 [hep-ex]} \BibitemShut
  {NoStop}%
\bibitem [{\citenamefont {Andriamirado}\ \emph {et~al.}(2020)\citenamefont
  {Andriamirado} \emph {et~al.}}]{PROSPECT:2020raz}%
  \BibitemOpen
  \bibfield  {author} {\bibinfo {author} {\bibfnamefont {M.}~\bibnamefont
  {Andriamirado}} \emph {et~al.} (\bibinfo {collaboration} {PROSPECT,
  STEREO}),\ }\href@noop {} {\  (\bibinfo {year} {2020})},\ \Eprint
  {http://arxiv.org/abs/2006.13147} {arXiv:2006.13147 [hep-ex]} \BibitemShut
  {NoStop}%
\bibitem [{\citenamefont {Giunti}\ \emph {et~al.}(2021)\citenamefont {Giunti},
  \citenamefont {Li}, \citenamefont {Ternes},\ and\ \citenamefont
  {Zhang}}]{Giunti:2021iti}%
  \BibitemOpen
  \bibfield  {author} {\bibinfo {author} {\bibfnamefont {C.}~\bibnamefont
  {Giunti}}, \bibinfo {author} {\bibfnamefont {Y.~F.}\ \bibnamefont {Li}},
  \bibinfo {author} {\bibfnamefont {C.~A.}\ \bibnamefont {Ternes}}, \ and\
  \bibinfo {author} {\bibfnamefont {Y.~Y.}\ \bibnamefont {Zhang}},\ }\href
  {\doibase 10.1016/j.physletb.2021.136214} {\bibfield  {journal} {\bibinfo
  {journal} {Phys. Lett. B}\ }\textbf {\bibinfo {volume} {816}},\ \bibinfo
  {pages} {136214} (\bibinfo {year} {2021})},\ \Eprint
  {http://arxiv.org/abs/2101.06785} {arXiv:2101.06785 [hep-ph]} \BibitemShut
  {NoStop}%
\bibitem [{\citenamefont {Acero}\ \emph {et~al.}(2008)\citenamefont {Acero},
  \citenamefont {Giunti},\ and\ \citenamefont {Laveder}}]{Acero:2007su}%
  \BibitemOpen
  \bibfield  {author} {\bibinfo {author} {\bibfnamefont {M.~A.}\ \bibnamefont
  {Acero}}, \bibinfo {author} {\bibfnamefont {C.}~\bibnamefont {Giunti}}, \
  and\ \bibinfo {author} {\bibfnamefont {M.}~\bibnamefont {Laveder}},\ }\href
  {\doibase 10.1103/PhysRevD.78.073009} {\bibfield  {journal} {\bibinfo
  {journal} {Phys. Rev. D}\ }\textbf {\bibinfo {volume} {78}},\ \bibinfo
  {pages} {073009} (\bibinfo {year} {2008})},\ \Eprint
  {http://arxiv.org/abs/0711.4222} {arXiv:0711.4222 [hep-ph]} \BibitemShut
  {NoStop}%
\bibitem [{\citenamefont {Giunti}\ and\ \citenamefont
  {Laveder}(2011{\natexlab{a}})}]{Giunti:2010zu}%
  \BibitemOpen
  \bibfield  {author} {\bibinfo {author} {\bibfnamefont {C.}~\bibnamefont
  {Giunti}}\ and\ \bibinfo {author} {\bibfnamefont {M.}~\bibnamefont
  {Laveder}},\ }\href {\doibase 10.1103/PhysRevC.83.065504} {\bibfield
  {journal} {\bibinfo  {journal} {Phys. Rev. C}\ }\textbf {\bibinfo {volume}
  {83}},\ \bibinfo {pages} {065504} (\bibinfo {year} {2011}{\natexlab{a}})},\
  \Eprint {http://arxiv.org/abs/1006.3244} {arXiv:1006.3244 [hep-ph]}
  \BibitemShut {NoStop}%
\bibitem [{\citenamefont {Barinov}\ \emph {et~al.}(2022)\citenamefont {Barinov}
  \emph {et~al.}}]{Barinov:2021asz}%
  \BibitemOpen
  \bibfield  {author} {\bibinfo {author} {\bibfnamefont {V.~V.}\ \bibnamefont
  {Barinov}} \emph {et~al.},\ }\href {\doibase 10.1103/PhysRevLett.128.232501}
  {\bibfield  {journal} {\bibinfo  {journal} {Phys. Rev. Lett.}\ }\textbf
  {\bibinfo {volume} {128}},\ \bibinfo {pages} {232501} (\bibinfo {year}
  {2022})},\ \Eprint {http://arxiv.org/abs/2109.11482} {arXiv:2109.11482
  [nucl-ex]} \BibitemShut {NoStop}%
\bibitem [{\citenamefont {Goldhagen}\ \emph {et~al.}(2022)\citenamefont
  {Goldhagen}, \citenamefont {Maltoni}, \citenamefont {Reichard},\ and\
  \citenamefont {Schwetz}}]{Goldhagen:2021kxe}%
  \BibitemOpen
  \bibfield  {author} {\bibinfo {author} {\bibfnamefont {K.}~\bibnamefont
  {Goldhagen}}, \bibinfo {author} {\bibfnamefont {M.}~\bibnamefont {Maltoni}},
  \bibinfo {author} {\bibfnamefont {S.~E.}\ \bibnamefont {Reichard}}, \ and\
  \bibinfo {author} {\bibfnamefont {T.}~\bibnamefont {Schwetz}},\ }\href
  {\doibase 10.1140/epjc/s10052-022-10052-2} {\bibfield  {journal} {\bibinfo
  {journal} {Eur. Phys. J. C}\ }\textbf {\bibinfo {volume} {82}},\ \bibinfo
  {pages} {116} (\bibinfo {year} {2022})},\ \Eprint
  {http://arxiv.org/abs/2109.14898} {arXiv:2109.14898 [hep-ph]} \BibitemShut
  {NoStop}%
\bibitem [{\citenamefont {Giunti}\ \emph
  {et~al.}(2022{\natexlab{b}})\citenamefont {Giunti}, \citenamefont {Li},
  \citenamefont {Ternes}, \citenamefont {Tyagi},\ and\ \citenamefont
  {Xin}}]{Giunti:2022btk}%
  \BibitemOpen
  \bibfield  {author} {\bibinfo {author} {\bibfnamefont {C.}~\bibnamefont
  {Giunti}}, \bibinfo {author} {\bibfnamefont {Y.~F.}\ \bibnamefont {Li}},
  \bibinfo {author} {\bibfnamefont {C.~A.}\ \bibnamefont {Ternes}}, \bibinfo
  {author} {\bibfnamefont {O.}~\bibnamefont {Tyagi}}, \ and\ \bibinfo {author}
  {\bibfnamefont {Z.}~\bibnamefont {Xin}},\ }\href {\doibase
  10.1007/JHEP10(2022)164} {\bibfield  {journal} {\bibinfo  {journal} {JHEP}\
  }\textbf {\bibinfo {volume} {10}},\ \bibinfo {pages} {164} (\bibinfo {year}
  {2022}{\natexlab{b}})},\ \Eprint {http://arxiv.org/abs/2209.00916}
  {arXiv:2209.00916 [hep-ph]} \BibitemShut {NoStop}%
\bibitem [{\citenamefont {Giunti}\ \emph {et~al.}(2023)\citenamefont {Giunti},
  \citenamefont {Li}, \citenamefont {Ternes},\ and\ \citenamefont
  {Xin}}]{Giunti:2022xat}%
  \BibitemOpen
  \bibfield  {author} {\bibinfo {author} {\bibfnamefont {C.}~\bibnamefont
  {Giunti}}, \bibinfo {author} {\bibfnamefont {Y.~F.}\ \bibnamefont {Li}},
  \bibinfo {author} {\bibfnamefont {C.~A.}\ \bibnamefont {Ternes}}, \ and\
  \bibinfo {author} {\bibfnamefont {Z.}~\bibnamefont {Xin}},\ }\href {\doibase
  10.1016/j.physletb.2023.137983} {\bibfield  {journal} {\bibinfo  {journal}
  {Phys. Lett. B}\ }\textbf {\bibinfo {volume} {842}},\ \bibinfo {pages}
  {137983} (\bibinfo {year} {2023})},\ \Eprint
  {http://arxiv.org/abs/2212.09722} {arXiv:2212.09722 [hep-ph]} \BibitemShut
  {NoStop}%
\bibitem [{\citenamefont {Brdar}\ \emph {et~al.}(2023)\citenamefont {Brdar},
  \citenamefont {Gehrlein},\ and\ \citenamefont {Kopp}}]{Brdar:2023cms}%
  \BibitemOpen
  \bibfield  {author} {\bibinfo {author} {\bibfnamefont {V.}~\bibnamefont
  {Brdar}}, \bibinfo {author} {\bibfnamefont {J.}~\bibnamefont {Gehrlein}}, \
  and\ \bibinfo {author} {\bibfnamefont {J.}~\bibnamefont {Kopp}},\ }\href
  {\doibase 10.1007/JHEP05(2023)143} {\bibfield  {journal} {\bibinfo  {journal}
  {JHEP}\ }\textbf {\bibinfo {volume} {05}},\ \bibinfo {pages} {143} (\bibinfo
  {year} {2023})},\ \Eprint {http://arxiv.org/abs/2303.05528} {arXiv:2303.05528
  [hep-ph]} \BibitemShut {NoStop}%
\bibitem [{\citenamefont {Berryman}\ \emph {et~al.}(2022)\citenamefont
  {Berryman}, \citenamefont {Coloma}, \citenamefont {Huber}, \citenamefont
  {Schwetz},\ and\ \citenamefont {Zhou}}]{Berryman:2021yan}%
  \BibitemOpen
  \bibfield  {author} {\bibinfo {author} {\bibfnamefont {J.~M.}\ \bibnamefont
  {Berryman}}, \bibinfo {author} {\bibfnamefont {P.}~\bibnamefont {Coloma}},
  \bibinfo {author} {\bibfnamefont {P.}~\bibnamefont {Huber}}, \bibinfo
  {author} {\bibfnamefont {T.}~\bibnamefont {Schwetz}}, \ and\ \bibinfo
  {author} {\bibfnamefont {A.}~\bibnamefont {Zhou}},\ }\href {\doibase
  10.1007/JHEP02(2022)055} {\bibfield  {journal} {\bibinfo  {journal} {JHEP}\
  }\textbf {\bibinfo {volume} {02}},\ \bibinfo {pages} {055} (\bibinfo {year}
  {2022})},\ \Eprint {http://arxiv.org/abs/2111.12530} {arXiv:2111.12530
  [hep-ph]} \BibitemShut {NoStop}%
\bibitem [{\citenamefont {Dydak}\ \emph {et~al.}(1984)\citenamefont {Dydak}
  \emph {et~al.}}]{Dydak:1983zq}%
  \BibitemOpen
  \bibfield  {author} {\bibinfo {author} {\bibfnamefont {F.}~\bibnamefont
  {Dydak}} \emph {et~al.},\ }\href {\doibase 10.1016/0370-2693(84)90688-9}
  {\bibfield  {journal} {\bibinfo  {journal} {Phys. Lett. B}\ }\textbf
  {\bibinfo {volume} {134}},\ \bibinfo {pages} {281} (\bibinfo {year}
  {1984})}\BibitemShut {NoStop}%
\bibitem [{\citenamefont {Louis}(2018)}]{Louis:2018yeg}%
  \BibitemOpen
  \bibfield  {author} {\bibinfo {author} {\bibfnamefont {W.~C.}\ \bibnamefont
  {Louis}},\ }\href@noop {} {\  (\bibinfo {year} {2018})},\ \Eprint
  {http://arxiv.org/abs/1803.11488} {arXiv:1803.11488 [hep-ex]} \BibitemShut
  {NoStop}%
\bibitem [{\citenamefont {Eskut}\ \emph {et~al.}(2008)\citenamefont {Eskut}
  \emph {et~al.}}]{CHORUS:2007wlo}%
  \BibitemOpen
  \bibfield  {author} {\bibinfo {author} {\bibfnamefont {E.}~\bibnamefont
  {Eskut}} \emph {et~al.} (\bibinfo {collaboration} {CHORUS}),\ }\href
  {\doibase 10.1016/j.nuclphysb.2007.10.023} {\bibfield  {journal} {\bibinfo
  {journal} {Nucl. Phys. B}\ }\textbf {\bibinfo {volume} {793}},\ \bibinfo
  {pages} {326} (\bibinfo {year} {2008})},\ \Eprint
  {http://arxiv.org/abs/0710.3361} {arXiv:0710.3361 [hep-ex]} \BibitemShut
  {NoStop}%
\bibitem [{\citenamefont {Agafonova}\ \emph {et~al.}(2015)\citenamefont
  {Agafonova} \emph {et~al.}}]{OPERA:2015zci}%
  \BibitemOpen
  \bibfield  {author} {\bibinfo {author} {\bibfnamefont {N.}~\bibnamefont
  {Agafonova}} \emph {et~al.} (\bibinfo {collaboration} {OPERA}),\ }\href
  {\doibase 10.1007/JHEP06(2015)069} {\bibfield  {journal} {\bibinfo  {journal}
  {JHEP}\ }\textbf {\bibinfo {volume} {06}},\ \bibinfo {pages} {069} (\bibinfo
  {year} {2015})},\ \Eprint {http://arxiv.org/abs/1503.01876} {arXiv:1503.01876
  [hep-ex]} \BibitemShut {NoStop}%
\bibitem [{\citenamefont {Smithers}\ \emph {et~al.}(2022)\citenamefont
  {Smithers}, \citenamefont {Jones}, \citenamefont {Arg\"uelles}, \citenamefont
  {Conrad},\ and\ \citenamefont {Diaz}}]{Smithers:2021orb}%
  \BibitemOpen
  \bibfield  {author} {\bibinfo {author} {\bibfnamefont {B.~R.}\ \bibnamefont
  {Smithers}}, \bibinfo {author} {\bibfnamefont {B.~J.~P.}\ \bibnamefont
  {Jones}}, \bibinfo {author} {\bibfnamefont {C.~A.}\ \bibnamefont
  {Arg\"uelles}}, \bibinfo {author} {\bibfnamefont {J.~M.}\ \bibnamefont
  {Conrad}}, \ and\ \bibinfo {author} {\bibfnamefont {A.}~\bibnamefont
  {Diaz}},\ }\href {\doibase 10.1103/PhysRevD.105.052001} {\bibfield  {journal}
  {\bibinfo  {journal} {Phys. Rev. D}\ }\textbf {\bibinfo {volume} {105}},\
  \bibinfo {pages} {052001} (\bibinfo {year} {2022})},\ \Eprint
  {http://arxiv.org/abs/2111.08722} {arXiv:2111.08722 [hep-ph]} \BibitemShut
  {NoStop}%
\bibitem [{\citenamefont {Aiello}\ \emph {et~al.}(2021)\citenamefont {Aiello}
  \emph {et~al.}}]{KM3NeT:2021uez}%
  \BibitemOpen
  \bibfield  {author} {\bibinfo {author} {\bibfnamefont {S.}~\bibnamefont
  {Aiello}} \emph {et~al.} (\bibinfo {collaboration} {KM3NeT}),\ }\href
  {\doibase 10.1007/JHEP10(2021)180} {\bibfield  {journal} {\bibinfo  {journal}
  {JHEP}\ }\textbf {\bibinfo {volume} {10}},\ \bibinfo {pages} {180} (\bibinfo
  {year} {2021})},\ \Eprint {http://arxiv.org/abs/2107.00344} {arXiv:2107.00344
  [hep-ex]} \BibitemShut {NoStop}%
\bibitem [{\citenamefont {Adamson}\ \emph {et~al.}(2017)\citenamefont {Adamson}
  \emph {et~al.}}]{NOvA:2017geg}%
  \BibitemOpen
  \bibfield  {author} {\bibinfo {author} {\bibfnamefont {P.}~\bibnamefont
  {Adamson}} \emph {et~al.} (\bibinfo {collaboration} {NOvA}),\ }\href
  {\doibase 10.1103/PhysRevD.96.072006} {\bibfield  {journal} {\bibinfo
  {journal} {Phys. Rev. D}\ }\textbf {\bibinfo {volume} {96}},\ \bibinfo
  {pages} {072006} (\bibinfo {year} {2017})},\ \Eprint
  {http://arxiv.org/abs/1706.04592} {arXiv:1706.04592 [hep-ex]} \BibitemShut
  {NoStop}%
\bibitem [{\citenamefont {Acero}\ \emph {et~al.}(2021)\citenamefont {Acero}
  \emph {et~al.}}]{NOvA:2021smv}%
  \BibitemOpen
  \bibfield  {author} {\bibinfo {author} {\bibfnamefont {M.~A.}\ \bibnamefont
  {Acero}} \emph {et~al.} (\bibinfo {collaboration} {NOvA}),\ }\href {\doibase
  10.1103/PhysRevLett.127.201801} {\bibfield  {journal} {\bibinfo  {journal}
  {Phys. Rev. Lett.}\ }\textbf {\bibinfo {volume} {127}},\ \bibinfo {pages}
  {201801} (\bibinfo {year} {2021})},\ \Eprint
  {http://arxiv.org/abs/2106.04673} {arXiv:2106.04673 [hep-ex]} \BibitemShut
  {NoStop}%
\bibitem [{\citenamefont {Furmanski}\ and\ \citenamefont
  {Hilgenberg}(2021)}]{Furmanski:2020smg}%
  \BibitemOpen
  \bibfield  {author} {\bibinfo {author} {\bibfnamefont {A.~P.}\ \bibnamefont
  {Furmanski}}\ and\ \bibinfo {author} {\bibfnamefont {C.}~\bibnamefont
  {Hilgenberg}},\ }\href {\doibase 10.1103/PhysRevD.103.112011} {\bibfield
  {journal} {\bibinfo  {journal} {Phys. Rev. D}\ }\textbf {\bibinfo {volume}
  {103}},\ \bibinfo {pages} {112011} (\bibinfo {year} {2021})},\ \Eprint
  {http://arxiv.org/abs/2012.09788} {arXiv:2012.09788 [hep-ex]} \BibitemShut
  {NoStop}%
\bibitem [{\citenamefont {Aker}\ \emph {et~al.}(2022)\citenamefont {Aker} \emph
  {et~al.}}]{KATRIN:2022ith}%
  \BibitemOpen
  \bibfield  {author} {\bibinfo {author} {\bibfnamefont {M.}~\bibnamefont
  {Aker}} \emph {et~al.} (\bibinfo {collaboration} {KATRIN}),\ }\href {\doibase
  10.1103/PhysRevD.105.072004} {\bibfield  {journal} {\bibinfo  {journal}
  {Phys. Rev. D}\ }\textbf {\bibinfo {volume} {105}},\ \bibinfo {pages}
  {072004} (\bibinfo {year} {2022})},\ \Eprint
  {http://arxiv.org/abs/2201.11593} {arXiv:2201.11593 [hep-ex]} \BibitemShut
  {NoStop}%
\bibitem [{\citenamefont {Aghanim}\ \emph {et~al.}(2020)\citenamefont {Aghanim}
  \emph {et~al.}}]{Planck:2018vyg}%
  \BibitemOpen
  \bibfield  {author} {\bibinfo {author} {\bibfnamefont {N.}~\bibnamefont
  {Aghanim}} \emph {et~al.} (\bibinfo {collaboration} {Planck}),\ }\href
  {\doibase 10.1051/0004-6361/201833910} {\bibfield  {journal} {\bibinfo
  {journal} {Astron. Astrophys.}\ }\textbf {\bibinfo {volume} {641}},\ \bibinfo
  {pages} {A6} (\bibinfo {year} {2020})},\ \bibinfo {note} {[Erratum:
  Astron.Astrophys. 652, C4 (2021)]},\ \Eprint
  {http://arxiv.org/abs/1807.06209} {arXiv:1807.06209 [astro-ph.CO]}
  \BibitemShut {NoStop}%
\bibitem [{\citenamefont {Adams}\ \emph {et~al.}(2020)\citenamefont {Adams},
  \citenamefont {Bezrukov}, \citenamefont {Elvin-Poole}, \citenamefont {Evans},
  \citenamefont {Guzowski}, \citenamefont {Fearraigh},\ and\ \citenamefont
  {S\"oldner-Rembold}}]{Adams:2020nue}%
  \BibitemOpen
  \bibfield  {author} {\bibinfo {author} {\bibfnamefont {M.}~\bibnamefont
  {Adams}}, \bibinfo {author} {\bibfnamefont {F.}~\bibnamefont {Bezrukov}},
  \bibinfo {author} {\bibfnamefont {J.}~\bibnamefont {Elvin-Poole}}, \bibinfo
  {author} {\bibfnamefont {J.~J.}\ \bibnamefont {Evans}}, \bibinfo {author}
  {\bibfnamefont {P.}~\bibnamefont {Guzowski}}, \bibinfo {author}
  {\bibfnamefont {B.~O.}\ \bibnamefont {Fearraigh}}, \ and\ \bibinfo {author}
  {\bibfnamefont {S.}~\bibnamefont {S\"oldner-Rembold}},\ }\href {\doibase
  10.1140/epjc/s10052-020-8197-y} {\bibfield  {journal} {\bibinfo  {journal}
  {Eur. Phys. J. C}\ }\textbf {\bibinfo {volume} {80}},\ \bibinfo {pages} {758}
  (\bibinfo {year} {2020})},\ \Eprint {http://arxiv.org/abs/2002.07762}
  {arXiv:2002.07762 [hep-ph]} \BibitemShut {NoStop}%
\bibitem [{\citenamefont {Dasgupta}\ and\ \citenamefont
  {Kopp}(2014)}]{Dasgupta:2013zpn}%
  \BibitemOpen
  \bibfield  {author} {\bibinfo {author} {\bibfnamefont {B.}~\bibnamefont
  {Dasgupta}}\ and\ \bibinfo {author} {\bibfnamefont {J.}~\bibnamefont
  {Kopp}},\ }\href {\doibase 10.1103/PhysRevLett.112.031803} {\bibfield
  {journal} {\bibinfo  {journal} {Phys. Rev. Lett.}\ }\textbf {\bibinfo
  {volume} {112}},\ \bibinfo {pages} {031803} (\bibinfo {year} {2014})},\
  \Eprint {http://arxiv.org/abs/1310.6337} {arXiv:1310.6337 [hep-ph]}
  \BibitemShut {NoStop}%
\bibitem [{\citenamefont {Hannestad}\ \emph {et~al.}(2014)\citenamefont
  {Hannestad}, \citenamefont {Hansen},\ and\ \citenamefont
  {Tram}}]{Hannestad:2013ana}%
  \BibitemOpen
  \bibfield  {author} {\bibinfo {author} {\bibfnamefont {S.}~\bibnamefont
  {Hannestad}}, \bibinfo {author} {\bibfnamefont {R.~S.}\ \bibnamefont
  {Hansen}}, \ and\ \bibinfo {author} {\bibfnamefont {T.}~\bibnamefont
  {Tram}},\ }\href {\doibase 10.1103/PhysRevLett.112.031802} {\bibfield
  {journal} {\bibinfo  {journal} {Phys. Rev. Lett.}\ }\textbf {\bibinfo
  {volume} {112}},\ \bibinfo {pages} {031802} (\bibinfo {year} {2014})},\
  \Eprint {http://arxiv.org/abs/1310.5926} {arXiv:1310.5926 [astro-ph.CO]}
  \BibitemShut {NoStop}%
\bibitem [{\citenamefont {Chu}\ \emph {et~al.}(2018)\citenamefont {Chu},
  \citenamefont {Dasgupta}, \citenamefont {Dentler}, \citenamefont {Kopp},\
  and\ \citenamefont {Saviano}}]{Chu:2018gxk}%
  \BibitemOpen
  \bibfield  {author} {\bibinfo {author} {\bibfnamefont {X.}~\bibnamefont
  {Chu}}, \bibinfo {author} {\bibfnamefont {B.}~\bibnamefont {Dasgupta}},
  \bibinfo {author} {\bibfnamefont {M.}~\bibnamefont {Dentler}}, \bibinfo
  {author} {\bibfnamefont {J.}~\bibnamefont {Kopp}}, \ and\ \bibinfo {author}
  {\bibfnamefont {N.}~\bibnamefont {Saviano}},\ }\href {\doibase
  10.1088/1475-7516/2018/11/049} {\bibfield  {journal} {\bibinfo  {journal}
  {JCAP}\ }\textbf {\bibinfo {volume} {11}},\ \bibinfo {pages} {049} (\bibinfo
  {year} {2018})},\ \Eprint {http://arxiv.org/abs/1806.10629} {arXiv:1806.10629
  [hep-ph]} \BibitemShut {NoStop}%
\bibitem [{\citenamefont {Yaguna}(2007)}]{Yaguna:2007wi}%
  \BibitemOpen
  \bibfield  {author} {\bibinfo {author} {\bibfnamefont {C.~E.}\ \bibnamefont
  {Yaguna}},\ }\href {\doibase 10.1088/1126-6708/2007/06/002} {\bibfield
  {journal} {\bibinfo  {journal} {JHEP}\ }\textbf {\bibinfo {volume} {06}},\
  \bibinfo {pages} {002} (\bibinfo {year} {2007})},\ \Eprint
  {http://arxiv.org/abs/0706.0178} {arXiv:0706.0178 [hep-ph]} \BibitemShut
  {NoStop}%
\bibitem [{\citenamefont {Saviano}\ \emph {et~al.}(2013)\citenamefont
  {Saviano}, \citenamefont {Mirizzi}, \citenamefont {Pisanti}, \citenamefont
  {Serpico}, \citenamefont {Mangano},\ and\ \citenamefont
  {Miele}}]{Saviano:2013ktj}%
  \BibitemOpen
  \bibfield  {author} {\bibinfo {author} {\bibfnamefont {N.}~\bibnamefont
  {Saviano}}, \bibinfo {author} {\bibfnamefont {A.}~\bibnamefont {Mirizzi}},
  \bibinfo {author} {\bibfnamefont {O.}~\bibnamefont {Pisanti}}, \bibinfo
  {author} {\bibfnamefont {P.~D.}\ \bibnamefont {Serpico}}, \bibinfo {author}
  {\bibfnamefont {G.}~\bibnamefont {Mangano}}, \ and\ \bibinfo {author}
  {\bibfnamefont {G.}~\bibnamefont {Miele}},\ }\href {\doibase
  10.1103/PhysRevD.87.073006} {\bibfield  {journal} {\bibinfo  {journal} {Phys.
  Rev. D}\ }\textbf {\bibinfo {volume} {87}},\ \bibinfo {pages} {073006}
  (\bibinfo {year} {2013})},\ \Eprint {http://arxiv.org/abs/1302.1200}
  {arXiv:1302.1200 [astro-ph.CO]} \BibitemShut {NoStop}%
\bibitem [{\citenamefont {Giovannini}\ \emph {et~al.}(2002)\citenamefont
  {Giovannini}, \citenamefont {Kurki-Suonio},\ and\ \citenamefont
  {Sihvola}}]{Giovannini:2002qw}%
  \BibitemOpen
  \bibfield  {author} {\bibinfo {author} {\bibfnamefont {M.}~\bibnamefont
  {Giovannini}}, \bibinfo {author} {\bibfnamefont {H.}~\bibnamefont
  {Kurki-Suonio}}, \ and\ \bibinfo {author} {\bibfnamefont {E.}~\bibnamefont
  {Sihvola}},\ }\href {\doibase 10.1103/PhysRevD.66.043504} {\bibfield
  {journal} {\bibinfo  {journal} {Phys. Rev. D}\ }\textbf {\bibinfo {volume}
  {66}},\ \bibinfo {pages} {043504} (\bibinfo {year} {2002})},\ \Eprint
  {http://arxiv.org/abs/astro-ph/0203430} {arXiv:astro-ph/0203430} \BibitemShut
  {NoStop}%
\bibitem [{\citenamefont {Bezrukov}\ \emph {et~al.}(2017)\citenamefont
  {Bezrukov}, \citenamefont {Chudaykin},\ and\ \citenamefont
  {Gorbunov}}]{Bezrukov:2017ike}%
  \BibitemOpen
  \bibfield  {author} {\bibinfo {author} {\bibfnamefont {F.}~\bibnamefont
  {Bezrukov}}, \bibinfo {author} {\bibfnamefont {A.}~\bibnamefont {Chudaykin}},
  \ and\ \bibinfo {author} {\bibfnamefont {D.}~\bibnamefont {Gorbunov}},\
  }\href {\doibase 10.1088/1475-7516/2017/06/051} {\bibfield  {journal}
  {\bibinfo  {journal} {JCAP}\ }\textbf {\bibinfo {volume} {06}},\ \bibinfo
  {pages} {051} (\bibinfo {year} {2017})},\ \Eprint
  {http://arxiv.org/abs/1705.02184} {arXiv:1705.02184 [hep-ph]} \BibitemShut
  {NoStop}%
\bibitem [{\citenamefont {Farzan}(2019)}]{Farzan:2019yvo}%
  \BibitemOpen
  \bibfield  {author} {\bibinfo {author} {\bibfnamefont {Y.}~\bibnamefont
  {Farzan}},\ }\href {\doibase 10.1016/j.physletb.2019.134911} {\bibfield
  {journal} {\bibinfo  {journal} {Phys. Lett. B}\ }\textbf {\bibinfo {volume}
  {797}},\ \bibinfo {pages} {134911} (\bibinfo {year} {2019})},\ \Eprint
  {http://arxiv.org/abs/1907.04271} {arXiv:1907.04271 [hep-ph]} \BibitemShut
  {NoStop}%
\bibitem [{\citenamefont {Cline}(2020)}]{Cline:2019seo}%
  \BibitemOpen
  \bibfield  {author} {\bibinfo {author} {\bibfnamefont {J.~M.}\ \bibnamefont
  {Cline}},\ }\href {\doibase 10.1016/j.physletb.2019.135182} {\bibfield
  {journal} {\bibinfo  {journal} {Phys. Lett. B}\ }\textbf {\bibinfo {volume}
  {802}},\ \bibinfo {pages} {135182} (\bibinfo {year} {2020})},\ \Eprint
  {http://arxiv.org/abs/1908.02278} {arXiv:1908.02278 [hep-ph]} \BibitemShut
  {NoStop}%
\bibitem [{\citenamefont {Archidiacono}\ \emph {et~al.}(2020)\citenamefont
  {Archidiacono}, \citenamefont {Gariazzo}, \citenamefont {Giunti},
  \citenamefont {Hannestad},\ and\ \citenamefont
  {Tram}}]{Archidiacono:2020yey}%
  \BibitemOpen
  \bibfield  {author} {\bibinfo {author} {\bibfnamefont {M.}~\bibnamefont
  {Archidiacono}}, \bibinfo {author} {\bibfnamefont {S.}~\bibnamefont
  {Gariazzo}}, \bibinfo {author} {\bibfnamefont {C.}~\bibnamefont {Giunti}},
  \bibinfo {author} {\bibfnamefont {S.}~\bibnamefont {Hannestad}}, \ and\
  \bibinfo {author} {\bibfnamefont {T.}~\bibnamefont {Tram}},\ }\href {\doibase
  10.1088/1475-7516/2020/12/029} {\bibfield  {journal} {\bibinfo  {journal}
  {JCAP}\ }\textbf {\bibinfo {volume} {12}},\ \bibinfo {pages} {029} (\bibinfo
  {year} {2020})},\ \Eprint {http://arxiv.org/abs/2006.12885} {arXiv:2006.12885
  [astro-ph.CO]} \BibitemShut {NoStop}%
\bibitem [{\citenamefont {Di~Valentino}\ \emph {et~al.}(2022)\citenamefont
  {Di~Valentino}, \citenamefont {Gariazzo}, \citenamefont {Giunti},
  \citenamefont {Mena}, \citenamefont {Pan},\ and\ \citenamefont
  {Yang}}]{DiValentino:2021rjj}%
  \BibitemOpen
  \bibfield  {author} {\bibinfo {author} {\bibfnamefont {E.}~\bibnamefont
  {Di~Valentino}}, \bibinfo {author} {\bibfnamefont {S.}~\bibnamefont
  {Gariazzo}}, \bibinfo {author} {\bibfnamefont {C.}~\bibnamefont {Giunti}},
  \bibinfo {author} {\bibfnamefont {O.}~\bibnamefont {Mena}}, \bibinfo {author}
  {\bibfnamefont {S.}~\bibnamefont {Pan}}, \ and\ \bibinfo {author}
  {\bibfnamefont {W.}~\bibnamefont {Yang}},\ }\href {\doibase
  10.1103/PhysRevD.105.103511} {\bibfield  {journal} {\bibinfo  {journal}
  {Phys. Rev. D}\ }\textbf {\bibinfo {volume} {105}},\ \bibinfo {pages}
  {103511} (\bibinfo {year} {2022})},\ \Eprint
  {http://arxiv.org/abs/2110.03990} {arXiv:2110.03990 [astro-ph.CO]}
  \BibitemShut {NoStop}%
\bibitem [{\citenamefont {Gelmini}\ \emph {et~al.}(2004)\citenamefont
  {Gelmini}, \citenamefont {Palomares-Ruiz},\ and\ \citenamefont
  {Pascoli}}]{Gelmini:2004ah}%
  \BibitemOpen
  \bibfield  {author} {\bibinfo {author} {\bibfnamefont {G.}~\bibnamefont
  {Gelmini}}, \bibinfo {author} {\bibfnamefont {S.}~\bibnamefont
  {Palomares-Ruiz}}, \ and\ \bibinfo {author} {\bibfnamefont {S.}~\bibnamefont
  {Pascoli}},\ }\href {\doibase 10.1103/PhysRevLett.93.081302} {\bibfield
  {journal} {\bibinfo  {journal} {Phys. Rev. Lett.}\ }\textbf {\bibinfo
  {volume} {93}},\ \bibinfo {pages} {081302} (\bibinfo {year} {2004})},\
  \Eprint {http://arxiv.org/abs/astro-ph/0403323} {arXiv:astro-ph/0403323}
  \BibitemShut {NoStop}%
\bibitem [{\citenamefont {Giunti}\ and\ \citenamefont
  {Zavanin}(2015)}]{Giunti:2015mwa}%
  \BibitemOpen
  \bibfield  {author} {\bibinfo {author} {\bibfnamefont {C.}~\bibnamefont
  {Giunti}}\ and\ \bibinfo {author} {\bibfnamefont {E.~M.}\ \bibnamefont
  {Zavanin}},\ }\href {\doibase 10.1142/S0217732316500036} {\bibfield
  {journal} {\bibinfo  {journal} {Mod. Phys. Lett. A}\ }\textbf {\bibinfo
  {volume} {31}},\ \bibinfo {pages} {1650003} (\bibinfo {year} {2015})},\
  \Eprint {http://arxiv.org/abs/1508.03172} {arXiv:1508.03172 [hep-ph]}
  \BibitemShut {NoStop}%
\bibitem [{\citenamefont {Giunti}\ and\ \citenamefont
  {Laveder}(2011{\natexlab{b}})}]{Giunti:2011gz}%
  \BibitemOpen
  \bibfield  {author} {\bibinfo {author} {\bibfnamefont {C.}~\bibnamefont
  {Giunti}}\ and\ \bibinfo {author} {\bibfnamefont {M.}~\bibnamefont
  {Laveder}},\ }\href {\doibase 10.1103/PhysRevD.84.073008} {\bibfield
  {journal} {\bibinfo  {journal} {Phys. Rev. D}\ }\textbf {\bibinfo {volume}
  {84}},\ \bibinfo {pages} {073008} (\bibinfo {year} {2011}{\natexlab{b}})},\
  \Eprint {http://arxiv.org/abs/1107.1452} {arXiv:1107.1452 [hep-ph]}
  \BibitemShut {NoStop}%
\bibitem [{\citenamefont {Conrad}\ \emph {et~al.}(2013)\citenamefont {Conrad},
  \citenamefont {Ignarra}, \citenamefont {Karagiorgi}, \citenamefont
  {Shaevitz},\ and\ \citenamefont {Spitz}}]{Conrad:2012qt}%
  \BibitemOpen
  \bibfield  {author} {\bibinfo {author} {\bibfnamefont {J.~M.}\ \bibnamefont
  {Conrad}}, \bibinfo {author} {\bibfnamefont {C.~M.}\ \bibnamefont {Ignarra}},
  \bibinfo {author} {\bibfnamefont {G.}~\bibnamefont {Karagiorgi}}, \bibinfo
  {author} {\bibfnamefont {M.~H.}\ \bibnamefont {Shaevitz}}, \ and\ \bibinfo
  {author} {\bibfnamefont {J.}~\bibnamefont {Spitz}},\ }\href {\doibase
  10.1155/2013/163897} {\bibfield  {journal} {\bibinfo  {journal} {Adv. High
  Energy Phys.}\ }\textbf {\bibinfo {volume} {2013}},\ \bibinfo {pages}
  {163897} (\bibinfo {year} {2013})},\ \Eprint {http://arxiv.org/abs/1207.4765}
  {arXiv:1207.4765 [hep-ex]} \BibitemShut {NoStop}%
\bibitem [{\citenamefont {Kopp}\ \emph {et~al.}(2013)\citenamefont {Kopp},
  \citenamefont {Machado}, \citenamefont {Maltoni},\ and\ \citenamefont
  {Schwetz}}]{Kopp:2013vaa}%
  \BibitemOpen
  \bibfield  {author} {\bibinfo {author} {\bibfnamefont {J.}~\bibnamefont
  {Kopp}}, \bibinfo {author} {\bibfnamefont {P.~A.~N.}\ \bibnamefont
  {Machado}}, \bibinfo {author} {\bibfnamefont {M.}~\bibnamefont {Maltoni}}, \
  and\ \bibinfo {author} {\bibfnamefont {T.}~\bibnamefont {Schwetz}},\ }\href
  {\doibase 10.1007/JHEP05(2013)050} {\bibfield  {journal} {\bibinfo  {journal}
  {JHEP}\ }\textbf {\bibinfo {volume} {05}},\ \bibinfo {pages} {050} (\bibinfo
  {year} {2013})},\ \Eprint {http://arxiv.org/abs/1303.3011} {arXiv:1303.3011
  [hep-ph]} \BibitemShut {NoStop}%
\bibitem [{\citenamefont {Giunti}\ \emph {et~al.}(2013)\citenamefont {Giunti},
  \citenamefont {Laveder}, \citenamefont {Li},\ and\ \citenamefont
  {Long}}]{Giunti:2013aea}%
  \BibitemOpen
  \bibfield  {author} {\bibinfo {author} {\bibfnamefont {C.}~\bibnamefont
  {Giunti}}, \bibinfo {author} {\bibfnamefont {M.}~\bibnamefont {Laveder}},
  \bibinfo {author} {\bibfnamefont {Y.~F.}\ \bibnamefont {Li}}, \ and\ \bibinfo
  {author} {\bibfnamefont {H.~W.}\ \bibnamefont {Long}},\ }\href {\doibase
  10.1103/PhysRevD.88.073008} {\bibfield  {journal} {\bibinfo  {journal} {Phys.
  Rev. D}\ }\textbf {\bibinfo {volume} {88}},\ \bibinfo {pages} {073008}
  (\bibinfo {year} {2013})},\ \Eprint {http://arxiv.org/abs/1308.5288}
  {arXiv:1308.5288 [hep-ph]} \BibitemShut {NoStop}%
\bibitem [{\citenamefont {Cianci}\ \emph {et~al.}(2017)\citenamefont {Cianci},
  \citenamefont {Furmanski}, \citenamefont {Karagiorgi},\ and\ \citenamefont
  {Ross-Lonergan}}]{Cianci:2017okw}%
  \BibitemOpen
  \bibfield  {author} {\bibinfo {author} {\bibfnamefont {D.}~\bibnamefont
  {Cianci}}, \bibinfo {author} {\bibfnamefont {A.}~\bibnamefont {Furmanski}},
  \bibinfo {author} {\bibfnamefont {G.}~\bibnamefont {Karagiorgi}}, \ and\
  \bibinfo {author} {\bibfnamefont {M.}~\bibnamefont {Ross-Lonergan}},\ }\href
  {\doibase 10.1103/PhysRevD.96.055001} {\bibfield  {journal} {\bibinfo
  {journal} {Phys. Rev. D}\ }\textbf {\bibinfo {volume} {96}},\ \bibinfo
  {pages} {055001} (\bibinfo {year} {2017})},\ \Eprint
  {http://arxiv.org/abs/1702.01758} {arXiv:1702.01758 [hep-ph]} \BibitemShut
  {NoStop}%
\bibitem [{Pro(2019)}]{Proceedings:2019qno}%
  \BibitemOpen
  \href {\doibase 10.21468/SciPostPhysProc.2.001} {\emph {\bibinfo {title}
  {{Neutrino Non-Standard Interactions: A Status Report}}}},\ Vol.~\bibinfo
  {volume} {2}\ (\bibinfo {year} {2019})\ \Eprint
  {http://arxiv.org/abs/1907.00991} {arXiv:1907.00991 [hep-ph]} \BibitemShut
  {NoStop}%
\bibitem [{\citenamefont {Akhmedov}\ and\ \citenamefont
  {Schwetz}(2010)}]{Akhmedov:2010vy}%
  \BibitemOpen
  \bibfield  {author} {\bibinfo {author} {\bibfnamefont {E.}~\bibnamefont
  {Akhmedov}}\ and\ \bibinfo {author} {\bibfnamefont {T.}~\bibnamefont
  {Schwetz}},\ }\href {\doibase 10.1007/JHEP10(2010)115} {\bibfield  {journal}
  {\bibinfo  {journal} {JHEP}\ }\textbf {\bibinfo {volume} {10}},\ \bibinfo
  {pages} {115} (\bibinfo {year} {2010})},\ \Eprint
  {http://arxiv.org/abs/1007.4171} {arXiv:1007.4171 [hep-ph]} \BibitemShut
  {NoStop}%
\bibitem [{\citenamefont {Hu}(2021)}]{Hu:2020uvx}%
  \BibitemOpen
  \bibfield  {author} {\bibinfo {author} {\bibfnamefont {Z.}~\bibnamefont {Hu}}
  (\bibinfo {collaboration} {MINOS, MINOS+, Daya Bay, Bugey-3}),\ }\href
  {\doibase 10.22323/1.390.0201} {\bibfield  {journal} {\bibinfo  {journal}
  {PoS}\ }\textbf {\bibinfo {volume} {ICHEP2020}},\ \bibinfo {pages} {201}
  (\bibinfo {year} {2021})}\BibitemShut {NoStop}%
\bibitem [{\citenamefont {Schneider}\ \emph {et~al.}(2021)\citenamefont
  {Schneider}, \citenamefont {Skrzypek}, \citenamefont {Arg\"uelles},\ and\
  \citenamefont {Conrad}}]{Schneider:2021wzs}%
  \BibitemOpen
  \bibfield  {author} {\bibinfo {author} {\bibfnamefont {A.}~\bibnamefont
  {Schneider}}, \bibinfo {author} {\bibfnamefont {B.}~\bibnamefont {Skrzypek}},
  \bibinfo {author} {\bibfnamefont {C.~A.}\ \bibnamefont {Arg\"uelles}}, \ and\
  \bibinfo {author} {\bibfnamefont {J.~M.}\ \bibnamefont {Conrad}},\ }\href
  {\doibase 10.1103/PhysRevD.104.092015} {\bibfield  {journal} {\bibinfo
  {journal} {Phys. Rev. D}\ }\textbf {\bibinfo {volume} {104}},\ \bibinfo
  {pages} {092015} (\bibinfo {year} {2021})},\ \Eprint
  {http://arxiv.org/abs/2106.01508} {arXiv:2106.01508 [hep-ph]} \BibitemShut
  {NoStop}%
\bibitem [{\citenamefont {Esmaili}\ and\ \citenamefont
  {Nunokawa}(2019)}]{Esmaili:2018qzu}%
  \BibitemOpen
  \bibfield  {author} {\bibinfo {author} {\bibfnamefont {A.}~\bibnamefont
  {Esmaili}}\ and\ \bibinfo {author} {\bibfnamefont {H.}~\bibnamefont
  {Nunokawa}},\ }\href {\doibase 10.1140/epjc/s10052-019-6595-9} {\bibfield
  {journal} {\bibinfo  {journal} {Eur. Phys. J. C}\ }\textbf {\bibinfo {volume}
  {79}},\ \bibinfo {pages} {70} (\bibinfo {year} {2019})},\ \Eprint
  {http://arxiv.org/abs/1810.11940} {arXiv:1810.11940 [hep-ph]} \BibitemShut
  {NoStop}%
\bibitem [{\citenamefont {Liao}\ and\ \citenamefont
  {Marfatia}(2016)}]{Liao:2016reh}%
  \BibitemOpen
  \bibfield  {author} {\bibinfo {author} {\bibfnamefont {J.}~\bibnamefont
  {Liao}}\ and\ \bibinfo {author} {\bibfnamefont {D.}~\bibnamefont
  {Marfatia}},\ }\href {\doibase 10.1103/PhysRevLett.117.071802} {\bibfield
  {journal} {\bibinfo  {journal} {Phys. Rev. Lett.}\ }\textbf {\bibinfo
  {volume} {117}},\ \bibinfo {pages} {071802} (\bibinfo {year} {2016})},\
  \Eprint {http://arxiv.org/abs/1602.08766} {arXiv:1602.08766 [hep-ph]}
  \BibitemShut {NoStop}%
\bibitem [{\citenamefont {Liao}\ \emph {et~al.}(2019)\citenamefont {Liao},
  \citenamefont {Marfatia},\ and\ \citenamefont {Whisnant}}]{Liao:2018mbg}%
  \BibitemOpen
  \bibfield  {author} {\bibinfo {author} {\bibfnamefont {J.}~\bibnamefont
  {Liao}}, \bibinfo {author} {\bibfnamefont {D.}~\bibnamefont {Marfatia}}, \
  and\ \bibinfo {author} {\bibfnamefont {K.}~\bibnamefont {Whisnant}},\ }\href
  {\doibase 10.1103/PhysRevD.99.015016} {\bibfield  {journal} {\bibinfo
  {journal} {Phys. Rev. D}\ }\textbf {\bibinfo {volume} {99}},\ \bibinfo
  {pages} {015016} (\bibinfo {year} {2019})},\ \Eprint
  {http://arxiv.org/abs/1810.01000} {arXiv:1810.01000 [hep-ph]} \BibitemShut
  {NoStop}%
\bibitem [{\citenamefont {Denton}\ \emph {et~al.}(2019)\citenamefont {Denton},
  \citenamefont {Farzan},\ and\ \citenamefont {Shoemaker}}]{Denton:2018dqq}%
  \BibitemOpen
  \bibfield  {author} {\bibinfo {author} {\bibfnamefont {P.~B.}\ \bibnamefont
  {Denton}}, \bibinfo {author} {\bibfnamefont {Y.}~\bibnamefont {Farzan}}, \
  and\ \bibinfo {author} {\bibfnamefont {I.~M.}\ \bibnamefont {Shoemaker}},\
  }\href {\doibase 10.1103/PhysRevD.99.035003} {\bibfield  {journal} {\bibinfo
  {journal} {Phys. Rev. D}\ }\textbf {\bibinfo {volume} {99}},\ \bibinfo
  {pages} {035003} (\bibinfo {year} {2019})},\ \Eprint
  {http://arxiv.org/abs/1811.01310} {arXiv:1811.01310 [hep-ph]} \BibitemShut
  {NoStop}%
\bibitem [{\citenamefont {Moulai}(2021)}]{Moulai:2021zey}%
  \BibitemOpen
  \bibfield  {author} {\bibinfo {author} {\bibfnamefont {M.~H.}\ \bibnamefont
  {Moulai}},\ }\emph {\bibinfo {title} {{Light, Unstable Sterile Neutrinos:
  Phenomenology, a Search in the IceCube Experiment, and a Global Picture}}},\
  \href@noop {} {Ph.D. thesis},\ \bibinfo  {school} {MIT} (\bibinfo {year}
  {2021}),\ \Eprint {http://arxiv.org/abs/2110.02351} {arXiv:2110.02351
  [hep-ex]} \BibitemShut {NoStop}%
\bibitem [{\citenamefont {Abbasi}\ \emph {et~al.}(2022)\citenamefont {Abbasi}
  \emph {et~al.}}]{IceCubeCollaboration:2022tso}%
  \BibitemOpen
  \bibfield  {author} {\bibinfo {author} {\bibfnamefont {R.}~\bibnamefont
  {Abbasi}} \emph {et~al.} (\bibinfo {collaboration} {(IceCube Collaboration)*,
  IceCube}),\ }\href {\doibase 10.1103/PhysRevLett.129.151801} {\bibfield
  {journal} {\bibinfo  {journal} {Phys. Rev. Lett.}\ }\textbf {\bibinfo
  {volume} {129}},\ \bibinfo {pages} {151801} (\bibinfo {year} {2022})},\
  \Eprint {http://arxiv.org/abs/2204.00612} {arXiv:2204.00612 [hep-ex]}
  \BibitemShut {NoStop}%
\bibitem [{\citenamefont {Hardin}\ \emph {et~al.}(2022)\citenamefont {Hardin},
  \citenamefont {Martinez-Soler}, \citenamefont {Diaz}, \citenamefont {Jin},
  \citenamefont {Kamp}, \citenamefont {Arg\"uelles}, \citenamefont {Conrad},\
  and\ \citenamefont {Shaevitz}}]{Hardin:2022muu}%
  \BibitemOpen
  \bibfield  {author} {\bibinfo {author} {\bibfnamefont {J.~M.}\ \bibnamefont
  {Hardin}}, \bibinfo {author} {\bibfnamefont {I.}~\bibnamefont
  {Martinez-Soler}}, \bibinfo {author} {\bibfnamefont {A.}~\bibnamefont
  {Diaz}}, \bibinfo {author} {\bibfnamefont {M.}~\bibnamefont {Jin}}, \bibinfo
  {author} {\bibfnamefont {N.~W.}\ \bibnamefont {Kamp}}, \bibinfo {author}
  {\bibfnamefont {C.~A.}\ \bibnamefont {Arg\"uelles}}, \bibinfo {author}
  {\bibfnamefont {J.~M.}\ \bibnamefont {Conrad}}, \ and\ \bibinfo {author}
  {\bibfnamefont {M.~H.}\ \bibnamefont {Shaevitz}},\ }\href@noop {} {\
  (\bibinfo {year} {2022})},\ \Eprint {http://arxiv.org/abs/2211.02610}
  {arXiv:2211.02610 [hep-ph]} \BibitemShut {NoStop}%
\bibitem [{\citenamefont {Arg\"uelles}\ \emph
  {et~al.}(2023{\natexlab{a}})\citenamefont {Arg\"uelles}, \citenamefont
  {Bert\'olez-Mart\'\i{}nez},\ and\ \citenamefont
  {Salvado}}]{Arguelles:2022bvt}%
  \BibitemOpen
  \bibfield  {author} {\bibinfo {author} {\bibfnamefont {C.~A.}\ \bibnamefont
  {Arg\"uelles}}, \bibinfo {author} {\bibfnamefont {T.}~\bibnamefont
  {Bert\'olez-Mart\'\i{}nez}}, \ and\ \bibinfo {author} {\bibfnamefont
  {J.}~\bibnamefont {Salvado}},\ }\href {\doibase 10.1103/PhysRevD.107.036004}
  {\bibfield  {journal} {\bibinfo  {journal} {Phys. Rev. D}\ }\textbf {\bibinfo
  {volume} {107}},\ \bibinfo {pages} {036004} (\bibinfo {year}
  {2023}{\natexlab{a}})},\ \Eprint {http://arxiv.org/abs/2201.05108}
  {arXiv:2201.05108 [hep-ph]} \BibitemShut {NoStop}%
\bibitem [{\citenamefont {Akhmedov}\ and\ \citenamefont
  {Smirnov}(2022{\natexlab{a}})}]{Akhmedov:2022bjs}%
  \BibitemOpen
  \bibfield  {author} {\bibinfo {author} {\bibfnamefont {E.}~\bibnamefont
  {Akhmedov}}\ and\ \bibinfo {author} {\bibfnamefont {A.~Y.}\ \bibnamefont
  {Smirnov}},\ }\href {\doibase 10.1007/JHEP11(2022)082} {\bibfield  {journal}
  {\bibinfo  {journal} {JHEP}\ }\textbf {\bibinfo {volume} {11}},\ \bibinfo
  {pages} {082} (\bibinfo {year} {2022}{\natexlab{a}})},\ \Eprint
  {http://arxiv.org/abs/2208.03736} {arXiv:2208.03736 [hep-ph]} \BibitemShut
  {NoStop}%
\bibitem [{\citenamefont {Jones}\ \emph {et~al.}(2023)\citenamefont {Jones},
  \citenamefont {Marzec},\ and\ \citenamefont {Spitz}}]{Jones:2022hme}%
  \BibitemOpen
  \bibfield  {author} {\bibinfo {author} {\bibfnamefont {B.~J.~P.}\
  \bibnamefont {Jones}}, \bibinfo {author} {\bibfnamefont {E.}~\bibnamefont
  {Marzec}}, \ and\ \bibinfo {author} {\bibfnamefont {J.}~\bibnamefont
  {Spitz}},\ }\href {\doibase 10.1103/PhysRevD.107.013008} {\bibfield
  {journal} {\bibinfo  {journal} {Phys. Rev. D}\ }\textbf {\bibinfo {volume}
  {107}},\ \bibinfo {pages} {013008} (\bibinfo {year} {2023})},\ \Eprint
  {http://arxiv.org/abs/2211.00026} {arXiv:2211.00026 [hep-ph]} \BibitemShut
  {NoStop}%
\bibitem [{\citenamefont {Jones}(2022)}]{Jones:2022cvh}%
  \BibitemOpen
  \bibfield  {author} {\bibinfo {author} {\bibfnamefont {B.~J.~P.}\
  \bibnamefont {Jones}},\ }\href@noop {} {\  (\bibinfo {year} {2022})},\
  \Eprint {http://arxiv.org/abs/2209.00561} {arXiv:2209.00561 [hep-ph]}
  \BibitemShut {NoStop}%
\bibitem [{\citenamefont {Akhmedov}\ and\ \citenamefont
  {Smirnov}(2022{\natexlab{b}})}]{Akhmedov:2022mal}%
  \BibitemOpen
  \bibfield  {author} {\bibinfo {author} {\bibfnamefont {E.}~\bibnamefont
  {Akhmedov}}\ and\ \bibinfo {author} {\bibfnamefont {A.~Y.}\ \bibnamefont
  {Smirnov}},\ }\href@noop {} {\  (\bibinfo {year} {2022}{\natexlab{b}})},\
  \Eprint {http://arxiv.org/abs/2210.01547} {arXiv:2210.01547 [hep-ph]}
  \BibitemShut {NoStop}%
\bibitem [{\citenamefont {Schwetz}(2008)}]{Schwetz:2007cd}%
  \BibitemOpen
  \bibfield  {author} {\bibinfo {author} {\bibfnamefont {T.}~\bibnamefont
  {Schwetz}},\ }\href {\doibase 10.1088/1126-6708/2008/02/011} {\bibfield
  {journal} {\bibinfo  {journal} {JHEP}\ }\textbf {\bibinfo {volume} {02}},\
  \bibinfo {pages} {011} (\bibinfo {year} {2008})},\ \Eprint
  {http://arxiv.org/abs/0710.2985} {arXiv:0710.2985 [hep-ph]} \BibitemShut
  {NoStop}%
\bibitem [{\citenamefont {Babu}\ \emph {et~al.}(2023)\citenamefont {Babu},
  \citenamefont {Brdar}, \citenamefont {de~Gouv\^ea},\ and\ \citenamefont
  {Machado}}]{Babu:2022non}%
  \BibitemOpen
  \bibfield  {author} {\bibinfo {author} {\bibfnamefont {K.~S.}\ \bibnamefont
  {Babu}}, \bibinfo {author} {\bibfnamefont {V.}~\bibnamefont {Brdar}},
  \bibinfo {author} {\bibfnamefont {A.}~\bibnamefont {de~Gouv\^ea}}, \ and\
  \bibinfo {author} {\bibfnamefont {P.~A.~N.}\ \bibnamefont {Machado}},\ }\href
  {\doibase 10.1103/PhysRevD.107.015017} {\bibfield  {journal} {\bibinfo
  {journal} {Phys. Rev. D}\ }\textbf {\bibinfo {volume} {107}},\ \bibinfo
  {pages} {015017} (\bibinfo {year} {2023})},\ \Eprint
  {http://arxiv.org/abs/2209.00031} {arXiv:2209.00031 [hep-ph]} \BibitemShut
  {NoStop}%
\bibitem [{\citenamefont {Pas}\ \emph {et~al.}(2005)\citenamefont {Pas},
  \citenamefont {Pakvasa},\ and\ \citenamefont {Weiler}}]{Pas:2005rb}%
  \BibitemOpen
  \bibfield  {author} {\bibinfo {author} {\bibfnamefont {H.}~\bibnamefont
  {Pas}}, \bibinfo {author} {\bibfnamefont {S.}~\bibnamefont {Pakvasa}}, \ and\
  \bibinfo {author} {\bibfnamefont {T.~J.}\ \bibnamefont {Weiler}},\ }\href
  {\doibase 10.1103/PhysRevD.72.095017} {\bibfield  {journal} {\bibinfo
  {journal} {Phys. Rev. D}\ }\textbf {\bibinfo {volume} {72}},\ \bibinfo
  {pages} {095017} (\bibinfo {year} {2005})},\ \Eprint
  {http://arxiv.org/abs/hep-ph/0504096} {arXiv:hep-ph/0504096} \BibitemShut
  {NoStop}%
\bibitem [{\citenamefont {Hollenberg}\ \emph {et~al.}(2009)\citenamefont
  {Hollenberg}, \citenamefont {Micu}, \citenamefont {Pas},\ and\ \citenamefont
  {Weiler}}]{Hollenberg:2009ws}%
  \BibitemOpen
  \bibfield  {author} {\bibinfo {author} {\bibfnamefont {S.}~\bibnamefont
  {Hollenberg}}, \bibinfo {author} {\bibfnamefont {O.}~\bibnamefont {Micu}},
  \bibinfo {author} {\bibfnamefont {H.}~\bibnamefont {Pas}}, \ and\ \bibinfo
  {author} {\bibfnamefont {T.~J.}\ \bibnamefont {Weiler}},\ }\href {\doibase
  10.1103/PhysRevD.80.093005} {\bibfield  {journal} {\bibinfo  {journal} {Phys.
  Rev. D}\ }\textbf {\bibinfo {volume} {80}},\ \bibinfo {pages} {093005}
  (\bibinfo {year} {2009})},\ \Eprint {http://arxiv.org/abs/0906.0150}
  {arXiv:0906.0150 [hep-ph]} \BibitemShut {NoStop}%
\bibitem [{\citenamefont {Carena}\ \emph {et~al.}(2017)\citenamefont {Carena},
  \citenamefont {Li}, \citenamefont {Machado}, \citenamefont {Machado},\ and\
  \citenamefont {Wagner}}]{Carena:2017qhd}%
  \BibitemOpen
  \bibfield  {author} {\bibinfo {author} {\bibfnamefont {M.}~\bibnamefont
  {Carena}}, \bibinfo {author} {\bibfnamefont {Y.-Y.}\ \bibnamefont {Li}},
  \bibinfo {author} {\bibfnamefont {C.~S.}\ \bibnamefont {Machado}}, \bibinfo
  {author} {\bibfnamefont {P.~A.~N.}\ \bibnamefont {Machado}}, \ and\ \bibinfo
  {author} {\bibfnamefont {C.~E.~M.}\ \bibnamefont {Wagner}},\ }\href {\doibase
  10.1103/PhysRevD.96.095014} {\bibfield  {journal} {\bibinfo  {journal} {Phys.
  Rev. D}\ }\textbf {\bibinfo {volume} {96}},\ \bibinfo {pages} {095014}
  (\bibinfo {year} {2017})},\ \Eprint {http://arxiv.org/abs/1708.09548}
  {arXiv:1708.09548 [hep-ph]} \BibitemShut {NoStop}%
\bibitem [{\citenamefont {D\"oring}\ and\ \citenamefont
  {P\"as}(2019)}]{Doring:2018ncz}%
  \BibitemOpen
  \bibfield  {author} {\bibinfo {author} {\bibfnamefont {D.}~\bibnamefont
  {D\"oring}}\ and\ \bibinfo {author} {\bibfnamefont {H.}~\bibnamefont
  {P\"as}},\ }\href {\doibase 10.1140/epjc/s10052-019-7122-8} {\bibfield
  {journal} {\bibinfo  {journal} {Eur. Phys. J. C}\ }\textbf {\bibinfo {volume}
  {79}},\ \bibinfo {pages} {604} (\bibinfo {year} {2019})},\ \Eprint
  {http://arxiv.org/abs/1808.07734} {arXiv:1808.07734 [hep-ph]} \BibitemShut
  {NoStop}%
\bibitem [{\citenamefont {D\"oring}\ \emph {et~al.}(2020)\citenamefont
  {D\"oring}, \citenamefont {P\"as}, \citenamefont {Sicking},\ and\
  \citenamefont {Weiler}}]{Doring:2018cob}%
  \BibitemOpen
  \bibfield  {author} {\bibinfo {author} {\bibfnamefont {D.}~\bibnamefont
  {D\"oring}}, \bibinfo {author} {\bibfnamefont {H.}~\bibnamefont {P\"as}},
  \bibinfo {author} {\bibfnamefont {P.}~\bibnamefont {Sicking}}, \ and\
  \bibinfo {author} {\bibfnamefont {T.~J.}\ \bibnamefont {Weiler}},\ }\href
  {\doibase 10.1140/epjc/s10052-020-08720-2} {\bibfield  {journal} {\bibinfo
  {journal} {Eur. Phys. J. C}\ }\textbf {\bibinfo {volume} {80}},\ \bibinfo
  {pages} {1202} (\bibinfo {year} {2020})},\ \Eprint
  {http://arxiv.org/abs/1808.07460} {arXiv:1808.07460 [hep-ph]} \BibitemShut
  {NoStop}%
\bibitem [{\citenamefont {Kostelecky}\ and\ \citenamefont
  {Mewes}(2004)}]{Kostelecky:2004hg}%
  \BibitemOpen
  \bibfield  {author} {\bibinfo {author} {\bibfnamefont {V.~A.}\ \bibnamefont
  {Kostelecky}}\ and\ \bibinfo {author} {\bibfnamefont {M.}~\bibnamefont
  {Mewes}},\ }\href {\doibase 10.1103/PhysRevD.70.076002} {\bibfield  {journal}
  {\bibinfo  {journal} {Phys. Rev. D}\ }\textbf {\bibinfo {volume} {70}},\
  \bibinfo {pages} {076002} (\bibinfo {year} {2004})},\ \Eprint
  {http://arxiv.org/abs/hep-ph/0406255} {arXiv:hep-ph/0406255} \BibitemShut
  {NoStop}%
\bibitem [{\citenamefont {Diaz}\ and\ \citenamefont
  {Kostelecky}(2011)}]{Diaz:2010ft}%
  \BibitemOpen
  \bibfield  {author} {\bibinfo {author} {\bibfnamefont {J.~S.}\ \bibnamefont
  {Diaz}}\ and\ \bibinfo {author} {\bibfnamefont {V.~A.}\ \bibnamefont
  {Kostelecky}},\ }\href {\doibase 10.1016/j.physletb.2011.04.049} {\bibfield
  {journal} {\bibinfo  {journal} {Phys. Lett. B}\ }\textbf {\bibinfo {volume}
  {700}},\ \bibinfo {pages} {25} (\bibinfo {year} {2011})},\ \Eprint
  {http://arxiv.org/abs/1012.5985} {arXiv:1012.5985 [hep-ph]} \BibitemShut
  {NoStop}%
\bibitem [{\citenamefont {Diaz}\ and\ \citenamefont
  {Kostelecky}(2012)}]{Diaz:2011ia}%
  \BibitemOpen
  \bibfield  {author} {\bibinfo {author} {\bibfnamefont {J.~S.}\ \bibnamefont
  {Diaz}}\ and\ \bibinfo {author} {\bibfnamefont {A.}~\bibnamefont
  {Kostelecky}},\ }\href {\doibase 10.1103/PhysRevD.85.016013} {\bibfield
  {journal} {\bibinfo  {journal} {Phys. Rev. D}\ }\textbf {\bibinfo {volume}
  {85}},\ \bibinfo {pages} {016013} (\bibinfo {year} {2012})},\ \Eprint
  {http://arxiv.org/abs/1108.1799} {arXiv:1108.1799 [hep-ph]} \BibitemShut
  {NoStop}%
\bibitem [{\citenamefont {Barenboim}\ \emph {et~al.}(2020)\citenamefont
  {Barenboim}, \citenamefont {Mart\'\i{}nez-Mirav\'e}, \citenamefont {Ternes},\
  and\ \citenamefont {T\'ortola}}]{Barenboim:2019hso}%
  \BibitemOpen
  \bibfield  {author} {\bibinfo {author} {\bibfnamefont {G.~A.}\ \bibnamefont
  {Barenboim}}, \bibinfo {author} {\bibfnamefont {P.}~\bibnamefont
  {Mart\'\i{}nez-Mirav\'e}}, \bibinfo {author} {\bibfnamefont {C.~A.}\
  \bibnamefont {Ternes}}, \ and\ \bibinfo {author} {\bibfnamefont {M.~A.}\
  \bibnamefont {T\'ortola}},\ }\href {\doibase 10.1007/JHEP03(2020)070}
  {\bibfield  {journal} {\bibinfo  {journal} {JHEP}\ }\textbf {\bibinfo
  {volume} {03}},\ \bibinfo {pages} {070} (\bibinfo {year} {2020})},\ \Eprint
  {http://arxiv.org/abs/1911.02329} {arXiv:1911.02329 [hep-ph]} \BibitemShut
  {NoStop}%
\bibitem [{\citenamefont {Aguilar-Arevalo}\ \emph
  {et~al.}(2021{\natexlab{b}})\citenamefont {Aguilar-Arevalo} \emph
  {et~al.}}]{MiniBooNE:2021bgc}%
  \BibitemOpen
  \bibfield  {author} {\bibinfo {author} {\bibfnamefont {A.~A.}\ \bibnamefont
  {Aguilar-Arevalo}} \emph {et~al.} (\bibinfo {collaboration} {MiniBooNE}),\
  }\href@noop {} {\  (\bibinfo {year} {2021}{\natexlab{b}})},\ \Eprint
  {http://arxiv.org/abs/2110.15055} {arXiv:2110.15055 [hep-ex]} \BibitemShut
  {NoStop}%
\bibitem [{\citenamefont {Arg\"uelles}\ \emph
  {et~al.}(2022{\natexlab{b}})\citenamefont {Arg\"uelles}, \citenamefont
  {Foppiani},\ and\ \citenamefont {Hostert}}]{Arguelles:2021dqn}%
  \BibitemOpen
  \bibfield  {author} {\bibinfo {author} {\bibfnamefont {C.~A.}\ \bibnamefont
  {Arg\"uelles}}, \bibinfo {author} {\bibfnamefont {N.}~\bibnamefont
  {Foppiani}}, \ and\ \bibinfo {author} {\bibfnamefont {M.}~\bibnamefont
  {Hostert}},\ }\href {\doibase 10.1103/PhysRevD.105.095006} {\bibfield
  {journal} {\bibinfo  {journal} {Phys. Rev. D}\ }\textbf {\bibinfo {volume}
  {105}},\ \bibinfo {pages} {095006} (\bibinfo {year} {2022}{\natexlab{b}})},\
  \Eprint {http://arxiv.org/abs/2109.03831} {arXiv:2109.03831 [hep-ph]}
  \BibitemShut {NoStop}%
\bibitem [{\citenamefont {Shoemaker}\ and\ \citenamefont
  {Wyenberg}(2019)}]{Shoemaker:2018vii}%
  \BibitemOpen
  \bibfield  {author} {\bibinfo {author} {\bibfnamefont {I.~M.}\ \bibnamefont
  {Shoemaker}}\ and\ \bibinfo {author} {\bibfnamefont {J.}~\bibnamefont
  {Wyenberg}},\ }\href {\doibase 10.1103/PhysRevD.99.075010} {\bibfield
  {journal} {\bibinfo  {journal} {Phys. Rev. D}\ }\textbf {\bibinfo {volume}
  {99}},\ \bibinfo {pages} {075010} (\bibinfo {year} {2019})},\ \Eprint
  {http://arxiv.org/abs/1811.12435} {arXiv:1811.12435 [hep-ph]} \BibitemShut
  {NoStop}%
\bibitem [{\citenamefont {Dasgupta}\ \emph {et~al.}(2021)\citenamefont
  {Dasgupta}, \citenamefont {Kang},\ and\ \citenamefont {Kim}}]{Kim:2021lun}%
  \BibitemOpen
  \bibfield  {author} {\bibinfo {author} {\bibfnamefont {A.}~\bibnamefont
  {Dasgupta}}, \bibinfo {author} {\bibfnamefont {S.~K.}\ \bibnamefont {Kang}},
  \ and\ \bibinfo {author} {\bibfnamefont {J.~E.}\ \bibnamefont {Kim}},\ }\href
  {\doibase 10.1007/JHEP11(2021)120} {\bibfield  {journal} {\bibinfo  {journal}
  {JHEP}\ }\textbf {\bibinfo {volume} {11}},\ \bibinfo {pages} {120} (\bibinfo
  {year} {2021})},\ \Eprint {http://arxiv.org/abs/2108.12998} {arXiv:2108.12998
  [hep-ph]} \BibitemShut {NoStop}%
\bibitem [{\citenamefont {Bolton}\ \emph {et~al.}(2022)\citenamefont {Bolton},
  \citenamefont {Deppisch}, \citenamefont {Fridell}, \citenamefont {Harz},
  \citenamefont {Hati},\ and\ \citenamefont {Kulkarni}}]{Bolton:2021pey}%
  \BibitemOpen
  \bibfield  {author} {\bibinfo {author} {\bibfnamefont {P.~D.}\ \bibnamefont
  {Bolton}}, \bibinfo {author} {\bibfnamefont {F.~F.}\ \bibnamefont
  {Deppisch}}, \bibinfo {author} {\bibfnamefont {K.}~\bibnamefont {Fridell}},
  \bibinfo {author} {\bibfnamefont {J.}~\bibnamefont {Harz}}, \bibinfo {author}
  {\bibfnamefont {C.}~\bibnamefont {Hati}}, \ and\ \bibinfo {author}
  {\bibfnamefont {S.}~\bibnamefont {Kulkarni}},\ }\href {\doibase
  10.1103/PhysRevD.106.035036} {\bibfield  {journal} {\bibinfo  {journal}
  {Phys. Rev. D}\ }\textbf {\bibinfo {volume} {106}},\ \bibinfo {pages}
  {035036} (\bibinfo {year} {2022})},\ \Eprint
  {http://arxiv.org/abs/2110.02233} {arXiv:2110.02233 [hep-ph]} \BibitemShut
  {NoStop}%
\bibitem [{\citenamefont {Coloma}\ \emph {et~al.}(2017)\citenamefont {Coloma},
  \citenamefont {Machado}, \citenamefont {Martinez-Soler},\ and\ \citenamefont
  {Shoemaker}}]{Coloma:2017ppo}%
  \BibitemOpen
  \bibfield  {author} {\bibinfo {author} {\bibfnamefont {P.}~\bibnamefont
  {Coloma}}, \bibinfo {author} {\bibfnamefont {P.~A.~N.}\ \bibnamefont
  {Machado}}, \bibinfo {author} {\bibfnamefont {I.}~\bibnamefont
  {Martinez-Soler}}, \ and\ \bibinfo {author} {\bibfnamefont {I.~M.}\
  \bibnamefont {Shoemaker}},\ }\href {\doibase 10.1103/PhysRevLett.119.201804}
  {\bibfield  {journal} {\bibinfo  {journal} {Phys. Rev. Lett.}\ }\textbf
  {\bibinfo {volume} {119}},\ \bibinfo {pages} {201804} (\bibinfo {year}
  {2017})},\ \Eprint {http://arxiv.org/abs/1707.08573} {arXiv:1707.08573
  [hep-ph]} \BibitemShut {NoStop}%
\bibitem [{\citenamefont {Atkinson}\ \emph {et~al.}(2022)\citenamefont
  {Atkinson}, \citenamefont {Coloma}, \citenamefont {Martinez-Soler},
  \citenamefont {Rocco},\ and\ \citenamefont {Shoemaker}}]{Atkinson:2021rnp}%
  \BibitemOpen
  \bibfield  {author} {\bibinfo {author} {\bibfnamefont {M.}~\bibnamefont
  {Atkinson}}, \bibinfo {author} {\bibfnamefont {P.}~\bibnamefont {Coloma}},
  \bibinfo {author} {\bibfnamefont {I.}~\bibnamefont {Martinez-Soler}},
  \bibinfo {author} {\bibfnamefont {N.}~\bibnamefont {Rocco}}, \ and\ \bibinfo
  {author} {\bibfnamefont {I.~M.}\ \bibnamefont {Shoemaker}},\ }\href {\doibase
  10.1007/JHEP04(2022)174} {\bibfield  {journal} {\bibinfo  {journal} {JHEP}\
  }\textbf {\bibinfo {volume} {04}},\ \bibinfo {pages} {174} (\bibinfo {year}
  {2022})},\ \Eprint {http://arxiv.org/abs/2105.09357} {arXiv:2105.09357
  [hep-ph]} \BibitemShut {NoStop}%
\bibitem [{\citenamefont {Plestid}(2021)}]{Plestid:2020vqf}%
  \BibitemOpen
  \bibfield  {author} {\bibinfo {author} {\bibfnamefont {R.}~\bibnamefont
  {Plestid}},\ }\href {\doibase 10.1103/PhysRevD.104.075027} {\bibfield
  {journal} {\bibinfo  {journal} {Phys. Rev. D}\ }\textbf {\bibinfo {volume}
  {104}},\ \bibinfo {pages} {075027} (\bibinfo {year} {2021})},\ \Eprint
  {http://arxiv.org/abs/2010.04193} {arXiv:2010.04193 [hep-ph]} \BibitemShut
  {NoStop}%
\bibitem [{\citenamefont {Huang}\ \emph {et~al.}(2022)\citenamefont {Huang},
  \citenamefont {Jana}, \citenamefont {Lindner},\ and\ \citenamefont
  {Rodejohann}}]{Huang:2021mki}%
  \BibitemOpen
  \bibfield  {author} {\bibinfo {author} {\bibfnamefont {G.-y.}\ \bibnamefont
  {Huang}}, \bibinfo {author} {\bibfnamefont {S.}~\bibnamefont {Jana}},
  \bibinfo {author} {\bibfnamefont {M.}~\bibnamefont {Lindner}}, \ and\
  \bibinfo {author} {\bibfnamefont {W.}~\bibnamefont {Rodejohann}},\ }\href
  {\doibase 10.1088/1475-7516/2022/02/038} {\bibfield  {journal} {\bibinfo
  {journal} {JCAP}\ }\textbf {\bibinfo {volume} {02}},\ \bibinfo {pages} {038}
  (\bibinfo {year} {2022})},\ \Eprint {http://arxiv.org/abs/2112.09476}
  {arXiv:2112.09476 [hep-ph]} \BibitemShut {NoStop}%
\bibitem [{\citenamefont {Ismail}\ \emph {et~al.}(2022)\citenamefont {Ismail},
  \citenamefont {Jana},\ and\ \citenamefont {Abraham}}]{Ismail:2021dyp}%
  \BibitemOpen
  \bibfield  {author} {\bibinfo {author} {\bibfnamefont {A.}~\bibnamefont
  {Ismail}}, \bibinfo {author} {\bibfnamefont {S.}~\bibnamefont {Jana}}, \ and\
  \bibinfo {author} {\bibfnamefont {R.~M.}\ \bibnamefont {Abraham}},\ }\href
  {\doibase 10.1103/PhysRevD.105.055008} {\bibfield  {journal} {\bibinfo
  {journal} {Phys. Rev. D}\ }\textbf {\bibinfo {volume} {105}},\ \bibinfo
  {pages} {055008} (\bibinfo {year} {2022})},\ \Eprint
  {http://arxiv.org/abs/2109.05032} {arXiv:2109.05032 [hep-ph]} \BibitemShut
  {NoStop}%
\bibitem [{\citenamefont {Brdar}\ \emph
  {et~al.}(2021{\natexlab{a}})\citenamefont {Brdar}, \citenamefont {Greljo},
  \citenamefont {Kopp},\ and\ \citenamefont {Opferkuch}}]{Brdar:2020quo}%
  \BibitemOpen
  \bibfield  {author} {\bibinfo {author} {\bibfnamefont {V.}~\bibnamefont
  {Brdar}}, \bibinfo {author} {\bibfnamefont {A.}~\bibnamefont {Greljo}},
  \bibinfo {author} {\bibfnamefont {J.}~\bibnamefont {Kopp}}, \ and\ \bibinfo
  {author} {\bibfnamefont {T.}~\bibnamefont {Opferkuch}},\ }\href {\doibase
  10.1088/1475-7516/2021/01/039} {\bibfield  {journal} {\bibinfo  {journal}
  {JCAP}\ }\textbf {\bibinfo {volume} {01}},\ \bibinfo {pages} {039} (\bibinfo
  {year} {2021}{\natexlab{a}})},\ \Eprint {http://arxiv.org/abs/2007.15563}
  {arXiv:2007.15563 [hep-ph]} \BibitemShut {NoStop}%
\bibitem [{\citenamefont {Voloshin}(1988)}]{Voloshin:1987qy}%
  \BibitemOpen
  \bibfield  {author} {\bibinfo {author} {\bibfnamefont {M.~B.}\ \bibnamefont
  {Voloshin}},\ }\href@noop {} {\bibfield  {journal} {\bibinfo  {journal} {Sov.
  J. Nucl. Phys.}\ }\textbf {\bibinfo {volume} {48}},\ \bibinfo {pages} {512}
  (\bibinfo {year} {1988})}\BibitemShut {NoStop}%
\bibitem [{\citenamefont {Barbieri}\ and\ \citenamefont
  {Mohapatra}(1989)}]{Barbieri:1988fh}%
  \BibitemOpen
  \bibfield  {author} {\bibinfo {author} {\bibfnamefont {R.}~\bibnamefont
  {Barbieri}}\ and\ \bibinfo {author} {\bibfnamefont {R.~N.}\ \bibnamefont
  {Mohapatra}},\ }\href {\doibase 10.1016/0370-2693(89)91423-8} {\bibfield
  {journal} {\bibinfo  {journal} {Phys. Lett. B}\ }\textbf {\bibinfo {volume}
  {218}},\ \bibinfo {pages} {225} (\bibinfo {year} {1989})}\BibitemShut
  {NoStop}%
\bibitem [{\citenamefont {Babu}\ and\ \citenamefont
  {Mohapatra}(1990)}]{Babu:1989px}%
  \BibitemOpen
  \bibfield  {author} {\bibinfo {author} {\bibfnamefont {K.~S.}\ \bibnamefont
  {Babu}}\ and\ \bibinfo {author} {\bibfnamefont {R.~N.}\ \bibnamefont
  {Mohapatra}},\ }\href {\doibase 10.1103/PhysRevLett.64.1705} {\bibfield
  {journal} {\bibinfo  {journal} {Phys. Rev. Lett.}\ }\textbf {\bibinfo
  {volume} {64}},\ \bibinfo {pages} {1705} (\bibinfo {year}
  {1990})}\BibitemShut {NoStop}%
\bibitem [{\citenamefont {Babu}\ and\ \citenamefont
  {Mohapatra}(1989)}]{Babu:1989wn}%
  \BibitemOpen
  \bibfield  {author} {\bibinfo {author} {\bibfnamefont {K.~S.}\ \bibnamefont
  {Babu}}\ and\ \bibinfo {author} {\bibfnamefont {R.~N.}\ \bibnamefont
  {Mohapatra}},\ }\href {\doibase 10.1103/PhysRevLett.63.228} {\bibfield
  {journal} {\bibinfo  {journal} {Phys. Rev. Lett.}\ }\textbf {\bibinfo
  {volume} {63}},\ \bibinfo {pages} {228} (\bibinfo {year} {1989})}\BibitemShut
  {NoStop}%
\bibitem [{\citenamefont {Leurer}\ and\ \citenamefont
  {Marcus}(1990)}]{Leurer:1989hx}%
  \BibitemOpen
  \bibfield  {author} {\bibinfo {author} {\bibfnamefont {M.}~\bibnamefont
  {Leurer}}\ and\ \bibinfo {author} {\bibfnamefont {N.}~\bibnamefont
  {Marcus}},\ }\href {\doibase 10.1016/0370-2693(90)90466-J} {\bibfield
  {journal} {\bibinfo  {journal} {Phys. Lett. B}\ }\textbf {\bibinfo {volume}
  {237}},\ \bibinfo {pages} {81} (\bibinfo {year} {1990})}\BibitemShut
  {NoStop}%
\bibitem [{\citenamefont {Babu}\ \emph {et~al.}(2020)\citenamefont {Babu},
  \citenamefont {Jana},\ and\ \citenamefont {Lindner}}]{Babu:2020ivd}%
  \BibitemOpen
  \bibfield  {author} {\bibinfo {author} {\bibfnamefont {K.~S.}\ \bibnamefont
  {Babu}}, \bibinfo {author} {\bibfnamefont {S.}~\bibnamefont {Jana}}, \ and\
  \bibinfo {author} {\bibfnamefont {M.}~\bibnamefont {Lindner}},\ }\href
  {\doibase 10.1007/JHEP10(2020)040} {\bibfield  {journal} {\bibinfo  {journal}
  {JHEP}\ }\textbf {\bibinfo {volume} {10}},\ \bibinfo {pages} {040} (\bibinfo
  {year} {2020})},\ \Eprint {http://arxiv.org/abs/2007.04291} {arXiv:2007.04291
  [hep-ph]} \BibitemShut {NoStop}%
\bibitem [{\citenamefont {McKeen}\ and\ \citenamefont
  {Pospelov}(2010)}]{McKeen:2010rx}%
  \BibitemOpen
  \bibfield  {author} {\bibinfo {author} {\bibfnamefont {D.}~\bibnamefont
  {McKeen}}\ and\ \bibinfo {author} {\bibfnamefont {M.}~\bibnamefont
  {Pospelov}},\ }\href {\doibase 10.1103/PhysRevD.82.113018} {\bibfield
  {journal} {\bibinfo  {journal} {Phys. Rev. D}\ }\textbf {\bibinfo {volume}
  {82}},\ \bibinfo {pages} {113018} (\bibinfo {year} {2010})},\ \Eprint
  {http://arxiv.org/abs/1011.3046} {arXiv:1011.3046 [hep-ph]} \BibitemShut
  {NoStop}%
\bibitem [{\citenamefont {Park}\ \emph {et~al.}(2016)\citenamefont {Park} \emph
  {et~al.}}]{MINERvA:2015nqi}%
  \BibitemOpen
  \bibfield  {author} {\bibinfo {author} {\bibfnamefont {J.}~\bibnamefont
  {Park}} \emph {et~al.} (\bibinfo {collaboration} {MINERvA}),\ }\href
  {\doibase 10.1103/PhysRevD.93.112007} {\bibfield  {journal} {\bibinfo
  {journal} {Phys. Rev. D}\ }\textbf {\bibinfo {volume} {93}},\ \bibinfo
  {pages} {112007} (\bibinfo {year} {2016})},\ \Eprint
  {http://arxiv.org/abs/1512.07699} {arXiv:1512.07699 [physics.ins-det]}
  \BibitemShut {NoStop}%
\bibitem [{\citenamefont {Valencia}\ \emph {et~al.}(2019)\citenamefont
  {Valencia} \emph {et~al.}}]{MINERvA:2019hhc}%
  \BibitemOpen
  \bibfield  {author} {\bibinfo {author} {\bibfnamefont {E.}~\bibnamefont
  {Valencia}} \emph {et~al.} (\bibinfo {collaboration} {MINERvA}),\ }\href
  {\doibase 10.1103/PhysRevD.100.092001} {\bibfield  {journal} {\bibinfo
  {journal} {Phys. Rev. D}\ }\textbf {\bibinfo {volume} {100}},\ \bibinfo
  {pages} {092001} (\bibinfo {year} {2019})},\ \Eprint
  {http://arxiv.org/abs/1906.00111} {arXiv:1906.00111 [hep-ex]} \BibitemShut
  {NoStop}%
\bibitem [{\citenamefont {Zazueta}\ \emph {et~al.}(2023)\citenamefont {Zazueta}
  \emph {et~al.}}]{MINERvA:2022vmb}%
  \BibitemOpen
  \bibfield  {author} {\bibinfo {author} {\bibfnamefont {L.}~\bibnamefont
  {Zazueta}} \emph {et~al.} (\bibinfo {collaboration} {MINERvA}),\ }\href
  {\doibase 10.1103/PhysRevD.107.012001} {\bibfield  {journal} {\bibinfo
  {journal} {Phys. Rev. D}\ }\textbf {\bibinfo {volume} {107}},\ \bibinfo
  {pages} {012001} (\bibinfo {year} {2023})},\ \Eprint
  {http://arxiv.org/abs/2209.05540} {arXiv:2209.05540 [hep-ex]} \BibitemShut
  {NoStop}%
\bibitem [{\citenamefont {Arg\"uelles}\ \emph {et~al.}(2019)\citenamefont
  {Arg\"uelles}, \citenamefont {Hostert},\ and\ \citenamefont
  {Tsai}}]{Arguelles:2018mtc}%
  \BibitemOpen
  \bibfield  {author} {\bibinfo {author} {\bibfnamefont {C.~A.}\ \bibnamefont
  {Arg\"uelles}}, \bibinfo {author} {\bibfnamefont {M.}~\bibnamefont
  {Hostert}}, \ and\ \bibinfo {author} {\bibfnamefont {Y.-D.}\ \bibnamefont
  {Tsai}},\ }\href {\doibase 10.1103/PhysRevLett.123.261801} {\bibfield
  {journal} {\bibinfo  {journal} {Phys. Rev. Lett.}\ }\textbf {\bibinfo
  {volume} {123}},\ \bibinfo {pages} {261801} (\bibinfo {year} {2019})},\
  \Eprint {http://arxiv.org/abs/1812.08768} {arXiv:1812.08768 [hep-ph]}
  \BibitemShut {NoStop}%
\bibitem [{\citenamefont {Henry}\ \emph {et~al.}(2024)\citenamefont {Henry}
  \emph {et~al.}}]{MINERvA:2023ner}%
  \BibitemOpen
  \bibfield  {author} {\bibinfo {author} {\bibfnamefont {S.}~\bibnamefont
  {Henry}} \emph {et~al.} (\bibinfo {collaboration} {MINERvA}),\ }\href
  {\doibase 10.1103/PhysRevD.109.092008} {\bibfield  {journal} {\bibinfo
  {journal} {Phys. Rev. D}\ }\textbf {\bibinfo {volume} {109}},\ \bibinfo
  {pages} {092008} (\bibinfo {year} {2024})},\ \Eprint
  {http://arxiv.org/abs/2312.16631} {arXiv:2312.16631 [hep-ex]} \BibitemShut
  {NoStop}%
\bibitem [{\citenamefont {Abe}\ \emph {et~al.}(2019)\citenamefont {Abe} \emph
  {et~al.}}]{T2K:2019jwa}%
  \BibitemOpen
  \bibfield  {author} {\bibinfo {author} {\bibfnamefont {K.}~\bibnamefont
  {Abe}} \emph {et~al.} (\bibinfo {collaboration} {T2K}),\ }\href {\doibase
  10.1103/PhysRevD.100.052006} {\bibfield  {journal} {\bibinfo  {journal}
  {Phys. Rev. D}\ }\textbf {\bibinfo {volume} {100}},\ \bibinfo {pages}
  {052006} (\bibinfo {year} {2019})},\ \Eprint
  {http://arxiv.org/abs/1902.07598} {arXiv:1902.07598 [hep-ex]} \BibitemShut
  {NoStop}%
\bibitem [{\citenamefont {Brdar}\ \emph
  {et~al.}(2021{\natexlab{b}})\citenamefont {Brdar}, \citenamefont {Fischer},\
  and\ \citenamefont {Smirnov}}]{Brdar:2020tle}%
  \BibitemOpen
  \bibfield  {author} {\bibinfo {author} {\bibfnamefont {V.}~\bibnamefont
  {Brdar}}, \bibinfo {author} {\bibfnamefont {O.}~\bibnamefont {Fischer}}, \
  and\ \bibinfo {author} {\bibfnamefont {A.~Y.}\ \bibnamefont {Smirnov}},\
  }\href {\doibase 10.1103/PhysRevD.103.075008} {\bibfield  {journal} {\bibinfo
   {journal} {Phys. Rev. D}\ }\textbf {\bibinfo {volume} {103}},\ \bibinfo
  {pages} {075008} (\bibinfo {year} {2021}{\natexlab{b}})},\ \Eprint
  {http://arxiv.org/abs/2007.14411} {arXiv:2007.14411 [hep-ph]} \BibitemShut
  {NoStop}%
\bibitem [{\citenamefont {Arg\"uelles}\ \emph
  {et~al.}(2023{\natexlab{b}})\citenamefont {Arg\"uelles}, \citenamefont
  {Foppiani},\ and\ \citenamefont {Hostert}}]{Arguelles:2022lzs}%
  \BibitemOpen
  \bibfield  {author} {\bibinfo {author} {\bibfnamefont {C.~A.}\ \bibnamefont
  {Arg\"uelles}}, \bibinfo {author} {\bibfnamefont {N.}~\bibnamefont
  {Foppiani}}, \ and\ \bibinfo {author} {\bibfnamefont {M.}~\bibnamefont
  {Hostert}},\ }\href {\doibase 10.1103/PhysRevD.107.035027} {\bibfield
  {journal} {\bibinfo  {journal} {Phys. Rev. D}\ }\textbf {\bibinfo {volume}
  {107}},\ \bibinfo {pages} {035027} (\bibinfo {year} {2023}{\natexlab{b}})},\
  \Eprint {http://arxiv.org/abs/2205.12273} {arXiv:2205.12273 [hep-ph]}
  \BibitemShut {NoStop}%
\bibitem [{\citenamefont {Bertuzzo}\ \emph {et~al.}(2019)\citenamefont
  {Bertuzzo}, \citenamefont {Jana}, \citenamefont {Machado},\ and\
  \citenamefont {Zukanovich~Funchal}}]{Bertuzzo:2018ftf}%
  \BibitemOpen
  \bibfield  {author} {\bibinfo {author} {\bibfnamefont {E.}~\bibnamefont
  {Bertuzzo}}, \bibinfo {author} {\bibfnamefont {S.}~\bibnamefont {Jana}},
  \bibinfo {author} {\bibfnamefont {P.~A.~N.}\ \bibnamefont {Machado}}, \ and\
  \bibinfo {author} {\bibfnamefont {R.}~\bibnamefont {Zukanovich~Funchal}},\
  }\href {\doibase 10.1016/j.physletb.2019.02.023} {\bibfield  {journal}
  {\bibinfo  {journal} {Phys. Lett. B}\ }\textbf {\bibinfo {volume} {791}},\
  \bibinfo {pages} {210} (\bibinfo {year} {2019})},\ \Eprint
  {http://arxiv.org/abs/1808.02500} {arXiv:1808.02500 [hep-ph]} \BibitemShut
  {NoStop}%
\bibitem [{\citenamefont {Abi}\ \emph {et~al.}(2021)\citenamefont {Abi} \emph
  {et~al.}}]{Muong-2:2021ojo}%
  \BibitemOpen
  \bibfield  {author} {\bibinfo {author} {\bibfnamefont {B.}~\bibnamefont
  {Abi}} \emph {et~al.} (\bibinfo {collaboration} {Muon g-2}),\ }\href
  {\doibase 10.1103/PhysRevLett.126.141801} {\bibfield  {journal} {\bibinfo
  {journal} {Phys. Rev. Lett.}\ }\textbf {\bibinfo {volume} {126}},\ \bibinfo
  {pages} {141801} (\bibinfo {year} {2021})},\ \Eprint
  {http://arxiv.org/abs/2104.03281} {arXiv:2104.03281 [hep-ex]} \BibitemShut
  {NoStop}%
\bibitem [{\citenamefont {Pospelov}(2009)}]{Pospelov:2008zw}%
  \BibitemOpen
  \bibfield  {author} {\bibinfo {author} {\bibfnamefont {M.}~\bibnamefont
  {Pospelov}},\ }\href {\doibase 10.1103/PhysRevD.80.095002} {\bibfield
  {journal} {\bibinfo  {journal} {Phys. Rev. D}\ }\textbf {\bibinfo {volume}
  {80}},\ \bibinfo {pages} {095002} (\bibinfo {year} {2009})},\ \Eprint
  {http://arxiv.org/abs/0811.1030} {arXiv:0811.1030 [hep-ph]} \BibitemShut
  {NoStop}%
\bibitem [{\citenamefont {Lees}\ \emph {et~al.}(2017)\citenamefont {Lees} \emph
  {et~al.}}]{BaBar:2017tiz}%
  \BibitemOpen
  \bibfield  {author} {\bibinfo {author} {\bibfnamefont {J.~P.}\ \bibnamefont
  {Lees}} \emph {et~al.} (\bibinfo {collaboration} {BaBar}),\ }\href {\doibase
  10.1103/PhysRevLett.119.131804} {\bibfield  {journal} {\bibinfo  {journal}
  {Phys. Rev. Lett.}\ }\textbf {\bibinfo {volume} {119}},\ \bibinfo {pages}
  {131804} (\bibinfo {year} {2017})},\ \Eprint
  {http://arxiv.org/abs/1702.03327} {arXiv:1702.03327 [hep-ex]} \BibitemShut
  {NoStop}%
\bibitem [{\citenamefont {Banerjee}\ \emph {et~al.}(2019)\citenamefont
  {Banerjee} \emph {et~al.}}]{Banerjee:2019pds}%
  \BibitemOpen
  \bibfield  {author} {\bibinfo {author} {\bibfnamefont {D.}~\bibnamefont
  {Banerjee}} \emph {et~al.},\ }\href {\doibase 10.1103/PhysRevLett.123.121801}
  {\bibfield  {journal} {\bibinfo  {journal} {Phys. Rev. Lett.}\ }\textbf
  {\bibinfo {volume} {123}},\ \bibinfo {pages} {121801} (\bibinfo {year}
  {2019})},\ \Eprint {http://arxiv.org/abs/1906.00176} {arXiv:1906.00176
  [hep-ex]} \BibitemShut {NoStop}%
\bibitem [{\citenamefont {Mohlabeng}(2019)}]{Mohlabeng:2019vrz}%
  \BibitemOpen
  \bibfield  {author} {\bibinfo {author} {\bibfnamefont {G.}~\bibnamefont
  {Mohlabeng}},\ }\href {\doibase 10.1103/PhysRevD.99.115001} {\bibfield
  {journal} {\bibinfo  {journal} {Phys. Rev. D}\ }\textbf {\bibinfo {volume}
  {99}},\ \bibinfo {pages} {115001} (\bibinfo {year} {2019})},\ \Eprint
  {http://arxiv.org/abs/1902.05075} {arXiv:1902.05075 [hep-ph]} \BibitemShut
  {NoStop}%
\bibitem [{\citenamefont {Abdullahi}\ \emph {et~al.}(2023)\citenamefont
  {Abdullahi}, \citenamefont {Hostert}, \citenamefont {Massaro},\ and\
  \citenamefont {Pascoli}}]{Abdullahi:2023tyk}%
  \BibitemOpen
  \bibfield  {author} {\bibinfo {author} {\bibfnamefont {A.~M.}\ \bibnamefont
  {Abdullahi}}, \bibinfo {author} {\bibfnamefont {M.}~\bibnamefont {Hostert}},
  \bibinfo {author} {\bibfnamefont {D.}~\bibnamefont {Massaro}}, \ and\
  \bibinfo {author} {\bibfnamefont {S.}~\bibnamefont {Pascoli}},\ }\href@noop
  {} {\  (\bibinfo {year} {2023})},\ \Eprint {http://arxiv.org/abs/2302.05410}
  {arXiv:2302.05410 [hep-ph]} \BibitemShut {NoStop}%
\bibitem [{\citenamefont {Tumasyan}\ \emph {et~al.}(2022)\citenamefont
  {Tumasyan} \emph {et~al.}}]{CMS:2022qva}%
  \BibitemOpen
  \bibfield  {author} {\bibinfo {author} {\bibfnamefont {A.}~\bibnamefont
  {Tumasyan}} \emph {et~al.} (\bibinfo {collaboration} {CMS}),\ }\href
  {\doibase 10.1103/PhysRevD.105.092007} {\bibfield  {journal} {\bibinfo
  {journal} {Phys. Rev. D}\ }\textbf {\bibinfo {volume} {105}},\ \bibinfo
  {pages} {092007} (\bibinfo {year} {2022})},\ \Eprint
  {http://arxiv.org/abs/2201.11585} {arXiv:2201.11585 [hep-ex]} \BibitemShut
  {NoStop}%
\bibitem [{\citenamefont {Aaboud}\ \emph {et~al.}(2019)\citenamefont {Aaboud}
  \emph {et~al.}}]{ATLAS:2019cid}%
  \BibitemOpen
  \bibfield  {author} {\bibinfo {author} {\bibfnamefont {M.}~\bibnamefont
  {Aaboud}} \emph {et~al.} (\bibinfo {collaboration} {ATLAS}),\ }\href
  {\doibase 10.1103/PhysRevLett.122.231801} {\bibfield  {journal} {\bibinfo
  {journal} {Phys. Rev. Lett.}\ }\textbf {\bibinfo {volume} {122}},\ \bibinfo
  {pages} {231801} (\bibinfo {year} {2019})},\ \Eprint
  {http://arxiv.org/abs/1904.05105} {arXiv:1904.05105 [hep-ex]} \BibitemShut
  {NoStop}%
\bibitem [{\citenamefont {de~Gouv\^ea}\ \emph {et~al.}(2019)\citenamefont
  {de~Gouv\^ea}, \citenamefont {Fox}, \citenamefont {Harnik}, \citenamefont
  {Kelly},\ and\ \citenamefont {Zhang}}]{deGouvea:2018cfv}%
  \BibitemOpen
  \bibfield  {author} {\bibinfo {author} {\bibfnamefont {A.}~\bibnamefont
  {de~Gouv\^ea}}, \bibinfo {author} {\bibfnamefont {P.~J.}\ \bibnamefont
  {Fox}}, \bibinfo {author} {\bibfnamefont {R.}~\bibnamefont {Harnik}},
  \bibinfo {author} {\bibfnamefont {K.~J.}\ \bibnamefont {Kelly}}, \ and\
  \bibinfo {author} {\bibfnamefont {Y.}~\bibnamefont {Zhang}},\ }\href
  {\doibase 10.1007/JHEP01(2019)001} {\bibfield  {journal} {\bibinfo  {journal}
  {JHEP}\ }\textbf {\bibinfo {volume} {01}},\ \bibinfo {pages} {001} (\bibinfo
  {year} {2019})},\ \Eprint {http://arxiv.org/abs/1809.06388} {arXiv:1809.06388
  [hep-ph]} \BibitemShut {NoStop}%
\bibitem [{\citenamefont {D\"obrich}\ \emph {et~al.}(2016)\citenamefont
  {D\"obrich}, \citenamefont {Jaeckel}, \citenamefont {Kahlhoefer},
  \citenamefont {Ringwald},\ and\ \citenamefont
  {Schmidt-Hoberg}}]{Dobrich:2015jyk}%
  \BibitemOpen
  \bibfield  {author} {\bibinfo {author} {\bibfnamefont {B.}~\bibnamefont
  {D\"obrich}}, \bibinfo {author} {\bibfnamefont {J.}~\bibnamefont {Jaeckel}},
  \bibinfo {author} {\bibfnamefont {F.}~\bibnamefont {Kahlhoefer}}, \bibinfo
  {author} {\bibfnamefont {A.}~\bibnamefont {Ringwald}}, \ and\ \bibinfo
  {author} {\bibfnamefont {K.}~\bibnamefont {Schmidt-Hoberg}},\ }\href
  {\doibase 10.1007/JHEP02(2016)018} {\bibfield  {journal} {\bibinfo  {journal}
  {JHEP}\ }\textbf {\bibinfo {volume} {02}},\ \bibinfo {pages} {018} (\bibinfo
  {year} {2016})},\ \Eprint {http://arxiv.org/abs/1512.03069} {arXiv:1512.03069
  [hep-ph]} \BibitemShut {NoStop}%
\bibitem [{\citenamefont {Aguilar-Arevalo}\ \emph {et~al.}(2012)\citenamefont
  {Aguilar-Arevalo} \emph {et~al.}}]{MiniBooNE:2012maf}%
  \BibitemOpen
  \bibfield  {author} {\bibinfo {author} {\bibfnamefont {A.~A.}\ \bibnamefont
  {Aguilar-Arevalo}} \emph {et~al.} (\bibinfo {collaboration} {MiniBooNE})\
  }(\bibinfo {year} {2012})\ \Eprint {http://arxiv.org/abs/1207.4809}
  {arXiv:1207.4809 [hep-ex]} \BibitemShut {NoStop}%
\bibitem [{\citenamefont {Fern\'andez-Mart\'\i{}nez}\ \emph
  {et~al.}(2023)\citenamefont {Fern\'andez-Mart\'\i{}nez}, \citenamefont
  {Gonz\'alez-L\'opez}, \citenamefont {Hern\'andez-Garc\'\i{}a}, \citenamefont
  {Hostert},\ and\ \citenamefont
  {L\'opez-Pav\'on}}]{Fernandez-Martinez:2023phj}%
  \BibitemOpen
  \bibfield  {author} {\bibinfo {author} {\bibfnamefont {E.}~\bibnamefont
  {Fern\'andez-Mart\'\i{}nez}}, \bibinfo {author} {\bibfnamefont
  {M.}~\bibnamefont {Gonz\'alez-L\'opez}}, \bibinfo {author} {\bibfnamefont
  {J.}~\bibnamefont {Hern\'andez-Garc\'\i{}a}}, \bibinfo {author}
  {\bibfnamefont {M.}~\bibnamefont {Hostert}}, \ and\ \bibinfo {author}
  {\bibfnamefont {J.}~\bibnamefont {L\'opez-Pav\'on}},\ }\href {\doibase
  10.1007/JHEP09(2023)001} {\bibfield  {journal} {\bibinfo  {journal} {JHEP}\
  }\textbf {\bibinfo {volume} {09}},\ \bibinfo {pages} {001} (\bibinfo {year}
  {2023})},\ \Eprint {http://arxiv.org/abs/2304.06772} {arXiv:2304.06772
  [hep-ph]} \BibitemShut {NoStop}%
\bibitem [{\citenamefont {Aguilar-Arevalo}\ \emph
  {et~al.}(2009{\natexlab{b}})\citenamefont {Aguilar-Arevalo} \emph
  {et~al.}}]{MiniBooNE:2008hfu}%
  \BibitemOpen
  \bibfield  {author} {\bibinfo {author} {\bibfnamefont {A.~A.}\ \bibnamefont
  {Aguilar-Arevalo}} \emph {et~al.} (\bibinfo {collaboration} {MiniBooNE}),\
  }\href {\doibase 10.1103/PhysRevD.79.072002} {\bibfield  {journal} {\bibinfo
  {journal} {Phys. Rev. D}\ }\textbf {\bibinfo {volume} {79}},\ \bibinfo
  {pages} {072002} (\bibinfo {year} {2009}{\natexlab{b}})},\ \Eprint
  {http://arxiv.org/abs/0806.1449} {arXiv:0806.1449 [hep-ex]} \BibitemShut
  {NoStop}%
\bibitem [{\citenamefont {Patterson}\ \emph {et~al.}(2009)\citenamefont
  {Patterson}, \citenamefont {Laird}, \citenamefont {Liu}, \citenamefont
  {Meyers}, \citenamefont {Stancu},\ and\ \citenamefont
  {Tanaka}}]{Patterson:2009ki}%
  \BibitemOpen
  \bibfield  {author} {\bibinfo {author} {\bibfnamefont {R.~B.}\ \bibnamefont
  {Patterson}}, \bibinfo {author} {\bibfnamefont {E.~M.}\ \bibnamefont
  {Laird}}, \bibinfo {author} {\bibfnamefont {Y.}~\bibnamefont {Liu}}, \bibinfo
  {author} {\bibfnamefont {P.~D.}\ \bibnamefont {Meyers}}, \bibinfo {author}
  {\bibfnamefont {I.}~\bibnamefont {Stancu}}, \ and\ \bibinfo {author}
  {\bibfnamefont {H.~A.}\ \bibnamefont {Tanaka}},\ }\href {\doibase
  10.1016/j.nima.2009.06.064} {\bibfield  {journal} {\bibinfo  {journal} {Nucl.
  Instrum. Meth. A}\ }\textbf {\bibinfo {volume} {608}},\ \bibinfo {pages}
  {206} (\bibinfo {year} {2009})},\ \Eprint {http://arxiv.org/abs/0902.2222}
  {arXiv:0902.2222 [hep-ex]} \BibitemShut {NoStop}%
\bibitem [{\citenamefont {Wilson}()}]{SBNpots}%
  \BibitemOpen
  \bibfield  {author} {\bibinfo {author} {\bibfnamefont {R.}~\bibnamefont
  {Wilson}} (\bibinfo {collaboration} {SBN program}),\ }\href
  {{https://indico.fnal.gov/event/22303/contributions/244923/}} {\enquote
  {\bibinfo {title} {{Fermilab Short-Baseline Neutrino Program}},}\ }\bibinfo
  {note} {{Snowmass Community Summer Study 2022}}\BibitemShut {NoStop}%
\bibitem [{\citenamefont {Isaacson}\ \emph {et~al.}(2023)\citenamefont
  {Isaacson}, \citenamefont {Jay}, \citenamefont {Lovato}, \citenamefont
  {Machado},\ and\ \citenamefont {Rocco}}]{Isaacson:2022cwh}%
  \BibitemOpen
  \bibfield  {author} {\bibinfo {author} {\bibfnamefont {J.}~\bibnamefont
  {Isaacson}}, \bibinfo {author} {\bibfnamefont {W.~I.}\ \bibnamefont {Jay}},
  \bibinfo {author} {\bibfnamefont {A.}~\bibnamefont {Lovato}}, \bibinfo
  {author} {\bibfnamefont {P.~A.~N.}\ \bibnamefont {Machado}}, \ and\ \bibinfo
  {author} {\bibfnamefont {N.}~\bibnamefont {Rocco}},\ }\href {\doibase
  10.1103/PhysRevD.107.033007} {\bibfield  {journal} {\bibinfo  {journal}
  {Phys. Rev. D}\ }\textbf {\bibinfo {volume} {107}},\ \bibinfo {pages}
  {033007} (\bibinfo {year} {2023})},\ \Eprint
  {http://arxiv.org/abs/2205.06378} {arXiv:2205.06378 [hep-ph]} \BibitemShut
  {NoStop}%
\bibitem [{\citenamefont {Andreopoulos}\ \emph {et~al.}(2010)\citenamefont
  {Andreopoulos} \emph {et~al.}}]{Andreopoulos:2009rq}%
  \BibitemOpen
  \bibfield  {author} {\bibinfo {author} {\bibfnamefont {C.}~\bibnamefont
  {Andreopoulos}} \emph {et~al.},\ }\href {\doibase 10.1016/j.nima.2009.12.009}
  {\bibfield  {journal} {\bibinfo  {journal} {Nucl. Instrum. Meth. A}\ }\textbf
  {\bibinfo {volume} {614}},\ \bibinfo {pages} {87} (\bibinfo {year} {2010})},\
  \Eprint {http://arxiv.org/abs/0905.2517} {arXiv:0905.2517 [hep-ph]}
  \BibitemShut {NoStop}%
\bibitem [{\citenamefont {Patterson}(2007)}]{Patterson:2007zz}%
  \BibitemOpen
  \bibfield  {author} {\bibinfo {author} {\bibfnamefont {R.~B.}\ \bibnamefont
  {Patterson}},\ }\emph {\bibinfo {title} {{A Search for Muon Neutrino to
  Electron Neutrino Oscillations at $\delta(m^2) >0.1$ eV$^2$}}},\ \href
  {\doibase 10.2172/917855} {Ph.D. thesis},\ \bibinfo  {school} {Princeton U.}
  (\bibinfo {year} {2007})\BibitemShut {NoStop}%
\end{thebibliography}%

\end{document}